\def\fileversion{v1.0n} \def\filedate{1997/09/15}
\documentclass[11pt,oneside]{book}
\usepackage[fullpage]{uiucthesis}
\usepackage[matrix,frame,arrow]{xypic}
\usepackage{graphics,epsfig,amsfonts,amsmath,amssymb,psfrag}

\def\openone{\leavevmode\hbox{\small1\kern-3.8pt\normalsize1}}

\def\ketbraS#1{\vert#1\rangle\langle#1\vert}
\def\ipr#1#2{\langle#1\vert#2\rangle}

\def\Longarrow{\protect\@lra}
\def\@lra{\relbar\joinrel\relbar\joinrel\relbar\joinrel%
          \relbar\joinrel\rightarrow}
\def\wstate{{\rm W}}
\def\wtilde{{\widetilde{\rm W}}}

\def\coe#1{{({\rm co}\,E_{#1})}}
\def\Chi{X}
\def\ketbra#1{\vert#1\rangle\langle#1\vert}
\def\Real{{\rm Re\,}}
\newcommand{\ind}{\hspace{\parindent}}
\input{Qcircuit}

\begin{document}

\title{QUANTUM ENTANGLEMENT: GEOMETRIC QUANTIFICATION AND
APPLICATIONS TO MULTI-PARTITE STATES AND QUANTUM PHASE TRANSITIONS}
\author{Tzu-Chieh Wei}
\department{Physics}
\schools{B.S., National Taiwan University, 1994 \\
         M.S., National Taiwan University, 1996}
\phdthesis
\degreeyear{2004}
\maketitle

\frontmatter


\singlespace
\begin{center}
QUANTUM ENTANGLEMENT: GEOMETRIC QUANTIFICATION AND\\
APPLICATIONS TO MULTI-PARTITE STATES AND QUANTUM PHASE\\ TRANSITIONS\\

\vspace{.2in} \normalsize
Tzu-Chieh Wei, Ph.D. \\
Department of Physics\\
University of Illinois at Urbana-Champaign, 2004\\
Prof. Paul M. Goldbart, Advisor \\
\end{center}
\doublespace

\ind The degree to which a pure quantum state is entangled can be
characterized by the distance or angle to the nearest unentangled state.  This
geometric measure of entanglement is explored for bi-partite and multi-partite
pure and mixed states. It is determined analytically for arbitrary two-qubit
mixed states, generalized Werner,  and isotropic states, and is also applied
to certain multi-partite mixed states, including two distinct multi-partite
bound entangled states. Moreover, the ground-state entanglement of the XY
model in a transverse field is calculated and shown to exhibit singular
behavior near the quantum critical line. Along the way, connections are
pointed out between the geometric measure of entanglement, the Hartree
approximation, entanglement witnesses, correlation functions, and the relative
entropy of entanglement.

\newpage

\leavevmode\vfill
\begin{center}
To my parents
\end{center}
\vfill

\chapter*{Acknowledgments}
The person to whom I owe the greatest thanks is my advisor, Prof.~Paul Goldbart.
This thesis originated back to the early days of QUISS seminars here
in UIUC, when once he described a simple geometric picture of entanglement.
Under his insight and guidance, I was able to carry out his ideas
further and they evolved into this thesis. In addition to working
with him on quantum information science, I was and still am
very fortunate to learn a lot of condensed matter physics from him.

Paul is not only a mentor for my scientific career, but is also
a very caring friend in my everyday life. He and his wife, Jenny,
treat me like a member of their family. While I was visiting
him in Boulder, Colorado, where he spent his sabatical leave,
they invited me to their house for dinner every night. I enjoyed
staying with them and playing with their children Ollie and Greta,
and I felt like at home. I also want to thank Jenny and the children for sharing
their family time as Paul often invited me to their
house for a cup of tea and then we worked till very late, sometimes
after midnight.

Another mentor that I am very fortunate to have is Prof.~Paul Kwiat.
I want to thank him for bringing the science of quantum information to
this campus and for building up a forum---the Quantum Information
Science Seminar---to discuss quantum information science.
The discussions with him and his group---Kwiat's Clan---have been
very stimulating and enlightening.
The Kwiat Clan has included Dave Branning, Daryl Achilles, Joe Altepeter, Julio Barreiro, Marie Ericsson, Mike Goggin, Onur Hosten, Evan Jeffrey, Nick Peters, Matt Rakher, and Aaron VanDevender.
They are the people responsible for my involvement and interest
in experimental quantum information processing. They  taught me
a lot about their experiments on quantum optics, and they  even
allowed me to adjust their waveplates! It is my privilege to work with
 them.

In addition to Paul and Paul,
I would like to express my earnest thanks to  Prof.~Tony Leggett and Prof.~Mike Weissman
for serving in both my Preliminary Exam and thesis committees and
for many inspiring and stimulating discussions, which
have greatly
helped to shape my scientitific and intellectual outlook.

I am grateful to Smitha Vishveshwara, who is another source of insipration
for me.
She has educated me on a lot of science, including Luttinger liquids, bosonization,
superfluid and Mott insulator transitions, shot noise, and critical dynamics.
Her exuberant spirit is always an encouragement to me. She is also a very
caring friend.

I would like to thank Prof.~Brian DeMarco for educating me on optical lattices and atomic physics, as well as for showing me the coolest stuff in the Midwest,
and  Prof.~Jim Eckstein for educating me on electro-optic frequency
shifters. It is also a great pleasure for me to thank
Prof.~Alexey Bezryadin and Ulas Coskun, who kindly
taught me a lot about carbon nanotube physics and
shared with me many results of their beautiful research.

I am very grateful to Prof.~Mike Stone and Prof.~Eduado Fradkin for many insightful and enlightening discussions, which have greatly helped some of my
projects.

I would like to thank other people in ESB: my office-mate and collaborator Swagatam Mukhopadyay for many stimulating discussions; David Pekker for teaching me the physics of Fiske modes; Dyutiman Das for collaboration and
for helping me with numerical work; and Eun-Ah Kim, my former office-mate
in Loomis Laboratory, for many useful suggestions and references.

I am also thankful to many people who have contributed to our Quantum Information
Science Study Group:   Julio Barreiro, Bryan Clark, David Ferguson, Richard Kassman, Nick Peters, Yu Shi, Smitha Vishveshwara,  Xiangjun Xing,  Guojun Zhu, and the many others who have particpated in our discussions.

I thank Yuli Lyanda-Geller for educating me
on mesoscopic physics, when he was in Urbana.
I also thank Mohit Randeria and Nandini Trivedi for useful
discussions during their stay here.

It is a pleasure of me to thank several colleagues outside
the University of Illinois:
Marie Ericsson, Daniel James, Bill Munro, Frank Verstraete, and
Andrew White, with whom I have had fruitful collaboration.

I also benefited a great deal from many enlightening discussions
with Ivan Deutsch, Rosario Fazio, Pawel Horodecki, Gerardo Ortiz, Vlatko Vedral,
and Lorenza Viola.
I  acknowledge useful discussions with people
that I met in Benasque: Howard Barnum,
Dagmar Bru\ss, Ignacio Cirac,  Wolfgang D\"ur, Jens Eisert,
Otfried G\"uhne, Artur Ekert,
L.-C. Kwek, Debbie Leung,  Chiara Macchiavello, Kiran Manne, and G\"ufre Vidal,
some of whom I have had the fortune to meet on several occasions.

I would also like to thank the directors, organizers, and participants of
the Les Houches summer school on  Nanoscopic Quantum Physics,
where I enjoyed the intellectual stimulation  of the scientific activities,
as well as the breath-taking views
of nature explored during my stay at the school.

During my graduate career here at the University of Illinois at Urbana-Champaign I have been supported in part by the Department of Physics, the  U.S. Department of Energy, Division of
Material Sciences under Award No.~DEFG02-96ER45434,
through the Federick Seitz Materials Research Laboratory at the University of Illinois
at Urbana-Champaign, and the National Science Foundation under
Grant No.~EIA01-21568. I also acknowledge the receipt of a Mavis Memorial Fund Scholarship (2002),
a Harry G. Drickamer Graduate Fellowship (2003), and a John Bardeen Award for graduate research (2004).

Finally, I dedicate this thesis to my parents, who have always
been the strongest support behind me.

\tableofcontents
\listoffigures


\chapter*{List of Abbreviations}
\begin{description}
\item[CHSH] Clauser-Horne-Simony-Holt
\item[CNOT] Controlled-NOT
\item[EPR] Einstein-Podolsky-Rosen
\item[GHZ] Greenberger-Horne-Zeilinger
\item[GME] Geometric Measure of Entanglement
\item[LHV] Local Hidden Variables
\item[LOCC] Local Operations and Classical Communication
\item[MREGS] Minimal Reversible Entanglement Generating Set
\item[NPT] Negative Partial Transpose
\item[PPT] Positive Partial Transpose
\item[QDC] Quantum Data Compression
\item[REE] Relative Entropy of Entanglement
\item[UPB] Unextendible Product Bases
\end{description}

\mainmatter

\def\mygamma{{r}}
\chapter{Introduction to quantum entanglement}
\label{chap:intro} The superposition principle---one of the several postulates
of quantum mechanics---produces consequences that deviate from the predictions
of classical mechanics. Consider, e.g., a pair of spin-1/2 particles, each of
which can inhabit an up ($\ket{\!\!\uparrow}$) or down
($\ket{\!\!\downarrow}$) spin state relative to some spin quantization axis or
any linear combination of $\ket{\!\uparrow}$ and $\ket{\!\downarrow}$. Then
two spin-1/2 particles can, e.g., be in a state $\ket{\!\uparrow\downarrow}$
or $\ket{\!\downarrow\uparrow}$, in which each spin has a definite direction.
However, they are also allowed to be in the singlet state
\begin{equation}
\ket{\Psi^-}\equiv\frac{1}{\sqrt{2}}\left(\ket{\!\uparrow\downarrow}-\ket{\!\downarrow\uparrow}\right),
\end{equation}
in which neither spin possesses a definite direction. States such as
$\ket{\Psi^-}$ are said to be entangled, a term coined by Erwin Schr\"odinger
in 1935. The two spins, which we shall call A and B, can be spatially far from
each other, even---in principle---at  galactic separations.

Suppose that a Stern-Gerlach measurement is performed on A and the result
$\ket{\!\uparrow}$ is obtained. Then the spin state of B is in a definite
state $\ket{\!\downarrow}$. However, if the result $\ket{\!\downarrow}$ is
obtained for A then B, as required by quantum mechanics, is in the state
$\ket{\!\uparrow}$. Quantum mechanics appears to imply a nonlocal correlation
between the two spins. More specifically, if one consider observables for the
state $\ket{\Psi^-}$, such as $\sigma_A^z$ and $\sigma_B^z$,  one obtains the
following predictions from quantum mechanics:
\begin{equation}
\label{eqn:sigz} \langle\sigma_A^z\otimes\sigma_B^z\rangle=-1, \ \
\langle\sigma_A^z\rangle=\langle\sigma_B^z\rangle=0.
\end{equation}
Therefore, quantum mechanics can have the consequence that the expectation
value of the product of the two observables is not necessarily equal to the
product of the expectation values of the two observables, e.g.,
\begin{equation}
\langle\sigma_A^z\otimes\sigma_B^z\rangle\ne
\langle\sigma_A^z\rangle\langle\sigma_B^z\rangle.
\end{equation}

Actually, the results in Eq.~(\ref{eqn:sigz}) can be easily explained by a
classical theory, e.g., the ensemble of the pairs of spins in the measurement
is a statistical equal mixture of $\uparrow\downarrow$ and
$\downarrow\uparrow$. The question of whether quantum mechanics can predict
something that cannot be explained by classical theories remained unanswered
until in 1964, when John S. Bell came up with an inequality  that all
classical local theories (usually called local hidden variable [LHV] theories;
local in the sense that the measurement outcome on one side should not
influence the other side) obey, whereas it can be violated by quantum
mechanics. Clauser, Horne, Shimony, and Holt
(CHSH)~\cite{ClauserHorneShimonyHolt69} later derived an inequality based on
correlations, such as that in Eq.~(\ref{eqn:sigz}), that all classical local
theories must satisfy:
\begin{equation}
\label{eqn:CHSH}
\Big|\langle\vec{\sigma}\cdot\vec{a}\otimes\vec{\sigma}\cdot\vec{b}\rangle
+\langle\vec{\sigma}\cdot\vec{a}\otimes\vec{\sigma}\cdot\vec{b'}\rangle+
\langle\vec{\sigma}\cdot\vec{a'}\otimes\vec{\sigma}\cdot\vec{b}\rangle
-\langle\vec{\sigma}\cdot\vec{a'}\otimes\vec{\sigma}\cdot\vec{b'}\rangle\Big|\le
2,
\end{equation}
where $\vec{a}$ and $\vec{a'}$ ($\vec{b}$ and $\vec{b'}$) are unit vectors,
representing pairs of different orientations of the Stern-Gerlach apparatus at
A (B). This Bell-CHSH inequality can be violated by the singlet state
$\ket{\Psi^-}$, with the maximum value of the left-hand side being $2\sqrt{2}$
(see Appendix~\ref{app:Bell1} for more details). The violation has been
demonstrated experimentally, e.g., using photon polarization
states~\cite{Aspect82,Kwiat99}.

So far, we have been discussing the singlet state. What about other entangled
states?
 Gisin~\cite{Gisin91} 
showed that any non-product two-spin-1/2 state indeed violates the CHSH
inequality. This non-product, unfactorizable property of wavefunctions is
entanglement, the characteristic trait of quantum mechanics,  and is
responsible for the deviation from classical theories.

Although entanglement was initially associated with a rather philosophical
debate over the foundations of quantum mechanics, it has recently been
discovered to be a useful resource. For example, for quantum cryptography
Ekert~\cite{Ekert91} found that  violation of the CHSH inequality can be used
as a test of security in the process of the random key distribution via a spin
singlet state. Bennett and Wiesner~\cite{BennettWiesner92} found that
transmission of a two-level quantum state (i.e., qubit), which is initially
maximally entangled with another two-level system at the receiving end, can
encode classical information of two bits (``super dense-coding''). Bennett and
co-workers~\cite{BennettBrassardCrepeauJozsaPeresWootters93} further found
that with the shared maximal entanglement (i.e., a singlet state), an unknown
two-level state can be faithfully reconstructed via communication of only two
classical bits (``quantum teleportation''). All these tasks are made possible
by entanglement. Although it is not yet clear whether entanglement is
necessary for the speed-up of quantum algorithms, such as Shor's
factoring~\cite{Shor94} and Grover's searching~\cite{Grover97}, it has been
established that entanglement does enable quantum computation. This is well
illustrated by the so-called one-way quantum computer, due to Rossendorf and
Brigel~\cite{RaussendorfBriegel01}, in which an initial highly entangled state
(specifically, a ``cluster'' state), together with subsequent local
measurements alone, allows efficient execution of quantum computation.

Although violation of Bell inequalities is a necessary and sufficient
signature of entanglement in the pure-state setting, the situation is more
subtle in the setting of mixed states.
What do we mean by a mixed state? When is a mixed state entangled? Let us
begin with a familiar setting for density matrices, quantum statistical
mechanics, where a Hamiltonian ${\cal H}$ leads to the density matrix $\rho$,
given by
\begin{equation}
\rho=Z^{-1}e^{-\beta {\cal H}}=\sum_n \frac{e^{-\beta E_n}}{Z}\ketbra{n},
\end{equation}
where $Z={\rm Tr}(e^{-\beta {\cal H}})$, $\beta=1/k_B T$, and $\ket{n}$ is the
energy eigenstate with energy eigenvalue $E_n$. A more general description of
a system than by  wavefunctions is thus provided by a density matrix,
which can be regarded as a probabilistic mixture of pure states, hence, the
name mixed state:
\begin{equation}
\label{eqn:decomp} \rho=\sum_i p_i\ketbra{\psi_i},
\end{equation}
with $0\le p_i\le 1$ and $\sum_i p_i=1$. Both a wavefunction $\ket{\psi}$ and
a density matrix $\rho$ will hereafter be simply referred to as a state. The
density a matrix description is necessary
 when the system interacts with an environment or when we only have access to
part of a larger system. We remark that the decomposition in
Eq.~(\ref{eqn:decomp}) is by no means unique. For example, the ``white''
distribution $\frac{1}{2}\openone$ can be expressed as
\begin{equation}
\frac{1}{2}\ketbra{\uparrow}+\frac{1}{2}\ketbra{\downarrow}
\end{equation}
or
\begin{equation}
\frac{1}{2}\ketbra{+}+\frac{1}{2}\ketbra{-},
\end{equation}
where $\ket{\pm}\equiv (\ket{\!\uparrow}\pm\ket{\!\downarrow})/\sqrt{2}$.
Furthermore, there is no requirement that the pure states in the decomposition
be orthogonal to one another.

The idea of a {\it mixed\/} entangled state of two or more parties is
naturally extended from the pure-state case as a state that allows {\it no\/}
decomposition into a mixture of factorizable pure states. If a mixed state
does have such a decomposition, it is said to be separable or unentangled.
Despite its seemingly innocuous definition, the question of whether or not a
mixed state is entangled turns out to present deep  mathematical challenges.
There have been proposed several useful criteria (usually called separability
criteria) that can (but not always) determine whether or not a given state is
entangled. It has been found that certain mixed states, although entangled, do
not violate any Bell inequality and, even more surprisingly, allow a
classical, local description~\cite{Werner89}. Thus we see that entanglement in
the mixed-state scenario is a much richer and subtler phenomenon than it is
for pure states.

As Bell inequalities  do not, in general, completely reveal entanglement for
mixed entangled states,
other approaches to quantifying entanglement have emerged. These include (i)
entanglement as a quantifiable resource: entanglement of distillation and
entanglement cost; (ii) information-theoretic considerations: relative entropy
of entanglement and related measures; and (iii) other, mathematical
approaches, including the central theme of this dissertation. Approach (i) is
perhaps the most natural way to quantify entanglement. However, as we shall
see later, it has so far been limited to the settings of two parties (i.e.,
bi-partite systems), and there are major difficulties in extending it to
multi-paritite systems. Approaches (ii) and (iii) are thus seen as
indispensible for providing a better understanding of entanglement in various
settings, especially multi-partite ones. Although the measure that we shall
focus on for  most of this dissertation belongs to the mathematical approach,
later we shall show that it is nevertheless related to other entanglement and
physical properties.

As one of the central themes in the entanglement theory is to quantity the
degree of entanglement, in the following, we shall introduce several
important, standard measures of entanglement.  We shall conclude this chapter
with an overview of the dissertation.

\section{Separability and entanglement}
\label{sec:negativity} Let us define precisely whether a state is entangled or
not. A state is entangled if it is not separable. A bi-partite state,
describing parties A and B, $\rho^{\rm AB}_{\rm s}$ is separable (or
unentangled) if and only if $\rho^{\rm AB}_{\rm s}$ can be expressed as
\begin{equation}
\label{eqn:sep} \rho_{\rm s}^{\rm AB}=\sum_{i}p_i\,\rho^{\rm
A}_i\otimes\rho^{\rm B}_i,
\end{equation}
where the $\{\rho_i^{A(B)}\}$'s are local states of A(B), which can be either
pure or mixed, and $\{p_i\}$'s are probabilities with $0\le p_i\le 1$ and
$\sum_i p_i=1$. Such a sum, in which the weights are non-negative, is called a
convex sum. The generalization to multi-partite states involves including more
parties:
\begin{equation}
\rho_{\rm s}^{\rm AB\cdots K}=\sum_{i}p_i\,\rho^{\rm A}_i\otimes\rho^{\rm
B}_i\otimes\cdots\otimes\rho^{\rm K}_i.
\end{equation}
In the present chapter we shall focus on bi-partite states. If a bi-partite
state cannot be written as a convex sum of direct products of density matrices
then it is entangled. However, this definition does not offer a practical way
of determining separability or entanglement.

Peres~\cite{Peres96} proposed a very simple but useful criterion for
separability.  As a density matrix is Hermitian and positive semi-definite,
its transpose is still a valid density matrix. If we take the transpose of the
matrices $\{ \rho^{\rm B}_i\}$'s in Eq.~(\ref{eqn:sep}), the resulting matrix,
denoted by $\rho^{\rm T_B}_{\rm s}$, still contains non-negative eigenvalues.
The operation is called partial transpose and can be defined for any
bi-partite state:
\begin{equation}
\rho=\sum_{i,j,k,l}\rho_{ij;kl}|e^{\rm A}_i\otimes e^{\rm B}_j\rangle\langle
e^{\rm A}_k\otimes e^{\rm B}_l|\ {\longrightarrow} \ \rho^{\rm
T_B}\equiv\sum_{i,j,k,l}\rho_{i\underline{l};k\underline{j}}|e^{\rm
A}_i\otimes e^{\rm B}_j\rangle\langle e^{\rm A}_k\otimes e^{\rm B}_l|,
\end{equation}
where $|e^{\rm A}_i\otimes e^{\rm B}_j\rangle\equiv\ket{e^{\rm A}_i}\otimes
\ket{e^{\rm B}_j}$ is the product basis used to represent the density matrix,
and the underscores are used to highlight the changes under the partial
transpose. Thus we have that if the state is separable, its partially
transposed matrix has non-negative eigenvalues (usually called PPT). Said
equivalently, if the state is not PPT under the partial transpose, the state
must be entangled\footnote{In general, if a state has PPT, no entanglement can
be distilled out from it~\cite{Horodecki398}. But the state can be either
unentangled or entangled. When the state has PPT and is also entangled, it is
called a bound entangled state.}.
 This is the Peres
positive partial transpose (PPT) criterion for separability~\cite{Peres96}.

Let us examine the example of a singlet state $\ket{\Psi^-}$. When written in
the form of density matrix in the basis $\{\ket{\!\uparrow\uparrow},
\ket{\!\uparrow\downarrow},\ket{\!\downarrow\uparrow},\ket{\!\downarrow\downarrow}\}$,
it corresponds to the density matrix
\begin{equation}
\ketbra{\Psi^-}\longleftrightarrow\begin{pmatrix} 0 & 0 & 0 & 0\cr 0 &
\frac{1}{2} & -\frac{1}{2} & 0\cr 0& -\frac{1}{2} & \frac{1}{2} & 0\cr 0 & 0 &
0 & 0
\end{pmatrix}.
\end{equation}
The partial transpose takes it to
\begin{equation}
\begin{pmatrix}
0 & 0 & 0 & -\frac{1}{2}\cr 0 & \frac{1}{2} & 0 & 0\cr 0& 0 & \frac{1}{2} &
0\cr -\frac{1}{2} & 0 & 0 & 0
\end{pmatrix},
\end{equation}
which has one negative eigenvalue, $-1/2$. Thus, via the PPT criterion we see
that $\ket{\Psi^-}$ is entangled.

In general, this PPT criterion is necessary but not sufficient for
establishing separability. However, it was shown by Horodecki and
co-workers~\cite{Horodecki396} that PPT is sufficient in the cases of
$C^2\otimes C^2$ (two-qubit) and $C^2\otimes C^3$ (qubit-qutrit) systems: if
two-qubit or qubit-qutrit states obey PPT, they are separable. On the other
hand, if PPT is violated, the state is entangled. The extent to which a state
violates PPT is manifested in the negative eigenvalues of the partially
transposed density matrix, and can be used as a measure (not just
an identifier) of entanglement; this measure is called the {\it negativity\/}~\cite{ZyczkowskiHorodeckiSanperaLewenstein98,VidalWerner02}. 
Following \.Zyczkowski and
co-workers~\cite{ZyczkowskiHorodeckiSanperaLewenstein98}, we define the
negativity  ${\cal N}$ to be twice the absolute value of the sum of the
negative eigenvalues:
\begin{equation}
{\cal N}(\rho) = 2\max(0,-\lambda_{\rm neg}),
\end{equation}
where $\lambda_{\rm neg}$ is the sum of the negative eigenvalues of $\rho^{\rm
T_B}$ and the factor of two is a normalization chosen such that the singlet
state $\ket{\Psi^-}$ has ${\cal N}=1$.

\section{Entanglement of distillation}
The notion of the entanglement of distillation was introduced by Bennett and
co-workers~\cite{BennettBernsteinPopescuSchumacher96,BennettBrassardPopescuSchumacherSmolinWootters96}
to give an operational definition of the degree of entanglement. Suppose
$\rho$ represents the state of two particles possessed by two parties (usually
referred to as Alice and Bob) separated by some distance. A way to envisage
the degree of entanglement that $\rho$ has is to ask how useful $\rho$ is
compared to a standard state, such as any of the four Bell states:
\begin{equation}
\ket{\Psi^\pm}\equiv\frac{1}{\sqrt{2}}(\ket{01}\pm \ket{10}), \ \ \
\ket{\Phi^\pm}\equiv\frac{1}{\sqrt{2}}(\ket{00}\pm \ket{11}).
\end{equation}
Here $\{\ket{0},\ket{1}\}$ represents an orthonormal basis of a two-level
system, for instance, the $z$-component of the spin of a spin-1/2 particle, or
the polarization of a photon. More specifically, given $n$ copies of the state
$\rho$ shared between Alice and Bob, how many pairs, say $k$, of Bell states
can be obtained if each of Alice and Bob is allowed to (i) perform any local
operations (including measurement) on the particles he or she possesses and
(ii) share with the other party classical information, e.g., the outcome of
some measurements. These operations are called local operations and classical
communication (LOCC). The asymptotic limit
\begin{eqnarray}
 E_{\rm D}(\rho)\equiv\lim_{n\rightarrow\infty}({k}/{n}),
\end{eqnarray}
is called the entanglement of distillation~\cite{BennettBernsteinPopescuSchumacher96,BennettBrassardPopescuSchumacherSmolinWootters96}. 
In it, $k$ is the average number of Bell states taken over different
possibilities (due to measurement) of an optimal procedure. $E_{\rm D}$
quantifies the entanglement as a resource, using Bell states as a standard
ruler.

Let us illustrate the idea with an example. Suppose $\rho$ is a pure state
corresponding to the ket
\begin{equation}
\label{eqn:psitheta}
|\psi_\theta\rangle=\cos\theta\ket{00}+\sin\theta\ket{11},
\end{equation}
 with $\theta\in[0,\pi/2]$. For $\theta=0$ or $\pi/2$, it is not entangled.
For the intermediate range of $\theta$, the state is entangled, and maximally
so at $\theta=\pi/4$. Suppose that two copies of non-maximally entangled
$\ket{\psi_\theta}$ are shared between Alice and Bob:
\begin{eqnarray}
\ket{\psi_\theta}_{12}\otimes\ket{\psi_\theta}_{34} & = & (\cos\theta|0_1
0_2\rangle +\sin\theta| 1_1 1_2\rangle )
    \otimes (\cos\theta|0_3 0_4\rangle +\sin\theta| 1_3 1_4\rangle ) \nonumber \\
\!\!\!\!\!     & = & \cos^2\theta |0_1 0_2 0_3 0_4\rangle  +
\sqrt{2}\cos\theta\sin\theta \frac{1}{\sqrt{2}}(|0_1 0_2 1_3 1_4\rangle +|1_1
1_2 0_3 0_4\rangle ) + \sin^2\theta | 1_1 1_2 1_3 1_4\rangle, \nonumber
\end{eqnarray}
where Alice has particles $1$ and $3$, whereas Bob has $2$ and $4$. Their
joint goal is to extract a Bell state under LOCC.

Both parties can perform any local operations allowed by quantum mechanics. A
possible operation is to measure the number of his/her particles in state
$\ket{1}$ (e.g., the $z$-component of total angular momentum). If Alice
measures the number of $1$'s of her particles to be 0 or 2 then the resulting
state ($|0_1 0_2 0_3 0_4\rangle $ or $| 1_1 1_2 1_3 1_4\rangle$) is
unentangled. She needs to tell Bob to abort the operation, as there is now no
entanglement to extract. But with probability $2 \cos^2\theta\sin^2\theta$ she
gets the state
\begin{eqnarray}
|\psi\rangle =\frac{1}{\sqrt{2}}(|0_1 0_2 1_3 1_4\rangle +|1_1 1_2 0_3
0_4\rangle ),
\end{eqnarray}
which is evidently entangled. But how do they establish from this a Bell
state, say, $\ket{\Phi^+}$?

This time, Alice proceeds to perform a unitary transformation $U$ on her
particles and contacts Bob (which is when the classical communication takes
place) and asks him to perform the same unitary transformation on {\it his\/}
particles. Suppose that the unitary transformation they agree to perform is
(also known as a CNOT operation)
\begin{eqnarray}
U=\left( \begin{array}{rrrr}  1  & 0  &  0 &  0 \\
                              0  & 1  &  0 &  0 \\
                               0  & 0  &  0 &  1 \\
                               0  & 0  &  1 &  0 \end{array}\right)
\end{eqnarray}
in the basis of \{$| 0 0\rangle , |0 1\rangle ,| 1 0\rangle ,|11\rangle$\}.
 In particular, for it we have
\begin{eqnarray}
 U_{13}|0_1 1_3\rangle  =  |0_1 1_3\rangle , \
 U_{13}|1_1 0_3\rangle  =  |1_1 1_3\rangle , \
 U_{24}|0_2 1_4\rangle  =  |0_2 1_4\rangle , \
 U_{24}|1_2 0_4\rangle  =  |1_2 1_4\rangle .
\end{eqnarray}
Then the joint state after the transformations becomes
\begin{eqnarray}
U_{13}U_{24}|\psi\rangle
        & = &U_{13}U_{24}\left(\frac{1}{\sqrt{2}}(|0_1 0_2 1_3 1_4\rangle +|1_1 1_2 0_3 0_4\rangle)\right)  \nonumber \\
       & = &\frac{1}{\sqrt{2}}(|0_1 0_2 1_3 1_4\rangle +|1_1 1_2 1_3 1_4\rangle )
 \nonumber \\
       & = &\frac{1}{\sqrt{2}}(|0_1 0_2\rangle +|1_1 1_2\rangle )\otimes |1_3 1_4\rangle .
\end{eqnarray}
As particles $3$ and $4$ are not entangled with $1$ and $2$, what now needs to
be done is that Alice throws away her particle $3$ and Bob throws away his
particle $4$. Finally, they have distilled one maximally entangled pair
$\frac{1}{\sqrt{2}}(|0_1 0_2\rangle +|1_1 1_2\rangle )$ out of two
non-maximally entangled pairs. The probability $P$ of success is $2
\cos^2\theta\sin^2\theta$, i.e., on average they can distill $k/n=\frac{1}{2}P
=\cos^2\theta\sin^2\theta\,$ Bell pairs per initial pair.
\begin{figure}
\[
\Qcircuit @C=2em @R=1.5em {
  & &\mbox{Alice}  & & & & \mbox{Entangled} & &\mbox{ sources}& &  & \mbox{Bob} &
   \\
  & \lstick{1}& \ctrl{1} & \qw& \qw & \qw & \qw & \qw & \qw &
    \qw &   \qw  & \ctrl{1}
  & \qw& \lstick{2}\\
  & \lstick{3}& \targ & \qw &\qw & \qw & \qw & \qw & \qw &
    \qw & \qw & \targ &
    \qw & \lstick{4} \gategroup{2}{5}{3}{10}{.8em}{--}
}
\]
\caption[Entanglement of distillation]{Entanglement of distillation. }
\label{fig:ED}
\end{figure}
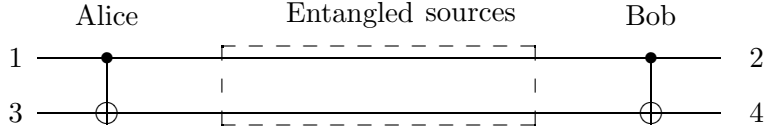

The above discussion involves Alice and Bob dealing with two pairs at a time.
In fact, it can be extended to the case where they can manipulate $n$ copies
at a time~\cite{BennettBernsteinPopescuSchumacher96}. The average number of
Bell pairs per initial pair can be dervied to be
\begin{subequations}
\begin{eqnarray}
&&\overline E(n)=\frac{1}{n}\sum_{k=0}^{n}P(k)E_k =\frac{1}{n}\sum_{k=0}^{n}P(k)\log_2(C^n_k), \\
&&P(k) \equiv (\cos^2\theta)^{n-k}(\sin^2\theta)^k C^n_k, \label{eq:AveEntS}
\end{eqnarray}
\end{subequations}
where $C_k^n\equiv n!/[k!(n-k)!]$. As the number $n$ of copies approaches
infinity,
\begin{equation}
\lim_{n\rightarrow\infty}\overline E(n)\rightarrow E_{\rm D}=S(\rho_{\rm A}),
\end{equation}
where $\rho_A\equiv{\rm Tr}_{\rm B}\ketbra{\psi_\theta}$, and
$S(\rho)\equiv-{\rm Tr}\rho\log_2\rho$ is the von Neumann entropy of $\rho$.
In the case of $\ket{\psi_\theta}$, its entanglement of distillation is
$E_{\rm D}=h(\cos^2\theta)$, where $h(x)\equiv -x\log_2(x)-(1-x)\log_2(1-x)$,
i.e., is the Shannon entropy. The result
\begin{eqnarray}
\label{eqn:EDentropy} E_{\rm D}=-{\rm Tr}\big(\rho_A\log_2(\rho_A)\big),
\end{eqnarray}
is valid for {\it any\/} bi-partite pure state.

\begin{figure}
\centerline{\psfig{figure=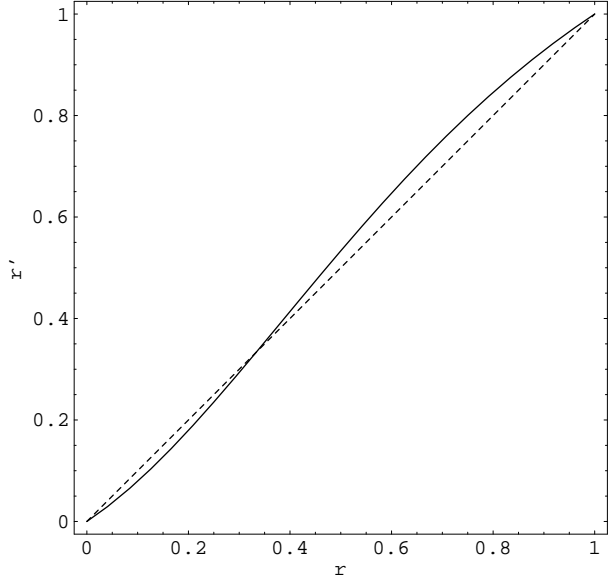,width=8cm}} \caption[Mixed state
distillation for Werner state $\rho_{\rm W+}(r)$]{\label{Fig:rrp}Mixed state
distillation for Werner state $\rho_{\rm W+}(r)$. States above the dashed line
(i.e., $r>1/3$) can be distilled by the procedure. } \label{fig:rrp}
\end{figure}

In Fig.~\ref{fig:ED} we show a  distillation scheme slightly modified from the
two-pair example. In this modified scheme, Alice and Bob both perform the CNOT
operation {\it before\/} the measurement. This transforms the initial state
$\ket{\psi_\theta}_{12}\otimes\ket{\psi_\theta}_{34}$ as follows:
\begin{equation}
\ket{\psi_\theta}_{12}\otimes\ket{\psi_\theta}_{34}  \rightarrow
(\cos^2\theta |0_1 0_2\rangle+ \sin^2\theta | 1_1 1_2 \rangle) |0_3 0_4\rangle
+ \sqrt{2}\cos\theta\sin\theta \frac{1}{\sqrt{2}}(|0_1 0_2 \rangle +|1_1
1_2\rangle) |1_3 1_4\rangle.
\end{equation}
If  Alice and/or Bob then measures the third and/or fourth qubit,
respectively, and the outcome is $\ket{1}$, they immediately obtain a Bell
state shared between particles 1 and 2. If the outcome is $\ket{0}$, they get
a slightly less entangled state, which they can store for a second trial of
distillation. What we mean by this is that two pairs of the states
\begin{equation}
\frac{1}{\sqrt{\cos^4\theta+\sin^4\theta}}\big(\cos^2\theta |0 0\rangle+
\sin^2\theta | 11 \rangle\big),
\end{equation}
although less entangled than the original pairs of Eq.~(\ref{eqn:psitheta}),
is distillable.  Thus, this modified scheme performs slightly better than the
original two-pair scheme.

For mixed entangled states, there are very few cases for which $E_{\rm D}$ is
known. No general optimal distillation procedure is known for generic states.
But a similar set-up to the one shown in Fig.~\ref{fig:ED} (except that the
measurement is performed in a different basis) does provide a way (although
not optimal) to distill very general two-qubit states. For example, Bennett
and co-workers~\cite{BennettBrassardPopescuSchumacherSmolinWootters96} have
shown that  after one step of the mixed-state distillation procedure, two
initial pairs of the state (which is usually called the Werner state)
\begin{equation}
\rho_{\rm W+}(r)\equiv r \ketbra{\Psi^+}+ \frac{1-r}{4}\openone,
\end{equation}
will be transformed into one pair with  a new parameter (see
Fig.~\ref{fig:rrp})
\begin{equation}
r'=\frac{2r(1+2r)}{3(1+r^2)}.
\end{equation}
Note that the larger the parameter $r$ is, the higher entanglement the Werner
state possesses. If $r'>r$, i.e., when $r> 1/3$ or equivalently the fidelity
$F\equiv\bra{\Psi^+}\rho_{\rm W}\ket{\Psi^+}>1/2$, the entanglement is said to
be increased. Horodecki and co-workers~\cite{HorodeckiHorodeckiHorodecki97}
further showed that any entangled two-qubit state can be transformed into a
state $\rho_{\rm W+}(r)$ with $r>1/3$, and hence can be distilled via the
scheme of Bennett and co-workers (also known as the BBPSSW
scheme)~\cite{BennettBrassardPopescuSchumacherSmolinWootters96}.

However, the procedure is not optimal for an arbitrary state $\rho$, and it is
generally rather difficult to compute $E_{\rm D}(\rho)$. Nevertheless, if
$E_{\rm D}(\rho)>0$ we say that the state $\rho$ is {\it distillable\/}. We
remark that there is a connection between the PPT criterion and the
distillability of a bi-partite state. Horodecki and
co-workers~\cite{Horodecki398} found that if a state has PPT then it cannot be
distilled. But the converse is not generally true. In fact, there is still no
simple criterion to determine whether  or not a state is distillable.

\section{Entanglement cost and entanglement of formation}
The distillation is a process for concentrating entanglement from a large
number of pairs with less entanglement into a small number of pairs with more
(and ultimately maximal) entanglement. On the other hand, we can consider the
converse process, which is usually called {\it dilution\/}. Given $k$ pairs of
Bell states shared between Alice and Bob, how many pairs $n$ of a given state
$\rho$ can be obtained by local operations (including adding unentangled
particles) and classical communication?
The goal is to maximize the number $n$ of copies of the output state $\rho$. The optimal ratio  defines the {\it entanglement cost\/}~\cite{BennettDiVincenzoSmolinWootters96}: 
\begin{eqnarray}
 E_{\rm C}(\rho)\equiv\lim_{k\rightarrow\infty}({k}/{n}).
\end{eqnarray}
As with $E_{\rm D}$, $E_{\rm C}$ is very difficult to calculate for general
mixed states, and is only known for a very few special cases.

However, for pure states such as the state $\ket{\psi_\theta}$ discussed
previously, $E_{\rm C} =-{\rm Tr}\big(\rho_A\log_2(\rho_A)\big)$, which equals
$E_{\rm D}$. The optimal way to realize this dilution process for the pure
state is to utilize two techniques: (i) quantum teleportation, which we have
introduced at the beginning and which simply says that a Bell state shared
between two parties can be used to tranfer an unknown qubit state with
certainty, and (ii) {\it quantum data compression\/}~\cite{Schumacher95},
which basically states that  a large message consistive of say $n$ qubits,
with each qubit on average being described by a density matrix $\rho_{\rm A}$,
can be compressed into a possibly smaller number $k=n S(\rho_{\rm A})\le n$ of
qubits; and one can faithfully recover the whole message, as long as $n$ is
large enough. For more detail of quantum data compression, see
Appendix~\ref{app:compression}.

With these two tools in hand, Alice can first prepare $n$ copies of
$\ket{\psi_\theta}$
 ($2n$ qubits in total) locally, compress the $n$ qubits to $k$ qubits that she will ``send'' to Bob, and teleport the compressed $k$ qubits to Bob using the shared $k$ Bell
states. Bob then decompresses the $k$ qubits back to the uncompressed $n$
qubits, which belong to half of the $n$ copies of the entangled state
$\ket{\psi_\theta}$. Thus, Alice and Bob establish $n$ pairs of
$\ket{\psi_\theta}$. This describes the optimal procedure for the dilution
process for a pure state.

The entanglement of distillation and entanglement cost are defined
asymptotically, i.e., both processes involve an infinite number of copies of
the initial states. For pure states, $E_{\rm C}=E_{\rm
D}$~\cite{BennettBernsteinPopescuSchumacher96}, which means that the two
processes are reversible asymptotically. Yet, for mixed states, both
quantities are very difficult to calculate. Nevertheless, it is expected that
$E_{\rm C}(\rho)\ge E_{\rm D}(\rho)$, viz. that one can not distill more
entanglement than is put in.

However, as we now explain, there is a modification of $E_{\rm C}$, obtained
by averaging $E_{\rm C}$ over pure states, and it is called the {\it
entanglement of formation\/} $E_{\rm
F}$~\cite{BennettDiVincenzoSmolinWootters96,Wootters98}. Any mixed state
$\rho$ can be decomposed into mixture of pure states
$\{p_i,|\psi_i\rangle\langle\psi_i|\}$ as in Eq.~(\ref{eqn:decomp}), although
the decomposition is far from unique. To construct the mixed state via mixing
pure states in this way will cost, on average,
 $\sum_i p_iE(|\psi_i\rangle\langle\psi_i|)$ pairs of Bell states. The entanglement of formation for a mixed state $\rho$ is thus defined as the {\it minimal\/} average number of Bell states needed to realize an ensemble described by $\rho$, i.e.,
\begin{eqnarray}
\label{eqn:Ef} E_{\rm F}(\rho)\equiv\min_{\{p_i,\psi_i\}}\sum_i p_i \, E_{\rm
C}(|\psi_i\rangle\langle\psi_i|),
\end{eqnarray}
where the minimization is taken over those probabilities $\{p_i\}$ and pure
states $\{\psi_i\}$ that, taken together, reproduce the density matrix
$\rho=\sum_i p_i |\psi_i\rangle\langle\psi_i|$. Such a construction is usually
called a {\it convex hull\/} construction. Furthermore, the quantity $E_{\rm
C}(|\psi_i\rangle\langle\psi_i|)$ is the entropy of entanglement of pure state
$|\psi_i\rangle$, viz. the expression in the right-hand side of
Eq.~(\ref{eqn:EDentropy}). However, $E_{\rm F}$ is, in general, also difficult
to calculate for mixed states, as it involves a minimization over all possible
decompositions. So far, there has been more analytic progress for $E_{\rm F}$
than forn $E_{\rm C}$ and $E_{\rm D}$. Notable cases include (i) Wootters'
formula for arbitary two qubits~\cite{Wootters98} (or see
Appendix~\ref{app:Wootters}); (ii) Terhal and Vollbrecht's
formula~\cite{TerhalVollbrecht00} for {\it isotropic\/} states for two qu-{\it
dit}s ($d$-level parties); and (iii) Vollbrecht and Werner's
formula~\cite{VollbrechtWerner01} for generalized Werner states of two
qu-dits.

One of the central issues in entanglement theory is the so-called {\it
additivity\/} of entanglement, i.e., whether the entanglement of formation,
defined as an average quantity, equals the entanglement cost, which is defined
asymptotically. Recently, Shor~\cite{Shor03} has established that the
additivity problem of entanglement of formation is equivalent to three other
additivity problems: the strong superadditivity of the entanglement of
formation, the additivity of the minimum output entropy of a quantum channel,
and the additivity of the Holevo classical capacity of a quantum channel.
However, further discussion of these additivity problems is beyond the scope
of this dissertation.

\section{Entanglement via a distance measure}
\label{sec:EntDist} As any mixture of separable density matrices is still, by
definition, separable, any separable state can be expressed as a sum of two
separable states
\begin{equation}
\rho_s=p\,\rho_s^1+(1-p)\rho_s^2,
\end{equation}
unless it is the extremal point, viz. a pure product state. Thus we see that
 the set of separable states is a {\it convex\/} set. This leads to another
type of entanglement measure: the shortest ``distance'' $E(\rho)$ from an
entangled state to the convex set $D_{\rm s}$ of separable mixed
states~\cite{VedralPlenio98}, i.e.,
$E(\rho)\equiv\min_{\sigma\in D_{\rm s}} {d}(\rho||\sigma)$.
One example of such an entanglement measure is the relative entropy of
entanglement,
\begin{equation}
E_{\rm R}(\rho)\equiv\min_{\sigma\in D_{\rm s}}{\rm
Tr}\left(\rho\log\rho-\rho\log\sigma\right),
\end{equation}
where the distance measure ${d}$ is defined to be the relative entropy of two
states:
\begin{equation}
 {d}(\rho||\sigma)\equiv {\rm Tr}\big[\rho\log\rho-\rho\log\sigma\big].
 \end{equation}
We remark that the relative entropy is non-negative, but it is also not
symmetric, i.e., ${d}(\rho||\sigma)\ne{d}(\sigma||\rho)$. For pure states this
definition of entanglement reduces to the entropy of entanglement.

Another example is the Bures metric of entanglement $E_{\rm B}(\rho)$, defined
via
\begin{eqnarray}
E_{\rm B}(\rho)\equiv \min_{\sigma\in D_{\rm s}}\big[2-2{F(\rho,\sigma)}\big],
\end{eqnarray}
where $F(\rho,\sigma)\!\equiv\! \big({\rm
Tr}\sqrt{\sqrt{\sigma}\rho\sqrt{\sigma}}\big)^2$ is called the {\it fidelity}
and {\it is\/} symmetric.  For two pure states $\rho=\ketbra{\psi}$ and
$\sigma=\ketbra{\phi}$, the distance
${d}(\rho||\sigma)\equiv\big(2-2{F(\rho,\sigma)}\big)$ reduces to
$2(1-|\langle\psi|\phi\rangle|^2)$.

We shall give more discussion of the relative entropy of entanglement later
on.
\def\test1{\section{Entanglement criteria}
From the discussions above, we know that entanglement is a nonlocal quantum
characteristic which cannot be generated by local operations and classical
communication between two distant parties. Local unitary transformations are
simply changes of local basis, and cannot change the amount of entanglement.
These, as well as other considerations lead to postulates for entanglement
measures. For a given state $\rho$
a good entanglement measure should, at least, satisfy the following conditions~\cite{VedralPlenio98,VedralPlenioRippinKnight97,Horodecki300}:\\
C1. (a) $E(\rho)\ge 0$; (b) $E(\rho)=0$ if $\rho$ is not entangled. \\
C2.  For any state $\rho$ and any local unitary transformation (i.e., a
unitary
transformation of the form $U_A \otimes U_B$), the entanglement should remain unchanged. \\
C3. Local operations, classical communication and postselection (i.e., keeping certain measurement outcomes and discarding the rest) should not increase the expectation value of the entanglement. \\
C4. Entanglement is convex under discarding information: $\sum_i p_iE(\rho_i)\ge E(\sum_i p_i\rho_i)$. \\
There are additional postulates such as continuity and additivity, but C1-C4
are those widely accepted in the literature. Note that in C1b we do not say
\lq\lq$E(\rho)=0$ {\it iff\/} $\rho$ is not entangled\rlap". This is because
cases exist in which $E_{\rm F}(\rho)>0$ although $E_{\rm D}(\rho)=0$,
i.e.,~there can exist {\it bound entanglement\/}~\cite{Horodecki398},
entanglement that cannot be distilled. }

\section{A simple model}
\begin{figure}
\vspace{-0.6cm} \psfrag{T}{{\bf T} $[k_{\rm B}/|J|]$}
\centerline{\psfig{figure=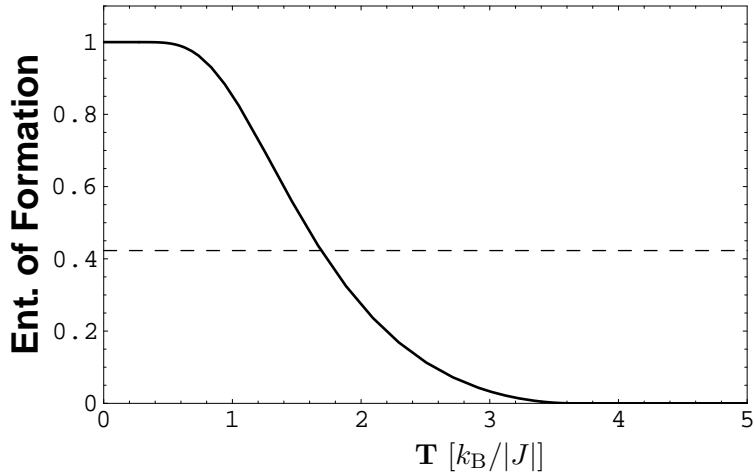,width=10cm}} \caption[Entanglement
vs.~temperature]{\label{Fig:EfT}Entanglement vs.~temperature. The dashed line
represents the threshold of entanglement above which there is a violation of
Bell's inequality. }
\end{figure}
\def\test2{As a simple example, let us consider a system of two spins in the absense of external fields.
Suppose the Hamiltonian of the system is ${\cal H}=-J\sigma_z^1\sigma_z^2$. It
is straightforward to solve the eigenproblem. The eigenstates are
$|++\rangle$, $|+-\rangle$, $|-+\rangle$, and $|--\rangle$ with respective
eigenvalues $-J$, $+J$, $+J$, and $-J$. The density matrix, when expressed in
terms of the four separable bases, is diagonal, and
is clearly separable and contains no entanglement.} Let us illustrate how
mixed states arise naturally and study the entanglement properties of the
mixed states. Consider the Hamiltonian ${\cal
H}=-J\vec{\sigma}^1\cdot\vec{\sigma}^2$. The eigenstates are
$|\!\uparrow\uparrow\rangle$, $|\!\downarrow\downarrow\rangle$,
$|\Psi^+\rangle=(|\!\uparrow\downarrow\rangle+|\!\downarrow\uparrow\rangle)/\sqrt{2}$,
and
$|\Psi^-\rangle=(|\!\uparrow\downarrow\rangle-|\!\downarrow\uparrow\rangle)/\sqrt{2}$
with respective eigenvalues $-J$, $-J$, $-J$ and $+3J$.

At a temperature $T=1/(k_{\rm B}\beta)$ the two-spin system  is described by a
density matrix
\begin{equation}
\rho=\frac{e^{\beta
J}}{Z}(\ketbra{\Phi^+}+\ketbra{\Phi^-}+\ketbra{\Psi^+})+\frac{e^{-3\beta
J}}{Z}\ketbra{\Psi^-} =\mygamma\ketbra{\Psi^-}+\frac{1-\mygamma}{4}\openone,
\end{equation}
where $Z=3e^{\beta J}+e^{-3\beta J}$ and $\mygamma\equiv(e^{-3\beta
J}-e^{\beta J})/(3e^{\beta J}+e^{-3\beta J})$. This happens to be a Werner
state. Using the Peres-Horodecki separability criterion, we find that if $J>0$
the state is separable, as $\mygamma<1/3$. On the other hand, if $J<0$, there
is a transition from being entangled to separable as the temparature is
increased~\cite{Nielsen98}. This can be seen from  the entanglement of
formation for the state as a function of temperature; see Fig.~\ref{Fig:EfT}.
In fact, Werner~\cite{Werner89} found that such a state violates the Bell-CHSH
inequality if $r>1/\sqrt{2}$. Thus, there is a finite region between being
entangled ($r>1/3$) and violating the Bell-CHSH inequality. Moreover, Werner
found that for $r\le 1/2$ the state can be described by a classical local
theory. The interval has recently been extended to $r\le 2/3$ by Terhal and
co-workers~\cite{TerhalDohertySchwab03}. Even more interestingly, as we have
seen previously, two pairs of Werner states with $r>1/2$ can be distilled into
a single Werner state of higher $r$, viz. higher entanglement. Thus, given a
sufficient supply of Werner states, each of which does not violate any Bell
inequality, a highly entangled Werner state can be distilled to violate a Bell
inequality. This nonlocal quantum feature hidden in the initial states is thus
revealed by the distillation
process~\cite{BennettBrassardPopescuSchumacherSmolinWootters96,KwiatNature01}.

This toy system demonstrates an interesting behavior  of entanglement of
canonical ensembles as the temperature is varied. Later in this dissertation
we shall study a more realistic model of large number of spins but at zero
temperature. The model that we shall discuss exhibits quantum phase
transitions as some system parameters vary.

\section{Overview of the dissertation}
One of the goals of this thesis is to develop an entanglement theory
applicable to many-body (multi-partite) systems. But before we apply to more
realistic models of many bodies, in Chapter~\ref{chap:GME} we examine the
theory in the context of more familiar bi-partite systems, where corresponding
results for other entanglement measures are already known.  We then apply our
entanglement measure to several nontrivial families of multi-partite states.
Connections of our entanglement measure  to other topics, such as the Hartree
approximation, entanglement witnesses, and correlation functions are also
discussed along the way. In Chapter~\ref{chap:ReEnt} we discuss in detail the
connection between the geometric measure of entanglement to the relative
entropy of entanglement. In Chapter~\ref{chap:Bound} we compute the
multi-partite measure for two distinct peculiar states, i.e., the so-called
bound entangled states. Finally, in Chapter~\ref{chap:QPT} we apply our
measure to the ground state of the XY quantum spin chain model in a transverse
magnetic field. The behavior of the entanglement near the quantum critical
points is found to be dictated by the universality classes of the model.

\chapter{Geometric measure of entanglement for multi-partite states}
\label{chap:GME}
\section{Introduction}
\label{sec:IntroGME} Only recently, after more than half a century of
existence, has the notion of entanglement become recognized as central to
quantum information processing.  As a result, the task of characterizing and
quantifying entanglement have emerged as one of the prominent themes of
quantum information theory.  There have been many achievements in this
direction, primarily in the setting of {\it bi-partite\/} systems.  Among
these, one highlight is Wootters' formula~\cite{Wootters98} for the
entanglement of formation for arbitrary two-qubit mixed states. This formula
enables discussions of entanglement between any pair of two-level systems,
which are quite common in
 various
physical systems (or idealizations of them). Other achivements include
corresponding results for highly symmetrical states of higher-dimensional
systems~\cite{TerhalVollbrecht00,VollbrechtWerner01}.

The success of bi-partite entanglement theories, such as the entanglement of
distillation and formation, hinges on the reversible interconvertibility of
pure entangled states. To be more precise, any bi-partite pure entangled
states can be, via local operations and classical communication, transformed
into  Bell states, asymptotically and reversibly. This means that there is
only one type of pure entangled state and one can use Bell states as a
``standard ruler'' to quantify the degree of entanglement. As a result. the
entanglement cost $E_{\rm C}$ equals the entanglement of distillation $E_{\rm
D}$ for pure states.

However, things become more delicate in the multi-partite settings. For
example, in the case of three qubits, two of the qubits can be entangled
whilst the third one is separable from (not entangled with) with them. This
kind of entanglement is nothing more than the bi-partite entanglement we
already know. But there are two other extreme types of entangled states. One
is the so-called GHZ state: $\ket{\rm
GHZ}\equiv(\ket{000}+\ket{111})/\sqrt{2}$, and the other is the W state:
$\ket{W}\equiv (\ket{001}+\ket{010}+\ket{100})/\sqrt{3}$. It was found by
D\"ur and co-workers~\cite{DurVidalCirac00} 
that, given a single copy of GHZ state, there is no way
 via local operations and classical communication (LOCC), not even probabilistically, that it can be transformed
into a W state, or vice versa. GHZ and W states have different types of
entanglement.

The GHZ state can be rewritten as
\begin{equation}
\ket{\rm
GHZ}=\frac{1}{2}\ket{00}\otimes(\ket{+}+\ket{-})+\frac{1}{2}\ket{11}\otimes(\ket{+}-\ket{-})
=\frac{1}{2}(\ket{00}+\ket{11})\otimes\ket{+}+\frac{1}{2}(\ket{00}-\ket{11})\otimes\ket{-},
\end{equation}
where $\ket{\pm}\equiv(\ket{0}\pm\ket{1})/\sqrt{2}$. If one of the parties,
e.g., the third one, performs measurement in the $\{\pm\}$ basis, and if the
outcome is $\ket{+}$, the other two parties have the entangled state
$\ket{\Phi^+}$. If the outcome is $\ket{-}$, the other two parties have the
entangled state $\ket{\Phi^-}$. Therefore, two of the three parties can
establish a Bell state. Thus, given two copies of GHZ states, they can
establish a Bell state shared between say A and B, and another Bell state
shared between A and C.  A can locally prepares a W state
\begin{equation}
\ket{{\rm
W}}=\frac{1}{\sqrt{3}}(\ket{0_10_21_3}+\ket{0_11_20_3}+\ket{1_10_20_3}).
\end{equation}
As she has one Bell state shared with B and the other with C, she can use
quantum teleportation to teleport the state of particle 2 of the W state to B
and that of particle 3 to C. Thus two copies of GHZ states can achieve a copy
of the W state.

On the other hand, if the three parties share a W state and one of them makes
a measurement in the $\{0/1\}$ basis, 2/3 of the time the other two parties
can establish a Bell state. The other 1/3, they fail to do so. If they are
given 2 copies of W states, 4/9 of the time they can establish one Bell state
between, say, A and B, and the other Bell state between A and C. A can use the
same trick of teleportation such that they end up with a GHZ state. So two
copies of W states can achieve 4/9 copy of GHZ state.

However, it is still not yet clear whether or not given some number of copies
of GHZ states, they can be transformed into as many copies of W states and
then transformed back to the original number of GHZ states. Namely, it is not
yet known that whether the process is asymptotically reversible.

Bennett and co-workers have come up with the notion of a finite minimal
reversible entanglement generating set
(MREGS)~\cite{BennettPopescuRohrlichSmolinThapliyal}. Such a set should
include different types of ``ruler'' states. But the numer of states in this
set should be finite, otherwise, it is not practical to use an infinite number
of different rulers to ``measure'' the property of entanglement. For example,
the minimal set might contain $\{\ket{\rm Bell}_{12},\ket{\rm
Bell}_{13},\ket{\rm Bell}_{23},\ket{\rm GHZ}\}$. Perhaps a large number of any
three-qubit pure state can be transformed reversibly into, say, $x_1$ copies
of $\ket{\rm Bell}_{12}$, $x_2$ copies of $\ket{\rm Bell}_{13}$, $x_3$ copies
of $\ket{\rm Bell}_{23}$, and $x_4$ copies of $\ket{\rm GHZ}$? Or perhaps the
minimal set might contain $\{\ket{\rm Bell}_{12},\ket{\rm Bell}_{13},\ket{\rm
Bell}_{23},\ket{\rm GHZ},\ket{W}\}$? Then the entanglement could be defined as
some kind of vector. However, this MREGS problem has not been solved yet. The
situation gets even worse beyond three qubits. Verstraete and
co-workers~\cite{VerstraeteDehaeneDeMoor02} have found that
there are nine inequivalent classes of four-qubit entangled states. As the
dimensions and the number of parties grow, the number of states in MREGS, if
such a set exists, is expected to grow considerably large.

The above considerations complicate the task of extending measures such as the
entanglement of distillation~\cite{BennettBernsteinPopescuSchumacher96} and
formation~\cite{BennettDiVincenzoSmolinWootters96,Wootters98} to multi-partite
systems.  Moreover, we have seen that the characterization of general
multi-partite entanglement remains incomplete, as the number of the {\it
types\/} of entanglement grows with the number of parties and the dimensions
of Hilbert space. The issue of entanglement for multi-partite states hence
poses an even greater challenge than bi-partite states. On the other hand, one
can quantify multi-partite entanglement via other measures, such as the
relative entropy of entanglement, the Bures
metric~\cite{VedralPlenio98,VedralPlenioRippinKnight97}, and the Schmidt
measure~\cite{EisertBriegel01}, which are naturally extendible to
multi-partite settings.

In this chapter, we present an attempt to quantify multi-partite entanglement
 by
developing and investigating a certain geometric measure of entanglement
(GME), first introduced by Shimony~\cite{Shimony95} in the setting of
bi-partite pure states and generalized to the multi-partite setting (via
projection operators of various ranks) by Barnum and
Linden~\cite{BarnumLinden01}. In Sec.~\ref{sec:Pure} we begin by examining
this geometric measure in pure-state settings and establishing a connection
with entanglement witnesses, Hartree approximations, and correlation
functions. In Sec.~\ref{sec:Mixed} we extend the measure to mixed states,
showing that it satisfies certain criteria required of good entanglement
measures. In Sec.~\ref{sec:Analytic} we examine the GME for several families
of mixed states of bi-partite systems: (i)~arbitrary two-qubit mixed,
(ii)~generalized Werner, and (iii)~isotropic states in bi-partite systems, as
well as (iv)~certain mixtures of multi-partite symmetric states. In
Sec.~\ref{sec:GWW} we give a detailed application of the GME to arbitrary
mixtures of three-qubit GHZ, W and inverted-W states. In
Sec.~\ref{sec:Conclude} we discuss some open questions and further directions.
The discussion in this chapter is based on Ref.~\cite{WeiGoldbart03}

It is not our intention to cast aspersions on existing approaches to
entanglement; rather we simply wish to add one further element to the
discussion. Our discussion focuses on quantifying multi-partite entanglement
rather than characterizing it.

\section{Basic geometric ideas and application to pure states}
\label{sec:Pure}
\begin{figure}[t]
\centerline{\psfig{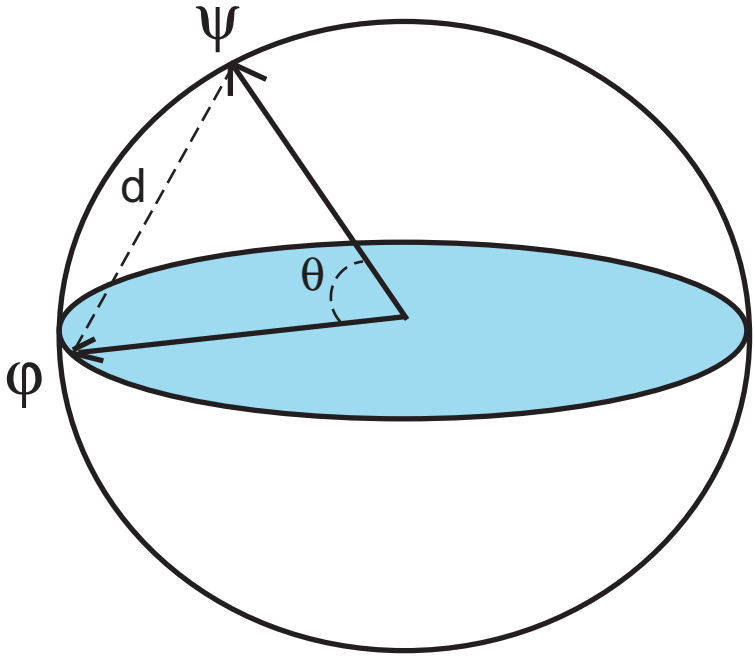}}
\caption[The schematic picture of the geometric measure.]{The schematic
picture of the geometric measure. Imagine all pure states lie on the sphere
and all separable pure states $\ket{\phi}$'s lie on the equator. The degree of
entanglement for $\ket{\psi}$ is reflected in the shortest distance to the set
of separable states.} \label{fig:sphereGME}
\end{figure}
We begin with an examination of entangled {\it pure\/} states, and of how one
might quantify their entanglement by making use of simple ideas of Hilbert
space geometry.  Let us start by developing a quite general formulation,
appropriate for multi-partite systems comprising $n$ parts, in which each part
can have a distinct Hilbert space.  Consider a general $n$-partite pure state
\begin{equation}
|\psi\rangle=\sum_{p_1\cdots p_n}\chi_{p_1p_2\cdots p_n}
|e_{p_1}^{(1)}e_{p_2}^{(2)}\cdots e_{p_n}^{(n)}\rangle,
\end{equation}
where $\{e_{p_k}^{(k)}\}$ is the local basis of the $k$-th party, e.g., the
spin $\uparrow$ or $\downarrow$. One can envisage a geometric definition of
its entanglement content via the distance
\begin{equation}
d=\min_{|\phi\rangle} \Vert\,|\psi\rangle-|\phi\rangle\Vert
\end{equation}
between $\ket{\psi}$ and the nearest of the separable states $\ket{\phi}$ (or
equivalently the angle between them). Here
\begin{equation}
\label{eqn:product}
|\phi\rangle\equiv\otimes_{i=1}^n|\phi^{(i)}\rangle=\ket{\phi^{(1)}}\otimes\ket{\phi^{(2)}}\otimes\cdots\otimes\ket{\phi^{(n)}}
\end{equation}
is an arbitrary separable (i.e., Hartree) $n$-partite pure state, the index
$i=1\ldots n$ labels the parties, and a state vector of part $i$ is written as
\begin{equation}
|\phi^{(i)}\rangle\equiv \sum_{p_i}c_{p_i}^{(i)}\,|e_{p_i}^{(i)}\rangle.
\end{equation}
It seems natural to assert that the more entangled a state is, the further
away it will be from its best unentangled approximant (and, correspondingly,
the wider will be the angle between them). We emphasize that we only compare
the entangled pure state to the set of {\it pure\/} unentangled state. We
shall extend to mixed states via the so-called convex-hull construction in
Sec.~\ref{sec:Mixed}. Another approach one might take is to compare any
entangled state to the set of unentalged states, including both pure and
mixed. The Bures measure, introduced in Sec.~\ref{sec:EntDist}, is such an
example.

To actually find the nearest separable state, it is convenient to minimize,
instead of $d$, the quantity $d^2$, i.e.,
\begin{equation}\Vert|\psi\rangle-|\phi\rangle\Vert^2,
\end{equation}
subject to the constraint $\langle\phi|\phi\rangle=1$. In fact, in solving the
resulting stationarity condition one may restrict one's attention to the
subset of solutions $\ket{\phi}$ that obey the further condition that each
factor $\ket{\phi^{(i)}}$ obeys its own normalization condition
$\ipr{\phi^{(i)}}{\phi^{(i)}}=1$. Thus, by introducing a Lagrange mulitplier
$\Lambda$ to enforce the constraint $\langle\phi|\phi\rangle=1$,
differentiating with respect to the independent amplitudes, and then imposing
the further condition $\ipr{\phi^{(i)}}{\phi^{(i)}}=1$, one arrives at the
{\it nonlinear eigenproblem\/} for the stationary $\ket{\phi}$:
\def\phast{\phantom{\ast}}
\begin{subequations}
\label{eqn:Eigen}
\begin{eqnarray}
\sum_{p_1\cdots\widehat{p_i}\cdots p_n} \chi_{p_1p_2\cdots
p_n}^*{c_{p_1}^{(1)\phast}}\cdots\widehat{c_{p_i}^{(i)\phast}}\cdots
c_{p_n}^{(n)\phast}&=&
\Lambda\,{c_{p_i}^{(i)*}}, \\
\sum_{p_1\cdots\widehat{p_i}\cdots p_n}\chi_{p_1p_2\cdots p_n}
{c_{p_1}^{(1)*}}\cdots\widehat{{c_{p_i}^{(i)*}}}\cdots {c_{p_n}^{(n)*}}&=&
\Lambda\,c_{p_i}^{(i)\phast}\,,
\end{eqnarray}
\end{subequations}
where the eigenvalue $\Lambda$ is associated with the Lagrange multiplier
enforcing the constraint $\ipr{\phi}{\phi}\!=\!1$, and the symbol
\,\,$\widehat{}$\,\, denotes the exclusion of the corresponding term or
factor. In a form independent of the choice of basis within each party,
Eqs.~(\ref{eqn:Eigen}) read
\begin{subequations}
\label{eqn:EigenForm}
\begin{eqnarray}
\label{eqn:EigenForma} \langle\psi|\Big(\mathop{\otimes}_{j(\ne
i)}^n|\phi^{(j)}\rangle\Big)
&=&\Lambda\bra{\phi^{(i)}}, \\
\Big(\mathop{\otimes}_{j(\ne i)}^n\langle\phi^{(j)}|\Big)|\psi\rangle
&=&\Lambda\ket{\phi^{(i)}}.
\end{eqnarray}
\end{subequations}
From Eqs.~(\ref{eqn:Eigen}) or (\ref{eqn:EigenForm}), e.g., by taking inner
product of both sides of Eq.~(\ref{eqn:EigenForma}) with $\ket{\phi^{(i)}}$
one readily sees that
\begin{equation}
\Lambda=\ipr{\psi}{\phi}=\ipr{\phi}{\psi}
\end{equation}
and thus the eigenvalues $\Lambda$ are real, in $[-1,1]$, and independent of
the choice of the local basis $\big\{|e_{p_i}^{(i)}\rangle\big\}$. Hence, the
spectrum  $\Lambda$ is the cosine of the angle between $|\psi\rangle$ and
$\ket{\phi}$; the largest, $\Lambda_{\max}$, which we call the {\it
entanglement eigenvalue\/}, corresponds to the closest separable state and is
equal to the maximal overlap
\begin{equation}
\label{eqn:LambdaMax} \Lambda_{\max}=\max_{\phi} ||\ipr{\phi}{\psi}||,
\end{equation}
where $\ket{\phi}$ is an arbitrary separable pure state.

Although, in determining the closest separable state, we have used the squared
distance between the states, there are alternative (basis-independent)
candidates for entanglement measures which are related to it in an elementary
way: the distance, the sine, or the sine squared of the angle $\theta$ between
them (with $\cos\theta\equiv\Real\ipr{\psi}{\phi}$). We shall adopt
$E_{\sin^2}\equiv 1-\Lambda_{\max}^2$ as our entanglement measure because, as
we shall see, when generalizing $E_{\sin^2}$ to mixed states we have been able
to show that it satisfies a set of criteria demanded of entanglement measures.

In bi-partite applications, the eigenproblem~(\ref{eqn:Eigen}) is in fact {\it
linear\/}, and solving it is actually equivalent to finding the Schmidt
decomposition~\cite{Shimony95}. Moreover, the entanglement eigenvalue is equal
to the maximal Schmidt coefficient. To be more precise, in bi-partite systems
the stationarity conditions~(\ref{eqn:Eigen}) reduce to the linear form
\begin{subequations}
\begin{eqnarray}
\label{eqn:bi-eigen-a} &&\sum_{p_1} \chi_{p_1p_2}^*
c_{p_1}^{(1)\phantom{\ast}}
=\Lambda \,{c_{p_2}^{(2)}}^*, \\
\label{eqn:bi-eigen-b} &&\sum_{p_1}
\chi_{p_1p_2} {c_{p_1}^{(1)}}^*=\Lambda\,{c_{p_2}^{(2)}}, \\
\label{eqn:bi-eigen-c} &&\sum_{p_2} \chi_{p_1p_2}^*
c_{p_2}^{(2)\phantom{\ast}}
=\Lambda \,{c_{p_1}^{(1)}}^*, \\
\label{eqn:bi-eigen-d} &&\sum_{p_2} \chi_{p_1p_2} {c_{p_2}^{(2)}}^*=\Lambda
\,{c_{p_1}^{(1)}}.
\end{eqnarray}
\end{subequations}
Eliminating $c_{p}^{(2)}$ between Eqs.~(\ref{eqn:bi-eigen-a}) and
(\ref{eqn:bi-eigen-d}) and, similarly, eliminating $c_{p}^{(1)}$ between
Eqs.~(\ref{eqn:bi-eigen-b}) and (\ref{eqn:bi-eigen-c}) gives
\begin{subequations}
\begin{eqnarray}
\label{eqn:bi-merge-eigen-ad} \sum_{p_1'\,p_2}\chi_{p_1p_2}\chi_{p_1'p_2}^*
c_{p_1'}^{(1)}
&=&\Lambda^2  \,c_{p_1}^{(1)}, \\
\label{eqn:bi-merge-eigen-bc} \sum_{p_1\,p_2'}\chi_{p_1p_2}\chi_{p_1p_2'}^*
{c_{p_2'}^{(2)}} &=&\Lambda^2  \,{c_{p_2}^{(2)}},
\end{eqnarray}
\end{subequations}
or equivalently
\begin{subequations}
\label{eqn:bipart}
\begin{eqnarray}
{\rm Tr}_2\big(|\psi\rangle\!\langle\psi|\big)| \phi^{{(1)}}\rangle &=&
\Lambda^2\,|\phi^{{(1)}}\rangle,\\
{\rm Tr}_1\big(|\psi\rangle\!\langle\psi|\big)| \phi^{{(2)}}\rangle &=&
\Lambda^2\,|\phi^{{(2)}}\rangle.
\end{eqnarray}
\end{subequations}

Now, solving the above equations is equivalent to finding the Schmidt
decomposition for $|\psi\rangle$.  To see this, first recall that Schmidt
decomposability guarantees that an arbitrary pure bi-partite state
\begin{equation}
|\psi\rangle= \sum_{p_1p_2}\chi_{p_1p_2} \ket{e_{p_1}^{(1)}}\otimes
\ket{e_{p_2}^{(2)}}
\end{equation}
in the (product) Hilbert space ${\cal H}_1\otimes{\cal H}_2$ (with factor
dimensions $d_1$ and $d_2$) can always be expressed in the form
\begin{equation}
|\psi\rangle= \sum_{k=1}^{\min\{d_1,d_2\}}{\lambda_k}\,
\ket{\tilde{e}_k^{(1)}}\otimes \ket{\tilde{e}_k^{(2)}}.
\end{equation}
Here, $\lambda_k\ge 0$, $\sum_k \lambda_k^2=1$, the
$|\tilde{e}_k^{(1)}\rangle$'s and $|\tilde{e}_k^{(2)}\rangle$'s are
orthonormal, respectively, in ${\cal H}_1$ and ${\cal H}_2$. Moreover, the new
(tilde) bases, as well as the $\lambda_k$'s, follow as the solution of the
eigenproblems of the reduced density matrix that one obtains by tracing
$\ket{\psi}\bra{\psi}$ over party 1 or 2:
\begin{subequations}
\begin{eqnarray}
\label{eqn:} {\rm
Tr}_2\big(|\psi\rangle\langle\psi|\big)|\tilde{e}_k^{(1)}\rangle
&=&\lambda_k^2|\tilde{e}_k^{(1)}\rangle,
\\
{\rm Tr}_1\big(|\psi\rangle\langle\psi|\big)|\tilde{e}_k^{(2)}\rangle
&=&\lambda_k^2|\tilde{e}_k^{(2)}\rangle.
\end{eqnarray}
\end{subequations}
These are Eqs.~(\ref{eqn:bipart}); thus we see that determining entanglement
for bi-partite pure states is equivalent to finding their Schmidt
decomposition, except that one only needs the {\it largest\/} Schmidt
coefficient $\Lambda_{\rm max}=\lambda_{\rm max}$.

By contrast, for the case of three or more parts, the eigenproblem is a {\it
nonlinear\/} one. For example, in the setting of tri-partite systems, the
stationarity conditions~(\ref{eqn:Eigen}) associated with the pure state
$|\psi\rangle=\sum_{p_1p_2p_3} \chi_{p_1p_2p_3}
|e_{p_1}^{(1)}e_{p_2}^{(2)}e_{p_3}^{(3)}\rangle$ become
\def\mysum#1{\sum\limits_{#1}}
\begin{subequations}
\begin{eqnarray}
&&\mysum{p_2p_3}\chi_{p_1p_2p_3}^* c_{p_2}^{{(2)}}c_{p_3}^{{(3)}} = \Lambda
\,{c_{p_1}^{{(1)}*}},
\\
&&\mysum{p_1p_3}\chi_{p_1p_2p_3}^* c_{p_1}^{{(1)}}c_{p_3}^{{(3)}} =
\Lambda\, {c_{p_2}^{{(2)}*}}, \\
&& \mysum{p_1p_2}\chi_{p_1p_2p_3}^* c_{p_1}^{{(1)}}c_{p_2}^{{(2)}}
 =
\Lambda \,{c_{p_3}^{{(3)}*}}.
\end{eqnarray}
\end{subequations}
Note the nonlinear structure of this tri-partite (and, in general, any $n\ge
3$-partite) eigenproblem.  As far as we are aware, for nonlinear eigenproblems
such as these, one cannot take advantage of the simplicity of linear
eigenproblems, for which one can address the task of determining the
eigenvalues directly (via the characteristic polynomial), without having to
address the eigenvectors.  Hence, even for systems comprising qubits, one is
forced to proceed numerically. This is consistent with the notion that no
Schmidt decomposition exists beyond bi-partite systems.
 Yet, as we shall
illustrate shortly, there do exist certain families of pure states whose
entanglement eigenvalues can be determined analytically.

For $C^d\otimes C^d$ bi-partite systems, the equivalence between the geometric
approach and Schmidt decomposition immediately indicates why the maximally
entangled pure states have (up to local unitary transformations) the
well-known form $|\Phi^+\rangle\equiv \frac{1}{\sqrt{d}}
\sum_{i=1}^d|ii\rangle$. As the Schmidt coefficients are non-negative and sum
(when squared) to unity, any less symmetric state must have a larger
$\Lambda_{\max}$, i.e., a smaller entanglement.

\subsection{Illustrative examples}
Suppose we are already in possession of the Schmidt decomposition of some
two-qubit pure state:
\begin{equation}
|\psi\rangle=
 \sqrt{p}\,|00\rangle
+\sqrt{1-p}\,|11\rangle.
\end{equation}
Then we can read off the entanglement eigenvalue:
\begin{equation}
\label{eqn:2qubitLambda} \Lambda_{\max}= \max\{\sqrt{p},\sqrt{1-p}\}.
\end{equation}

Another approach to obtain $\Lambda_{\max}$ is to solve, e.g.,
Eqs.~(\ref{eqn:bipart}), which lead to solving the maximal eigenvalue of
\begin{equation}
\begin{pmatrix}
p & 0 \cr 0 & 1-p
\end{pmatrix}
\vec{v}=\Lambda^2 \, \vec{v},
\end{equation}
resulting in the solution~(\ref{eqn:2qubitLambda}).

Now, recall~\cite{Wootters98} that the concurrence $C$ for this state is
$2\sqrt{p(1-p)}$ [see Appendix~\ref{app:Wootters}].  Hence, one has
\begin{eqnarray}
\label{eqn:LamConc}
\Lambda_{\max}^{2}=\frac{1}{2}\left(1+\sqrt{1-C^{2}}\right),
\end{eqnarray}
a connection between the entanglement eigenvalue and the concurrence that
holds for arbitrary two-qubit pure states.

The possession of symmetry by a state can alleviate the difficulty associated
with solving the nonlinear eigenproblem.  To see this, consider a state
\begin{equation}
|\psi\rangle= \sum_{p_1\cdots p_n}\chi_{p_1p_2\cdots p_n}
|e_{p_1}^{(1)}e_{p_2}^{(2)}\cdots e_{p_n}^{(n)}\rangle
\end{equation}
that possesses the symmetry that the (nonzero) amplitudes $\chi$ are invariant
under permutations. What we mean by this is that, regardless of the dimensions
of the factor Hilbert spaces, the amplitudes are only nonzero when the indices
take on the first $\nu$ values (or can be arranged to do so by appropriate
relabeling of the basis in each factor) and, moreover, that these amplitudes
are invariant under permutations of the parties, i.e.,
$\chi_{\sigma_1\sigma_2\cdots\sigma_n}=
 \chi_{p_1p_2\cdots p_n}$,
where the $\sigma$'s are any permutation of the $p$'s. (This symmetry may be
obscured by arbitrary local unitary transformations.) For such states, it
seems reasonable to anticipate that the closest Hartree approximant retains
this permutation symmetry.  Assuming this to be the case---and numerical
experiments of ours support this assumption---in the task of determining the
entanglement eigenvalue one can start with the Ansatz that the closest
separable state has the form
\begin{equation}
|\phi\rangle\equiv\otimes_{i=1}^n\left(\sum\nolimits_j
c_j|e_j^{(i)}\rangle\right),
\end{equation}
i.e., is expressed in terms of copies of a single factor state, for which
$c^{(i)}_j=c_j$.  To obtain the entanglement eigenvalue it is thus only
necessary to maximize ${\rm Re}\,\langle\phi|\psi\rangle$ with respect to the
$\nu$ amplitudes $\{c_j\}_{j=1}^{\nu}$, a simpler task than maximization over
the $\sum_{i=1}^{n}d_{i}$ amplitudes of a generic product state.

To illustrate this symmetry-induced simplification, we consider several
examples involving permutation-invariant states, first restricting our
attention to the case $\nu=2$ (i.e., two-state parties). The most natural
realizations are $n$-qubit systems. One can label these symmetric states
according to the number of $0$'s, as follows~\cite{Stockton}:
\begin{equation}
|S(n,k)\rangle\equiv \sqrt{\frac{k!(n-k)!}{n!}} \sum_{\rm{\scriptstyle
permutations}} |\underbrace{0\cdots0}_{k}\underbrace{1\cdots1}_{n-k}\rangle.
\end{equation}
As the amplitudes in this state are all positive and the state is
permutationally invariant, one can assume that the closest Hartree state also
has the symmetry and has non-negative amplitudes, and is of the form
\begin{equation}
|\phi\rangle=\Big(\sqrt{p}\,|0\rangle+\sqrt{1-p}\,|1\rangle\Big)^{\otimes n}.
\end{equation}
Maximizing the overlap with respect to $p$, one obtains  the entanglement
eigenvalue for $|{\rm S}(n,k)\rangle$:
\begin{eqnarray}
\label{eqn:Lambda} \Lambda_{\max}(n,k)= \sqrt{\frac{n!}{k!(n\!-\!k)!}}
\left(\frac{k}{n}\right)^{\frac{k}{2}}
{\left(\frac{n-k}{n}\right)}^{\frac{n\!-\!k}{2}}.
\end{eqnarray}
For fixed $n$, the minimum $\Lambda_{\max}$ (and hence the maximum
entanglement) among the $\ket{{\rm S}(n,k)}$'s occurs for $k=n/2$ (for $n$
even) and $k=(n\pm1)/2$ (for $n$ odd). In fact, for fixed $n$ the general
permutation-invariant state can be expressed as $\sum_k \alpha_k\ket{{\rm
S}(n,k)}$ with $\sum_k |\alpha_k|^2=1$. The entanglement of such states can be
addressed via the strategy that we have been discussing, i.e., via the
maximization of a function of (at most) three real parameters. The simplest
example is provided by the $n$GHZ state:
\begin{equation}
|n{\rm GHZ}\rangle\equiv \frac{1}{\sqrt{2}}\big(\ket{{\rm S}(n,0)}+\ket{{\rm
S}(n,n)}\big).
\end{equation}
It is easy to show that (for all $n$) $\Lambda_{\max}(n{\rm GHZ})=1/\sqrt{2}$
and, equivalently, $E_{\sin^2}=1/2$. Note that one could have rescale the
definion of the GME by a factor of two such that Bell and N-GHZ states have
entanglement of unity. However, we do not make this rescaling.

We now focus our attention on three-qubit settings. Of these, the states
$\ket{S(3,0)}=\ket{000}$ and $\ket{S(3,3)}=\ket{111}$ are not entangled and
are, respectively, the components of the 3-GHZ state: $\ket{{\rm GHZ}}\equiv
\big(\ket{000}+\ket{111})/\sqrt{2}$. The states $\ket{{\rm S}(3,2)}$ and
$\ket{{\rm S}(3,1)}$, are the W state and the ``inverted W'' state,
respectively,
\begin{subequations}
\begin{eqnarray}
|{\rm W}\rangle&\equiv&\ket{{\rm S}(3,2)} =\frac{1}{\sqrt{3}}
\big(\ket{001}+\ket{010}+\ket{100}\big),\\
\ket{\widetilde{\rm W}}&\equiv&\ket{{\rm S}(3,1)}=
\frac{1}{\sqrt{3}}\big(\ket{110}+\ket{101}+\ket{011}\big),
\end{eqnarray}
\end{subequations}
are equally entangled, having $\Lambda_{\max}=2/3$ and
$E_{\sin^2}=1-\Lambda_{\max}^2=5/9$.

\begin{figure}[t]
\centerline{\psfig{figure=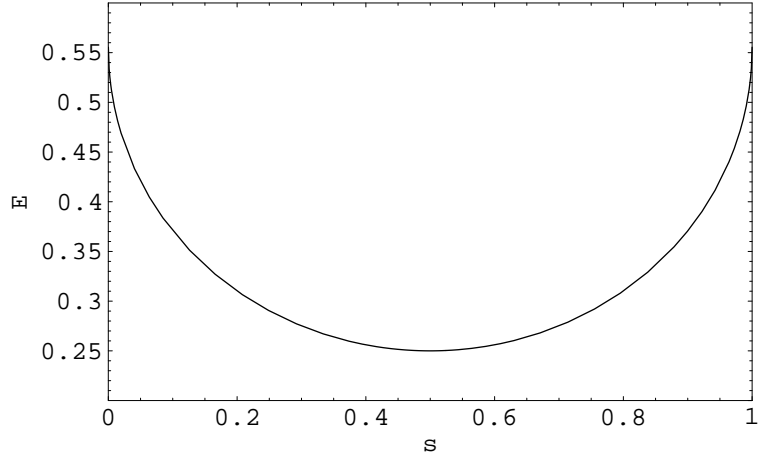,width=10cm,angle=0}}
\caption[Entanglement curve for the state $\sqrt{s}\,|{\rm
W}\rangle+\sqrt{1-s}\,|\widetilde{\rm W}\rangle$]{Entanglement of the pure
state $\sqrt{s}\,|{\rm W}\rangle+\sqrt{1-s}\,|\widetilde{\rm W}\rangle$
vs.~$s$. This also turns out to be the entanglement curve for the mixed state
$s\,\ketbra{\rm W}+(1-s)\ketbra{\widetilde{\rm W}}$.} \label{fig:WW}
\end{figure}
Next, consider a superposition of the ${\rm W}$ and $\widetilde{\rm W}$
states:
\begin{equation}
\ket{\wstate\wtilde(s,\phi)} \equiv \sqrt{s}\, \ket{\wstate}+\sqrt{1-s}\,{\rm
e}^{i\phi}\ket{\wtilde}.
\end{equation}
It is straightforward to see that its entanglement is independent of $\phi$:
the transformation $\big\{\ket{0},\ket{1}\big\}\to \big\{\ket{0},{\rm
e}^{-i\phi}\ket{1}\big\}$ induces $\ket{\wstate\wtilde(s,\phi)} \rightarrow
{\rm e}^{-i\phi}\ket{\wstate\wtilde(s,0)}$, and the entanglement is
independent of the overall phase of a state. To calculate $\Lambda_{\max}$, we
assume that the separable state is of the form
$(\cos\theta\ket{0}+\sin\theta\ket{1})^{\otimes 3}$, and maximize its overlap
with $\ket{\wstate\wtilde(s,0)}$ with respect to $\theta$. Thus we find that
$\theta(s)$ is determined via its tangent $t\equiv\tan\theta$, which is the
particular root of the polynomial equation
\begin{equation}
\sqrt{1-s}\,t^3+2\sqrt{s}\,t^2-2\sqrt{1-s}\,t-\sqrt{s}=0
\end{equation}
that lies in the range $t\in [\sqrt{{1}/{2}},\sqrt{2}]$. Via $\theta(s)$,
$\Lambda_{\max}$ (and thus $E_{\sin^2}=1-\Lambda_{\max}^2$) can be expressed
as
\begin{eqnarray}
\Lambda_{\max}(s) =\frac{\sqrt{3}}{2}\big[\sqrt{s}\cos\theta(s)\!
+\sqrt{1\!-\!s}\sin\theta(s)\big] \sin2\theta(s).
\end{eqnarray}
In Fig.~\ref{fig:WW}, we show
$E_{\sin^{2}}\big(\ket{\wstate\wtilde(s,\phi)}\big)$ vs.~$s$.

In fact,  for the more general superposition
\begin{equation}
\label{eqn:SSnk} \ket{{\rm SS}_{n;k_1k_2}(r,\phi)}\equiv \sqrt{r}\,
\ket{{\rm S}(n,k_1)}+ \sqrt{1\!-\!r}\,{\rm e}^{i\phi}\,\ket{{\rm S}(n,k_2)}
\end{equation}
(with $k_1\ne k_2$) $\Lambda_{\max}$ also turns out to be independent of
$\phi$, and can be computed in the same way as was used for
$\ket{\wstate\wtilde(s,\phi)}$. We note that although the curve in
Fig.~\ref{fig:WW} is convex, convexity does not hold uniformly over $k_1$ and
$k_2$. This adds some complexity to calculation of mixed-state entanglement,
as we shall see later.

As our last pure-state example in the qubit setting, we consider
superpositions of W and GHZ states:
\begin{equation}
\ket{\Psi_{\rm GHZ+W}(s,\phi)} \equiv\sqrt{s}\, \ket{{\rm
GHZ}}+\sqrt{1-s}\,\,{\rm e}^{i\phi}\ket{{\rm W}}.
\end{equation}
For these, the phase $\phi$ cannot be ``gauged'' away and, hence, $E_{\sin^2}$
depends on $\phi$.
\begin{figure}[t]
\centerline{\psfig{figure=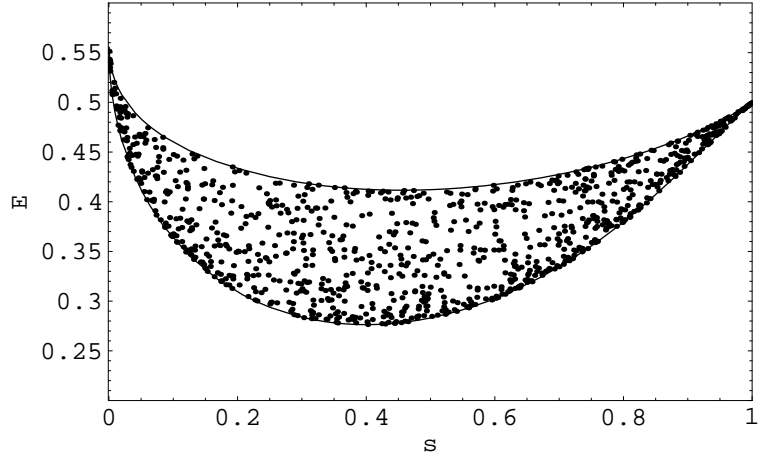,width=10cm,angle=0}}
\caption[Entanglement curve for the state $\ket{\Psi_{\rm
GHZ+W}(s,\phi)}$]{Entanglement of $\ket{\Psi_{\rm GHZ+W}(s,\phi)}$ vs.~$s$.
The upper curve is for $\phi=\pi$ whereas the lower one is for $\phi=0$. Dots
represent states with randomly generated $s$ and $\phi$.} \label{fig:WGHZ}
\end{figure}
In Fig.~\ref{fig:WGHZ} we show $E_{\sin^2}$ vs.~$s$ at $\phi=0$ and $\phi=\pi$
(i.e., the bounding curves), as well as $E_{\sin^2}$ for randomly generated
values of $s\in[0,1]$ and $\phi\in[0,2\pi]$ (dots).  We observe that the
\lq$\pi$\rq\ state has higher entanglement than the \lq$0$\rq\ does.  As the
numerical results suggest, the ($\phi$-parametrized) $E_{\sin^2}$ vs.~$s$
curves of the states $\ket{\Psi_{\rm GHZ+W}(s,\phi)}$ lie between the
\lq$\pi$\rq\ and \lq$0$\rq\ curves.

We remark that, more generally, for systems comprising $n$ parts, each being a
$d$-level system, the symmetric state
\begin{equation}
\ket{S(n;\{k\})}\equiv \sqrt{\frac{\prod_i k_i!}{n!}} \sum_{\rm{\scriptstyle
P}}
|\underbrace{0\ldots0}_{k_0}\underbrace{1\ldots1}_{k_1}\dots\underbrace{(d\!-\!1)
\ldots (d\!-\!1)\,}_{k_{d\!-\!1}}\rangle
\end{equation}
(with $\sum_i k_i=n$) has entanglement eigenvalue
\begin{equation}
\Lambda_{\max}(n;\{k\})=\sqrt{\frac{n!}{\prod_i (k_i!)}}
\,\,\prod_{i=0}^{d-1}\left(\frac{k_i}{n}\right)^{\frac{k_i}{2}}.
\end{equation}

One can also consider other symmetries. Consider the totally {\it
anti}-symmetric (viz.~determinant) state of $n$ parts, each with $n$ levels,
\begin{equation}
\ket{{\rm Det}_n}\equiv \frac{1}{\sqrt{n!}}\sum_{i_1,\dots,i_n=1}^{n}
\epsilon_{i_1,\dots,i_n}\ket{i_1,\dots,i_n}.
\end{equation}
It has been shown by Bravyi~\cite{Bravyi02} that the maximal squared overlap
of this state $\Lambda^2_{\max}$ is $1/n!$. Bravyi has also generalized the
anti-symmetric state to the $n=p\,d^p$-partite determinant state, via the
construction
\begin{eqnarray}
\phi(1)&=&(0,0,\dots,0,0), \nonumber \\
\phi(2)&=&(0,0,\dots,0,1), \nonumber \\
&\vdots& \nonumber \\
\phi(d^p\!-\!1)&=&(d\!-\!1,d\!-\!1,\dots,d\!-\!1,d\!-\!2), \nonumber \\
\phi(d^p)&=&(d\!-\!1,d\!-\!1,\dots,d\!-\!1,d\!-\!1),\nonumber
\end{eqnarray}
and
\begin{equation}
\label{eqn:DetndGME} \ket{{\rm
Det}_{n,d}}\equiv\frac{1}{\sqrt{(d^p!)}}\sum_{i_1,\dots,i_{d^p}}
\epsilon_{i_1,\dots,i_{d^p}}\ket{\phi(i_1),\dots,\phi(i_{d^p})}.
\end{equation}
In this case, we can show that $\Lambda^2_{\max}=[(d^p)!]^{-1}$.
\subsection{Connection with the Hartree approximation}
Recall from Eq.~(\ref{eqn:LambdaMax}) that the maximal overlap or the
entanglement eigenvalue of a pure state $\ket{\psi}$ is defined as
$\Lambda_{\max}(\psi)=\max_{\phi} ||\ipr{\phi}{\psi}||$, where the
maximization is taken over arbitrary product state $\ket{\phi}$. On the other
hand, consider the Hamiltonian constructed from the entangled state
$\ket{\psi}$
\begin{equation}
\label{eqn:Hpsi} {\cal H_\psi}=-\ketbra{\psi},
\end{equation}
which has the form of a projector on to the state $\ket{\psi}$ and which has
the minimum eigen-energy $-1$. The Hartree approximation is to take some
product state $\ket{\psi}$ such that it minimzes the expectation value of the
Hamiltonian
\begin{equation}
\label{eqn:Hartree} \min_{\phi}\bra{\phi}{\cal H}_\psi\ket{\phi}=-\max_{\phi}
|\ipr{\phi}{\psi}|^2=-\Lambda_{\max}^2(\psi).
\end{equation}
It thus arises that the difference between the true ground-state energy of
${\cal H}_\psi$ and the one given by the best Hartree
approximation~(\ref{eqn:Hartree}) is
\begin{equation}
1-\Lambda_{\max}^2(\psi),
\end{equation}
i.e., the GME of $\ket{\psi}$ and that that the best Hartree approximant
$\ket{\phi^*}$ is the closest separable state to $\ket{\psi}$.

This connection, although elementary, can be generalized to the Hartree-Fock
approximation when we wish to extend the idea of entanglement to fermionic
systems.

If we are interested in the entanglement of a nondegenerate ground state (with
energy $E_0$) of a Hamiltonian, in principle, we can express the maximal
overlap as
\begin{equation}
\Lambda^2_{\max}(\ket{E_0})=\lim_{\beta\rightarrow\infty}
\min_\phi\bra{\phi}{e^{-\beta ({\cal H}-E_0)}}\ket{\phi}.
\end{equation}

\subsection{Connection with entanglement witnesses}
We now digress to discuss a relationship between the geometric measure of
entanglement and another entanglement property---entanglement witnesses. The
entanglement witness ${\cal W}$ for an entangled state $\rho$ is defined to be
an operator that is (a)~Hermitian and (b)~obeys the following
conditions~\cite{Horodecki396}:\\
 (i)~${\rm Tr}({\cal W}\sigma)\ge 0$ for all separable states $\sigma$, and\\
(ii)~${\rm Tr}({\cal W}\rho)< 0$.\\
Here, we wish to establish a correspondence between $\Lambda_{\max}$ for the
entangled pure state $\ket{\psi}$ and the optimal element of the set of
entanglement witnesses ${\cal W}$ for $\ket{\psi}$ that have the specific form
\begin{equation}
\label{eqn:classW} {\cal W}=\lambda^{2}\openone-\ketbra{\psi},
\end{equation}
where $\openone$ is the identity operator. This set is parametrized by the
real, non-negative number $\lambda^{2}$. By {\it optimal\/} we mean that, for
this specific form of witnesses, the value of the expression ${\rm
Tr}\big({\cal W}\ketbra{\psi}\big)$ is as negative as can be.

In order to satisfy condition~(i) we must ensure that, for any {\it
separable\/} state $\sigma$, we have ${\rm Tr}\big({\cal W}\sigma\big)\ge 0$.
As the density matrix for any separable state can be decomposed into a mixture
of {\it separable pure\/} states [i.e., $\sigma=\sum_i\ketbra{\phi_i}$ where
$\{\ket{\phi_i}\}$ are separable pure states], condition~(i) will be satisfied
as long as ${\rm Tr}\big({\cal W}\ketbra{\phi}\big)\ge 0$ for all separable
{\it pure} states $\ket{\phi}$. This condition is equivalent to
\begin{equation}
\lambda^{2} -||\ipr{\psi}{\phi}||^2\ge 0\ (\mbox{for all separable}\
\ket{\phi}),
\end{equation}
which leads to
\begin{equation}
\lambda^{2} \ge \max_{\ket{\phi}}||\ipr{\psi}{\phi}||^2=
\Lambda^2_{\max}(\ket{\psi}).
\end{equation}

Condition~(ii) requires that ${\rm Tr}\big({\cal W}\ketbra{\psi}\big)<0$, in
order for ${\cal W}$ to be a valid entanglement witness for $\ket{\psi}$; this
gives $\lambda^{2} - 1<0$. Thus, we have established the range of $\lambda$
for which $\lambda^{2}\openone-\ketbra{\psi}$ is a valid entanglement witness
for $\ket{\psi}$:
\begin{equation}
\Lambda^2_{\max}(\ket{\psi}) \le \lambda^{2} < 1.
\end{equation}

With these preliminaries in place, we can now establish the connection we are
seeking.  Of the specific family~(\ref{eqn:classW}) of entanglement witnesses
for $\ket{\psi}$, the one of the form ${\cal W}_{\rm
opt}=\Lambda^2_{\max}(\ket{\psi})\openone -\ketbra{\psi}$ is optimal, in the
sense that it achieves the most negative value for the detector ${\rm
Tr}\big({\cal W}_{\rm opt}\ketbra{\psi}\big)$:
\begin{equation}
\min_{{\cal W}}{\rm Tr}\big({\cal W}\ketbra{\psi}\big)= {\rm Tr}\big({\cal
W}_{\rm opt}\ketbra{\psi}\big)= -E_{\sin^2}(\ket{\psi}),
\end{equation}
where ${\cal W}$ runs over the class~(\ref{eqn:classW}) of witnesses.

We now look at some examples. For the GHZ state the optimal witness is
\begin{equation}
{\cal W}_{\rm GHZ}=\frac{1}{2}\openone-\ketbra{\rm GHZ}
\end{equation}
and ${\rm Tr}\big({\cal W}_{\rm GHZ}\ketbra{{\rm GHZ}}\big)=
-E_{\sin^2}(\ket{{\rm GHZ}})=-1/2$. Similarly, for the W and inverted-W states
we have
\begin{equation}
{\cal W}_{\rm W}=\frac{4}{9}\openone-\ketbra{\rm W} \ \ \mbox{and} \ \ {\cal
W}_{\rm \widetilde{\rm W}}=\frac{4}{9}\openone-\ketbra{\widetilde{\rm W}}
\end{equation}
and ${\rm Tr}\big({\cal W}_{\rm W}\ketbra{{\rm W}}\big)= -E_{\sin^2}(\ket{{\rm
W}})=-5/9$, and similarly for $\ket{\widetilde{\rm W}}$. For the four-qubit
state
\begin{equation}
\ket{\Psi}\equiv (\ket{0011}+\ket{0101}+\ket{0110}+
 \ket{1001}+\ket{1010}+\ket{1100})/\sqrt{6}
\end{equation}
the optimal witness is
\begin{equation}
{\cal W}_{\Psi}=\frac{3}{8}\openone-\ketbra{\Psi}
\end{equation}
and ${\rm Tr}\big({\cal W}_{\Psi}\ketbra{\Psi}\big)=
-E_{\sin^2}(\ket{\Psi})=-5/8$.

Although the observations we have made in this section are, from a technical
standpoint, elementary, we nevertheless find it intriguing that two distinct
aspects of entanglement---the geometric measure of entanglement and
entanglement witnesses---are so closely related.  Furthermore, this connection
sheds new light on the content of the geometric measure of entanglement.  In
particular, as entanglement witnesses are Hermitian operators, they can, at
least in principle, be realized and measured locally~\cite{Guhne}. Their
connection with the geometric measure of entanglement ensures that the
geometric measure of entanglement can, at least in principle, be verified
experimentally.

\subsection{Connection with correlation functions}
\label{sec:correlations} In this section we explore the connection of the
entanglement eigenvalue to the correlation functions. To illustrate the
connection, we shall mainly focus on $N$-qubit systems. Recall that a single
pure qubit state, when written in the form of  a density matrix, can be
expressed in terms of Pauli matrices plus the identity:
\begin{equation}
\frac{1}{2}(\openone + \vec{r}\cdot \vec{\sigma}),
\end{equation}
with the constraint on the real vector parameter $\vec{r}$ that $|\vec{r}|=1$.
The density matrix for an $N$-qubit separable pure state
$\ket{\phi}$~(\ref{eqn:product}) is hence a direct product of $N$ such terms:
\begin{eqnarray}
\ketbra{\phi}=\frac{1}{2}(\openone + \vec{r}_1\cdot
\vec{\sigma}^{(1)})\otimes\frac{1}{2}(\openone + \vec{r}_2\cdot
\vec{\sigma}^{(2)})\otimes\cdots\otimes\frac{1}{2}(\openone + \vec{r}_N\cdot
\vec{\sigma}^{(N)}),
\end{eqnarray}
and expanding out the products gives the decomposition of $\ketbra{\phi}$
\begin{eqnarray}
\frac{1}{2^N}\left\{\openone+\sum_{j=1}^N \vec{r}_j\cdot
\vec{\sigma}^{(j)}+\sum_{j< k}\vec{r}_j\cdot
\vec{\sigma}^{(j)}\otimes\vec{r}_k\cdot \vec{\sigma}^{(k)}+\sum_{j< k<
l}\vec{r}_j\cdot \vec{\sigma}^{(j)}\otimes\vec{r}_k\cdot
\vec{\sigma}^{(k)}\otimes\vec{r}_l\cdot \vec{\sigma}^{(l)} +\cdots\right\}\!.
\end{eqnarray}
The overlap squared of any $N$-qubit state $\ket{\psi}$ with the separable
state $\ket{\phi}$ is
\begin{equation}
\Lambda^2=\bra{\psi}\big(\ketbra{\phi}\big)\ket{\psi},
\end{equation}
which can in turn be expressed in terms of the correlations of the operators
of the form $\sigma\otimes\cdots\otimes\sigma\otimes\cdots$. Therefore, the
maximal overlap is a maximization over all combinations of $M$-point (with
$M\le N$) correlation functions:
\begin{eqnarray}
\Lambda_{\max}^2&=&\max_{|\vec{r}|=1}\frac{1}{2^N}\left\{1+\sum_{j=1}^N \langle\vec{r}_j\cdot \vec{\sigma}^{(j)}\rangle+\sum_{j< k}\langle\vec{r}_j\cdot \vec{\sigma}^{(j)}\otimes\vec{r}_k\cdot \vec{\sigma}^{(k)}\rangle\right.\nonumber\\
&&\left. +\sum_{j< k< l}\langle\vec{r}_j\cdot
\vec{\sigma}^{(j)}\otimes\vec{r}_k\cdot
\vec{\sigma}^{(k)}\otimes\vec{r}_l\cdot \vec{\sigma}^{(l)}\rangle +\cdots+
\langle\cdots\rangle+\cdots\right\}.
\end{eqnarray}
The average is taken with respect to the $N$-qubit pure state $\ket{\psi}$
whose entanglement we are to quantify. This result echos the fact mentioned in
the previous section that the geometric measure (or here the maximal overlap)
can be measured, via the correlations, which are physical quantities.
\section{Extension to mixed states}
\label{sec:Mixed} The extension of the geometric measure to mixed states
$\rho$ can be made via the use of the {\it convex roof\/} (or {\it hull\/})
construction [indicated by ``co''], as was done for the entanglement of
formation [c.f. Eq.~(\ref{eqn:Ef})]~\cite{Wootters98}.  The essence is a
minimization over all decompositions $\rho=\sum_i
p_i\,|\psi_i\rangle\langle\psi_i|$ into pure states, i.e.,
\begin{eqnarray}
\label{eqn:Emixed} E(\rho) \equiv \coe{\rm pure}(\rho) \equiv
{\min_{\{p_i,\psi_i\}}} \sum\nolimits_i p_i \, E_{\rm pure}(|\psi_i\rangle).
\end{eqnarray}
This convex hull construction ensures that the measure gives zero for
separable states; however, in general it also complicates the task of
determining mixed-state entanglement.

Now, any good entanglement measure $E$ should, at least, satisfy the following
criteria (c.f.~Refs.~\cite{VedralPlenio98,Horodecki300,Vidal00})
\begin{itemize}
\item[C1.](a)~$E(\rho)\!\ge\! 0$;
(b)~$E(\rho)\!=\!0$ if $\rho$ is not entangled.
\item[C2.]Local unitary transformations do not change $E$.
\item[C3.]Local operations and classical communication (LOCC)
(as well as post-selection) do not increase the expectation value of
$E$~\footnote{This requirement does not contradict with distillation, as it
takes into account the cases when distillation fails.}.
\item[C4.]Entanglement is convex under the discarding of
information, i.e., $\sum_i p_i\,E(\rho_i)\ge E(\sum_i p_i\,\rho_i)$.
\end{itemize}
The first requirement simply states that the entanglement is a non-negative
quantity, and it vanishes for unentangled states. Local unitary
transformations are simply a change of local basis, and entanglement should be
invariant under such a change. The third criterion simply requires that the
average entanglement cannot be increased during manipulations that are local,
which reflects the fact that entanglement is a nonlocal resource. The last
criterion states the fact that, for a set of states, entanglement can never be
increased if the information about which state is which is lost. The issue of
the desirability of additional features, such as continuity and additivity,
requires further investigation, but C1-C4 are regarded as the minimal set. If
a measure satisfies C1-C4, it is called an {\it entanglement
monotone\/}~\cite{Vidal00}.

Does the geometric measure of entanglement obey C1-4? The answer is
affirmative but we shall relegate the proof to Appendix~\ref{app:Monotone}. We
remark that from the definition~(\ref{eqn:Emixed}) it is evident that C1 and
C2 are satisfied, provided that $E_{\rm pure}$ satisfies them. Furthermore,
the convex-hull construction automatically fulfills C4. The consideration of
C3 seems to be more delicate and whether or not it holds depends on the
explicit form of $E_{\rm pure}$. In Appendix~\ref{app:Monotone} we show that
the choice of taking $E_{\sin^2}$ as the entanglement measure {\it does\/}
satisfy C3.

\def\ProofMonotone{
as it does for $E_{\rm pure}$ being any function of $\Lambda_{\max}$
consistent with C1.  It is straightforward to check that C4 holds, by the
convex hull construction. First, consider the case is which $\rho=\sum_i
p_i|\psi_i\rangle\langle\psi_i|$. From the definition~(\ref{eqn:Emixed}) of
$E(\rho)$, which is the {\it minimum\/} over all decompositions, we have that
$E(\rho)\le\sum_i p_iE_{\rm pure}(|\psi_i\rangle)$. Hence we have that
$E(\sum_i p_i|\psi_i\rangle\langle\psi_i|)\le \sum_i  p_i
E(|\psi_i\rangle\langle\psi_i|)$, i.e., C4 is obeyed whenever the deomposition
is into {\it pure\/} states. Second, allow $\rho_i$ to be mixed.  To deal with
this case, express $\rho_i$ as its optimal decomposition: $\rho_i=\sum_k
q_{ik}|\psi_{ik}\rangle\langle\psi_{ik}|$, for which $E(\rho_i)=\sum_k
q_{ik}E_{\rm pure}(|\psi_{ik}\rangle)$. Inserting the above expression for
$\rho_i$ and $E(\rho_i)$ into the left hand side of the sought criterion, and
using the pure-state result just proved, we find $\sum_i p_i E(\rho_i)=
  \sum_{ik} p_i\,q_{ik} E(|\psi_{ik}\rangle\langle\psi_{ik}|)\ge
E(\sum_{ik} p_i\,q_{ik}|   \psi_{ik}\rangle\langle\psi_{ik}|)= E(\sum_i
p_i\rho_i)$. Thus we see that C4 is indeed obeyed.

 The consideration of C3 seems to be more delicate.
The reason is that our analysis of whether or not it holds depends on the
explicit form of $E_{\rm pure}$. For C3 to hold, it is sufficient to show that
the average entanglement is non-increasing under any trace-preserving,
unilocal operation: $\rho\rightarrow \sum_k V_k\rho V_k^\dagger$, where the
Kraus operator has the form $V_k=\openone\otimes\cdots \openone\otimes
V_k^{(i)}\otimes\openone\cdots\otimes\openone$ and $\sum_k V_k^{\dagger}
V_k=\openone$. Furthermore, it suffices to show that C3 holds for the case of
a pure  initial state, i.e., $\rho=\ketbra{\psi}$.

We now prove that for the particular (and by no means un-natural) choice
$E_{\rm pure}=E_{\sin^2}$, C3 holds. To be precise, for any quantum operation
on a pure initial state, i.e.,
\begin{equation}
|\psi\rangle\langle\psi| \rightarrow\sum\nolimits_k
V_k|\psi\rangle\langle\psi|V_k^\dagger,
\end{equation}
we aim to show that
\begin{equation}
\sum_k p_k \,E_{\sin^{2}} \left({V_k|\psi\rangle}/\!{\sqrt{p_k}}\right) \le
E_{\sin^{2}}(|\psi\rangle),
\end{equation}
where $p_k\!\equiv\! {\rm Tr}\,V_k|\psi\rangle\langle\psi|V_k^\dagger
=\langle\psi|V_k^\dagger V_k|\psi\rangle$, regardless of whether the operation
$\{V_k\}$ is state-to-state or state-to-ensemble. Let us respectively denote
by $\Lambda$ and $\Lambda_k$ the entanglement eigenvalues corresponding to
$\ket{\psi}$ and the (normalized) pure state
${V_k|\psi\rangle}/{\sqrt{p_k}}\,$. Then our task is to show that $\sum_k
p_k\,\Lambda_k^2 \ge \Lambda^2$, of which the left hand side is, by the
definition of $\Lambda_k$, equivalent to
\begin{equation}
\sum_k p_k \max_{\scriptscriptstyle\xi_k\in D_s} \Vert {\langle\xi_k|
V_k|\psi\rangle}/\!{\sqrt{p_k}} \Vert^2=\sum_k
\max_{\scriptscriptstyle\xi_k\in D_s}\Vert \langle\xi_k| V_k|
\psi\rangle\Vert^2.
\end{equation}
Without loss of generality, we may assume that it is the first party who
performs the operation. Recall that the condition~(\ref{eqn:EigenForm}) for
the closest separable state
\begin{equation}
\ket{\phi}\equiv|\tilde{\alpha}\rangle_1\otimes|\tilde{\gamma}
\rangle_{2\cdots n}
\end{equation}
can be recast as
\begin{equation}
{}_{2\cdots n}\langle\tilde{\gamma}|\psi\rangle_{1\cdots n}=
\Lambda|\tilde{\alpha}\rangle_1.
\end{equation}
Then, by making the specific choice
\begin{equation}
\langle\xi_k|= ({\langle\tilde{\alpha}|V_k^{(1)\dagger}/\!{\sqrt{q_k}}})
\otimes\langle\tilde{\gamma}|,
\end{equation}
where $q_k \equiv \langle\tilde{\alpha}| V_k^{(1)\dagger}V_k^{(1)}
|\tilde{\alpha}\rangle$, we have the sought result
\begin{eqnarray}
\sum_k p_k\Lambda_k^2= \sum_k \max_{\xi_k\in D_s} \Vert\langle\xi_k|
V_k|\psi\rangle\Vert^2 \ge\Lambda^2\sum_k ({\langle\tilde{\alpha}|
V_k^{(1)\dagger}V_k^{(1)}| \tilde{\alpha}\rangle/\!\sqrt{q_k}} )^2=\Lambda^2.
\end{eqnarray}
  Hence, the form $1-\Lambda^2$, when generalized to
mixed states, is an entanglement monotone. We note that a different approach
to establishing this result has been used by Barnum and
Linden~\cite{BarnumLinden01}. Moreover, using the result that $\sum_k
p_k\Lambda_k^2\ge \Lambda^2$, one can further show that for any convex
increasing function $f_c(x)$ with $x\in[0,1]$,
\begin{equation}
\sum_k p_k\,f_c(\Lambda_k^{2})\ge f_c(\Lambda^{2}).
\end{equation}
Therefore, the quantity ${const.}-f_c(\Lambda^{2})$ (where the ${const.}$ is
to ensure the whole expression is non-negative), when extended to mixed
states, is also an entanglement monotone, hence a good entanglement measure.
For the following discussion we simply take $E=1-\Lambda^2$. }
\section{Analytic results for mixed states}
\label{sec:Analytic} Before moving on to the {\it terra incognita\/} of mixed
{\it multi-partite\/} entanglement, we test the geometric approach in the
setting of mixed {\it bi-partite\/} states, by computing $E_{\sin^2}$ for
three classes of states for which $E_{\rm F}$ is known.
\subsection{Arbitrary two-qubit mixed states}
For arbitrary two-qubit states we can show that
\begin{eqnarray}
\label{eqn:EC} E_{\sin^2}(\rho)= \frac{1}{2}\big(1-\sqrt{1-C(\rho)^2}\,\big),
\end{eqnarray}
where $C(\rho)$ is the Wootters concurrence of the state $\rho$. The essential
point is that, in his derivation of the formula for $E_{\rm F}$, Wootters
showed that there exists an optimal decomposition
$\rho=\sum_{i}p_i\,|\psi_i\rangle\langle\psi_i|$ in which every
$|\psi_i\rangle$ has the concurrence of $\rho$ itself. (More explicitly, every
$|\psi_i\rangle$ has the identical concurrence, that concurrence being the
infimum over all decompositions.) By using Eq.~(\ref{eqn:LamConc}) we see that
Eq.~({\ref{eqn:EC})holds for any two-qubit {\it pure\/} state. As $E_{\sin^2}$
is a monotonically increasing convex function of $C\in[0,1]$, the optimal
decomposition for $E_{\sin^2}$ is identical to that for the entanglement of
formation $E_{\rm F}$.  Thus, we see that Eq.~(\ref{eqn:EC}) holds for {\it
any two-qubit mixed state}.

The fact that $E_{\sin^2}$ is  related to $E_{\rm F}$ via the concurrence $C$
is inevitable for two-qubit systems, as both are fully determined by the one
independent Schmidt coefficient. We note that Vidal~\cite{Vidalb00} has
derived this expression when he considered the probability of success for
converting a single copy of some pure state $\ket{\psi}$ into the desired
mixed state $\rho$. The probability of the conversion is
\begin{equation}
P(\psi\rightarrow
\rho)=\min\left\{1,\frac{E_{\sin^2}(\psi)}{E_{\sin^2}(\rho)}\right\},
\end{equation}
 which gives a physical
interpretation of the geometric measure of entanglement. Unfortunately, this
connection only holds for two-qubit states.

\subsection{Generalized Werner states}
Any state $\rho_{\rm W}$ of a $C^d\otimes C^d$ system is called a generalized
Werner state if it is invariant under
\begin{equation}
\label{eqn:P1} {\rm\bf P}_{1}:\rho\rightarrow \int dU(U\otimes U)\rho\,
(U^\dagger\otimes U^\dagger),
\end{equation}
where $U$ is any element of the unitary group ${\cal U}(d)$ and $dU$ is the
corresponding normalized Haar measure. Both parties perform the same unitary
transformation $U$ on their own system but they  choose $U$ randomly according
to the measure $dU$.

 Generalized Werner states
can be expressed as a linear combination of two operators: the {\it
identity\/} $\hat{\openone}$, and the {\it swap\/} $\hat{\rm
F}\equiv\sum_{ij}|ij\rangle\langle ji|$, i.e., $\rho_{\rm W}\equiv
a\hat{\openone}+b\hat{\rm F}$, where $a$ and $b$ are real parameters related
via the constraint ${\rm Tr}\rho_{\rm W}=1$. This one-parameter family of
states can be neatly expressed in terms of the single parameter $f\equiv{\rm
Tr}(\rho_{\rm W}\hat{\rm F})$:
\begin{equation}
\label{eqn:Wernerf} \rho_{\rm
W}(f)=\frac{d^2-fd}{d^4-d^2}\openone\otimes\openone+
\frac{fd^2-d}{d^4-d^2}\hat{\rm F}.
\end{equation}
By applying to $E_{\sin^2}$ the technique by developed by Vollbrecht and
Werner for $E_{\rm F}(\rho_{\rm W})$ [to be briefly reviewed shortly; or see
Ref.~\cite{VollbrechtWerner01}], one arrives at the geometric entanglement
function for Werner states:
\begin{eqnarray}
 \label{eqn:Werner}
 E_{\sin^2}\big(\rho_{\rm W}(f)\big)=
 \frac{1}{2}\big({1-\sqrt{1-f^2}}\,\big)\quad {\rm for} \ f\le 0,
 \end{eqnarray}
 and zero otherwise.

In order to prepare for this, we now briefly review the technique developed by
Vollbrecht and Werner~\cite{VollbrechtWerner01} for computing the entanglement
of formationfor $\rho_{\rm W}$; this turns out to be applicable to the
computation of the sought quantity $E_{\sin^2}$.  We start by fixing some
notation.
Let \\
$K$ be a compact convex set (e.g., a set of states that includes both pure and
mixed ones);
\\
$M$ be a convex subset of $K$ (e.g., set of pure states);
\\
$E:M\rightarrow R\cup\{+\infty\}$ be function that maps elements of $M$ to the
real numbers (e.g., $E=E_{\sin^2}$); and
\\
$G$ be a compact group of symmetries, acting on $K$ (e.g., the group $U\times
U^\dagger$) as $\alpha_g:K\rightarrow K$ (where $\alpha_g$ is the
representation of the element $g\in G$) that preserve convex combinations.
\\
We assume that $\alpha_g M\subset M$ (e.g., pures states are mapped into pure
states), and that $E(\alpha_g m)=E(m)$ for all $m\in M$ and $g\in G$ (e.g.,
that the entanglement of a pure state is preserved under $\alpha_g$). We
denote by ${\rm\bf P}$ the invariant projection operator defined via
\begin{equation}
{\rm\bf P}k= \int dg\alpha_g(k),
\end{equation}
where $k\in K$. One example of ${\rm\bf P}_1$ is the operation in
Eq.~(\ref{eqn:P1}). Vollbrecht and Werner also defined the following
real-valued function $\epsilon$ on the invariant subset ${\rm\bf P}K$:
\begin{equation}
\epsilon(x)= {\rm inf} \left\{E(m)\vert m\in M, {\rm \bf P}m=x\right\}.
\end{equation}
They then showed that, for $x\in{\rm\bf P}K$,
\begin{equation}
{\rm co}\,E(x)= {\rm co}\,\epsilon(x),
\end{equation}
and provided the following recipe for computing the function ${\rm co}\,E$ for
$G$-invariant states:
\\
1.~For every invarinat state $\rho$ (i.e., obeying $\rho={\rm\bf P}\rho$),
find the set $M_\rho$ of pure states $\sigma$ such that ${\rm \bf
P}\sigma=\rho$.
\\
2.~Compute $\epsilon(\rho)\equiv{\rm inf} \left\{E(\sigma)\vert \sigma\in
M_\rho\right\}$.
\\
3.~Then ${\rm co}\,E$ is the convex hull of this function $\epsilon$.

Having reviewed the Vollbrecht-Werner technique, we now apply it to the
geometric measure of entanglement $E_{\sin^{2}}$ by applying their recipe.

The essential points of the derivation are as follows:\\
(i)~In order to find the set $M_{\rho_{\rm W}}$  it is sufficient, due to the
invariance of $\rho_{\rm W}$ under ${\rm\bf P_1}$, to consider any pure state
$|\Phi\rangle=\sum_{jk}\Phi_{jk} |e^{(1)}_j\rangle\otimes|e^{(2)}_k\rangle$
that has a diagonal reduced density matrix ${\rm Tr}_2\ketbra{\Phi}$ and the
value ${\rm Tr}(|\Phi\rangle\langle\Phi|\hat{F})$ equal to the parameter $f$
associated with the Werner state $\rho_{\rm W}(f)$.  It can be shown that
\begin{equation}
E_{\sin^2}(|\Phi\rangle\langle\Phi|) \ge\frac{1}{2}
\left({1-\sqrt{1-(f-\sum\nolimits_i\lambda_{ii})^2}}\,\right),
\end{equation}
where $\lambda_{ii}\equiv|\Phi_{ii}|^2$.
\\
(ii)~If $f>0$, we can set the only nonzero elements of $|\Phi\rangle$ to be
$\Phi_{i1}$, $\Phi_{i2},\ldots$, $\Phi_{ii},\ldots$, $\Phi_{id}$ such that
$|\Phi_{ii}|^2=f$, this state obviously being separable. Hence, for $f>0$ we
have $E_{\sin^2}(\rho_{\rm W}(f))=0$. On the other hand, if $f<0$ then any
nonzero $\lambda_{ii}$ would increase $(f-\sum_i\lambda_{ii})^2$ and, hence,
increase the value of $E(|\Phi\rangle\langle\Phi|)$, not conforming with the
convex hull. Thus, for a fixed value of $f$, the lowest possible value of the
entanglement $E(|\Phi\rangle\langle\Phi|)$ that can be achieved occurs when
$\lambda_{ii}=0$ and there are only two nonzero elements $\Phi_{ij}$ and
$\Phi_{ji}$ ($i\ne j$). This leads to
\begin{equation}
\min_{\ket{\Phi} {\rm \ at \ fixed\/}\ f} E(|\Phi\rangle\langle\Phi|)=
\frac{1}{2}\left({1-\sqrt{1-f^2}}\right).
\end{equation}
Thus, as a function of $f$, $\epsilon(f)$ is given by
\begin{equation}
\epsilon(f)=
\begin{cases}
\frac{1}{2}\left({1-\sqrt{1-f^2}}\right)&{\rm for} \ f\le 0,\cr 0&{\rm for} \
f\ge 0,
\end{cases}
\end{equation}
which, being convex for $f\in[-1,1]$, gives the entanglement
function~(\ref{eqn:Werner}) for Werner states.

\subsection{Isotropic states}
Instead of both performing $U$ on their system, one of the two parties
performs $U$ whereas the other performs $U^*$. The isotropic states are are
invariant under
\begin{equation}
{\rm \bf P}_{2}:\rho\rightarrow \int dU\,(U\otimes U^*)\rho\,
(U^\dagger\otimes {U^*}^\dagger),
\end{equation} and can be expressed as
\begin{eqnarray}
\rho_{\rm iso}(F) \equiv \frac{1-F}{d^2-1}
\left(\hat{\openone}-|\Phi^+\rangle\langle\Phi^+|\right)+
F|\Phi^+\rangle\langle\Phi^+|,
\end{eqnarray}
where $|\Phi^+\rangle\equiv\frac{1}{\sqrt{d}}\sum_{i=1}^{d}|ii\rangle$ and
$F\in[0,1]$. For $F\in[0,1/d]$ this state is known to be
separable~\cite{Horodecki299}. By following arguments similar to those applied
by Terhal and Vollbrecht~\cite{TerhalVollbrecht00} for $E_{\rm F}(\rho_{\rm
iso})$ one arrives at
\begin{eqnarray}
\label{eqn:Eiso}
  E_{\sin^2}\left(\rho_{\rm iso}(F)\right)=
1-\frac{1}{d}\big(\sqrt{{F}} +\sqrt{(\!1-\!F)(d\!-\!1)}\,\big)^2,
\end{eqnarray}
for $F\ge 1/d$.
The essential point of the derivation is the following Lemma (c.f.~Ref.~\cite{TerhalVollbrecht00}): \\
{\it Lemma 1\/}. The entanglement $E_{\sin^2}$ for isotropic states in
$C^d\otimes C^d$ for $F\in[1/d,1]$ is given by
\begin{equation}
E_{\sin^2}(\rho_{\rm iso}(F))={\rm co\/}(R(F)),
\end{equation}
where ${\rm co}(R(F))$ is the convex hull of the function $R$ and
\begin{equation}
R(F)=1-\max_{\{\mu_i\}}\big\{\mu_i\,|F=\Big(\sum_{i=1}^d\sqrt{\mu_i}\Big)^2/d
;\ \sum_{i=1}^d \mu_i=1\big\}.
\end{equation}
Straightforward extremization shows that
\begin{equation}
R(F)=1-\left(\sqrt{\frac{F}{d}}+\sqrt{\frac{F+d-1}{d}-F}\right)^2,
\end{equation}
which is convex, and hence ${\rm co}(R(F))=R(F)$. Thus we arrive at the
entanglement result for isotropic states given in Eq.~(\ref{eqn:Eiso}).

\begin{figure}[t]
\centerline{\psfig{figure=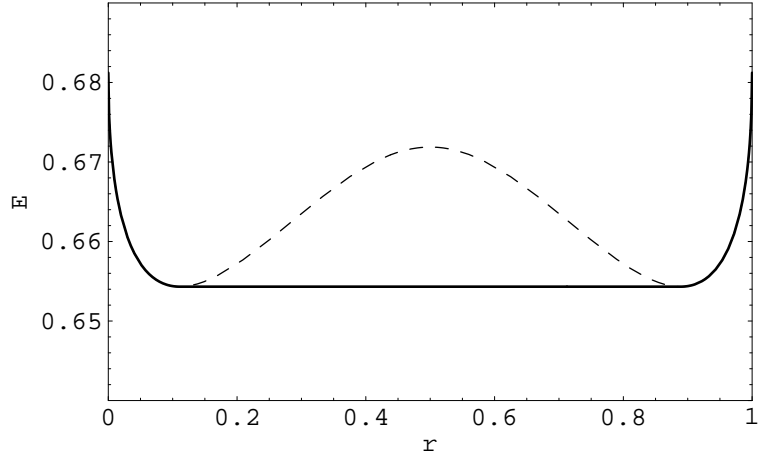,width=10cm,angle=0}}
\caption[Entanglement curve for the mixed state
$\rho_{7;2,5}(r)$]{Entanglement curve for the mixed state $\rho_{7;2,5}(r)$
(full line) constructed as the convex hull of the curve for the pure state
$\ket{{\rm S}{\rm S}_{7;2,5}(r,\phi)}$ (dashed in the middle; full at the
edges).} \label{fig:SS72}
\end{figure}

\subsection{Mixtures of multi-partite symmetric states}
\label{sec:MultiMixed} Before exploring more general mixed states, it is
useful to first examine states with high symmetry. With this in mind, we
consider states formed by mixing two distinct symmetric states (i.e., $k_1\ne
k_2$):
\begin{eqnarray}
\rho_{n;k_1\!k_2}(r)\equiv r\, \ket{{\rm S}(n,k_1)}\bra{{\rm
S}(n,k_1)}+(1-r)\ket{{\rm S}(n,k_2)}\bra{{\rm S}(n,k_2)}.
\end{eqnarray}
From the independence of $E_{\sin^2}\left(\ket{{\rm S}{\rm
S}_{n;k_1k_2}(r,\phi)}\right)$ on $\phi$ and the fact that the mixed state
$\rho_{n;k_1\!k_2}\!(r)$ is invariant under the projection
\begin{equation}
{\rm P_3}:\rho\rightarrow \int\frac{d\phi}{2\pi}\,U^{\otimes n}\rho\,
U^{\dagger\otimes n}
\end{equation}
with $U:\big\{\ket{0},\ket{1}\big\}\to \big\{\ket{0},{\rm
e}^{-i\phi}\ket{1}\big\}$, we have that
$E_{\sin^2}\left(\rho_{n;k_1k_2}(r)\right)$ vs.~$r$ can be constructed from
the convex hull of the entanglement function of $\ket{{\rm S}{\rm
S}_{n;k_1k_2}(r,0)}$ vs.~$r$.  An example, $(n,k_1,k_2)=(7,2,5)$, is shown in
Fig.~\ref{fig:SS72}. If the dependence of $E_{\sin^2}$ on $r$ is already
convex for the pure state, its mixed-state counterpart has precisely the same
dependence. Figure~\ref{fig:WW}, for which $(n,k_1,k_2)=(3,1,2)$, exemplifies
such behavior. More generally, one can consider mixed states of the form
\begin{equation}
\label{eqn:mixSnk} \rho(\{p\})=\sum_k p_k\ketbra{S(n,k)}.
\end{equation}
The entanglement $E_{\rm mixed}(\{p\})$ can then be obtained as a function of
the mixture $\{p\}$ from the convex hull of the entanglement function $E_{\rm
pure}(\{q\})$ for the {\it pure\/} state $\sum_k\sqrt{q_k}\ket{S(n,k)}$. That
is, $E_{\rm mix}(\{p\})={\rm co}\,E_{\rm pure}(\{q\})|_{\{q=p\}}$. Therefore,
the entanglement for a mixture of symmetric states $\ket{S(n,k)}$ is known
from $E_{\rm pure}(\{q\})$, up to some convexification.

\section{Application to arbitrary mixture of GHZ, W and inverted-W states}
\label{sec:GWW} Having warmed up in Sec.~\ref{sec:MultiMixed} by analyzing
mixtures of multi-partite symmetric states, we now turn our attention to
mixtures of three-qubit GHZ, W and inverted-W states.
\subsection{Symmetry and entanglement preliminaries}
These states---GHZ, W and inverted-W states---are important, in the sense that
all pure three-qubit entangled states can, under probabilistic LOCC, be
transformed either to GHZ or W (equivalently inverted-W) states. It is thus
interesting to determine the entanglement content (using any measure of
entanglement) for mixed states of the form:
\begin{equation}
\label{eqn:GWW} \rho(x,y)\equiv x\ketbra{{\rm GHZ}}+y\ketbra{{\rm
W}}+(1-x-y)\ketbra{\widetilde{\rm W}},
\end{equation}
where $x,y\ge 0$ and $x+y\le 1$.  This family of mixed states is not contained
in the family~(\ref{eqn:mixSnk}), as $\ket{\rm
GHZ}=\big(\ket{S(3,0)}+\ket{S(3,3)})/\sqrt{2}$. The property of $\rho(x,y)$
that facilitates the computation of its entanglement is a certain invariance,
which we now describe.  Consider the local unitary transformation on a single
qubit:
\begin{subequations}
\begin{eqnarray}
\ket{0}&\rightarrow& \ket{0}, \\
\ket{1}&\rightarrow& g^k\ket{1},
\end{eqnarray}
\end{subequations}
where $g=\exp{(2\pi i/3)}$, i.e., a relative phase shift. This transformation,
when applied simultaneously to all three qubits, is denoted by $U_k$.  It is
straightforward to see that $\rho(x,y)$ is invariant under the mapping
\begin{equation}
{\rm\bf P}_4:\rho\rightarrow \frac{1}{3}\sum_{k=1}^3 U_k\,\rho\,U_k^\dagger\,.
\end{equation}
Thus, we can apply Vollbrecht-Werner technique~\cite{VollbrechtWerner01} to
the compution of the entanglement of $\rho(x,y)$.

Now, the Vollbrecht-Werner procedure requires one to characterize the set
$S_{\rm inv}$ of all pure states that are invariant under the projection
${\rm\bf P}_4$. Then, the convex hull of $E_{\sin^2}(\rho)$ need only be taken
over $S_{\rm inv}$, instead of the set of {\it all\/} pure states. However, as
the state $\rho(x,y)$ is a mixture of three orthogonal pure states (viz.
$\ket{\rm GHZ}$, $\ket{\rm W}$ and $\ket{\widetilde{\rm W}}$) that are
themselves invariant under ${\rm\bf P}_4$, the pure states that can enter any
possible decomposition of $\rho$ must be of the restricted form:
\begin{equation}
\alpha \ket{{\rm GHZ}}+\beta\ket{{\rm W}}+\gamma\ket{\widetilde{\rm W}},
\end{equation}
with $|\alpha|^2+|\beta|^2+|\gamma|^2=1$. Thus, there is no need to
characterize $S_{\rm inv}$, but only to characterize the pure states that,
under ${\rm {\bf P}_4}$, are projected to $\rho(x,y)$. These states are
readily seen to be of the form:
\begin{equation}
\sqrt{x}\,e^{i\phi_1}\ket{{\rm GHZ}}+ \sqrt{y}\,e^{i\phi_2}\ket{{\rm W}}+
\sqrt{1-x-y}\,e^{i\phi_3}\ket{\widetilde{\rm W}}.
\end{equation}
Of these, the least entangled state, for given $(x,y)$, has all coefficients
non-negative (up to a global phase), i.e.,
\begin{equation}
\ket{\psi(x,y)}\equiv \sqrt{x}\ket{{\rm GHZ}}+ \sqrt{y}\ket{{\rm W}}+
\sqrt{1-x-y}\ket{\widetilde{\rm W}}.
\end{equation}
The entanglement eigenvalue of $\ket{\psi(x,y)}$ can then be readily
calculated, and one obtains
\begin{equation}
\Lambda(x,y)\!=\!\frac{1}{(1 \!+\! t^2)^{\frac{3}{2}}}
\left\{\sqrt{\frac{x}{2}}(1\! +\! t^3) + \sqrt{3y}\,t +
      \sqrt{3(1\!-\!x\!-\!y)}\,t^2\right\},
\end{equation}
where $t$ is the (unique) non-negative real root of the following third-order
polynomial equation:
\begin{eqnarray}
3\sqrt{\frac{x}{2}}(-t + t^2) + \sqrt{3y}(-2t^2 + 1) +\sqrt{3(1 - x-y)}(-t^3 +
2t) = 0.
\end{eqnarray}
Hence, the entanglement function for $\ket{\psi(x,y)}$, i.e.,
$E_\psi(x,y)\equiv 1-\Lambda(x,y)^2$, is determined (up to the numerically
straightforward task of root-finding).

\subsection{Finding the convex hull}
Recall that our aim is to determine the entanglement of the mixed state
$\rho(x,y)$.  As we already know the entanglement of the corresponding pure
state $\ket{\psi(x,y)}$, we may accomplish our aim via the Vollbrecht-Werner
technique~\cite{VollbrechtWerner01}, which gives the entanglement of
$\rho(x,y)$ in terms of that of $\ket{\psi(x,y)}$ via the convex hull
construction: $E_\rho(x,y)=({\rm co}\,E_\psi)(x,y)$.  Said in words, the
entanglement surface $z=E_{\rho}(x,y)$ is the convex surface constructed from
the surface $z=E_{\psi}(x,y)$.

The idea underlying the use of the convex hull is this.  Due to its linearity
in $x$ and $y$, the state $\rho(x,y)$~(\ref{eqn:GWW}) can [except when $(x,y)$
lies on the boundary] be decomposed into two parts:
\begin{equation}
\rho(x,y)=p \,\rho(x_1,y_1) +(1-p)\rho(x_2,y_2),
\end{equation}
with the (real, non-negative) weight $p$ and end-points $(x_1,y_1)$ and
$(x_2,y_2)$ related by
\begin{subequations}
\begin{eqnarray}
p\, x_1+(1-p)x_2=x, \\
p\, y_1+(1-p)y_2=y.
\end{eqnarray}
\end{subequations}
Now, if it should happen that
\begin{equation}
p   E_{\psi}(x_1,y_1)+ (1-p)E_{\psi}(x_2,y_2)<
     E_{\psi}(x,y)
\end{equation}
then the entanglement, averaged over the end-points, would give a value lower
than that at the interior point $(x,y)$; this conforms with the convex-hull
construction.

It should be pointed out that the convex hull should be taken with respect to
parameters on which the density matrix depends {\it linearly\/}, such as $x$
and $y$ in the example of $\rho(x,y)$. Furthermore, in order to obtain the
convex hull of a function, one needs to know the {\it global\/} structure of
the function---in the present case $E_{\psi}(x,y)$.  We note that efficient
numerical algorithms have been developed for constructing convex
hulls~\cite{QHull}.
\begin{figure}[t]
\centerline{\psfig{figure=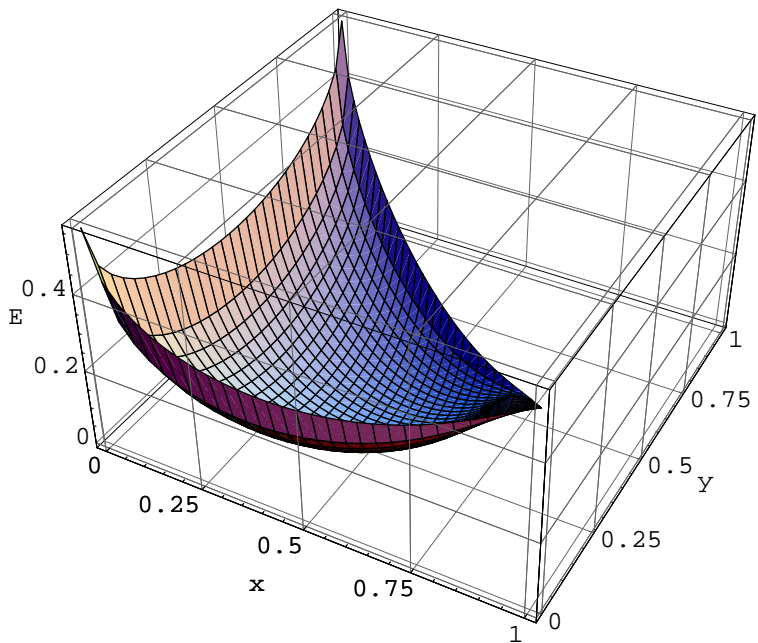,width=10cm,angle=0}}
\caption[Entanglement surface for the state $\ket{\psi(x,y)}$]{Entanglement
vs.~the composition of the pure state $\ket{\psi(x,y)}$. Although not obvious
from the plot, this entanglement surface is not convex near $(x,y)=(1,0)$, }
\label{fig:gGWW0}
\end{figure}
\begin{figure}[h!]
\vspace{0.5cm} \centerline{\psfig{figure=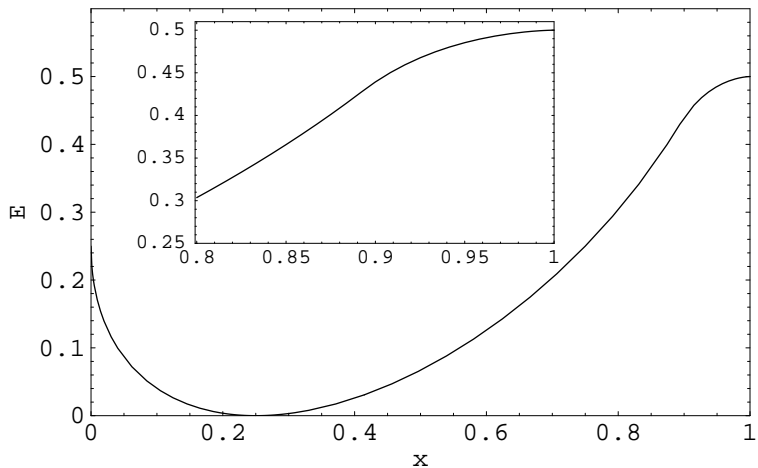,width=10cm,angle=0}}
\caption[Entanglement curve for  the  state $\sqrt{x}\,\ket{{\rm
GHZ}}+\sqrt{(1-x)/2}\,\ket{{\rm W}}+\sqrt{(1-x)/2}\,\ket{\widetilde{\rm
W}}$]{Entanglement of the pure state
$\ket{\psi\big(x,y=(1-x)/2\big)}=\sqrt{x}\,\ket{{\rm
GHZ}}+\sqrt{(1-x)/2}\,\ket{{\rm W}}+\sqrt{(1-x)/2}\,\ket{\widetilde{\rm W}}$
vs.~$x$. This shows the entanglement along the diagonal boundary $x+2y=1$.
Note the absence of convexity near $x=1$; this region is repeated in the
inset.} \label{fig:gGW}
\end{figure}

As we have discussed, our route to establishing the entanglement of
$\rho(x,y)$ involves the analysis of the entanglement of $\ket{\psi(x,y)}$,
which we show in Fig.~\ref{fig:gGWW0}.  Although it is not obvious from the
figure, the corresponding surface fails to be convex near the point
$(x,y)=(1,0)$, and therefore in this region we must suitably convexify in
order to obtain  the entanglement of $\rho(x,y)$.
As previously shown in Fig.~\ref{fig:WW}, where the entanglement of
$\ket{\psi(x,y)}$ is plotted along the line $(x,y)=(0,s)$, the behavior of the
entanglement curve is convex. By contrast, along the line $x+2y=1$ there is a
region in which the entanglement is not convex, as Fig.~\ref{fig:gGW} shows.
The nonconvexity of the entanglement of $\ket{\psi(x,y)}$ complicates the
calculation of the entanglement of $\rho(x,y)$, as it necessitates a procedure
for constructing the convex hull in the (as it happens, small) nonconvex
region. Elsewhere in the $xy$ plane the entanglement of $\rho(x,y)$ is given
directly by the entanglement of $\ket{\psi(x,y)}$.

At worst, convexification would have to be undertaken numerically. However, in
the present setting it turns out that one can determine the convex surface
essentially analytically, by performing the necessary convexifying surgery on
surface $z=E_{\psi}(x,y)$.  To do this, we make use of the fact that if we
parametrize $y$ via $(1-x)r$, i.e., we consider
\begin{eqnarray}
{\rho\big(x,(1-x)r\big)}=x\, \ketbra{{\rm GHZ}}+(1-x)r\,\ketbra{{\rm
W}}+(1-x)(1-r)\ketbra{\widetilde{\rm W}}, \label{eqn:Rhoxr}
\end{eqnarray}
where $0\le r\le 1$ [and similarly for $\ket{\psi(x,y)}$] then, as a function
of $(x,r)$, the entanglement would be symmetric with respect to $r=1/2$, as
Fig.~\ref{fig:gGWWxr0} makes evident. With this parametrization, the nonconvex
region of the entanglement of $\ket{\psi}$ can more clearly be identified.
\begin{figure}[t]
\vspace{0.2cm} \centerline{\psfig{figure=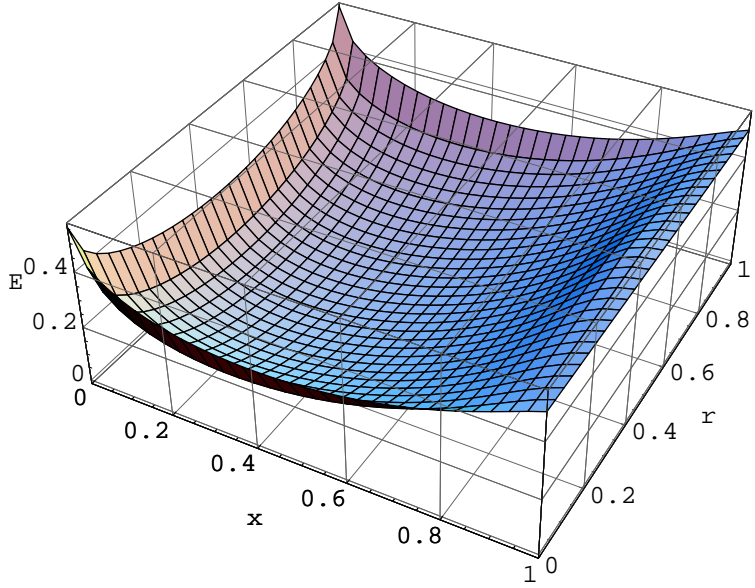,width=10cm,angle=0}}
\caption[Entanglement surface for the state $\sqrt{x}\,\ket{{\rm GHZ}}+
\sqrt{(1-x)r}\,\ket{{\rm W}}+\sqrt{(1-x)(1-r)}\ket{\widetilde{\rm
W}}$]{Entanglement of the pure state $\ket{\psi\big(x,(1-x)r\big)}=\sqrt{x}\,
\ket{{\rm GHZ}}+ \sqrt{(1-x)r}\,\ket{{\rm W}}+
\sqrt{(1-x)(1-r)}\ket{\widetilde{\rm W}}$ vs.~$x$ and $r$. Note the symmetry
of the surface with respect with $r=1/2$.} \label{fig:gGWWxr0}
\end{figure}
\begin{figure}[t]
\centerline{\psfig{figure=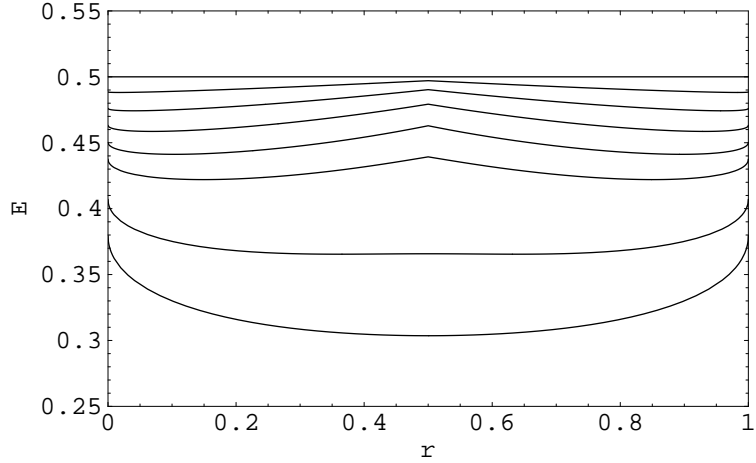,width=10cm,angle=0}}
\caption[Entanglement curves for the state $\sqrt{x}\, \ket{{\rm GHZ}}+
\sqrt{(1-x)r}\,\ket{{\rm W}}+ \sqrt{(1-x)(1-r)}\ket{\widetilde{{\rm W}}}$ for
various values of $x$]{Entanglement of the pure states
$\ket{\psi\big(x,(1-x)r\big)}= \sqrt{x}\, \ket{{\rm GHZ}}+
\sqrt{(1-x)r}\,\ket{{\rm W}}+ \sqrt{(1-x)(1-r)}\ket{\widetilde{{\rm W}}}$
vs.~$r$ for various values of $x$ (from the bottom: 0.8, 0.85, 0.9, 0.92,
0.94, 0.96, 0.98, 1). This reveals the nonconvexity in $r$ for intermediate
values of $x$.} \label{fig:gGWWxr1}
\end{figure}
\begin{figure}[h!]
\centerline{\psfig{figure=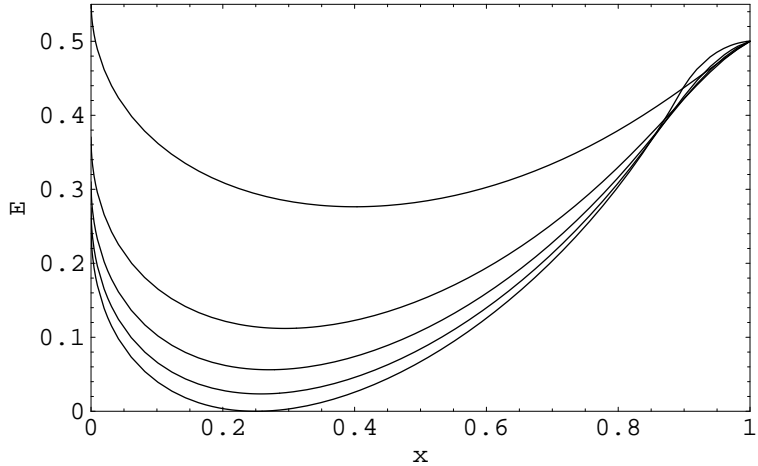,width=10cm,angle=0}}
\caption[Entanglement curves for the state $\sqrt{x}\,\ket{{\rm GHZ}}+
\sqrt{(1-x)r}\,\ket{{\rm W}}+ \sqrt{(1-x)(1-r)}\ket{\widetilde{\rm W}}$ for
various values of $r$]{Entanglement of the pure states
$\ket{\psi\big(x,(1-x)r\big)}= \sqrt{x}\,\ket{{\rm GHZ}}+
\sqrt{(1-x)r}\,\ket{{\rm W}}+ \sqrt{(1-x)(1-r)}\ket{\widetilde{\rm W}}$
vs.~$x$ for various values of $r$ (from the top: 0, 0.1, 0.2, 0.3, 0.5). This
reveals the nonconvexity in $x$ in the (approximate) interval $[0.85,1]$.}
\label{fig:gGWWxr2}
\end{figure}
To convexify this surface we adopt the following convenient strategy. First,
we reparametrize the coordinates, exchanging $y$ by $(1-x)r$. Now, owing to
the linearity, in $r$ at fixed $x$ and {\it vice versa\/}, of the coefficients
$x$, $(1-x)r$ and $(1-x)(1-r)$ in Eq.~(\ref{eqn:Rhoxr}), it is certainly
necessary for the entanglement of $\rho$ to be a convex function of $r$ at
fixed $x$ and {\it vice versa\/}.  Convexity is, however, not necessary in
other directions in the $(x,r)$ plane, owing to the nonlinearity of the
coefficients under simultaneous variations of $x$ and $r$.  Put more simply:
convexity is not necessary throughout the $(x,r)$ plane because straight lines
in the $(x,r)$ plane do not correspond to straight lines in the $(x,y)$ plane
(except along lines parallel either to the $r$ or the $x$ axis). Thus, our
strategy will be to convexify in a restricted sense: first along lines
parallel to the $r$ axis and then along lines parallel to the $x$ axis.
Having done this, we shall check to see that no further convexification is
necessary.

For each $x$, we convexify the curve $z=E_{\psi}\big(x,(1-x)r\big)$ as a
function of $r$, and then generate a new surface by allowing $x$ to vary.
More specifically, the nonconvexity in this direction has the form of a
symmetric pair of minima located on either side of a cusp, as shown in
Fig.~\ref{fig:gGWWxr1}.  Thus, to correct for it, we simply locate the minima
and connect them by a straight horizontal line.

What remains is to consider the issue of convexity along the $x$ (i.e., fixed
$r$) direction for the surface just constructed. In this direction,
nonconvexity occurs when $x$ is, roughly speaking, greater than $0.8$, as
Fig.~\ref{fig:gGWWxr2} suggests.  In contrast with the case of nonconvexity at
fixed $r$, this nonconvexity is due to an inflection point, at which the
second derivative vanishes. To correct for it, we locate the point $x=x_0$
such that the tangent at $x=x_0$ is equal to that of the line between the
point on the curve at $x_0$ and the end-point at $x=1$, and connect them with
a straight line.  This furnishes us with a surface convexified with respect to
$x$ (at fixed $r$) and {\it vice versa\/}.

\begin{figure}[t]
\centerline{\psfig{figure=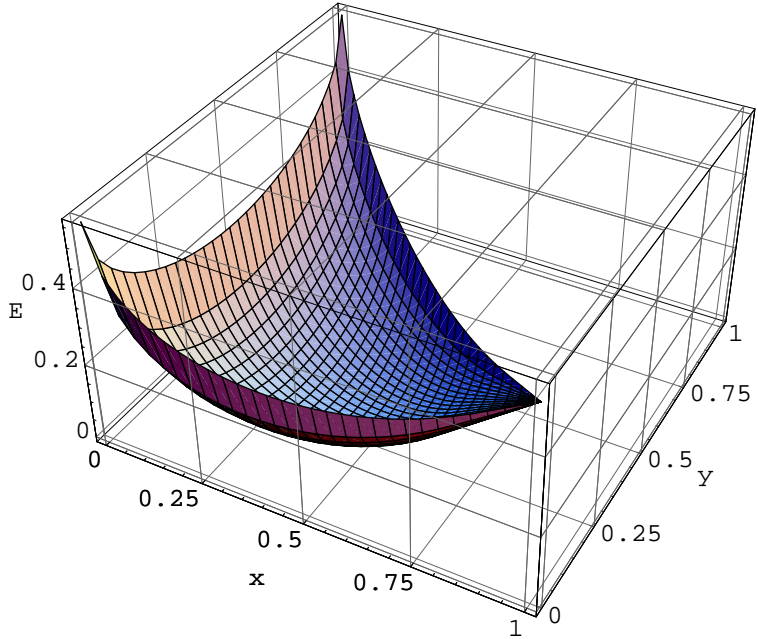,width=10cm,angle=0}}
\caption[Entanglement of the mixed state ${\rho(x,y)}$]{Entanglement of the
mixed state ${\rho(x,y)}$.} \label{fig:gGWW}
\end{figure}
Armed with this surface, we return to the $(x,y)$ parametrization, and ask
whether it is fully convex (i.e., convex along straight lines connecting {\it
any\/} pair of points).  Said equivalently, we ask whether or not any further
convexification is required.  Although we have not proven it, on the basis of
extensive numerical exploration we are confident that the resulting surface
is, indeed, convex.  The resulting convex entanglement surface for $\rho(x,y)$
is shown in Fig.~\ref{fig:gGWW}.  Figure~\ref{fig:gGWW05} exemplifies this
convexity along the line $2y+x=1$.  We have observed that, for the case at
hand, it is adequate to correct for nonconvexity only in the $x$ direction at
fixed $r$.
\begin{figure}[t]
\centerline{\psfig{figure=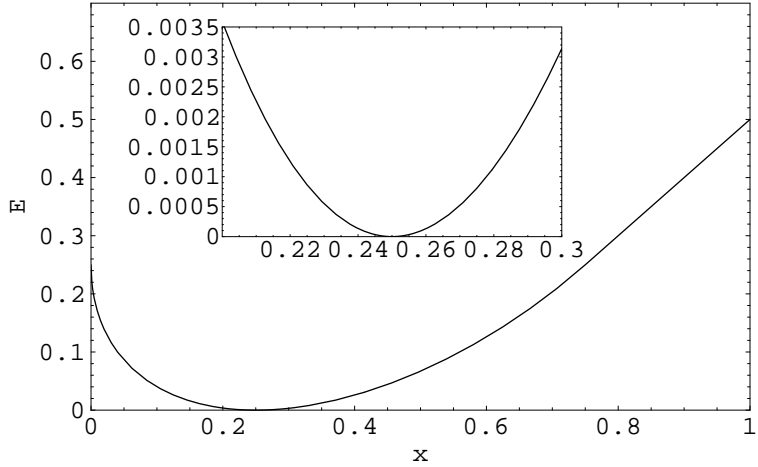,width=10cm,angle=0}}
\caption[Entanglement curve for the state $x\, \ketbra{{\rm GHZ}}+
\frac{1-x}{2}\big(\ketbra{{\rm W}}+ \ketbra{\widetilde{\rm
W}}\big)$]{Entanglement of the mixed state ${\rho\big(x,y=(1-x)/2\big)}=x\,
\ketbra{{\rm GHZ}}+ \frac{1-x}{2}\big(\ketbra{{\rm W}}+ \ketbra{\widetilde{\rm
W}}\big)$ vs.~$x$. Inset: enlargement of the region $x\in[0.2,0.3]$. This
contains the only point, $(x,y)=(1/4,3/8)$, at which $E_{\rho}(x,y)$
vanishes.} \label{fig:gGWW05}
\end{figure}

\subsection{Comparison with the negativity}
As introduced in Sec.~\ref{sec:negativity} the negativity measure of
entanglement is defined to be twice the absolute value of the sum of the
negative eigenvalues of the partial transpose of the density
matrix~\cite{ZyczkowskiHorodeckiSanperaLewenstein98,
    VidalWerner02,WeiNemotoGoldbartKwiatMunroVerstraete03}.
In the present setting, viz., the family $\rho(x,y)$ of three-qubit states,
the partial transpose may equivalently be taken with respect to any one of the
three parties, owing to the invariance of $\rho(x,y)$ under all permutations
of the parties.  Transposing with respect to the third party, one has
\begin{equation}
{\cal N}(\rho)\equiv-2\sum_{\lambda_i<0} \lambda_i,
\end{equation}
where the $\lambda$'s are the eigenvalues of the matrix $\rho^{\rm T_3}$,

\begin{figure}[t]
\centerline{\psfig{figure=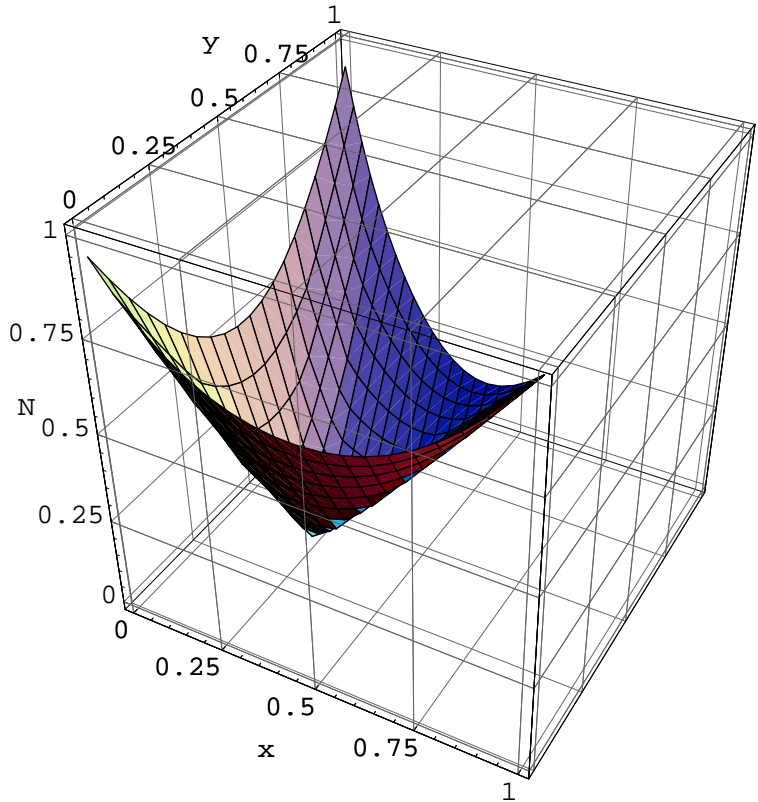,width=10cm,angle=0}}
\caption[Negativity of the mixed state ${\rho(x,y)}$]{Negativity of the mixed
state ${\rho(x,y)}$.} \label{fig:Nxy}
\end{figure}
It is straightforward to calculate the negativity of $\rho(x,y)$; the results
are shown in Fig.~\ref{fig:Nxy}. Interestingly, for all allowed values of
$(x,y)$, the state $\rho(x,y)$ has nonzero negativity, except at
$(x,y)=(1/4,3/8)$, at which the calculation of the GME shows that the density
matrix is indeed separable. One also sees  from that fact that it can be
obtained by applying the projection ${\rm \bf P}_4$ to the (un-normalized)
separable pure state $\big(\ket{0}+\ket{1}\big)^{\otimes 3}$ that
$\rho(1/4,3/8)$ is a separable state. The fact that the only
positve-partial-transpose (PPT) state is separable is the statement that there
are no entangled PPT states (i.e., no PPT bound entangled states) within this
family of three-qubit mixed states.  The negativity surface,
Fig.~\ref{fig:Nxy}, is qualitatively---but not quantitatively---the same as
that of GME. By inspecting the negativity and GME surfaces one can see that
they present ordering difficulties.  This means that one can find pairs of
states $\rho(x_1,y_1)$ and $\rho(x_2,y_2)$ that have respective negativities
${\cal N}_1$ and ${\cal N}_2$ and GMEs $E_1$ and $E_2$ such that, say, ${\cal
N}_1< {\cal N}_2$ but $E_1>E_2$.  Said equivalently, the negativity and the
GME do not necessarily agree on which of a pair of states is the more
entangled.  For two-qubit settings, such ordering difficulties do not show up
for pure states, but can for mixed
states~\cite{WeiNemotoGoldbartKwiatMunroVerstraete03,Ordering}. On the other
hand, for three qubits, such ordering difficulties already show up for pure
states, as the following example shows: ${\cal N}({\rm GHZ})=1>{\cal N}({\rm
W})=2\sqrt{2}/3$ whereas for the GME the order is reversed.  We note, however,
that for the relative entropy of entanglement $E_R$, one has $E_R({\rm
GHZ})=\log_2 2 < E_R({\rm W})=\log_2(9/4)$~\cite{PlenioVedral01}, which for
this particular case is in accord with the GME.

\section{Concluding remarks}
\label{sec:Conclude} We have considered a rather general, geometrically
motivated, measure of entanglement, applicable to pure and mixed quantum
states involving arbitrary numbers and structures of parties. In bi-partite
settings, this approach provides an alternative---and generally
inequivalent---measure to the entanglement of formation. For multi-partite
settings, there is, to date, no explicit generalization of entanglement of
formation. However, if such a generalization should emerge, and if it should
be based on the convex hull construction (as it is in the bi-partite case),
then one may be able to calculate the entanglement of formation for the
families of multi-partite mixed states considered in the present chapter.

As for explicit implementations, the geometric measure of entanglement yields
analytic results in several bi-partite cases for which the entanglement of
formation is already known. These cases include: (i)~arbitrary two-qubit
mixed, (ii)~generalized Werner, and (iii)~isotropic states. Furthermore, we
have obtained the geometric measure of entanglement for certain multi-partite
mixed states, such as mixtures of symmetric states. In addition, by making use
of the geometric measure, we have addressed the entanglement of a rather
general family of three-qubit mixed states analytically (up to root-finding).
This family consists of arbitrary mixtures of GHZ, W, and inverted-W states.
To the best of our knowledge, corresponding results have not, to date, been
obtained for other measures of entanglement, such as entanglement of formation
and relative entropy of entanglement. We have also obtained  corresponding
results for the negativity measure of entanglement. Among other things, we
have found that there are no PPT bound entangled states within this general
family.

A significant issue that we have not discussed is how to use the geometric
measure to provide a classification of entanglement of various multi-partite
entangled states, even in the pure-state setting. For example, given a
tri-partite state, is all the entanglement associated with pairs of parts, or
is some attributable only to the system as a whole? More generally, one can
envisage all possible partitionings of the parties, and for each, compute the
geometric measure of entanglement. This would provide a hierarchical
characterization of the entanglement of states, more refined than the {\it
global\/} characterization discussed here. Another extension would involve
augmenting the set of separable pure states with certain classes of entangled
pure states, such as bi-separable entangled, W-type and GHZ-type
states~\cite{AcinBrussLewensteinSanpera01}.

Although there is no generally valid analytic procedure for computing the
entanglement eigenvalue $\Lambda_{\max}$, one can give---and indeed we have
given---analytical results for several elementary cases.  Harder examples
require computation, but often this is (by today's computational standards)
trivial. We note that in order to find $\Lambda_{\max}$ for the state
$\ket{\psi}$ it is not necessary to solve the nonlinear
eigenproblem~(\ref{eqn:Eigen}); one can instead appropriately parametrize the
family of separable states $\ket{\phi}$ and then directly maximize their
overlap with the entangled state $\ket{\psi}$, i.e.,
$\Lambda_{\max}=\max_\phi||\ipr{\phi}{\psi}||$. Recently, Eisert and
co-workers~\cite{EisertHyllusGuhneCurty04} have reformulated the problem of
finding the entanglement eigenvalue as an efficient convex optimization.
Furthermore, there already exist numerical techniques for determing $E_{\rm
F}$ (see, e.g., Ref.~\cite{Zyczkowski99}), the construction of this measure
being also via the convex-hull construction. We believe that numerical
techniques for solving the geometric measure of entanglement for general
multi-partite mixed states can readily be developed.

The motivation for constructing the measure discussed in the present chapter
is that we wish to address the degree of entanglement from a geometric
viewpoint, regardless of the number of parties. Although the construction is
purely geometric, we have related this measure to entanglement witnesses,
which can in principle be measured locally~\cite{Guhne}. Moreover, the
geometric measure of entanglement is related to the probability of preparing a
single copy of a two-qubit mixed state from a certain pure
state~\cite{Vidalb00}. Yet it is still desirable to see whether, in general,
this measure can be associated with any physical process in quantum
information, as are the entanglement of formation and distillation.

There are further issues that remain to be explored, such as additivity and
ordering. The present form of entanglement for pure states, $E_{\sin^2}\equiv
1-\Lambda^2$, is not additive. However, one can consider a related form,
$E_{\log_2}\equiv -\log_2\Lambda^2$, which, e.g., is additive for
$\ket{\psi}_{AB}\otimes\ket{\psi}_{CD}$, i.e.,
\begin{equation}
E_{\log_2}\big( \ket{\psi_1}_{AB}\otimes\ket{\psi_2}_{CD}\big)=
E_{\log_2}\big(\ket{\psi_1}_{AB}\big)+E_{\log_2}\big(\ket{\psi_2}_{CD}\big).
\end{equation}
This suggests that it is more appropriate to use this logarithmic form of
entanglement to discuss additivity issues. However, $E_{\log_2}$ is not an
entanglement monotone when extended to mixed states by convex hull, as we
shall show later.

As regards the ordering issue, we first mention a result of bi-partite
entanglement measures, due to Virmani and Plenio~\cite{Ordering}, which states
that any two measures with continuity that give the same value as the
entanglement of formation for {\it pure\/} states are either \lq\lq identical
or induce different orderings in general\rlap.\rq\rq\ This result points out
that different entanglement measures will inevitably induce different
orderings if they are inequivalent. This result might still hold for
multi-partite settings, despite their discussion being based on the existence
of entanglement of formation and distillation, which have not been generalized
to multi-partite settings. Although the geometric measure gives the same
ordering as the entanglement of formation for two-qubit mixed states [see
Eq.~(\ref{eqn:EC})], the geometric measure will, in general, give a different
ordering. It is interesting to note that for bi-partite systems, even though
the relative entropy of entanglement coincides with entanglement of formation
for pure states, they can give different orderings for mixed states, as
pointed out by Verstraete and co-workers~\cite{Ordering}.

We conclude by remarking that the measure discussed in the present chapter is
not the same as the Bures measure~\cite{VedralPlenioRippinKnight97} (see also
Sec.~\ref{sec:EntDist}). The Bures measure of entanglement is based on the
minimal distance between the entangled state and the set of separable {\it
mixed\/} states. By contrast, the measure considered here is based upon the
minimal distance between the entangled pure state and the set of separable
pure states, and  is extended to mixed states by a convex hull construction.

\chapter{Connections between relative entropy of entanglement
         and geometric measure of entanglement}
     \label{chap:ReEnt}
\section{Introduction}
\label{sec:IntroReEnt} In Sec.~\ref{sec:IntroGME} we saw some difficulties
with generalization to multi-partite settings of entanglement measures such as
entanglement of formation and of distillation. These difficulties necessitate
the study of other measures, e.g., the relative entropy of entanglement
(REE)~\cite{VedralPlenioRippinKnight97,PlenioVedral01}. One of the reasons
that the REE is interesting in the bi-partite setting is that it provides a
lower bound on the distillable entanglement, the latter being  very difficult
to calculate generally. However, it is still non-trivial to calculate the REE
for generic states, and can be more challenging for multi-partite states. One
reason
 is the absence, in general, of Schmidt
decompositions for multi-partite pure states~\cite{Peres95}. This implies that
for multi-partite pure states the entropies of the reduced density matrices
can differ, in contrast to bi-partite pure states, as the following example
shows.

Consider a three-qubit pure state
$\ket{\psi}_{ABC}\equiv\alpha\ket{001}+\beta\ket{010}+\gamma\ket{100}$, where
$|\alpha|^2+|\beta|^2+|\gamma|^2=1$. The reduced density matrices for parties
A, B, and C are, respectively,
\begin{subequations}
\begin{eqnarray}
\rho_A&=&(|\alpha|^2+|\beta|^2)\ketbra{0}+|\gamma|^2\ketbra{1}, \\
\rho_B&=&(|\alpha|^2+|\gamma|^2)\ketbra{0}+|\beta|^2\ketbra{1},\\
\rho_C&=&(|\gamma|^2+|\beta|^2)\ketbra{0}+|\alpha|^2\ketbra{1},
\end{eqnarray}
\end{subequations}
which, in general, have different entropies.  Thus, for a multi-partite pure
state the entropy of the reduced density matrix does not give a consistent
entanglement measure.  However, even in the case in which all parties have the
identical entropy, e.g., $\alpha=\beta=\gamma=1/\sqrt{3}$~\cite{Note}, it is
in general nontrivial to obtain the relative entropy of entanglement for the
state.  More generally, for pure multi-partite states, it is not yet known how
to obtain their relative entropy of entanglement analytically. The situation
is even worse for {\it mixed\/} multi-partite states.

In Chapter~\ref{chap:GME} we have developed a multi-partite measure based on
a simple geometric picture. For pure states, this geometric measure of
entanglement depends on the maximal overlap between the entangled state and
unentangled states, and is easy to compute numerically. We have examined this
measure for several bi-partite and multi-partite pure and mixed
states~\cite{BarnumLinden01,WeiGoldbart03}. We shall see in the next chapter
the application to two distinct multi-partite bound entangled
states~\cite{WeiAltepeterGoldbartMunro03}. In the present chapter, we explore
connections between this measure and the relative entropy of entanglement. For
certain pure states, some bi-partite and some multi-partite, we find that this
lower bound is saturated, and thus their relative entropy of entanglement can
be found analytically, in terms of their known geometric measure of
entanglement. For certain mixed states, upper bounds on the relative entropy
of entanglement are also established.  Numerical evidence strongly suggests
that these upper bounds are tight, i.e., they are actually the relative
entropy of entanglement. These results, although not general enough to solve
the problem of calculating the relative entropy of entanglement for arbitrary
multi-partite states, may offer some insight into, and serve as a testbed for,
future analytic progress related to the relative entropy of entanglement.

The structure of the present chapter is as follows.  In
Sec.~\ref{sec:measures} we review the two entanglement measures considered in
the paper: the relative entropy of entanglement and the geometric measure of
entanglement.  In Sec.~\ref{sec:connection} we explore connections between the
two, in both pure- and mixed-state settings.  Examples are provided in which
bounds and exact values of the relative entropy of entanglement are obtained.
In Sec.~\ref{sec:summary} we give some concluding remarks. The discussion in
the present chapter is based on Ref.~\cite{WeiEricssonGoldbartMunro}.
\section{Entanglement measures}
\noindent%
\label{sec:measures}%
In this section we briefly review the two measures considered in the present
paper: the relative entropy of entanglement and the geometric measure of
entanglement.
\subsection{Relative entropy of entanglement}
\noindent
The relative entropy $S(\rho||\sigma)$ between two states $\rho$ and $\sigma$
is defined via
\begin{equation}
S(\rho||\sigma)\equiv {\rm Tr}\left(\rho\log_2\rho-\rho\log_2{\sigma}\right),
\end{equation}
which is evidently not symmetric under exchange of $\rho$ and $\sigma$, and is
non-negative, i.e., $S(\rho||\sigma)\ge 0$.  The relative REE for a mixed
state $\rho$ is defined to be the minimal relative entropy of $\rho$ over the
set of separable mixed
states~\cite{VedralPlenioRippinKnight97,VedralPlenio98}:
\begin{equation}
\label{eqn:ER} E_{\rm R}(\rho)\equiv \min_{\sigma\in {\cal
D}}S(\rho||\sigma)=\min_{\sigma\in {\cal D}}{\rm
Tr}\left(\rho\log_2\rho-\rho\log_2\sigma\right),
\end{equation}
where ${\cal D}$ denotes the set of all separable states.

In general, the task of finding the REE for arbitrary states $\rho$ involves a
minimization over all separable states, and this renders its computation very
difficult.  For bi-partite pure states, the REE is equal to entanglements of
formation and of distillation. But, despite recent progress~\cite{Ishizaka03},
for mixed states---even in the simplest setting of two qubits---no analog of
Wootters' formula~\cite{Wootters98} for the entanglement of formation has been
found. Things are even worse in multi-partite settings.  Even for pure states,
there has not been a systematic method for computing REE's. It is thus
worthwhile to seek cases in which one can explicitly obtain an expression for
the REE.  A trivial case arises when there exists a Schmidt decomposition for
a multi-partite pure state: in this case, the REE is the usual expression
\begin{equation}
-\sum_i \alpha_i^2 \log_2 \alpha_i^2\,,
\end{equation}
where the $\alpha_i$'s are Schmidt coefficients (with $\sum_i \alpha_i^2=1$).
We shall see that there exist other cases in which the REE can be determined
analytically, even though no Schmidt decomposition exists.

{We remark that an alternative definition of the REE is to replace the set of
separable states by the set of postive partial transpose (PPT) states. The REE
thus defined, as well as its regularized version, gives a tighter bound on
distillable entanglement. There has been important progress in calculating the
REE [and its regularized version, see Eq.~(\ref{eqn:regularized})] with
respect to PPT states for certain bi-partite mixed states; see
Refs.~\cite{AudenaertEtAl} for more detailed discussions. For multi-partite
settings one could
 define the set of states to optimize over
to be the set of states that are PPT with respect to all bi-partite
partitionings. However, throughout the discussion of the present chapter, we
shall use the first definition, i.e., optimization over the set of completely
separable states.}

\subsection{Geometric measure of entanglement}
\noindent We have introduced the geometric measure of entanglement (GME) in
Sec.~\ref{sec:Pure} in the pure-state settings, and generalized it to
mixed-state settings via the convex-hull construction in Sec.~\ref{sec:Mixed}.
The essential point is to find the maximal overlap (a.k.a. entanglement
eigenvalue) of the entangled state $\ket{\psi}$ with unentangled states:
\begin{equation}
\Lambda_{\max}(\ket{\psi})=\max_{\phi}|\ipr{\phi}{\psi}|,
\end{equation}
where $\ket{\phi}$ is an arbitrary unentangled pure state. The explicit form
of the measure we shall be concerned with in the present chapter is
$E_{\log_2}(\psi)\equiv-2\log_2\Lambda_{\max}(\ket{\psi})$. We shall later
show that it is a lower bound of the REE.

\def\remove1{
We have defined the particular form of the geometric measure $E_{\sin^2}\equiv
1-\Lambda^2_{\max}(\ket{\psi})=\sin^2\theta_{\min}$ for any pure state
$|\psi\rangle$. Here, we shall be concerned with the related quantity
$E_{\log_2}(\psi)\equiv-2\log_2\Lambda_{\max}(\ket{\psi})$, which we shall
show to be a lower bound on the relative entropy of entanglement for
$\ket{\psi}$.  Although this quantity is not, as we shall see later, an
entanglement monotone for mixed states, it is a good measure of {\it
pure-state\/} entanglement.

The extension to mixed states $\rho$ can be built upon pure states via the
{\it convex hull\/} construction (indicated by ``co''), as was done for the
entanglement of formation; see Ref.~\cite{Wootters98}.
 The essence is a minimization
over all decompositions $\rho=\sum_i p_i\,|\psi_i\rangle\langle\psi_i|$ into
pure states:
\begin{eqnarray}
E(\rho) \equiv \coe{\rm pure}(\rho) \equiv {\min_{\{p_i,\psi_i\}}}
\sum\nolimits_i p_i \, E_{\rm pure}(|\psi_i\rangle).
\end{eqnarray}
This convex hull construction ensures that the measure gives zero for
separable states; however, in general it also complicates the task of
determining mixed-state entanglement. }

Some of the examples we considered in previous chapter are relevant to the
discussion in the present chapter. We now briefly recap them here. The first
class contains the permutation-invariant pure states
\begin{equation}
\label{eqn:Snk} |S(n,k)\rangle\equiv \sqrt{\frac{k!(n-k)!}{n!}}
\sum_{\rm{\scriptstyle Permutations}} {\rm
P}|\underbrace{0\cdots0}_{k}\underbrace{1\cdots1}_{n-k}\rangle,
\end{equation}
which has the entanglement eigenvalue
%
\begin{eqnarray}
\label{eqn:Lambdank} \Lambda_{\max}(n,k)= \sqrt{\frac{n!}{k!(n\!-\!k)!}}
\left(\frac{k}{n}\right)^{\frac{k}{2}}
{\left(\frac{n-k}{n}\right)}^{\frac{n\!-\!k}{2}}.
\end{eqnarray}
More generally, for $n$ parties each a $(d+1)$-level system, the state
\begin{equation}
\label{eqn:Snkk} \ket{S(n;\{k\})}\equiv \sqrt{\frac{k_0!k_1!\cdots k_d!}{n!}}
\sum_{\rm{\scriptstyle Permutations}}{\rm P}\,
|\underbrace{0\ldots0}_{k_0}\,\underbrace{1\ldots1}_{k_1}\ldots
 \underbrace{d\ldots d\,}_{k_d}\,\rangle
\end{equation}
has the entanglement eigenvalue
\begin{equation}
\Lambda_{\max}(n;\{k\})=\sqrt{\frac{n!}{\prod_i (k_i!)}}
\,\prod_{i=0}^{d}\left(\frac{k_i}{n}\right)^{\frac{k_i}{2}}.
\end{equation}
The next is the totally antisymmetric state $\ket{{\rm Det}_n}$, defined via
\begin{equation}
\label{eqn:Detn} \ket{{\rm Det}_n}\equiv \frac{1}{\sqrt{n!}}
\sum_{i_1,\dots,i_n=1}^{n} \epsilon_{i_1,\dots,i_n}\ket{i_1,\dots,i_n},
\end{equation}
which has $\Lambda^2_{\max}=1/n!$. The generalization of the antisymmetric
state to the $n=p\,d^p$-partite determinant state is
\begin{equation}
\label{eqn:Detnd} \ket{{\rm
Det}_{n,d}}\equiv\frac{1}{\sqrt{(d^p!)}}\sum_{i_1,\dots,i_{d^p}}
\epsilon_{i_1,\dots,i_{d^p}}\ket{\phi(i_1),\dots,\phi(i_{d^p})},
\end{equation}
with the $\phi$'s defined above Eq.~(\ref{eqn:DetndGME}). The state $\ket{{\rm
Det}_{n,d}}$ has $\Lambda^2_{\max}=1/(d^p)!$.

Although the above states were discussed in terms of the
GME~\cite{WeiGoldbart03}, we shall, in the following section, show the rather
surprising fact that the relative entropy of entanglement of these example
states, is given by the corresponding expression: $-2 \log_2\Lambda_{\max}$.

\section{Connection between the two measures}
\label{sec:connection} \noindent In bi-partite systems, due to the existence
of Schmidt decompositions, the relative entropy of entanglement of a pure
state is simply the von Neumann entropy of its reduced density matrix.
However, for multi-partite systems there is, in general, no such
decomposition, and how to calculate the relative entropy of entanglement for
an arbitrary pure state remains an open question. We now connect the relative
entropy of entanglement to the geometric measure of entanglement for arbitrary
pure states by giving a lower bound on the former in terms of the latter or,
more specifically, via the entanglement eigenvalue.

\subsection{Pure states:
lower bound on relative entropy of entanglement} \noindent Let us begin with
the following theorem:

\smallskip
\noindent {\it Theorem\/} 1. For any pure state $\ket{\psi}$ with entanglement
eigenvalue $\Lambda_{\max}(\psi)$ the quantity $-2\log_2\Lambda_{\max}(\psi)$
is a lower bound on the relative entropy of entanglement of $\ket{\psi}$,
i.e.,
\begin{equation}
\label{eqn:2log} E_{\rm R}(\ketbra{\psi})\ge -2\log_2\Lambda_{\max}(\psi).
\end{equation}

\smallskip\noindent
{\it Proof\/}: From the definition~(\ref{eqn:ER}) of the relative entropy of
entanglement we have, for a pure state $\ket{\psi}$,
\begin{equation}
E_{\rm R}(\ketbra{\psi})=
 \min_{\sigma\in {\cal D}}-\bra{\psi}\log_2\sigma\ket{\psi}=
-\max_{\sigma\in {\cal D}} \bra{\psi}\log_2\sigma\ket{\psi}.
\end{equation}
Using the concavity of the log function, we have
\begin{equation}
\label{eqn:inequalityLog} \bra{\psi}\log_2\sigma\ket{\psi}\le \log_2
(\bra{\psi}\sigma\ket{\psi})
\end{equation}
and, furthermore,
\begin{equation}
\label{eqn:inequalityMax} \max_{\sigma\in{\cal D}}
\bra{\psi}\log_2\sigma\ket{\psi}\le \max_{\sigma\in{\cal D}} \log_2
(\bra{\psi}\sigma\ket{\psi}),
\end{equation}
although the $\sigma$'s maximizing the left- and right-hand sides are not
necessarily identical.  We then conclude that
\begin{equation}
E_{\rm R}(\ketbra{\psi})\ge -\max_{\sigma\in{\cal D}} \log_2
(\bra{\psi}\sigma\ket{\psi}).
\end{equation}
As any $\sigma\in{\cal D}$ can be expanded as $\sigma=\sum_i
p_i\ketbra{\phi_i}$, where ${\ket{\phi_i}}$'s are separable pure states, one
has
\begin{equation}
\bra{\psi}\sigma\ket{\psi}= \sum_i p_i |\ipr{\phi_i}{\psi}|^2 \le
\Lambda^{2}_{\max}({\psi}),
\end{equation}
and hence we arrive at the sought result
\begin{equation}
E_{\rm R}(\ketbra{\psi})\ge -2\log_2\Lambda_{\max}({\psi}). \label{eq:theorem}
\end{equation}

\smallskip
\noindent {We wish to point out that such an inequality was previously
established and exploited in Refs.~\cite{VidalEtAl}.}

{When does the inequality becomes an equality? The demand that
Eq.~(\ref{eqn:inequalityLog}) hold as an equality implies that  $\sigma$
(un-normalized) can be decomposed into either (a)
\begin{subequations}
\begin{equation}
\label{eqn:case1} \sigma=\sum_{i} \ketbra{i},
\end{equation}
where $\{\ket{i}\}$ are mutually orthogonal but {\it not\/} orthogonal to
$\ket{\psi}$, or (b)
\begin{equation}
\label{eqn:case2} \sigma= \ketbra{\psi}+\tau^\perp,
\end{equation}
\end{subequations}
where $\tau^\perp$ (either pure or mixed) is orthogonal to $\psi$, i.e.,
$\bra{\psi}\tau^\perp\ket{\psi}=0$. However, the separable $\sigma$ that has
either property is not necessarily the one that maximizes both sides of the
inequality~(\ref{eqn:inequalityMax}), unless $\ket{\psi}$ (and hence $\sigma$)
has high symmetry. } On the other hand, a corollary arises from Thereom 1
which says that for any multi-partite pure state $\ket{\psi}$, if one can find
a separable mixed state $\sigma$ such that
$S(\rho||\sigma)\vert_{\rho=\ketbra{\psi}}=
-2\log_2\Lambda_{\max}\big(\ket{\psi}\big)$ then $E_{\rm
R}=-2\log_2\Lambda_{\max}\big(\ket{\psi}\big)$. This result follows directly
from the fact that when the lower bound on $E_{\rm R}$ given in
Eq.~(\ref{eqn:2log}) equals an upper bound, the relative entropy of
entanglement is immediate. In all the examples we shall consider for which
this lower bound is saturated, it turns out that
\begin{equation}
\label{eqn:sigma} \sigma^*\equiv\sum_i p_i\, \ketbra{\phi_i}
\end{equation}
is a closest separable mixed state, in which $\{\ket{\phi_i}\}$ are separable
pure states closest to $\ket{\psi}$. (The distribution $p_i$ is uniform, and
can be either discrete or continuous, and $\{\ket{\phi_i}\}$ are not
necessarily mutually orthogonal.)

We now examine several illustrative states in the light of the above
corollary, thus obtaining $E_{\rm R}$ for each of them.  We begin with the
permutation-invariant states $\ket{S(n,k)}$ of Eq.~(\ref{eqn:Snk}), for which
$\Lambda_{\max}$ was given in Eq.~(\ref{eqn:Lambdank}). The above theorem
guarantees that $E_{\rm R}\big(\ket{S(n,k)}\big)\ge -2\log_2
\Lambda_{\max}(n,k)$. To find an {\it upper\/} bound we construct a separable
mixed state
\begin{subequations}
\begin{eqnarray}
\sigma^* &\equiv& \int \frac{d\phi}{2\pi}\ketbra{\xi(\phi)},
\\
\ket{\xi(\phi)} &\equiv& \left(\sqrt{p}\ket{0}+
e^{i\phi}\sqrt{1-p}\ket{1}\right)^{\otimes n},
\end{eqnarray}
\end{subequations}
with $p$ chosen to maximize $||\ipr{\xi}{S(n,k)}||=\sqrt{C_k^n
\,p^k(1-p)^{n-k}}$, which gives $p=k/n$. Direct evaluation then gives
\begin{equation}
\label{eqn:sigmastar} \sigma^*=\sum_{k=0}^n C^n_k
p^{k}(1-p)^{(n\!-\!k)}\ketbra{S(n,k)},
\end{equation}
and $S(\rho||\sigma)=-2\log_2\Lambda_{\max}(n,k)$, where
$\rho=\ketbra{S(n,k)}$ and $\Lambda_{\max}(n,k)$ is given in
Eq.~(\ref{eqn:Lambdank}). The upper and lower bounds on $E_{\rm R}$ coincide,
and hence we have that
\begin{equation}
\label{eqn:rhoSnk} E_{\rm R}\big(\ket{S(n,k)}\big)=-2\log_2
\Lambda_{\max}(n,k).
\end{equation}
The closest separable mixed state $\sigma^*$ belongs to the case (b),
i.e.,~Eq.~(\ref{eqn:case2}). Similar equalities can be established for the
generalized permutation-invariant $n$-party $(d+1)$-dit states
$\ket{S(n,\{k\})}$ of Eq.~(\ref{eqn:Snkk}). We remark that the entanglements
of the symmetric states $\ket{S(n,k)}$ (which are also known as {\it Dicke\/}
states) have been analyzed via other approaches; see Ref.~\cite{Stockton}.

For our next example we consider the totally anti-symmetric states $\ket{{\rm
Det}_n}$ of Eq.~(\ref{eqn:Detn}).  It was shown in Ref.~\cite{Bravyi02} that
for these states $\Lambda^2_{\max}=1/n!$, and hence it is straightforward to
see that each of the $n!$ basis states $\ket{i_1,\dots,i_n}$ is a closest
separable pure state. Thus, one can construct a separable mixed state from
these separable pure states [cf.~Eq.~(\ref{eqn:sigma})]:
\begin{equation}
\sigma_1\equiv\frac{1}{n!}\sum_{i_1,\dots,i_n} \ketbra{i_1,\dots,i_n}.
\end{equation}
Then, by direct calculation one gets $S(\rho_{{\rm
Det}_n}||\sigma_1)=\log_2(n!)$, which is identical to
$-2\log_2\Lambda_{\max}$, as mentioned above.  As in our previous examples,
upper and lower bounds on $E_{\rm R}$ coincide, and hence we have that $E_{\rm
R}(\ket{{\rm Det}_n})=\log_2(n!)$. The closest separable mixed state
$\sigma_1$ belongs to the case (a), i.e.,~Eq.~(\ref{eqn:case1}). Similarly,
for the generalized determinant state~(\ref{eqn:Detnd}) one can show that
$E_{\rm R}=\log_2(d^p!)$.

We now focus our attention on three-qubit settings. Of these, the states
$\ket{S(3,0)}=\ket{000}$ and $\ket{S(3,3)}=\ket{111}$ are not entangled and
are, respectively, the components of the the 3-GHZ state: $\ket{{\rm
GHZ}}\equiv \big(\ket{000}+\ket{111})/\sqrt{2}$. Although the GHZ state is not
of the form $\ket{S(n,k)}$, it has $\Lambda_{\max}=1/\sqrt{2}$, and two of its
closest separable pure states are $\ket{000}$ and
$\ket{111}$~\cite{WeiGoldbart03}. From these one can construct a separable
mixed state
\begin{eqnarray}
\sigma_2&=&\frac{1}{2}\big(\ketbra{000}+\ketbra{111}\big),
\end{eqnarray}
From the discussion given after Eq.~(\ref{eq:theorem}), one concludes that
$E_{\rm R}({\rm GHZ})=-2 \log_{2} \Lambda_{\max}=1$ and that $\sigma_2$ is one
of the closest separable mixed states to $\ket{\rm GHZ}$. This closest
separable mixed state $\sigma_2$ belongs to the case (a),
i.e.,~Eq.~(\ref{eqn:case1}). With some rewriting, it can also be classified as
case (b), i.e.,
\begin{equation}
\sigma_2=\frac{1}{2}\ketbra{\rm GHZ}+\frac{1}{2}\ketbra{\rm GHZ^-},
\end{equation}
where $\ket{{\rm GHZ^-}}\equiv \big(\ket{000}-\ket{111})/\sqrt{2}$.

The states
\begin{subequations}
\begin{eqnarray}
|{\rm W}\rangle &\equiv& \ket{{S}(3,2)}=
\big(\ket{001}+\ket{010}+\ket{100}\big)/\sqrt{3},
\\
\ket{\widetilde{\rm W}} &\equiv& \ket{{S}(3,1)}=
\big(\ket{110}+\ket{101}+\ket{011}\big)/\sqrt{3},
\end{eqnarray}
\end{subequations}
are equally entangled, and have $\Lambda_{\max}=2/3$~\cite{WeiGoldbart03}.
Again, from the discussion after Eq.~(\ref{eq:theorem}) we have $E_{\rm
R}=\log_{2}(9/4)$, and one of the closest separable mixed states to the W
state can be constructed from
\begin{eqnarray}
\sigma_{3}&\equiv& \int\frac{d\phi}{2\pi}\ketbra{\psi(\phi)}, \quad{\rm with}
\\
\ket{\psi(\phi)}&\equiv& \big(\sqrt{2/3}\ket{0}+
e^{i\phi}\sqrt{1/3}\ket{1}\big)^{\otimes 3},
\end{eqnarray}
which gives the result
\begin{eqnarray}
\sigma_{3}= \frac{4}{9}\ketbra{{\rm W}}+\frac{2}{9}\ketbra{\widetilde{\rm W}}
+ \frac{8}{27}\ketbra{000}+\frac{1}{27}\ketbra{111}.
\end{eqnarray}
We remark that the mixed state $\sigma_{3}$ is not the only closest separable
mixed state to the W state; the following state $\sigma_{4}$ is another
example (as would be any mixture of $\sigma_{3}$ and $\sigma_{4}$):
\begin{subequations}
\begin{eqnarray}
\sigma_4 \equiv\frac{1}{3}\sum_{k=0}^{2} \ketbra{\psi(2\pi k/3)}
=\frac{4}{9}\ketbra{{\rm W}}+\frac{2}{9}\ketbra{\widetilde{\rm W}}+
\frac{1}{3}\ketbra{\xi},
\end{eqnarray}
\end{subequations}
where $3\ket{\xi}\equiv {2\sqrt{2}}\ket{000}+\ket{111}$.  These closest
separable mixed states of W state belong to the case (b),
i.e.,~Eq.~(\ref{eqn:case2}).

\begin{figure}
\centerline{\psfig{figure=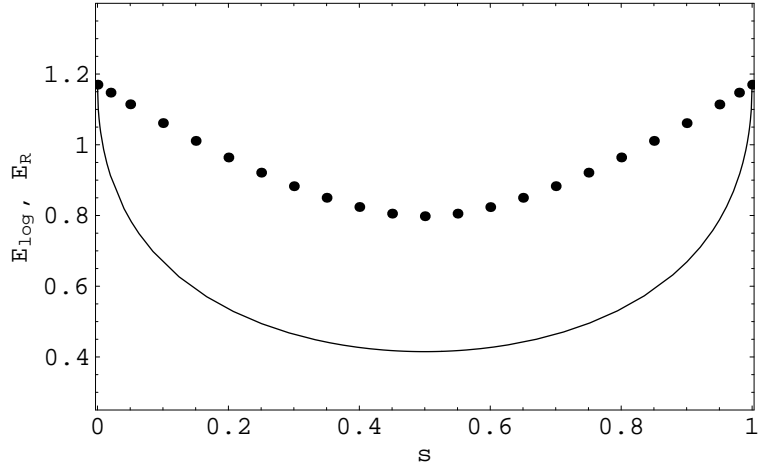,width=10cm,angle=0}}
\caption[Entanglement curve for the state  state $\sqrt{s}\,|{\rm
W}\rangle+\sqrt{1-s}\,|\widetilde{\rm W}\rangle$]{The solid curve represents
$E_{\log_{2}}(s)$ of the pure state $\sqrt{s}\,|{\rm
W}\rangle+\sqrt{1-s}\,|\widetilde{\rm W}\rangle$ vs.~$s$. The dots are the
corresponding relative entropies of entanglement, obtained numerically.}
\label{fig:ElogWW}
\end{figure}
Having obtained REE for ${\rm W}$ and $\widetilde{\rm W}$, it is interesting
to examine the REE of the following superposition of the two:
$\ket{\wstate\widetilde{\rm W}(s)} \equiv \sqrt{s}\,
\ket{\wstate}+\sqrt{1-s}\,\ket{\widetilde{\rm W}}$. We  have not been able to
find an analytical result for REE, but we can compare the analytical
expression for $-2\log_2\Lambda_{\max}(\wstate\widetilde{\rm W}(s))$ with the
numerical evaluation of $E_{\rm R}(\wstate\widetilde{\rm W}(s))$, and we do
this in Fig.~\ref{fig:ElogWW}.  As we see in this figure, the qualitative
behavior of the two functions is similar, but $-2\log_2\Lambda_{\max}$ and
$E_{\rm R}$ only coincide at the two end-points, $s=0$ and $s=1$.

\subsection{Mixed states: upper bound on relative entropy of entanglement}
\noindent In Chapter~\ref{chap:GME} the procedure was given to find the
geometric measure of entanglement, $E_{\sin^2}$, for the mixed state
comprising symmetric states:
\begin{equation}
\label{eqn:mixSnk1} \rho(\{p\})=\sum_k p_k\,\ketbra{S(n,k)}.
\end{equation}
Here, we focus instead on the quantity $E_{\log_{2}}$, but the basic procedure
is the same.  The first step is to find the entanglement eigenvalue
$\Lambda_{n}(\{q\})$ for the pure state
\begin{equation}
\sum_k \sqrt{q_k}\,\ket{S(n,k)},
\end{equation}
thus arriving at the quantity
\begin{equation}
\label{eqn:epsilon} {\cal E}(\{q\})\equiv -2 \log_2 \Lambda_{n}(\{q\}).
\end{equation}
Then the quantity $E_{\log_{2}}$ for the mixed state~(\ref{eqn:mixSnk1}) is
actually the convex hull of the expression~(\ref{eqn:epsilon}):
\begin{equation}
E_{\log_2}\left(\rho(\{p\})\right)={\rm co}\,{\cal E}(\{p\}).
\end{equation}

This prompts us to ask the question: Can we find REE for the mixture of
$\ket{S(n,k)}$ in Eq.~(\ref{eqn:mixSnk1})? To answer it, we shall first
construct an upper bound to REE, and then compare this bound with the
numerically evaluated REE.  To accomplish the first step, bearing in mind the
fact that any separable mixed state will yield an upper bound, we consider the
state formed by mixing the separable pure states $\ket{\xi(\theta,\phi)}$
[cf.~Eq.~(\ref{eqn:sigmastar})]:
\begin{eqnarray}
\label{eqn:sigmatheta} \sigma(\theta)=\int \frac{d\phi}{2\pi}
\ketbra{\xi(\theta,\phi)}=\sum_{k=0}^n
C^n_k\cos^{2k}\theta\sin^{2(n\!-\!k)}\theta\ketbra{S(n,k)},
\end{eqnarray}
where
\begin{equation}
\label{eqn:xi} \ket{\xi(\theta,\phi)}\equiv \left(\cos\theta \ket{0}+
e^{i\phi}\sin\theta\ket{1}\right)^{\otimes n}.
\end{equation}
We then minimize the relative entropy between $\rho(\{p\})$ and
$\sigma(\theta)$,
\begin{equation}
S\left(\rho(\{p\})\vert\vert\sigma(\theta)\right)=\sum_k p_k
\log\frac{p_k}{C^n_k\cos^{2k}\theta\sin^{2(n\!-\!k)}\theta},
\end{equation}
with respect to $\theta$, obtaining the stationarity condition
\begin{equation}
\label{eqn:theta} \tan^2\theta\equiv \frac{\sum_k p_k\,(n-k)}{\sum p_k \,k}.
\end{equation}
Due to the convexity of the relative entropy,
\begin{equation}
S\left(\sum_i q_i \rho_i\Vert\sum_i q_i \sigma_i\right) \le \sum_i q_i
S(\rho_i||\sigma_i),
\end{equation}
we can further tighten the expression of the relative entropy by taking its
convex hull. (Via the convexification process, i.e., the convex hull
construction, the corresponding separable state can also be
obtained.)\thinspace\ Therefore, we arrive at an upper bound for the relative
entropy of entanglement of the mixed state $\rho(\{p\})$:
\begin{equation}
\label{eqn:conjecture} E_{\rm R}\left(\rho(\{p\})\right)\le {\rm co}F(\{p\}),
\end{equation}
where
\begin{eqnarray}
\label{eqn:F} F(\{p\})\equiv\sum_k p_k \log_2 \frac{p_k}{C_k^n \cos^{2k}\theta
\sin^{2(n-k)}\theta}=\sum_k p_k \log_2 \frac{p_k\, n^n}{C_k^n \alpha^{k}
(n-\alpha)^{n-k}},
\end{eqnarray}
where the angle $\theta$ satisfies Eq.~(\ref{eqn:theta}), $C_k^n\equiv
n!/\big(k!(n-k)!\big)$, and $\alpha\equiv \sum_k p_k\, k$.

Having established an upper bound for REE for the state $\rho(\{p\})$, we now
make the restriction to mixtures of two distinct $n$-qubit states
$\ket{S(n,k_1)}$ and $\ket{S(n,k_2)}$ (with $k_1\ne k_2$):
\begin{eqnarray}
\rho_{n;k_1,k_2}(s)\equiv s\ketbra{S(n,k_1)} 
+(1-s)\ketbra{S(n,k_2)}.
\end{eqnarray}
One trivial example is $\rho_{n;0,n}(s)$, which is obviously unentangled as it
is the mixture of two separable pure states $\ket{0^{\otimes n}}$ and
$\ket{1^{\otimes n}}$.  Other mixtures are generally entangled, except
possibly at the end-points $s=0$ or $s=1$ when the mixture contains either
$\ket{S(n,0)}$ or $\ket{S(n,n)}$.  We first investigate the two-qubit (i.e.\
$n=2$) case.  Besides the trivial mixture, $\rho_{2;0,2}$, there is only one
inequivalent mixture, $\rho_{2;0,1}(s)$ [which is equivalent to
$\rho_{2;2,1}(s)$], which is---up to local basis change---the so-called {\it
maximally entangled mixed
state\/}~\cite{MunroEtAl,WeiNemotoGoldbartKwiatMunroVerstraete03} (for a
certain range of $s$)
\begin{equation}
\rho_{2;0,1}=s\,\ketbra{11}+(1-s)\ketbra{\Psi^+},
\end{equation}
where $\ket{\Psi^+}\equiv(\ket{01}+{10})/\sqrt{2}$. The function $F$ for this
state [denoted by $F_{2;0,1}(s)$] is
\begin{equation}
\label{eqn:rho201} F_{2;0,1}(s)=s
\,\log_2\frac{4s}{(1+s)^2}+(1-s)\log_2\frac{2}{1+s}\,,
\end{equation}
which is convex in $s$.  It is exactly the expression for the relative entropy
of entanglement for the state $\rho_{2;0,1}$ found by Vedral and
Plenio~\cite{VedralPlenio98} (see their Eq.~(56) with $\lambda$ replaced by
$1-s$).

For $n=3$ there are three other inequivalent mixtures: $\rho_{3;0,1}(s)$
[equivalent to $\rho_{3;3,2}(s)$], $\rho_{3;0,2}(s)$ [to $\rho_{3;3,1}(s)$],
and $\rho_{3;1,2}(s)$ [to $\rho_{3;2,1}(s)$]. In Fig.~\ref{fig:Er3} we compare
the function $F$ in Eq.~(\ref{eqn:F}), its convex hull  ${\rm co}\,F$, and
numerical values of $E_{\rm R}$ obtained using the general scheme described in
Ref.~\cite{VedralPlenio98} extended beyond the two-qubit case. The agreement
between ${\rm co}\,F$ and the numerical values of $E_{\rm R}$ appears to be
exact.

For $n=4$ there are five inequivalent nontrivial mixtures: $\rho_{4;0,1}(s)$,
$\rho_{4;0,2}(s)$, $\rho_{4;0,3}(s)$, $\rho_{4;1,2}(s)$, and
$\rho_{4;1,3}(s)$. In Figs.~\ref{fig:Er4A} and \ref{fig:Er4B} we again compare
the function $F$ in Eq.~(\ref{eqn:F}), its convex hull ${\rm co}\,F$, and
numerical values of $E_{\rm R}$. Again the agreement between ${\rm co}\,F$ and
the numerical values of $E_{\rm R}$ appears to be exact.

\begin{figure}
\centerline{\psfig{figure=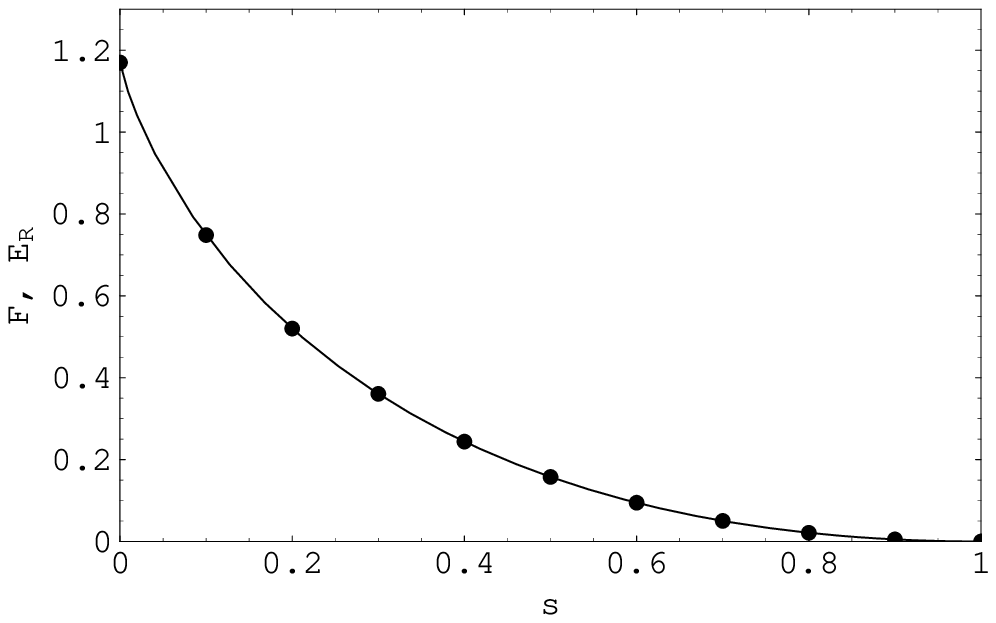,width=8cm,angle=0}} \vspace{0.2cm}
\centerline{\psfig{figure=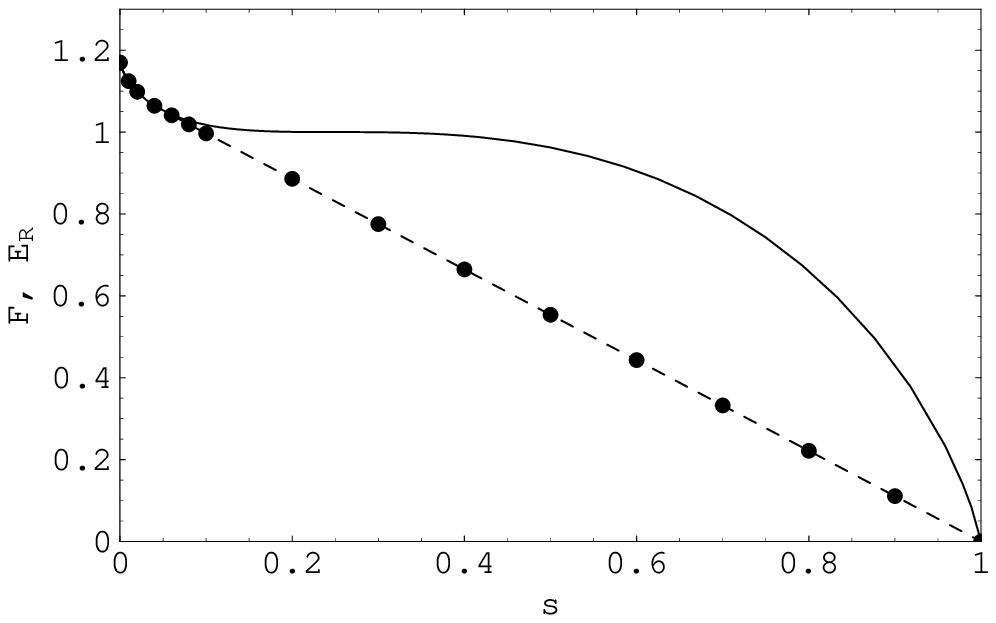,width=8cm,angle=0}} \vspace{0.2cm}
\centerline{\psfig{figure=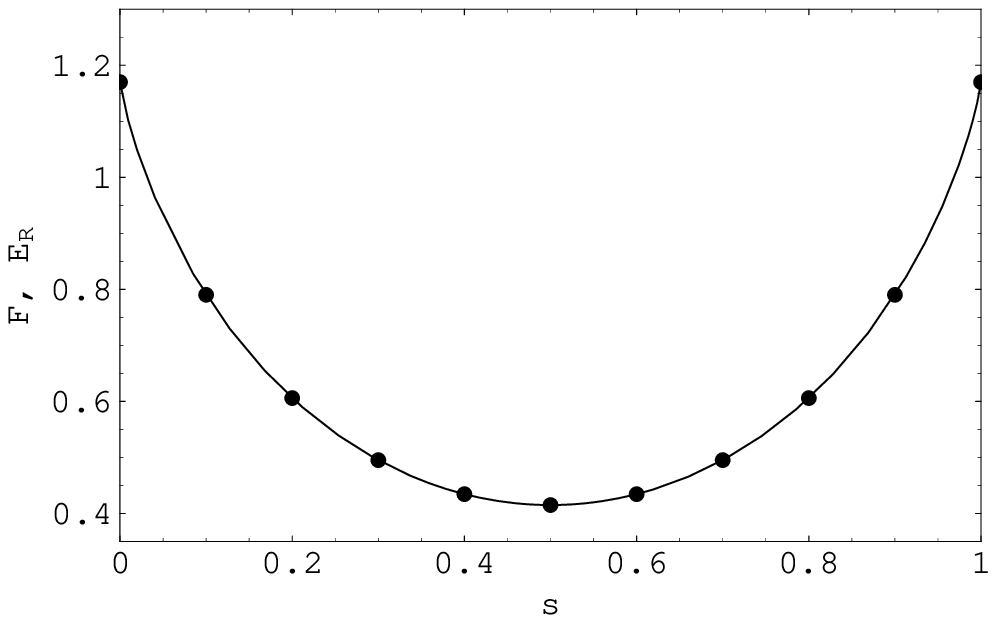,width=8cm,angle=0}}
\caption[Comparison of $F$, ${\rm co}\,F$, and the numerical value of $E_{\rm
R}$ for the states $\rho_{3;0,1}(s)$, $\rho_{3;0,2}(s)$, and
$\rho_{3;1,2}(s)$]{Comparison of $F$ (solid curve), ${\rm co}\,F$
(convexification, indicated by dashed line) and the numerical value of $E_{\rm
R}$ (dots) for the states $\rho_{3;0,1}(s)$, $\rho_{3;0,2}(s)$, and
$\rho_{3;1,2}(s)$ (from top to bottom). Note that the $\log$ function is
implicitly base-2. } \label{fig:Er3}
\end{figure}
\begin{figure}
\centerline{\psfig{figure=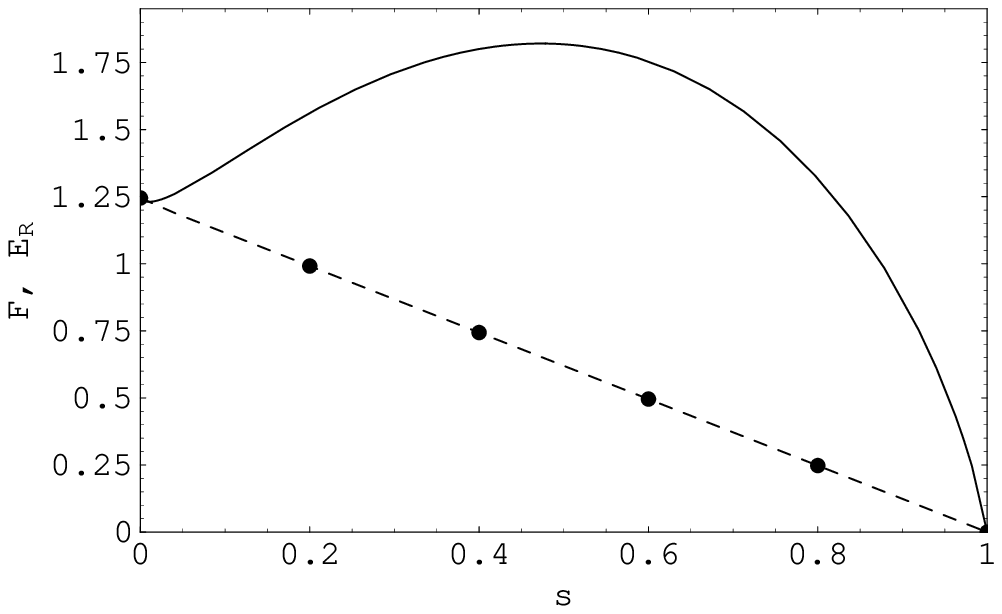,width=10cm,angle=0}} \vspace{0.2cm}
\centerline{\psfig{figure=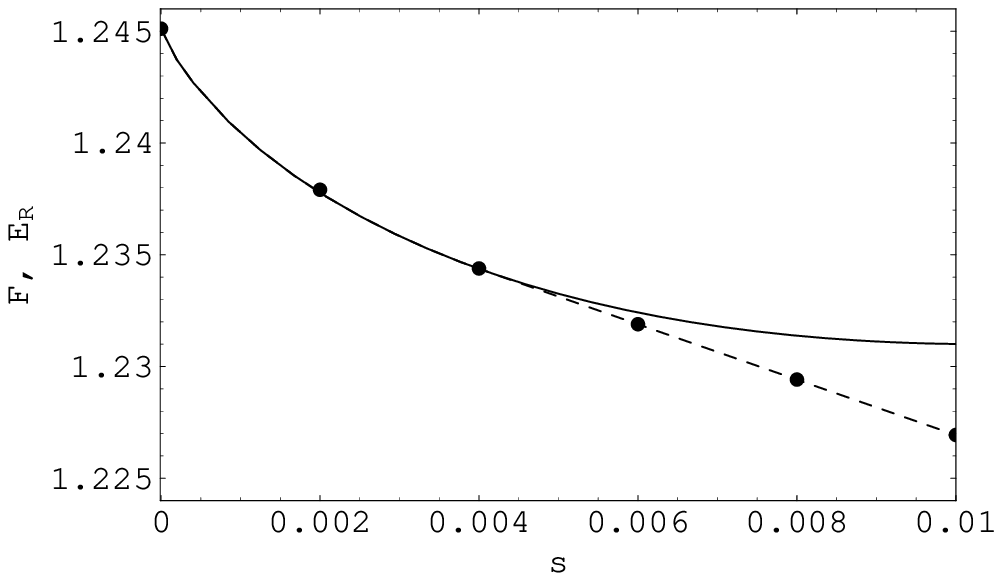,width=10.5cm,angle=0}}
\caption[Comparison of $F$, ${\rm co}\,F$, and the numerical value of $E_{\rm
R}$ for the state  $\rho_{4;0,3}(s)$]{Comparison of $F$ (solid curve), its
convex hull (dash line), and the numerical value of $E_{\rm R}$ for the state
 $\rho_{4;0,3}(s)$. Upper panel shows the whole range $s\in[0,1]$,
 whereas the lower panel shows a blow-up of the range $s\in[0,0.01]$.
} \label{fig:Er4A}
\end{figure}
\begin{figure}
\centerline{\psfig{figure=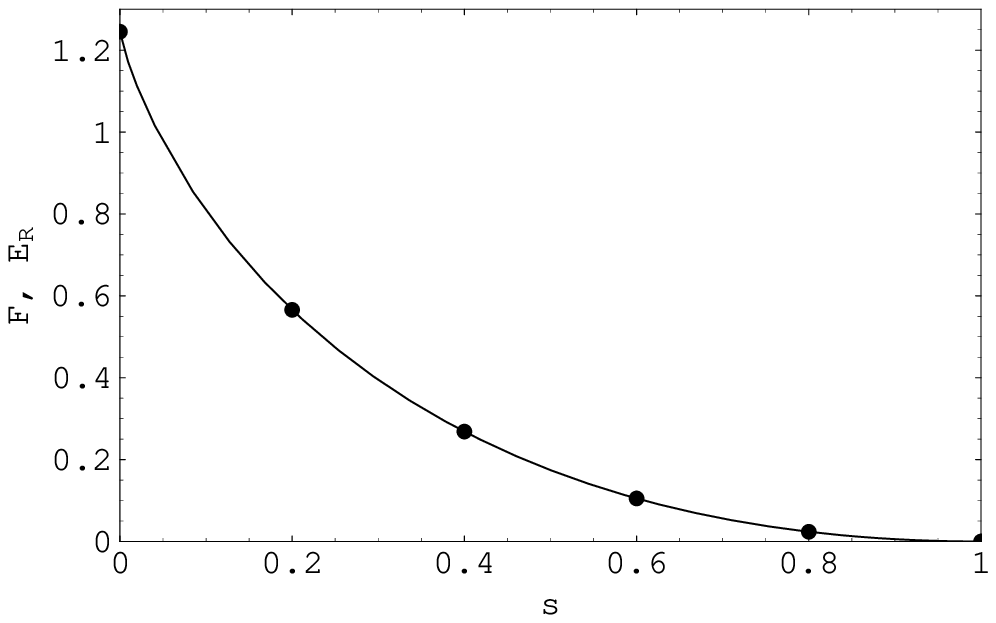,width=8cm,angle=0}} \vspace{0.2cm}
\centerline{\psfig{figure=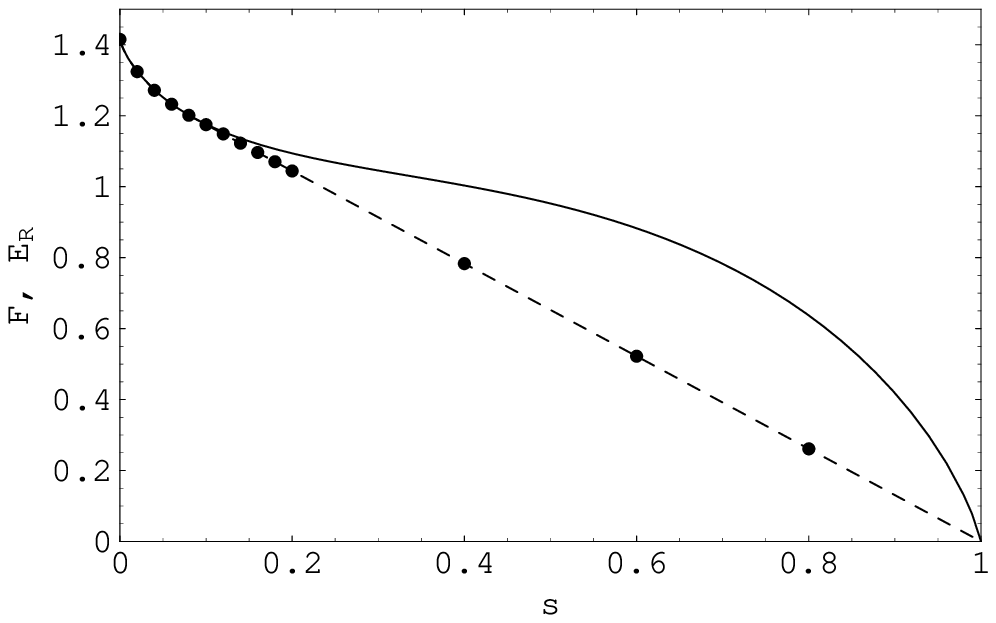,width=8cm,angle=0}} \vspace{0.2cm}
\centerline{\psfig{figure=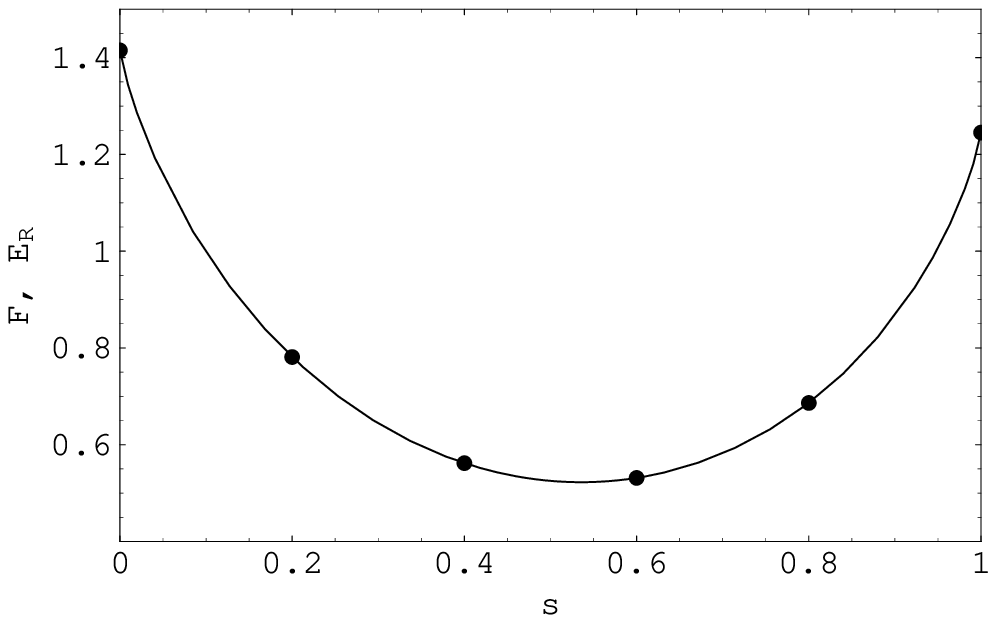,width=8cm,angle=0}} \vspace{0.2cm}
\centerline{\psfig{figure=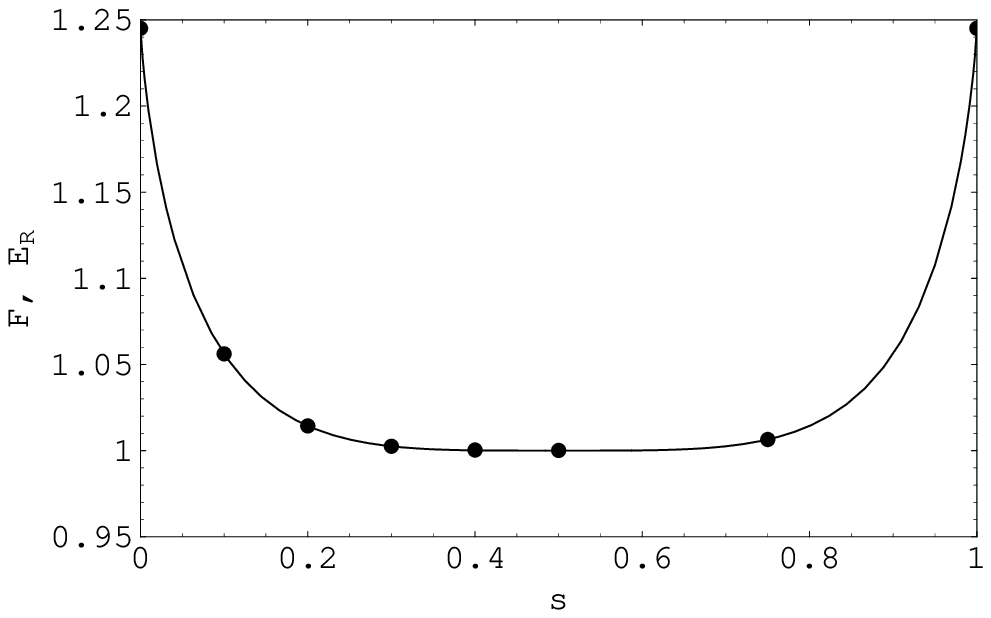,width=8.3cm,angle=0}}
\caption[Comparison of $F$ , ${\rm co}\,F$, and the numerical value of $E_{\rm
R}$ for the states $\rho_{4;0,1}(s)$, $\rho_{4;0,2}(s)$, $\rho_{4;1,2}(s)$,
and $\rho_{4;1,3}(s)$]{Comparison of $F$ (solid curve), its convex hull
(dashed line), and the numerical value of $E_{\rm R}$ for the states
$\rho_{4;0,1}(s)$, $\rho_{4;0,2}(s)$, $\rho_{4;1,2}(s)$, and $\rho_{4;1,3}(s)$
(from top to bottom). } \label{fig:Er4B}
\end{figure}
From these agreements, we are led to the following conjecture:
\\
\noindent {\it Conjecture}~1: The relative entropy of entanglement $E_{\rm
R}\left(\rho(\{p\})\right)$ for the mixed states $\rho(\{p\})$ is given
exactly by ${\rm co}F(\{p\})$.

\begin{figure}
\centerline{\psfig{figure=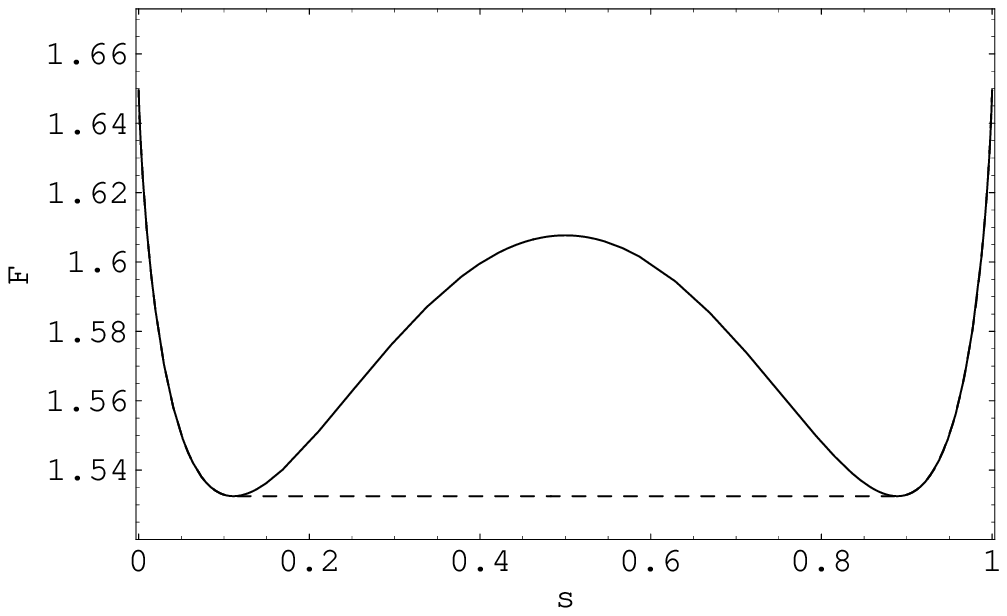,width=10cm,angle=0}}
\caption[Comparison of $F$ and ${\rm co}\,F$ for the seven-qubit mixed state
$\rho_{7;2,5}(s)$]{The function $F$ (solid curve) and its convex hull (dashed
line indicates convexification) for  the seven-qubit mixed state
$\rho_{7;2,5}(s)$. } \label{fig:Er725}
\end{figure}
For the states that we have just considered, we now pause to give the formulas
for $E_{\rm R}$ suggested by the conjecture.  For the three-qubit mixed state
$\rho_{3;2,1}(s)$, its conjectured $E_{\rm R}$ is
\begin{subequations}
\begin{equation}
\label{eqn:ErWW} s\log_2\frac{9s}{(1+s)^2(2-s)}+
(1-s)\log_2\frac{9(1-s)}{(2-s)^2(1+s)}.
\end{equation}
For $\rho_{3;0,1}(s)$, it is
\begin{equation}
\label{eqn:rho301} s\log_2\frac{27s}{(2+s)^3}+ (1-s)\log_2\frac{9}{(2+s)^2}.
\end{equation}
\end{subequations}
For $\rho_{4;0,1}(s)$, it is
\begin{subequations}
\begin{equation}
s\log_2\frac{256s}{(3+s)^4}+ (1-s)\log_2\frac{64}{(3+s)^3}.
\end{equation}
For $\rho_{4;1,2}(s)$, it is
\begin{equation}
s\log_2\frac{64s}{(2\!-\!s)(2\!+\!s)^3}+
(1\!-\!s)\log_2\frac{128(1-s)}{3(2\!-\!s)^2(2\!+\!s)^2}.
\end{equation}
For $\rho_{4;1,3}(s)$, it is
\begin{equation}
s\log_2\frac{64s}{(3\!-\!2s)(1\!+\!2s)^3}+
(1\!-\!s)\log_2\frac{64(1-s)}{(3\!-\!2s)^3(1\!+\!2s)}.
\end{equation}
\end{subequations}
For states such as $\rho_{3;0,2}$, $\rho_{4;0,2}$, and $\rho_{4;0,3}$,
convexifications (i.e. convex hull constructions) are needed; see
Figs.~\ref{fig:Er3}, \ref{fig:Er4A}, and \ref{fig:Er4B}. In
Fig.~\ref{fig:Er725} we give an example of a seven-qubit state, viz.,
$\rho_{7;2,5}(s)$.

Although we have not been able to prove our conjecture, we have observed some
supporting evidence, in addition to the numerical evidence presented above. We
begin by noting that the states $\rho(\{p\})$ are invariant under the
projection
\begin{equation}
\label{eqn:Projection} {\bf \rm P}:\rho\rightarrow \int\frac{d\phi}{2\pi}\,
U(\phi)^{\otimes n}\rho\, U(\phi)^{\dagger\otimes n}
\end{equation}
with $U(\phi)\big\{\ket{0},\ket{1}\big\}\to \big\{\ket{0},{\rm
e}^{-i\phi}\ket{1}\big\}$. Vollbrecht and Werner~\cite{VollbrechtWerner01}
have shown that in order to find the closest separable mixed state for a state
that is invariant under projections such as ${\bf \rm P}$, it is only
necessary to search within the separable states that are also invariant under
the projection.  We can further reduce the set of separable states to be
searched by invoking another symmetry property possessed by $\rho(\{p\})$:
these states are also, by construction, invariant under permutations of all
parties.  Let us denote by $\Pi_i$ one of the permutations of parties, and by
$\Pi_i(\rho)$ the state obtained from $\rho$ by permuting the parties under
$\Pi_i$.  We now show that the set of separable states to be searched can be
reduced to the separable states that are invariant under the permutations. To
see this, suppose that $\rho$ is a mixed state in the
family~(\ref{eqn:mixSnk1}), and that $\sigma^*$ is one of the closest
separable states to $\rho$, i.e.,
\begin{equation}
E_{\rm R}(\rho)\equiv\min_{\sigma\in {\cal D}}
S(\rho||\sigma)=S(\rho||\sigma^*). \label{eqn:extreme}
\end{equation}
As $\rho$ is invariant under all $\Pi_i$, we have
\begin{equation}
E_{\rm R}(\rho)=\frac{1}{N_\Pi}\sum_i
S\left(\rho\big\Vert\Pi_i(\sigma^*)\right),
\end{equation}
where $N_\Pi$ is the number of permutations. By using the convexity of the
relative entropy we have
\begin{equation}
E_{\rm R}(\rho)\ge S\left(\rho\big\Vert\big[\sum_i
\Pi_i(\sigma^*)/N_\Pi\big]\right).
\end{equation}
However, because of the extremal property, Eq.~(\ref{eqn:extreme}), the
inequality must be saturated, as the left-hand side is already minimal. This
shows that
\begin{equation}
\sigma^{**}\equiv \frac{1}{N_\Pi}\sum_i \Pi_i(\sigma^*)
\end{equation}
also a closest separable mixed state to $\rho$, and is manifestly invariant
under all permutations.  Thus, we only need to search within this restricted
family of separable states.

It is not difficult to see that the set ${\cal D}_S$ of all separable mixed
states that are diagonal in the basis of $\{\ket{S(n,k)}\}$ can be constructed
from a convex mixture of separable states in Eq.~(\ref{eqn:sigmatheta}). That
is, for any $\sigma_s\in {\cal D}_S$ we have a decomposition
\begin{equation}
\label{eqn:sigmas} \sigma_s=\sum_i t_i \,\sigma(\theta_i),
\end{equation}
where $t_i\ge 0$, $\sum_i t_i=1$, and $\sigma(\theta_i)$ is of the
form~(\ref{eqn:sigmatheta}). This is because the separability of the
states~(\ref{eqn:mixSnk1}) implies that there exists a decomposition into pure
states such that each pure state is a separable state. Furthermore,
 because $\{\ket{S(n,k)}\}$ are eigenstates of $\rho(\{p\})$,
the most general form of the pure state in its decomposition is
\begin{equation}
\sum_k\sqrt{q_k}\,e^{i\phi_k}\ket{S(n,k)}.
\end{equation}
This pure state is separable if and only if it is of the form~(\ref{eqn:xi}),
up to an overall irrelevant phase. As $\rho(\{p\})$ is invariant under the
projection ${\rm P}$~(\ref{eqn:Projection}), a pure state in
Eq.~(\ref{eqn:xi}) will be projected to the mixed state in
Eq.~(\ref{eqn:sigmatheta}) under ${\rm P}$. Thus, every separable state that
is diagonal in $\{\ket{S(n,k)}\}$ basis can be expressed in the
form~(\ref{eqn:sigmas}).

\begin{figure}
\psfrag{E}{${\cal E}$} \psfrag{F}{$F$}
\centerline{\psfig{figure=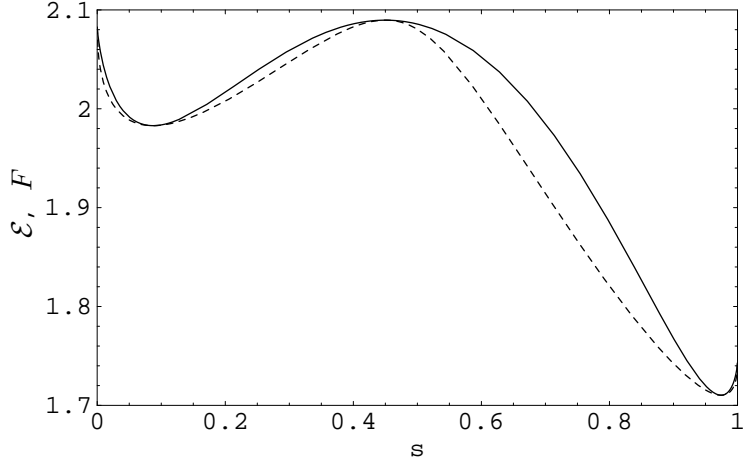,width=10cm,angle=0}}
\caption[Comparision of ${\cal E}$ and $F$  for the eleven-qubit mixed state
$\rho_{11;2,6}(s)$]{Comparision of ${\cal E}$ (dashed curve) and $F$ (solid
curve) for the eleven-qubit mixed state $\rho_{11;2,6}(s)$. }
\label{fig:ElogEr1126}
\end{figure}
Hence, our conjecture~(\ref{eqn:conjecture})  ensures (via any necessary
convexification) that it is at least the minimum (of the relative entropy)
when the separable mixed states are restricted to ${\cal D}_S$. However, in
order to prove the conjecture, one would still need to show that the
expression is also the minimum when the restriction to ${\cal D}_S$ is
relaxed.

We remark that our conjecture is consistent with the results of
Ishizaka~\cite{Ishizaka02}, in that our conjectured $\sigma^*$ satisfies the
condition that $[\rho,\sigma^*]=0$ and that $\sigma^*$ has the same reduction
as $\rho$ for every party. { Furthermore, suppose $\sigma^*$ (diagonal in the
basis $\{\ket{S(n,k)}\}$) represents the separable state that gives the
conjectured value of REE:
\begin{equation}
\sigma^*=\sum_k r_k \ket{S(n,k)}\bra{S(n,k)},
\end{equation}
where the $r$'s can be obtained by finding the convex hull of the function $F$
in Eq.~(\ref{eqn:F}). Now consider any separable state $\tau$ in the Hilbert
space {\it orthogonal\/} to the subspace spanned by $\{S(n,k)\}$. We need to
show that the separable state $\sigma(x)\equiv x\sigma^*+(1-x) \tau$, for any
$x\in [0,1]$, gives greater relative entropy with $\rho(\{p\})$ in
Eq.~(\ref{eqn:mixSnk1}) than $\sigma^*$ does with $\rho(\{p\})$, i.e.,
\begin{equation}
\label{eqn:Sinequality} S\left(\rho(\{p\})\Vert\sigma(x)\right)\ge
S\left(\rho(\{p\})\Vert\sigma^*\right).
\end{equation}
Writing out the expression explicitly, we have that
\begin{eqnarray}
S\left(\rho(\{p\})\Vert\sigma(x)\right)=\sum_k p_k \log\frac{p_k}{x\,r_k} 
\ge\sum_k p_k \log\frac{p_k}{ r_k}= S\left(\rho(\{p\})\Vert\sigma^*\right).
\end{eqnarray}
Note that $\tau$ gives no contribution in the relative entropy, as it is
orthogonal to $\rho(\{p\})$, and that we have not used the fact that $\tau$ is
separable. But to prove Conjecture 1 we need to show that
Eq.~(\ref{eqn:Sinequality}) holds if separable $\tau$ is not orthogonal to the
subspace spanned by $\{S(n,k)\}$. }

Recall that for pure states we found the inequality $E_{\log_2}\le E_{\rm R}$.
Does this inequality hold for mixed states? We do not know the complete answer
to this question, but for the mixed state $\rho(\{p\})$ we shall at least find
that this inequality would hold if Conjecture~1 holds.  To see this, we first
establish that ${\cal E}(\{q\})$ is a lower bound on $F(\{q\})$; see the
example in Fig.~\ref{fig:ElogEr1126}. The proof is as follows. Recall that
\begin{subequations}
\begin{equation}
{\cal E}(\{p\})= -2\log_2\left[\max_\theta \sum_k \sqrt{p_k}\, \sqrt{C_k^n}
\cos^k\theta\sin^{n-k}\theta\right].
\end{equation}
By the concavity of $\log$, we then have
\begin{eqnarray}
-2\log_2\left[\sum_k \sqrt{p_k}\, \sqrt{C_k^n} \cos^k\theta\sin^{n-k}\theta\right] 
\le \sum_k p_k \log_2\frac{p_k}{C_k^n \cos^{2k}\theta \sin^{2(n-k)}\theta}.
\end{eqnarray}
Hence
\begin{eqnarray}
\min_\theta -2\log_2\left[\sum_k \sqrt{p_k}\, \sqrt{C_k^n} \cos^k\theta\sin^{n-k}\theta\right] 
\le \min_\theta\sum_k p_k \log_2\frac{p_k}{C_k^n \cos^{2k}\theta
\sin^{2(n-k)}\theta},
\end{eqnarray}
or equivalently
\begin{equation}
{\cal E}(\{p\})\le F(\{p\}).
\end{equation}
If Conjecture~1 is correct then by taking the convex hull of both sides of
this inequality we would have
\begin{equation}
E_{\log_2}\le E_{\rm R}
\end{equation}
for the family of states~(\ref{eqn:mixSnk1}).
\end{subequations}
Notice that we have also shown that this relation holds for arbitrary pure
states. It would be interesting to know whether it also holds for arbitrary
mixed states.

\section{Concluding remarks}
\noindent \label{sec:summary} We have provided a lower bound on the relative
entropy of entanglement for arbitrary multi-partite pure states in terms of
their geometric measure of entanglement.  For several families of pure states
we have shown that the bound is in fact saturated, and thus provides the exact
value of the relative entropy of entanglement. For mixtures of certain
permutation-invariant states we have conjectured analytic expressions for the
relative entropy of entanglement.

It is possible that our results on the relative entropy of entanglement might
be applicable to the checking of the consistency of some equalities and
inequalities~\cite{PlenioVedral01,WuZhang00,GalvaoPlenioVirmani00} regarding
minimal reversible entanglement generating sets (MREGSs). Consider, e.g., the
particular family of $n$-qubit pure states $\{\ket{S(n,k)}\}$, the relative
entropy of entanglement of which we have given in Eq.~(\ref{eqn:rhoSnk}). Now,
if we trace over one party we get a mixed $(n-1)$-qubit state:
\begin{equation}
\label{eqn:Tr1} {\rm Tr}_1
\ketbra{S(n,k)}=\frac{n\!-\!k}{n}\ketbra{S(n\!-\!1,k)}+\frac{k}{n}\ketbra{S(n\!-\!1,k\!-\!1)}.
\end{equation}
We have also given a conjecture for the relative entropy of entanglement for
this mixed state.  If we trace over $m$ parties, the reduced mixed state would
be a mixture of $\{\ket{S(n-m,q)}\}$ [with $q\le (n-m)$], and again we have
given a conjecture for its relative entropy of entanglement.  For example, if
we start with $\ket{S(4,1)}$, and trace over one party and then another, we
get the sequence:
\begin{equation}
\ket{S(4,1)}\rightarrow \rho_{3;0,1}({1}/{4}) \rightarrow
\rho_{2;0,1}({1}/{2}),
\end{equation}
for which we have given the corresponding relative entropies of entanglement
in Eqs.~(\ref{eqn:rhoSnk}), (\ref{eqn:rho301}) and (\ref{eqn:rho201}). (To be
precise, the second formula is a conjecture; the others are
proven.)\thinspace\ The afore-mentioned equalities and inequalities concerning
MREGS usually involve only the von Neumann entropy and the regularized
(i.e.~asymptotic) relative entropy of entanglement of the pure state and its
reduced density matrices. The regularized relative entropy of entanglement is
defined as
\begin{equation}
\label{eqn:regularized} E_{\rm R}^\infty(\rho)\equiv \lim_{n\rightarrow
\infty}\frac{1}{n}E_{\rm R}(\rho^{\otimes n}).
\end{equation}
The calculation of the regularized relative entropy of entanglement is, in
general, much more difficult than for the non-regularized case, and the
(in)equalities involving the regularized relative entropy of entanglement are
thus difficult to check.  Nevertheless, it is known that $E_{\rm R}^\infty\le
E_{\rm R}$, so we can check their weaker forms by replacing $E_{\rm R}^\infty$
by $E_{\rm R}$, and the corresponding (in)equalities by weaker inequalities.

{ Plenio and Vedral~\cite{PlenioVedral01} have derived a lower bound on the
REE of a tripartite pure state $\rho_{\rm ABC}=\ketbra{\psi}$ in terms of the
the entropies and REE's of the reduced states of two parties:
\begin{equation}
\max\{E_{\rm R}(\rho_{\rm AB})+S(\rho_{\rm AB}),E_{\rm R}(\rho_{\rm
AC})+S(\rho_{\rm AC}),E_{\rm R}(\rho_{\rm BC})+S(\rho_{\rm BC})\}\le E_{\rm
R}(\rho_{\rm ABC}),
\end{equation}
where $\rho_{\rm AB}={\rm Tr}_C(\rho_{\rm ABC})$ (and similarly for $\rho_{\rm
AC}$ and $\rho_{\rm BC}$) and $S(\rho)\equiv - {\rm Tr}\rho\log_2\rho$ is the
von Neumann entropy. They have further found that this lower bound is
saturated by $\ket{\rm GHZ}$ and $\ket{\rm W}$. This raises an interesting
question: is the above lower bound (for $n$-partite pure states) saturated by
the states that saturate the lower bound
$E_{\log_2}=-2\log_2{\Lambda_{\max}}(\psi)\le E_{\rm R}(\psi)$? Numerical
tests seem to suggest that the Plenio-Vedral bound is tighter than
$E_{\log_2}$. If this is the case then all states that saturate the lower
bound $E_{\log_2}$ on $E_{\rm R}$ will saturate the Plenio-Vedral bound. Based
on Conjecture~1, we can show that for $\rho_{12\ldots n}=\ketbra{S(n,k)}$ the
inequality
\begin{equation}
\label{eqn:CRE} \max_i \{E_{\rm R}(\rho_{12\ldots\hat{i}\ldots n})+S(\rho_{
12\ldots\hat{i}\ldots n})\}\le E_{\rm R}(\rho_{12\ldots n})
\end{equation}
is saturated, where $\rho_{12\ldots\hat{i}\ldots n}\equiv{\rm
Tr}_i(\rho_{12\ldots n})$ is the reduced density matrix obtained from
$\rho_{12\ldots n}$ by tracing out the $i$-th party. The proof is as follows.
As $\ket{S(n,k)}$ is permutation-invariant, there is no need to maximize over
all parties, and we can simply take $i=1$, obtaining the reduced state
$\rho_{n-1;k-1,k}(k/n)$ as in Eq.~(\ref{eqn:Tr1}). As the corresponding
function $F_{n-1;k-1,k}(s)$ of $\rho_{n-1;k-1,k}(s)$ is convex for
$s\in[0,1]$, we immediately obtain from Conjecture~1 that, for
$\rho_{n-1;k-1,k}(k/n)$,
\begin{subequations}
\begin{eqnarray}
\!\!\!\!\!\!\!\!\!\!\!E_{\rm
R}\left(\rho_{n\!-\!1;k\!-\!1,k}(k/n)\right)&=&\log_2\left[C^n_k
\left(\frac{k}{n}\right)^k\left(\frac{n\!-\!k}{n}\right)^{n\!-\!k}\right]
+\frac{k}{n}\log_2\frac{k}{n}+\frac{n\!-\!k}{n}\log_2\frac{n\!-\!k}{n}\\
&=&E_{\rm R}\left(\ket{S(n,k)}\right)-
S\left(\rho_{n\!-\!1;k\!-\!1,k}(k/n)\right).
\end{eqnarray}
\end{subequations}
Therefore, the bound in Eq.~(\ref{eqn:CRE}) is saturated for $\rho_{12\ldots
n}=\ketbra{S(n,k)}$. }

A major challenge is to extend the ideas contained in the present Paper from
the relative entropy of entanglement to its regularized version, the latter in
fact giving tighter upper bound on the entanglement of distillation than the
former in the bi-partite settings. {The alternative way of defining the
relative entropy via the optimization over PPT states may also been used, in
view of the recent progress on the bi-partite regularized relative entropy of
entanglement~\cite{AudenaertEtAl}. }

We now explore the possibility that the geometric measures can provide lower
bounds on yet another entanglement measure---the entanglement of formation. If
the relationship $ E_{\rm R} \le E_{\rm F}$ between the two measures of
entanglement---the relative entropy of entanglement $E_{\rm R}$ and the
entanglement of formation $E_{\rm F}$---should continue to hold for {\it
multi-partite\/} states (at least for pure states), and if $E_{\rm F}$ should
remain a convex hull construction for mixed states, then we would be able to
construct a lower bound on the entanglement of formation:
\begin{eqnarray}
E_{\log_2}(\rho)& \equiv&\min_{p_i,\psi_i} \sum_i p_i E_{\log_2}(\ket{\psi_i})
\le \min_{p_i,\psi_i} \sum_i p_i E_{\rm R}(\ket{\psi_i}) \nonumber \\
&\le& \min_{p_i,\psi_i} \sum_i p_i E_{\rm F}(\ket{\psi_i})\equiv E_{\rm
F}(\rho),
\end{eqnarray}
where $\{p_i\}$ and $\{\psi_i\}$ are such that $\rho=\sum_i p_i
\ketbra{\psi_i}$. Thus, $E_{\log_2}(\rho)$ is a lower bound on $E_{\rm
F}(\rho)$. By using the inequality $(1-x^2)\log_2e\le -2\log_2 x$  (for $0\le
x\le 1$), one further has has that $(\log_2e)E_{\sin^2}(\rho)\le
{E}_{\log_2}(\rho)\le E_{\rm F}(\rho)$.

We remark that $E_{\sin^2}$ has been shown to be an entanglement
monotone~\cite{BarnumLinden01,WeiGoldbart03}, i.e., it is not increasing under
local operations and classical communication (LOCC). However, $E_{\log_2}$ is
{\it not\/} a monotone, as the following example shows. Consider the
bi-partite pure state
\begin{equation}
\ket{\psi}\equiv\frac{1}{\sqrt{1+N x^2}}\ket{00}+
\frac{x}{\sqrt{1+Nx^2}}\big(\ket{11}+\ket{22}+\dots+\ket{NN}\big),
\end{equation}
with $|x|\le 1$, for which $E_{\log_{2}}=\log_{2}(1+N x^2)$. Suppose that one
party makes the following measurement:
\begin{equation}
{\cal M}_1\equiv\ket{0}\bra{0}, \ \ \ {\cal
M}_2\equiv\ket{1}\bra{1}+\ket{2}\bra{2}+\dots+\ket{N}\bra{N}.
\end{equation}
With probability $P_1=1/(1+N x^2)$ the output state becomes
$\ket{\psi_1}=\ket{00}$; with probability $P_2=N x^2/(1+N x^2)$ the output
state becomes
$\ket{\psi_2}=\big(\ket{11}+\ket{22}+\dots+\ket{NN}\big)/\sqrt{N}$, for which
$E_{\log_{2}}=\log_{2}N$. For $E_{\log_{2}}$ to be a monotone it would be
necessary that
\begin{equation}
E_{\log_{2}}(\psi)\ge P_1 E_{\log_{2}}(\psi_1)+ P_2 E_{\log_{2}}(\psi_2).
\end{equation}
Putting in the corresponding values for the $P$'s and $E_{\log_{2}}$'s, we
find that this inequality is  equivalent to
\begin{equation}
\label{eqn:violatemono} f(N,x)\equiv\log_2(1+N x^2)- \frac{N x^2}{1+N
x^2}\log_2 N\ge 0.
\end{equation}
As this is violated for certain values of $x$ with $N>2$, as exemplified in
Fig.~\ref{fig:violate} for the plot of $f(4,x)$, we arrive at the conclusion
that $E_{\log_{2}}$ is, in general, not a monotone.

\begin{figure}
\centerline{\psfig{figure=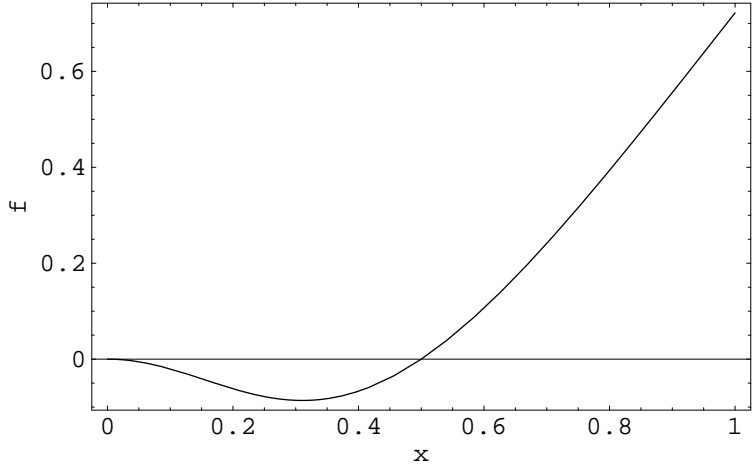,width=10cm,angle=0}}
\caption[Violation of the monotone condition]{ The function $f(4,x)$. It shows
the violation of monotone condition~(\ref{eqn:violatemono}) when the function
is negative. } \label{fig:violate}
\end{figure}
We conclude by mentioning that certain results reported in the present chapter
have recently been applied by Vedral~\cite{Vedral04} to the macroscopic
entanglement of $\eta$-paired superconductivity.

\chapter{Bound entanglement}
\label{chap:Bound}
\section{Introduction}
As we have discussed above, we are motivated to study the quantification of
entanglement for the basic reason that entanglement has been identified as a
{\it resource\/} central to much of quantum information processing.  As we
have also discussed above, to date, progress in the quantification of
entanglement for mixed states has resided primarily in the domain of
bi-partite systems. For multi-partite systems in pure and mixed states, the
characterization and quantification of entanglement presents even greater
challenges.

We have introduced previously the notion of the entanglement of distillation,
entanglement cost, and entanglement of formation. The entanglement of
distillation ($E_{\rm D}$) is the optimal asymptotic {\it yield\/} of Bell
states, given an infinite supply of replicas of an identical quantum state
shared between two distant parties. The entanglement cost ($E_{\rm C}$) is the
minimum asymptotic cost ratio of Bell states {\it consumed\/} to create an
large number of copies of a certain quantum state shared between two distant
parties. For pure states shared between two distant parties the two processes
are reversible, i.e., $E_{\rm D}=E_{\rm C}$. For mixed states it is expected
that $E_{\rm D}\le E_{\rm C}$; otherwise recycling the two processes would
churn out more entanglement than there was initially. The quantity
entanglement of formation $E_{\rm F}$ is an averaged version of $E_{\rm C}$.

All these three quantities, $E_{\rm D}$, $E_{\rm C}$, and $E_{\rm F}$,
especially the first two, are very difficult to calculate. The calculation of
these quantities for general quantum states remains a challenge in quantum
information theory. Another challenge regarding $E_{\rm F}$ concerns the
question of whether or not $E_{\rm F}$, defined in an average sense, is equal
to $E_{\rm C}$, which is defined for asymptotically large numbers of copies.
This is the so-called additivity problem for entanglement of formation,
mentioned earlier.


There is, however, some progress in small dimensions, especially for
two-qubit systems. Wootters' formula~\cite{Wootters98} for two-qubit
entanglement of formation is the most prominent example. It sets an upper
bound on the entanglement cost. It is obvious that an entangled state has
nonzero $E_{\rm C}$ and $E_{\rm F}$. It was also shown by Horodecki and
co-workers~\cite{HorodeckiHorodeckiHorodecki97} that all two-qubit entangled
states can be distilled into Bell states, and hence have nonzero $E_{\rm D}$.
It was then thought that all entangled states, however small the entanglement,
could be distilled. But shortly after, Horodecki and
co-workers~\cite{Horodecki398,Horodecki97} showed that in dimensions higher
than two-qubit ($C^2\otimes C^2$) and qubit-qubtrit ($C^2\otimes C^3$) [such
as $C^2\otimes C^4$ and $C^3\otimes C^3$] there exist entangled states that
have {\it zero\/} entanglement of distillation. These states, however
entangled, cannot be distilled into any pure entangled states. The
entanglement used to create them is somehow bound and inextractable! Such
states are called {\it bound entangled states\/}, and their entanglement is
called bound entanglement. The bound entangled states that Horodecki
constructed are bi-partite and have zero negativity, i.e., have positve
partial transpose (PPT). Bound entanglement is not limited to bi-partite
states. Bennett and co-workers~\cite{BennettDiVincenzoMorShorSmolinTerhal99}
constructed both bi- and multi-partite bound entangled states using the
technique of unextendible product bases; see also Appendix~\ref{app:UPB}.
Other multi-partite bound entangled states have also been
found~\cite{BrussPeres00}, including the
two~\cite{Smolin00,ShorSmolinThapliyal03,Dur01} that we shall discuss in this
chapter.

Although bound entanglement seems useless, as it cannot be used alone for
quantum communication,  Horodecki and co-workers~\cite{Horodecki399} found
that, surprisingly, a certain threshold fidelity of a teleportation process
that cannot be achieved via a single copy of non-maximally entangled pure
state {\it can\/} actually be achieved if combined with a supply of bound
entanglement. Recently, Shor and co-workers~\cite{ShorSmolinThapliyal03} have
shown that two multi-partite bound entangled states (of Smolin's), tensored
together, can be be distilled (via LOCC) into a Bell state shared between two
of the parties. Bound entanglement clearly has richer properties than was
thought initially.

We have seen that bound entangled states are states that are entangled, but
from which no pure entangled state can be distilled, provided all parties are
allowed only local operations and classical communication (LOCC). The
distillable entanglement ($E_{\rm D}$) is thus zero by definition. Bound
entangled states can be either bi-partite or multi-partite, the latter
possibly exhibiting more structure than the former. However, it does take
nonzero entanglement to {\it create\/} bound entangled states under LOCC. But
exactly how much entanglement is bound in these strange states is not known
analytically. In the present chapter, we study the entanglement content of two
distinct types of bound entangled state: Smolin's four-party unlockable bound
entangled state~\cite{Smolin00,ShorSmolinThapliyal03} and D\"ur's $N$-party
Bell-inequality-violating bound entangled states~\cite{Dur01}. For each, we
determine analytically  their geometric measure of entanglement $E_{\sin^2}$
and the related quantity ${E}_{\log_2}$. We have shown in the previous chapter
that, under certain circumstances, these give lower bounds on their
multi-partite $E_{\rm F}$. In particular, we showed that
$E_{\sin^2}(\rho)\log_2e\le {E}_{\log_2}(\rho)\le E_{\rm F}(\rho)$. In
addition, we make conjectures concerning the relative entropies of
entanglement for these bound entangled states.  Although quantities such as
the geometric measure or the relative entropy of entanglement may not be able
to reveal the exact nature of bound entanglement, they nevertheless quantify
for these bound entangled states the content of entanglement that is
inextractable. The discussion in the present chapter is based on
Ref.~\cite{WeiAltepeterGoldbartMunro03}.

\def\NONE{\section{Geometric measure of entanglement}
We begin by  briefly reviewing its formulation. Consider a general $n$-partite
pure state (expanded in the local bases $\{e_{p_i}^{(i)}\}$):
\begin{equation}
|\psi\rangle=\sum_{p_1\cdots p_n}\chi_{p_1p_2\cdots p_n}
|e_{p_1}^{(1)}e_{p_2}^{(2)}\cdots e_{p_n}^{(n)}\rangle.
\end{equation}
As shown in Ref.~\cite{WeiGoldbart03}, its closest {\it separable} (i.e.
product) pure state (with $i$ being the party index),
\begin{equation}
\ket{\phi}\equiv\mathop{\otimes}_{i=1}^n|\phi^{(i)}\rangle=\mathop{\otimes}_{i=1}^{n}
\Big({\sum}_{p_i}c_{p_i}^{(i)}\,|e_{p_i}^{(i)}\rangle\Big),
\end{equation}
satisfies the condition (and its complex conjugate)
\begin{equation}
\sum_{p_1\cdots\widehat{p_i}\cdots p_n} \chi_{p_1p_2\cdots
p_n}^*c_{p_1}^{(1)}\cdots\widehat{c_{p_i}^{(i)}}\cdots c_{p_n}^{(n)}=
\Lambda\,{c_{p_i}^{(i)}}^*,
\end{equation}
where the eigenvalue $\Lambda\in[-1,1]$ is associated with the Lagrange
multiplier enforcing $\ipr{\phi}{\phi}\!=\!1$, and the symbol
\,\,$\widehat{}$\,\, denotes exclusion. Moreover, $\Lambda$ is the cosine of
the angle between $|\psi\rangle$ and $\ket{\phi}$; the largest one,
$\Lambda_{\max}$, which we call the {\it entanglement eigenvalue\/},
corresponds to the closest separable state, and is the maximal overlap:
$\Lambda_{\max}(\ket{\psi})=\max_{\phi}|\ipr{\phi}{\psi}|$,
where $\ket{\phi}$ is separable but otherwise arbitrary. $E_{\sin^2}\equiv
1-\Lambda^2_{\max}(\ket{\psi})$ was defined to be the geometric measure of
entanglement~\cite{WeiGoldbart03} for  state $|\psi\rangle$, and it measures
the degree of inseparability via the squared sine of the angle away from the
closest separable pure state.

The extension  to mixed states  can be built upon  the pure-state theory, and
is made via the use of the {\it convex hull\/} construction, as was done for
$E_{\rm F}$~\cite{Wootters98}.  The essence is a minimization over all
decompositions $\rho=\sum_i p_i\,|\psi_i\rangle\langle\psi_i|$ into pure
states:
\begin{eqnarray}
E_{\sin^2}(\rho) \equiv {\min_{\{p_i,\psi_i\}}} \sum\nolimits_i p_i \,
E_{\sin^2}(|\psi_i\rangle).
\end{eqnarray}
This convex hull construction ensures that the measure gives zero for
unentangled states; however, it also complicates the task of determining
mixed-state entanglement. $E_{\sin^2}$ was shown to be
 an {\it entanglement
monotone\/}~\cite{BarnumLinden01,WeiGoldbart03} (i.e. the measure does not
increase under LOCC), hence is a good measure of entanglement. As there is no
explicit generalization of $E_{\rm F}$ to multi-partite states, we shall
calculate analytically $E_{\sin^2}$ for two bound entangled states: Smolin's
and D\"ur's. Because $E_{\rm F}(\rho)$ is the minimum average {\it ebit\/} to
create a single copy of $\rho$, we can regard $E_{\sin^2}(\rho)$ as the
minimum average degree of pure inseparability needed to realize the state
$\rho$.

In bi-partite settings, it is known~\cite{VedralPlenio98} that $E_{\rm
R}(\rho)\le E_{\rm F}(\rho)$, and that for pure states $\ket{\psi}$, $E_{\rm
R}({\psi})=E_{\rm F}({\psi})$. It is also
known~\cite{WeiEricssonGoldbartMunro} that for any bi- and multi-partite {\it
pure\/} state $\ket{\psi}$, ${E}_{\log_2}(\psi)\equiv-
2\log_2\Lambda_{\max}(\psi) \le E_{\rm R}(\psi)$. Together with the inequality
$(1-x^2)\log_2e\le -2\log_2 x$  (for $0\le x\le 1$), one has
\begin{eqnarray}
\sum\nolimits_i\! p_i \, E_{\sin^2}(\psi_i)\log_2e\le\! \sum\nolimits_i\! p_i
\, {E}_{\log_2}(\psi_i)\le \!\sum\nolimits_i p_i \, {E}_{\rm F}(\psi_i),
\nonumber
\end{eqnarray}
and thus $(\log_2e)E_{\sin^2}(\rho)\le {E}_{\log_2}(\rho)\le E_{\rm F}(\rho)$
for any {\it bi-partite\/} state $\rho$. If the generalization of $E_{\rm F}$
to multi-partite states maintains the property that $E_{\rm F}(\psi)\ge E_{\rm
R}(\psi)$ then the inequality $(\log_2e)E_{\sin^2}\le {E}_{\log_2} \le E_{\rm
F}$ will continue to hold for multi-partite mixed states. We remark that
\begin{equation}
{E}_{\log_2}(\rho)\equiv {\min_{\{p_i,\psi_i\}}}\sum\nolimits_i p_i \Big(-2
\log_2\Lambda_{\max}(\psi_i)\Big)
\end{equation}
is not an entanglement monotone~\cite{WeiEricssonGoldbartMunro}. However, we
see that both $(\log_2e)E_{\sin^2}$ and ${E}_{\log_2}$ could serve as lower
bounds on multi-partite entanglement of formation. } We now turn to the
calculations of entanglement for the two bound entangled states: Smolin's and
D\"ur's.

\section{Smolin's four-party unlockable bound entangled state}
Consider the four-qubit mixed state
\begin{equation}
\label{eqn:rhoABCD}
\rho^{ABCD}\equiv\frac{1}{4}\sum_{i=0}^3\big(\ketbra{\Psi_i}\big)_{\rm
AB}\otimes \big(\ketbra{\Psi_i}\big)_{\rm CD},
\end{equation}
where the $\ket{\Psi}$'s are the four Bell states:
$(\ket{00}\pm\ket{11})/\sqrt{2}$ and $(\ket{01}\pm\ket{10})/\sqrt{2}$. Now,
the state $\rho^{ABCD}$ can be conveniently rewritten as
\begin{equation}
\label{eqn:rhoABCD2} \rho^{ABCD}=\frac{1}{4}\sum_{i=0}^3\ketbra{\Chi_i},
\end{equation}
where $\ket{\Chi}$'s are the four orthogonal GHZ-like states:
\begin{eqnarray}
&&\!\!\!\!\!\!
\ket{\Chi_0}\!\equiv\!\frac{1}{\sqrt{2}}\big(\ket{0000}\!+\!\ket{1111}\big),
\
\ket{\Chi_1}\!\equiv\!\frac{1}{\sqrt{2}}\big(\ket{0011}\!+\!\ket{1100}\big),\nonumber\\
&&\!\!\!\!\!\!
\ket{\Chi_2}\!\equiv\!\frac{1}{\sqrt{2}}\big(\ket{0101}\!+\!\ket{1010}\big), 
\
\ket{\Chi_3}\!\equiv\!\frac{1}{\sqrt{2}}\big(\ket{0110}\!+\!\ket{1001}\big).\nonumber
\end{eqnarray}
From the decomposition in Eq.~(\ref{eqn:rhoABCD2}) we immediately see that the
state $\rho^{ABCD}$ is invariant under any permutations of the four parties.

If any two of the four parties, say C and D, get together, they can do a
Bell-measurement (namely, measurement done in the basis of $\ket{\Psi_i}$'s).
Depending on the result $i=0,\ldots,3$, they broadcast the outcome to A and B,
from which A and B can then establish a definite Bell state. This shows that
the state $\rho^{ABCD}$ must be entangled. But if all the four parties are far
apart, they have no way of distilling any pure entangled states. This can be
seen from the fact proven in Ref.~\cite{BennettDiVincenzoSmolinWootters96}
that if two parties are on opposite sides of a separable cut, then they will
remain in a separable cut under any local quantum operations and classical
communication. From the form in Eq.~(\ref{eqn:rhoABCD}) we see that A cannot
establish any entanglement between either C or D, as A is in the opposite side
of a separable cut from C and D. But from Eq.~(\ref{eqn:rhoABCD2}) we know
that the state is invariant under any permutation of the parties, hence, A
cannot establish any entanglement with B (by exchanging C with B), either.
Therefore, the state $\rho^{ABCD}$ is bound entangled.

Our goal here is to calculate how much entanglement is bound in the state. As
the state is bound entangled, it is equivalent to ask how entangled is the
state? For the purpose of using GME to quantify entanglement, we need to
characterize all decompositions of the mixed state into pure states. The most
general decomposition of a mixed state $\rho$ into pure states can be
expressed as
\begin{equation}
\label{eqn:rhoDecomp} \rho=\sum_{k=1}^{\cal M} \ketbra{\tilde{\varphi}_k},\ \
{\rm with} \ \ket{\tilde{\varphi}_k}=\sum_{i=1}^n {\cal
U}_{ki}\sqrt{\lambda_i}\,\ket{\xi_i},
\end{equation}
where ${\cal M}$ is an integer not smaller than $n$, the number of orthonormal
eigenvectors $\{\ket{\xi_i}\}$ (with nonzero eigenvalues $\{\lambda_i\}$) of
$\rho$, $\ket{\tilde{\varphi}}$'s are {\it un-normalized\/}, and ${\cal U}$
satisfies
$\sum_{k=1}^{\cal M}{\cal U}_{ki}\,{\cal U}^*_{kj}=\delta_{ij}$.
Thus, the most general pure state that appears in the decomposition of
Smolin's state is
\begin{equation}
\label{eqn:varphi} \ket{\tilde{\varphi}_k}=\sum_{i=0}^3 \frac{1}{2}\,{\cal
U}_{ki}\ket{\Chi_i}.
\end{equation}
Our goal is to minimize
$\sum_{k} p_k \, E_{\rm pure}\big(\ket{{\varphi}_k}\big)$
over all possible ${\cal U}$'s, where $E_{\rm pure}$ is some pure-state
entanglement ($E_{\sin^2}$ or ${E}_{\log_2}$ in our considerations),
$p_k\equiv \ipr{\tilde{\varphi}_k}{\tilde{\varphi}_k}$, and $\ket{\varphi_k}$
is the normalized state $\ket{\varphi_k}\equiv
\ket{\tilde{\varphi}_k}/\sqrt{p_k}$. Making a general minimization for an
arbitrary mixed state is extremely difficult. However, for the mixed state
$\rho^{ABCD}$ we shall show that the decomposition in Eq.~(\ref{eqn:rhoABCD2})
does indeed minimize the average entanglement over pure-state decompositions.
As in Eq.~(\ref{eqn:varphi}), $\ket{\varphi}$ can be explicitly written as
$\ket{{\varphi}}=\sum_{i=0}^3 \sqrt{q_i}\, e^{i\,\phi_i}\ket{\Chi_i}$, 
where the $q$'s are non-negative, satisfying $\sum_iq_i=1$, and the $\phi$'s
are phases. For fixed $q$'s, the state has a maximal entanglement eigenvalue
when all phases are zero. We shall show shortly that its maximal entanglement
eigenvalue is $1/\sqrt{2}$, which is achieved by the $\ket{\Chi}$'s.

The entanglement eigenvalue of the state
$\ket{{\varphi}}=\sum_{i=0}^3 \sqrt{q_i}\, \ket{\Chi_i}$
is the maximal overlap with the separable state
$\ket{\Phi}=\otimes_{i=1}^4 \big(c_i\ket{0}+s_i\ket{1}\big)$,
where $c_i\equiv\cos\theta_i$ and $s_i\equiv\sin\theta_i$ with $0\le
\theta_i\le \pi/2$. Thus
\begin{eqnarray}
\ipr{\Phi}{{\varphi}}&=& \sqrt{{q_0}/{2}}\,(c_1c_2c_3c_4
                    +s_1s_2s_3s_4)+         \sqrt{{q_1}/{2}}\,(c_1c_2s_3s_4
                   +s_1s_2c_3c_4)\nonumber \\
&+&         \sqrt{{q_2}/{2}}\,(c_1s_2c_3s_4
                    +s_1c_2s_3c_4)
            +\sqrt{{q_3}/{2}}\,(c_1s_2s_3c_4
                   +s_1c_2c_3s_4),
\nonumber
\end{eqnarray}
which has maximum $1/\sqrt{2}$. To see this, use the Cauchy-Schwarz
inequality, treating as one vector
$\big\{\sqrt{q_0/2},\sqrt{q_1/2},\sqrt{q_2/2},\sqrt{q_3/2}\big\}$ (whose
modulus is $1/\sqrt{2}$\,), and the corresponding coefficients as another
vector, whose modulus can be shown to be no greater than 1:
\begin{eqnarray}
&&(c_1c_2c_3c_4
                 +s_1s_2s_3s_4)^2
+           (c_1c_2s_3s_4
                    +s_1s_2c_3c_4)^2 \nonumber \\
&&+         (c_1s_2c_3s_4
                    +s_1c_2s_3s_4)^2
+           (c_1s_2s_3c_4
                    +s_1c_2c_3s_4)^2\le 1\nonumber.
\end{eqnarray}
 By subtracting the left-hand side from 1
and making some algebraic manipulation,  we arrive at the non-negative
expression (hence the sought result):
\begin{eqnarray}
 &&(c_1c_2c_3s_4-s_1s_2s_3c_4)^2+(c_1c_2s_3s_4-s_1s_2c_3s_4)^2+\nonumber \\ &&(c_1s_2c_3c_4-
 s_1c_2s_3s_4)^2+(s_1c_2c_3c_4-c_1s_2s_3s_4)^2\ge0\nonumber.
\end{eqnarray}

The states $\ket{\Chi_i}$ are GHZ-like states and have
$\Lambda_{\max}=1/\sqrt{2}$ and they clearly saturate
$|\ipr{\Phi}{{\varphi}}|\le 1/\sqrt{2}$. Hence, we have
\begin{equation}
E_{\sin^2}(\rho^{ABCD})={1}/{2} , \ \ {E}_{\log_2}(\rho^{ABCD}) = 1.
\end{equation}
This suggests that although bound entangled, Smolin's state has a very high
degree of entanglement, the same as that of a 4-partite GHZ state. This high
degree of entanglement seems to manifest in some bi-partite partitioning,
e.g., \{A:BCD\} (as we discuss below).

To compare the results with other measures of entanglement, we conjecture (and
later prove) that $E_{\rm R}\!=\!1$ for this state and one of its closest
separable mixed states is
\begin{eqnarray}
&&\!\!\!\!\!\!\!\!\frac{1}{8}\!\big(\ketbraS{0000}\!+\!\ketbraS{1111}\!+\!\ketbraS{0011}
\!+\!\ketbraS{1100}\nonumber\\
&&\!\!\!\!\!\!\!\!+\ketbraS{0101}\!+\!\ketbraS{1010}\!+\!\ketbraS{0110}\!+\!\ketbraS{1001}\big).\nonumber
\end{eqnarray}
The negativity ${\cal N}$ (a value used to quantify the degree of bi-partite
inseparability of states and defined as twice the absolute sum of negative
eigenvalues of the partial transpose (PT) of the density matrix with respect
to some bi-partite partitioning) is zero for any 2/2 partitioning, e.g., \{AB
: CD\}, but nonzero for 1/3 partitioning, e.g.,\{A:BCD\}. (This nonzero
negativity also demonstrates that the state $\rho^{ABCD}$ is entangled.)
Specifically, ${\cal N}_{\rm A:BCD}=1$ but ${\cal N}_{\rm AB:CD}=0$.

Let us now turn to D\"ur's bound entangled states.
\section{D\"ur's $N$-party  bound entangled states}
\label{sec:DurBound} D\"ur~\cite{Dur01} found that for $N\ge 4$ the following
state is bound entangled:
\begin{equation}
\label{eqn:Dur}
\rho_N\equiv\frac{1}{N+1}\left(\ketbra{\Psi_G}+\frac{1}{2}\sum_{k=1}^N\big(P_k+\bar{P}_k\big)\right),
\end{equation}
where $\ket{\Psi_G}\equiv \big(\ket{0^{\otimes
N}}+e^{i\alpha_N}\ket{1^{\otimes N}}\big)/        {\sqrt{2}}$ is a $N$-partite
GHZ state; $P_k\equiv\ketbra{u_k}$ is a projector onto the state
$\ket{u_k}\equiv\ket{0}_1\ket{0}_2\ldots\ket{1}_k\ldots\ket{0}_N$; and
$\bar{P}_k\equiv\ketbra{v_k}$ projects onto
$\ket{v_k}\equiv\ket{1}_1\ket{1}_2\ldots\ket{0}_k\ldots\ket{1}_N$. For $N\ge
8$ this state violates the Mermin-Klyshko-Bell inequality~\cite{Dur01};
violation was pushed down to $N\ge 7$ by Kaszlikowski and
co-workers~\cite{Kwek02} for a three-setting Bell inequality; it was pushed
further down to $N\ge 6$ by Sen and co-workers~\cite{SenSenZukowski02} for a
functional Bell inequality. For these inequalities, see
Appendix~\ref{app:Bell2} for more detail. These results are interesting and
somewhat surprising, as one might expect that bound entangled states has low
entanglement that they could not violate any Bell inequality. So how entangled
are D\"ur's bound entangled states?

We remark that the phase $\alpha_N$ in $\ket{\Psi_G}$ can be eliminated by
local unitary transformations, and hence we shall take $\alpha_N=0$ in the
following discussion. In fact, if we consider the family of states
\begin{equation}
\rho_N(x)\equiv x\ketbra{\Psi_G}+\frac{1-x}{2N}
\sum_{k=1}^N\big(P_k+\bar{P}_k\big),
\end{equation}
we find that for $N\ge 4$ the state is bound entangled if $0<x\le 1/(N+1)$ and
is still entangled but not bound entangled if $x > 1/(N+1)$. This can be seen
from the fact that the negativities of $\rho_N(x)$ with respect to the two
different partitions $(1:2\cdots N)$ and $(12:3\cdots N)$ are
\begin{subequations}
\begin{eqnarray}
&&\!\!\!\!\!\!\!\!\!\!{\cal N}_{1:2\cdots
 N}\big(\rho_N(x)\big)=\max\left\{0,[{(N\!+\!1)\,x-1}\,]/{N}\,\right\}, \\
 &&\!\!\!\!\!\!\!\!\!\!{\cal N}_{12:3\cdots
 N}\big(\rho_N(x)\big)=x.
\end{eqnarray}
\end{subequations}

By applying arguments similar to those used to calculate entanglement for
Smolin's state, we have that the general pure state in the decomposition of
$\rho_N(x)$ is
\begin{equation}
\sqrt{y}\,e^{i\phi_0}\ket{\Psi_G}+\sqrt{1\!-\!y}\sum_{k=1}^N
\big(\sqrt{q_k}e^{i \phi_{i}}\ket{u_i}+ \sqrt{r_k}e^{i
\phi'_{i}}\ket{v_i}\big),\nonumber
\end{equation}
where $q$'s and $r$'s are non-negative and satisfy $\sum_k(q_k+r_k)=1$. In
this family, the state with the least entanglement (or maximum
$\Lambda_{\max}$) for fixed $\{y,q_k,r_k\}$ is the one with all phase factors
zero:
\begin{equation}
\ket{\Psi\big(y,\{q,r\}\big)}\equiv\sqrt{y}\ket{\Psi_G}+\sqrt{1\!-\!y}\sum_{k=1}^N
\big(\sqrt{q_k}\ket{u_i}+ \sqrt{r_k}\ket{v_i}\big).\nonumber
\end{equation}
Next, we ask: For fixed $y$, what is the least entanglement that the above
state can have? Take a separable state of the form
$\ket{\Phi}=\otimes_{i=1}^N
\big(c_i\ket{0}+s_i\ket{1}\big)$; 
its overlap with $\ket{\Psi\big(y,\{q,r\}\big)}$ is then
\begin{eqnarray}
\ipr{\Psi}{\Phi} =\sqrt{{y}/{2}}\, (c_1\cdots c_N+s_1\cdots s_N)
+\sqrt{1\!-\!y}\sum_{k=1}^N (\sqrt{q_k}\,c_1\cdots s_k\cdots c_N
+\sqrt{r_k}\,s_1\cdots c_k\cdots s_N).\nonumber
\end{eqnarray}
This can be shown to no greater than $\sqrt{(2-y)/2}$, again by a
Cauchy-Schwarz inequality, taking
\begin{equation}
\left\{ \sqrt{{y}/{2}}, \big\{\sqrt{(1-y)q_k}\big\},
 \big\{\sqrt{(1-y)r_k}\big\}\right\}\nonumber
\end{equation}
as the first $(2N\!+\!1)$-component vector (with modulus $\sqrt{(2\!-\!y)/2}$)
and the corresponding coefficients as the second one, whose modulus can be
shown to be no greater than 1 for $N\ge 4$:
\begin{eqnarray}
f_N\equiv\big(c_1\cdots c_N+s_1\cdots s_N\big)^2 + \sum_{k=1}^N\left\{
(c_1\cdots s_k\cdots c_N)^2 +(s_1\cdots c_k\cdots s_N)^2\right\}\le
1.\nonumber
\end{eqnarray}
First, making similar arguments as previously, one can show that $f_4\le 1$.
One can also show that $f_{N+1}\le f_N$. Thus by induction, we have proved the
inequality.

The bound can be saturated, e.g., by
\begin{subequations}
\label{eqn:psiuv}
\begin{eqnarray}
&&\ket{\psi_{\pm,u,k}(y)}\equiv\sqrt{y}\ket{\Psi_G}\pm\sqrt{1-y}\ket{u_k},\\
&&\ket{\psi_{\pm,v,k}(y)}\equiv\sqrt{y}\ket{\Psi_G}\pm\sqrt{1-y}\ket{v_k},
\end{eqnarray}
\end{subequations}
for which $\Lambda_{\max}(y)=\sqrt{(2-y)/2}$~\cite{footnote}. This can be seen
as follows.
As one can make local relative phase shifts to transform
$\sqrt{y}\ket{\Psi_G}+\sqrt{1-y}\ket{u_k}$ to
$\sqrt{y}\ket{\Psi_G}-\sqrt{1-y}\ket{u_k}$, they have the same entanglement.
The change from $\sqrt{y}\ket{\Psi_G}\pm\sqrt{1-y}\ket{u_k}$ to
$\sqrt{y}\ket{\Psi_G}\pm\sqrt{1-y}\ket{v_k}$ is simply a flipping of 0 to 1,
and vice versa. The mapping from $k$ to $k'$ is just a relabelling of parties.
Thus, we need only consider the state
\begin{equation}
\sqrt{y/{2}}\,(\ket{00\cdots0}+\ket{11\cdots1})+\sqrt{1-y}
\ket{10\cdots0}.\nonumber
\end{equation}
As this state is invariant under permutation of all parties except the first
one, and as the coefficients are non-negative, in order to find the maximal
overlap we can make the hypothesis that the closest separable state is of the
form
\begin{equation}
\left(\sqrt{p}\ket{0}+\sqrt{1-p}\ket{1}\right)\otimes
(\sqrt{q}\ket{0}+\sqrt{1-q}\ket{1})^{\otimes N-1}. \nonumber
\end{equation}
We further see that in order for the overlap to be maximal, $q$ must be either
1 or 0. For the former case, we can further maximize the overlap to get
$\sqrt{(2-y)/2}$. For the latter case, the maximum overlap is $\sqrt{y/2}$,
which is less than $\sqrt{(2-y)/2}$ (as $0\le y\le 1$). Hence, the state
$\sqrt{y}\ket{\Psi_G}\pm\sqrt{1-y}\ket{u_k}$ has the entanglement eigenvalue
$\sqrt{(2-y)/2}$.

As $1-\Lambda^2_{\max}(y)$ is linear in $y$ and $-2\log_2\Lambda_{\max}(y)$ is
convex in $y$, one gets
\begin{equation}
E_{\sin^2}(\rho_N(x)) =\frac{x}{2}, \ \
{E}_{\log_2}(\rho_N(x))=\log_2\frac{2}{2-x},
\end{equation}
and one of the optimal decompositions is
\begin{equation}
\rho_N(x)=\frac{1}{4N}\sum_{k=1}^{N}\sum_{\alpha=\pm}\sum_{\beta=u,v}\ketbra{\psi_{\alpha,\beta,k}(x)}.
\end{equation}
The above calculations show that for $\rho_N(x)$, the entanglement depends on
the portion $x$ of the GHZ in states $\ketbra{\psi_{\alpha,\beta,k}(x)}$ and
it never becomes zero unless there is no GHZ mixture.

We conjecture that, for $N\ge 4$, $\rho_N(x)$ has $E_R(x)=x$, with one closest
separable mixed state being
\begin{equation}
\frac{x}{2}\big(\ketbra{0..0}+\ketbra{1..1}\big)
+\frac{1-x}{2N}\sum_{k=1}^N\big(P_k +\bar{P}_k\big),\nonumber
\end{equation}
which seems plausible as $\big(\ketbra{0..0}+\ketbra{1..1}\big)$ is a closest
separable mixed state to  $\ket{\Psi_G}$.

\section{Concluding remarks}
\label{sec:concludeBound} We have presented analytical results on how much
entanglement is bounded in two distinct multi-partite bound entangled states.
The measure we have used to quantify their entanglement is the geometric
measure of entanglement (GME), whose construction, similiar to the
entanglement of formation ($E_{\rm F}$), is via convex hull. In contrast to
GME, $E_{\rm F}$ has not been explicitly generalized to multi-partite states,
and hence is still unavailable for these bound entangled states. However,
under the circumstances discussed previously, the results of $E_{\sin^2}$ as
well as a related quantity, ${E}_{\log_2}$, might provide lower bounds on
$E_{\rm F}$.
 For the Smolin state, its bound entanglement is
as large as that of a four-partite GHZ state, whereas that for D\"ur states is
related to the portion of the $N$-partite GHZ state. For each case, an optimal
decomposition is given. Furthermore, we have conjectured that the relative
entropy of entanglement ($E_{\rm R}$) for the Smolin state is unity (proved
below), whereas we conjecture that $E_{\rm R}$ for D\"ur's state is equal to
the portion that is $N$-GHZ.

For  Smolin's  state we can establish its $E_{\rm F}$, $E_{\rm D}$, $E_{\rm
R}$
and $E_{\sin^2}$ 
for certain bi-partite partitionings. For example, if we group the four
parties ABCD in two, A:BCD, we can write the state as
\begin{equation}
\rho^{A:BCD}=\frac{1}{4}\sum_{i=0}^3\ketbra{\bar{\Chi}_i},
\end{equation}
with the 3-qubit states of BCD mapped on to the 8-level system
($000\rightarrow \underline{0}, 001\rightarrow\underline{1},...,
111\rightarrow\underline{7}$), involving the locally orthogonal and
convertible states (by BCD)
\begin{eqnarray}
&&\!\!\!\!\!
\ket{\bar{\Chi}_0}=\big(\ket{0\underline{0}}\!+\!\ket{1\underline{7}}\big)/{\sqrt{2}},
\ \ \
\ket{\bar{\Chi}_1}=\big(\ket{0\underline{3}}\!+\!\ket{1\underline{4}}\big)/{\sqrt{2}},\nonumber\\
&&\!\!\!\!\!
\ket{\bar{\Chi}_2}=\big(\ket{0\underline{5}}\!+\!\ket{1\underline{2}}\big)/{\sqrt{2}},
\ \ \
\ket{\bar{\Chi}_3}=\big(\ket{0\underline{6}}\!+\!\ket{1\underline{1}}\big)/{\sqrt{2}}.\nonumber
\end{eqnarray}
In order to find the entanglement of this bi-partite state (in $C^{2}\otimes
C^{8}$), we need to consider the entanglement of the general  (properly
normalized) pure state
\begin{equation}
\ket{\psi}\equiv{\sum}_i\sqrt{x_i}\, e^{i\phi_i}\ket{\bar{\Chi}_i} \nonumber
\end{equation}
that appears in the pure-state decompositions. In fact, regardless of the
values of the ${x_i}$'s, this pure state  has a reduced density matrix
(tracing over BCD) of the form
$\left(\ketbra{0}+\ketbra{1}\right)/2$.
This shows that $\rho^{A:BCD}$ has $E_{\rm F}=1$, $E_{\sin^2}=1/2$, and
${E}_{\log_2}=1$. In fact, there is a general result due to Horodecki and
co-workers~\cite{Horodecki398a} that $E_{\rm D}=E_{\rm F}$ for mixture of
locally orthogonal bi-partite states, e.g., $C^2\otimes C^{2m}$ states that
are derived from mixing Bell-like states
\begin{equation}
\label{eqn:Belllike}
\ket{\Psi^\pm_k}\equiv(\ket{0,\underline{k}}\pm\ket{1,\underline{2m-k-1}})/{\sqrt{2}},
\end{equation}
having {\it distinct\/} $k$'s, where $k=0,1,\ldots, m-1$.
 As $E_{\rm D}\le E_{\rm R}\le E_{\rm F}$, we have that $E_{\rm R}(\rho^{\rm A:BCD})=1$ as well. What about the original four-partite state
$\rho^{\rm ABCD}$? As $E_{\rm R}(\rho^{\rm ABCD})\ge E_{\rm R}(\rho^{\rm
A:BCD})$, we have $E_{\rm R}(\rho^{\rm ABCD})\ge 1$. But we also have that
$E_{\rm R}(\rho^{\rm ABCD})\le 1$, as our previous conjecture gives at least
an upper bound; we thus have that $E_{\rm R}(\rho^{\rm ABCD})=1$ and the
conjecture is proved. Naively, we expect that any arbitrary $\rho^{\rm ABCD}$
has greater entanglement than $\rho^{\rm A:BCD}$. However, for the Smolin
state, they have the same entanglement as quantified by both GME and the
relative entropy of entanglement.

Although D\"ur's bound entangled state violates a Bell inequality, it has
nonzero negativity under certain partitionings. One may raise the question:
Does there exist a bound entangled state that has positive PT (PPT) under all
partitionings but that  still violates a Bell's inequality? For example, does
a UPB bound entangled state~\cite{BennettDiVincenzoMorShorSmolinTerhal99}
violate a Bell inequality? We shall see shortly that the answer is ``No'', at
least for the three different Bell
inequalities~\cite{Dur01,Kwek02,SenSenZukowski02} mentioned earlier. Ac\'\i n
has shown~\cite{Acin02} that if an $N$-qubit state violates a two-setting Bell
inequality then it is distillable under certain  bi-partite partitioning.
Using the results of Refs.~\cite{DurCirac00,DurCirac00B} regarding
distillability, we can repeat the same analysis for the other two
inequalities~\cite{Kwek02,SenSenZukowski02} and indeed obtain the same
conclusion. We analyze this as follows.

 It was shown by D\"ur and Cirac~\cite{DurCirac00B} that an arbitrary $N$-qubit state $\rho$ can be locally depolarized
into the form
\begin{eqnarray}
\rho_N =\lambda_0^+\ketbra{\Psi^+_0}+\lambda_0^-\ketbra{\Psi^-_0} +
\sum_{j=1}^{2^{N\!-\!1}-1}\lambda_j\big(\ketbra{\Psi^+_j}+\ketbra{\Psi^-_j}\big),\nonumber
\end{eqnarray}
while preserving $\lambda_0^\pm=\langle\Psi_0^\pm|\rho|\Psi_0^\pm\rangle$ and
$\lambda_j=\langle\Psi_j^+|\rho|\Psi_j^+\rangle+\langle\Psi_j^-|\rho|\Psi_j^-\rangle$,
where $\ket{\Psi^\pm_0}\equiv (\ket{0^{\otimes N}}\pm\ket{1^{\otimes
N}})/\sqrt{2}$, and the $\ket{\Psi^\pm_j}$'s are GHZ-like states, i.e., the
states in~(\ref{eqn:Belllike}), unfolded into qubit notation. Normalization
gives the condition
\begin{equation}
\lambda_0^++\lambda_0^-+2\sum_j\lambda_j=1.\nonumber
\end{equation}
Now define $\Delta\equiv \lambda_0^+-\lambda_0^-$, which we assume to be
non-negative (w.l.o.g). The condition that there is no bi-partite
distillability for some bi-partite partitioning $P_j$ is~\cite{DurCirac00}
\begin{equation}
2\lambda_j\ge\Delta.\nonumber
\end{equation}
Assuming non-distillability for {\it all\/} bi-partite splittings, we have
\begin{equation}
2\sum_j \lambda_j = 1-(\lambda_0^++\lambda_0^-) \ge
(2^{N\!-\!1}-1)\Delta.\nonumber
\end{equation}
As $\lambda_0^++\lambda_0^-\ge \Delta$, we have further that
\begin{equation}
\label{eqn:NoDistill} 1-\Delta \ge (2^{N\!-\!1}-1)\Delta.
\end{equation}
For the Mermin-Klyshko-Bell inequality, violation implies
$\Delta> 1/ 2^{(N\!-\!1)/2}$. 
For the three-setting Bell inequality considered in~\cite{Kwek02},
violation 
implies
$\Delta> \sqrt{3} \,(2^N/3^N).$ 
For the functional Bell inequality in~\cite{SenSenZukowski02},
violation 
implies
$\Delta >2 \,(2^N/\pi^N).$ 
One can easily check that the three Bell inequalities considered are
inconsistent with non-bipartite-distillability condition,
Eq.~(\ref{eqn:NoDistill}). Hence, the violating of these three Bell
inequalities implies the existence of some bi-partite distillability.

 This bi-partite distillability then implies a
negative PT (NPT) under that bi-partite partitioning according to Horodecki
and co-workers~\cite{Horodecki398}. Hence, violating these Bell inequalities
implies NPT under certain bi-partite partitioning. Said equivalently, if an
$N$-qubit state has PPT under all bi-partite partitionings then the state
never violates these Bell inequalities. This seems to suggest that PPT bound
entangled states are truly bound in nature that cannot give deviation from
local theories.

\chapter{Global entanglement and quantum criticality in spin chains}
\label{chap:QPT}
\section{Introduction}
Entanglement has been recognized in the past decade as a useful resource in
quantum information processing. Very recently, it has emerged as an actor on
the nearby stage of quantum many-body physics, especially for systems that
exhibit quantum phase
transitions~\cite{OsborneNielsen02,OsterlohAmicoFalciFazio02,VidalLatorreRicoKitaev03,CalabreseCardy04,SommaOrtizBarnumKnillViola04},
where it can play the role of a diagnostic of quantum correlations.  Quantum
phase transitions~\cite{Sachdev} are transitions between qualitatively
distinct phases of quantum many-body systems, driven by quantum fluctuations.
In view of the connection between entanglement and quantum correlations, one
anticipates that entanglement will furnish a dramatic signature of the quantum
critical point. On the other hand, the more entangled a state is, the more
useful it is likely to be as resource for quantum information processing.
 It is thus desirable to study and quantify
the degree of entanglement near quantum phase transitions.  By employing
entanglement to diagnose many-body quantum states one may obtain fresh insight
into the quantum many-body problem.

To date, progress in quantifying entanglement has taken place primarily in the
domain of bi-partite systems.  Much of the previous work on entanglement in
quantum phase transitions has been based on bi-partite measures, i.e., focus
has been on entanglement either between pairs of
parties~\cite{OsborneNielsen02,OsterlohAmicoFalciFazio02} or between a part
and the remainder of a system~\cite{VidalLatorreRicoKitaev03}.  For
multi-partite systems, however, the complete characterization of entanglement
requires the consideration of multi-partite entanglement, for which a
consensus measure has not yet emerged.

Singular and scaling behavior of entanglement near quantum critical points was
discovered in important work by Osterloh and
co-workers~\cite{OsterlohAmicoFalciFazio02}, who invoked Wootters' {\it
bi-partite\/} concurrence~\cite{Wootters98} as a measure of entanglement. The
drawback of concurrence is that it can deal with only two spins (each with
spin-1/2) even though the system may contain an infinite number of spins.
Although attempts have been made to generalize concurrence to many spin-1/2
systems via the time reversal operation, the generalized concurrence loses its
connection to the entanglement of formation~\cite{WongChristensen01}.

Another approach is to consider the von Neumann entropy of a subsystem of $L$
spins with the rest $N-L$ spins of the system. It is found that for critical
spin chains the entropy scales logarithmically with the subsystem size $L$ for
$N\rightarrow\infty$, with a prefactor that is related to the central charge
of the corresponding conformal
theory~\cite{VidalLatorreRicoKitaev03,CalabreseCardy04}. However, the
entanglement addressed in this case is not truly many-body, but only between a
subsystem and the rest of the system, although the connection to central
charge is interesting.

Quite recently, Barnum and co-workers~\cite{SommaOrtizBarnumKnillViola04} have
developed an entanglement measure, which they call generalized entanglement.
Instead of using subsystems they use different algebras and generalized
coherent states to define the entanglement. They have also applied the
generalized entanglement to systems exhibiting quantum phase transitions.
Their approach opens a new approach to multi-partite entanglement. However,
there is no {\it a priori\/} choice of which algebra, amongst all possible
ones, is the most natural one to use.

In addtion to spin chains other models that have been studied by using either
the von Neumann entropy or the concurrence as the entanglement measure
include: (i) the super-radiance model, in which many two-level atoms interact
with a single-mode photon field~\cite{LambertEmaryBrandes04}; and (ii) the
one-dimensional extended Hubbard model, in which electrons can hop between the
nearest neighbors and there are Coulomb interactions among electrons on the
same site and with nearest-neighbor electrons as well~\cite{GuDengLiLin04}.
Verstraete and co-workers~\cite{VerstraeteMartin-DelgadoCirac04} have recently
defined an entanglement length, viz., the distance at which two sites can
establish a pure-state entanglement at the cost of measuring all other sites.
They found that this entanglement length is usually greater than the
correlation length. All these, including the theme of the present chapter, are
aimed at approaching many-body problems from different, and hopefully fresh,
prespectives.

In the present chapter, we apply the {\it global\/} measure that we have
developed in previous chapters, based on a geometric picture; it provides a
{\it holistic\/}, rather than bi-partite, characterization of the entanglement
of quantum many-body systems.  Our focus is on one-dimensional spin systems,
specifically ones that are exactly solvable and exhibit quantum criticality.
For these systems, we are able to determine the entanglement analytically, and
to observe that it varies in a singular manner near the quantum critical line.
This supports the view that entanglement---the non-factorization of wave
functions---reflects quantum correlations. Moreover, the boundaries between
different phases can be detected by the  entanglement.

\section{Global measure of entanglement}
We quickly review the global measure that we shall use in the present chapter.
Consider a general, $n$-partite, normalized pure state:
$|\Psi\rangle=\sum_{p_1\cdots p_n}\Psi_{p_1p_2\cdots p_n}
|e_{p_1}^{(1)}e_{p_2}^{(2)}\cdots e_{p_n}^{(n)}\rangle$. If the parties are
all spin-1/2 then each can be taken to have the basis
$\{\ket{\!\uparrow},\ket{\!\downarrow}\}$. Our scheme for analyzing the
entanglement involves considering how well an entangled state can be
approximated by some unentangled (normalized) state (e.g.,~the state in which
every spin points in a definite direction):
$\ket{\Phi}\equiv\mathop{\otimes}_{i=1}^n|\phi^{(i)}\rangle$. The proximity of
$\ket{\Psi}$ to $\ket{\Phi}$ is captured by their overlap; the entanglement of
$\ket{\Psi}$ is revealed by the maximal overlap~\cite{WeiGoldbart03}
\begin{equation}
\label{eq:lambdamax}
\Lambda_{\max}({\Psi})\equiv\max_{\Phi}|\ipr{\Phi}{\Psi}|\,;
\end{equation}
the larger $\Lambda_{\max}$ is, the less entangled is $\ket{\Psi}$. (Note that
for a product state, $\Lambda_{\max}$ is unity.) If the entangled state
consists of two separate entangled pairs of subsystems, $\Lambda_{\max}$ is
the product of the maximal overlaps of the two. Hence, it makes sense to
quantify the entanglement of $\ket{\Psi}$
via the following {\it extensive\/} quantity
\begin{equation}
E_{\log_2}({\Psi})\equiv-\log_2\Lambda^2_{\max}(\Psi), \label{eq:Entrelate}
\end{equation}
This normalizes to unity the entanglement of EPR-Bell and $N$-party GHZ
states, as well as giving zero for unentangled states. Finite-$N$ entanglement
is interesting in the context of quantum information processing. To
characterize the properties of the quantum critical point we use the
thermodynamic quantity ${\cal E}$ defined by
\begin{subequations}
\begin{eqnarray}
&&{\cal E}\equiv\lim_{N\to\infty}{\cal E}_{N}, \\
&&{\cal E}_{N}\equiv {N}^{-1}E_{\log_2}(\Psi),
\end{eqnarray}
\end{subequations}
where ${\cal E}_{N}$ is the {\it entanglement density\/}, i.e., the
entanglement per particle.

\section{Quantum XY spin chains and entanglement}
We consider the family of models governed by the Hamiltonian
\begin{equation}
\label{eqn:HXY} {{\cal H}_{\rm XY}}= - \sum_{j=1}^N \left(\frac{1\!+\!r}{2}
\sigma_j^x\sigma_{j\!+\!1}^x+ \frac{1\!-\!r}{2}\sigma_j^y\sigma_{j\!+\!1}^y+ h
\,\sigma_j^z\right),
\end{equation}
where $r$ measures the anisotropy between $x$ and $y$ couplings, $h$ is the
transverse external field, lying along the $z$-direction, and we impose
periodic boundary conditions, namely, a ring geometry. At $r=0$ we have the
isotropic XY limit (also known as the XX model) and at $r=1$, the Ising limit.
All anisotropic XY models ($0<r\le 1$) belong to the same universality class,
i.e., the Ising class, whereas the isotropic XX model belongs to a different
universality class. XY models exhibit three phases (see
Fig.~\ref{fig:XYEnt10000}): oscillatory, ferromagnetic and paramagnetic. In
contrast to the paramagnetic phase, the first two are ordered phases, with the
oscillatory phase being associated with a characteristic wavevector,
reflecting the modulation of the spin correlation functions (see, e.g.,
Ref.~\cite{Henkel99}). We shall see that the global entanglement detects the
boundaries between these phases, and that the universality class dictates the
behavior of entanglement near quantum phase transitions.

Before we solve the entanglement of the XY model, we give perturbative
analysis of, as an illustration of how entanglement arises and vanishes, the
Ising model in a transverse field (viz. $r=1$)
\begin{equation}
\label{eqn:Hising} {\cal H}=- \sum_{i=1}^N \left(\sigma_i^x\sigma_{i+1}^x+h
\sigma_i^z\right).
\end{equation}
At $h=0$ the ground state is that with all spins pointing up in the
$x$-direction $\ket{\!\!\rightarrow\rightarrow\dots\rightarrow}$ or down
$\ket{\!\!\leftarrow\leftarrow\dots\leftarrow}$, which is manifestly
unentangled. The ground state can be any superposition of
$(\ket{\!\!\rightarrow\rightarrow\dots\rightarrow}$ and
$\ket{\!\!\leftarrow\leftarrow\dots\leftarrow}$ when the $Z_2$ symmetry is not
spontaneously broken. For example, the states
$(\ket{\!\!\rightarrow\rightarrow\dots\rightarrow}\pm\ket{\!\!\leftarrow\leftarrow\dots\leftarrow})/\sqrt{2}$
are actually the two lowest levels obtained from solving the models using
Jordan-Wigner and Bogoliubov transformations and they both have
$E_{\log_2}=1$. (For small $h$ the entanglement rises quadratically in the
case of unbroken symmetry instead of quartically, as we shall show shortly.)
We shall see later that whether or not we use a broken-symmetry state  has no
effect in the thermodynamic limit.
 For small $h$ (i.e., $h\ll1/\sqrt{N}$) one can
obtain the ground state by treating the $h \sigma_i^z$ terms as perturbations.
Take the ground state at $h=0$ to be
$\ket{\!\!\rightarrow\rightarrow\dots\rightarrow}$. Then first-order
perturbation theory for the ground state gives
\begin{equation}
\frac{1}{\sqrt{1+\frac{Nh^2}{4}}}\Big(\ket{\!\!\rightarrow\rightarrow\dots\rightarrow}+\frac{h}{2}\sum_i\ket{\!\!\rightarrow\dots\leftarrow_{i}\dots\rightarrow}\Big).
\end{equation}
Using the method described in Sec.~\ref{sec:Pure} we obtain $E_{\log_2}\approx
N(N-1) h^4/32$ to leading order in $h$.
 At $h=\infty$ the ground state is a quantum paramagnet with all
spins aligning along the external field:
$\ket{\!\!\uparrow\uparrow\dots\uparrow}$, and once more is
 unentangled.
To ${\cal O}(1/h)$ perturbation theory gives (treating
$\sigma^z_i\sigma^z_{i+1}$ terms small)
\begin{equation}
\frac{1}{\sqrt{1+\frac{N}{16h^2}}}\Big(\ket{\!\!\uparrow\uparrow\dots\uparrow}+\frac{1}{4h}\sum_i\ket{\!\!\uparrow\dots\downarrow_{i}\downarrow_{{i\!+\!1}}\dots\uparrow}\Big),
\end{equation}
for which $E_{\log_2}\approx N/(16 h^2)$, to leading order of $1/h$. The
quantum phase transition from a ferromagnetic to a paramagnetic phase occurs
at $h=1$~\cite{Sachdev}. The two lowest levels, which we denote by
$\ket{\Psi_{1/2}}$ and $\ket{\Psi_0}$ (for reasons to be explained later) are,
respectively, the ground and first excited states, and they are asymptotically
degenerate for $0\le h\le1$ when $N\rightarrow\infty$.

As is well known~\cite{Sachdev,Henkel99,LiebSchultzMattis61}, the energy
eigenproblem for the XY spin chain can be solved via a Jordan-Wigner
transformation, through which the spin degrees of freedom are recast as
fermionic ones, followed by a Bogoliubov transformation, which diagonalizes
the resulting quadratic Hamiltonian.

The Jordan-Wigner transformation that we shall make from spins ($\sigma$'s) to
fermion particles ($c$'s) is
\begin{subequations}
\begin{eqnarray}
\sigma_i^z&=&1-2c_i^\dagger\,c_i, \\
\sigma_i^x&=&\prod_{j=1}^{i-1}\big(1-2c_j^\dagger\,c_j\big)\big(c_i+c_i^\dagger\big),\\
\sigma_i^y&=&-i\prod_{j=1}^{i-1}\big(1-2c_j^\dagger\,c_j\big)\big(c_i-c_i^\dagger\big).
\end{eqnarray}
\end{subequations}
One has to pay attention to the boundary conditions that are to be imposed on
the $c$'s. Although periodic in the $\sigma$'s, one cannot simply take
\begin{equation}
\sigma_{N+1}^x=\prod_{j=1}^{N}\big(1-2c_j^\dagger\,c_j\big)
\big(c_{N+1}+c_{N+1}^\dagger\big)=\sigma_1=(c_{1}+c_{1}^\dagger\big),
\end{equation}
and conclude that either (i)~$\prod_{j=1}^{N}\big(1-2c_j^\dagger\,c_j\big)=1$
and $c_{N+1}=c_1$, or (ii)~$\prod_{j=1}^{N}\big(1-2c_j^\dagger\,c_j\big)=-1$
and $c_{N+1}=-c_1$, neither of which are correct if one wishes to obtain the
correct spectrum and eigenstates for arbitrary finite $N$. Instead one should
impose that (as the $(N+1)$-th site is identified as the first site)
\begin{equation}
\sigma_{N}^x\sigma_{N+1}^x=\sigma_{N}^x\sigma_{1}^x,
\end{equation}
which then leads to
\begin{equation}
\big(c_{N}+c_{N}^\dagger\big)\big(c_{N+1}+c_{N+1}^\dagger\big)
=-\prod_{j=1}^{N}\big(1-2c_j^\dagger\,c_j\big)\big(c_{N}+c_{N}^\dagger\big)\big(c_{1}+c_{1}^\dagger\big).
\end{equation}
The two possible conditions that satisfy this are either
(I)~$\prod_{j=1}^{N}\big(1-2c_j^\dagger\,c_j\big)=-1$ and $c_{N+1}=c_1$, or
(II)~$\prod_{j=1}^{N}\big(1-2c_j^\dagger\,c_j\big)=1$ and $c_{N+1}=-c_1$. The
operator
\begin{equation}
\prod_{j=1}^{N}\big(1-2c_j^\dagger\,c_j\big)=e^{i\pi\sum_j c_j^\dagger\,c_j}
\end{equation}
counts whether the total number of particles is even ($+1$) or odd ($-1$). For
$c$'s that are periodic, the number is odd, whereas for antiperiodic $c$'s,
this number is even.

To incorporate these two boundary conditions on the $c$'s, we take
\begin{equation}
c_j=\frac{1}{\sqrt{N}}\sum_{m=0}^{N-1}e^{i\frac{2\pi}{N}j (m+b)}
\tilde{c}_{m}^{(b)},
\end{equation}
where $b=0$ for periodic $c$'s; $b=1/2$ for anti-periodic $c$'s. (This
explains why we label the lowest two states by $\ket{\Psi_{1/2}}$ and
$\ket{\Psi_0}$.) The momentum index $m$ ranges from $0$ to $N-1$. In terms of
these fermion operators the Hamiltonian becomes
\begin{equation}
{\cal H}=-N h
-\sum_{m=0}^{N-1}\left\{\Big[2\cos\frac{2\pi}{N}(m+b)-2h\Big]{\tilde{c}_{m}^{(b)\dagger}}
\tilde{c}_{m}^{(b)}+i r \sin\frac{2\pi}{N}(m+b)\Big[{\tilde{c}_{m}^{(b)}}
\tilde{c}_{N-m-2b}^{(b)}+{\tilde{c}_{m}^{(b)\dagger}}
{\tilde{c}_{N-m-2b}^{(b)\dagger}}\Big]\right\}.
\end{equation}

Upon using the Bogoliubov transformation
\begin{equation}
\tilde{c}_m^{(b)}=\cos\theta_m^{(b)} \gamma_m^{(b)}+i \sin\theta_m^{(b)}
\gamma^{(b)\dagger}_{N-m-2b}\,,
\end{equation}
with
\begin{equation}
\tan2\theta_m^{(b)}={r\sin\frac{2\pi\,(m+b)}{N}}\Big/{\left(h-\cos\frac{2\pi\,(m+b)}{N}\right)},
\end{equation}
one arrives at the diagonal the Hamiltonian:
\begin{subequations}
\begin{equation}
{\cal
H}=-Nh+\sum_{m=0}^{N-1}\varepsilon_m^{(b)}\Big(\tilde{\gamma}_{m}^{(b)\dagger}
\tilde{\gamma}_{m}^{(b)}-\frac{1}{2}\Big),
\end{equation}
\begin{equation}
\varepsilon_m^{(b)}=2\sqrt{\Big(h-\cos\frac{2\pi\,(m+b)}{N}\Big)^2+r^2\sin^2\frac{2\pi\,(m+b)}{N}},
\end{equation}
\end{subequations}
except for the special case that $\varepsilon_0^{(0)}=2(h-1)$.

We remark that we have not left out any constant in diagonalizing the
Hamiltonian in either case, so the energy spectrum is exact. For each value of
$b$ the diagonalization gives $2^N$ energy eigenvalues, so there are $2^{N+1}$
in total. Half of them are spurious. In determining the correct $2^N$ states
from the $2^{N+1}$ solutions, one has to impose a constraint from the boundary
conditions. Namely, in case I there can be only odd number of fermions,
whereas in case II there can be only even number of fermions.

For $b=0$, viz, the odd-number-fermion case, the lowest state $\ket{\Psi_{0}}$
is such that $\langle \tilde{\gamma}_{m}^{(0)\dagger}
\tilde{\gamma}_{m}^{(0)}\rangle=0$ except that $\langle
\tilde{\gamma}_{0}^{(0)\dagger} \tilde{\gamma}_{0}^{(0)}\rangle=1$. Its energy
eigenvalue is
\begin{equation}
\label{eqn:E0}
E_0^{(0)}(r,h)=(h-1)-\sum_{m=1}^{N-1}\sqrt{\Big(h-\cos\frac{2\pi\,m}{N}\Big)^2+r^2\sin^2\frac{2\pi\,m}{N}}.
\end{equation}

For $b=1/2$, namely, the even-number-fermion case, the lowest state
$\ket{\Psi_{1/2}}$ is such that $\langle \tilde{\gamma}_{m}^{(1/2)\dagger}
\tilde{\gamma}_{m}^{(1/2)}\rangle=0$ for all $m$. Its eigen-energy is
\begin{equation}
\label{eqn:E1/2}
E_0^{(1/2)}(r,h)=-\sum_{m=0}^{N-1}\sqrt{\Big(h-\cos\frac{2\pi\,(m+1/2)}{N}\Big)^2+r^2\sin^2\frac{2\pi\,(m+1/2)}{N}}.
\end{equation}

We see that, as $N\rightarrow \infty$, the above two energy levels are
degenerate for $h\le 1$. Furthermore,  as $N\rightarrow\infty$ the difference
between the two energy levels becomes
\begin{equation}
E_0^{(0)}(r,h)-E_0^{(1/2)}(r,h)=2(h-1) \Theta(h-1),
\end{equation}
where $\Theta(x)=1$ if $x>0$ and zero otherwise. The way the energy gap
vanishes as $h\rightarrow 1^+$ gives a relation betweem two exponents
\begin{equation}
\label{eqn:znu} z\nu=1;
\end{equation}
$z$ is the dynamical exponent (defined via the vanishing of energy gap
$\Delta\sim|h-h_c|^{z\nu}$) and $\nu$ is the correlation-length exponent
(defined via $L_c\sim |h-h_c|^{-\nu}$).

Having found the lowest two eigenstates, the quantity $\Lambda_{\rm max}$ of
Eq.~(\ref{eq:lambdamax})---and hence the entanglement---can be found, at least
in principle. To do this, we parametrize the separable states via
\begin{equation}
\ket{\Phi}\equiv\mathop{\otimes}_{i=1}^{N}
\big[\cos(\xi_i/2)\ket{\!\!\uparrow}_i +
e^{i\phi_i}\sin(\xi_i/2)\ket{\!\!\downarrow}_i\big],
\end{equation}
where $\ket{\!\uparrow\!\!/\!\!\downarrow}$ denote spin states
parallel/antiparallel to the $z$-axis.  Instead of maximizing the overlap with
respect to the $2N$ real parameters $\{\xi_{i},\phi_{i}\}$, for the lowest two
states it is adequate to appeal to the translational symmetry of and the
reality of the ground-state wavefunctions.  Thus taking $\xi_i=\xi$ and
$\phi_i=0$ we make the Ansatz:
\begin{equation}
\label{eqn:PhiTh}
 \ket{\Phi(\xi)}
\equiv e^{-i\frac{\xi}{2}\sum_{j=1}^N\sigma_j^y}
\ket{\!\uparrow\uparrow\dots\uparrow}
\end{equation}
for searching for the maximal the overlap
$\Lambda_{\max}(\Psi)$~\cite{footnote:ansatz}. This form shows that this
separable state can be constructed as a global rotation of the ground state at
$h=\infty$, viz., the separable state $\ket{\!\uparrow\uparrow\dots\uparrow}$.
In this particular limit the boundary condition on the $c$'s is irrelevant, as
the dominant term in the Hamiltonian is  $\,-\sum_j h(1-2c_j^\dagger c_j)$.

The energy eigenstates are readily expressed in terms of the Jordan-Wigner
fermion operators, and so too is the  family of the Ansatz states
$\ket{\Phi(\xi)}$.  By working in this fermion basis we are able to evaluate
the overlaps between the two lowest states and the Ansatz states. With
$\ket{\Psi_0}$ ($\ket{\Psi_{1/2}}$) denoting the lowest state in the odd
(even) fermion-number sector, we arrive at the overlaps
\def\myaa{b}
\begin{equation}
\label{eqn:overlap} \ipr{\Psi_{\myaa}(r,h)}{\Phi(\xi)} =
f^{(\myaa)}_{N}(\xi)\prod_{m=1-2\myaa}^{m<\frac{N-1}{2}} \left[
\cos\theta^{(\myaa)}_{m}(r,h)\cos^2({\xi}/{2})+
\sin\theta^{(\myaa)}_{m}(r,h)\sin^2({\xi}/{2}) \cot(k_{m,N}^{(\myaa)}/2)
\right],
\end{equation}
with
\begin{subequations}
\label{eqn:f}
\begin{eqnarray}
&& k_{m,N}^{(\myaa)} \equiv \frac{2\pi}{N}(m+{\myaa}),
\ \ \ \tan2\theta^{(\myaa)}_{m}(r,h) \equiv r\sin k_{m,N}^{(\myaa)}\big/
(h\!-\!\cos k_{m,N}^{(\myaa)});
\\
&&
f^{(1/2)}_{N}(\xi) \equiv 1, \ \ \ f^{(0)}_{N}(\xi) \equiv
\sqrt{N}\sin({\xi}/{2})\cos({\xi}/{2}), \ (\mbox{$N$ even});
\\
&&
f^{(1/2)}_{N}(\xi) \equiv \cos({\xi}/{2}), \ \ \ f^{(0)}_{N}(\xi) \equiv
\sqrt{N}\sin({\xi}/{2}), \ \ \,\, (\mbox{$N$  odd});
\end{eqnarray}
\end{subequations}
where $\myaa=0,1/2$ and $m\in[0,N-1]$ is the (integer) momentum index. The
above results are {\it exact\/} for arbitrary $N$, obtained with periodic
boundary conditions on spins rather than in the so-called $c$-cyclic
approximation~\cite{LiebSchultzMattis61}. Given these overlaps, we can readily
obtain the entanglement of the ground state, the first excited state, and any
linear superposition, $\cos\alpha\ket{\Psi_0}+\sin\alpha\ket{\Psi_1}$ of the
two lowest states, for arbitrary $(r,h)$ and $N$, by maximizing the magnitude
of the overlap with respect to the single, real parameter $\xi$. For the
derivation of the above results, see Appendix~\ref{app:Derivation}.

\def\deriv1{
We now derive the results listed above. We first analyze $b=1/2$ case, namely,
the even-fermion case. The lowest state $\ket{\Psi_{1/2}(r,h)}$ has zero
number of Bogoliubov fermions. It is related to the state that has no
$c$-fermions, i.e., $\ket{\Omega}\equiv\ket{\!\uparrow\cdots\uparrow}$ via
\begin{subequations}
\begin{eqnarray}
\ket{\Psi_{1/2}(r,h)}&=&\prod_{m=0}^{m<\frac{N\!-\!1}{2}}\cos\theta_m^{(1\!/2)}(r,h)
e^{i\tan\theta_m^{(1\!/2)}(r,h)\,\tilde{c}_m^{(1\!/2)\dagger}
\tilde{c}_{N-m-1}^{(1\!/2)\dagger} }\ket{\Omega}\\
&=&\prod_{m=0}^{m<\frac{N\!-\!1}{2}}\Big[\cos\theta_m(r,h)
+i\sin\theta_m(r,h)\,\tilde{c}_m^{\dagger} \tilde{c}_{N-m-1}^{\dagger}
\Big]\ket{\Omega}
\end{eqnarray}
\end{subequations}
The Ansatz state is then
\begin{subequations}
\begin{eqnarray}
\ket{\Phi(\xi)} &=&\prod_{j=1}^N
\Big[ \cos\frac{\xi}{2}+\sin\frac{\xi}{2}\prod_{1\le l<j}(1-2c_l^\dagger c_l) (c_j^\dagger-c_j) \Big] \ket{\Omega} \\
&=&\prod_{j=1}^N\big(\cos\frac{\xi}{2}+\sin\frac{\xi}{2}\,c_j^\dagger\big)
\ket{\Omega}\\
&=&\cos^{N}\frac{\xi}{2}\, e^{\tan\frac{\xi}{2}c_1^\dagger}\cdots e^{\tan\frac{\xi}{2}c_N^\dagger}\ket{\Omega}\\
&=&\cos^{N}\frac{\xi}{2}\, e^{\tan\frac{\xi}{2}\sum_{j=1}^N c_j^\dagger}
e^{\tan^2\frac{\xi}{2}\sum_{j< l}c_j^\dagger c_l^\dagger}\ket{\Omega},
\end{eqnarray}
\end{subequations}
where we have suppressed the index $(1/2)$. The term $\sum_{j< l}c_j^\dagger
c_l^\dagger$
 can be rewritten in momentum space as
 \begin{equation}
 \sum_{1\le j< l\le N} c_j^\dagger c_l^\dagger=i\sum_{m=0}^{m<\frac{N-1}{2}}
 \cot\frac{\pi(m+\frac{1}{2})}{N}\tilde{c}^\dagger_m \tilde{c}^\dagger_{N\!-\!m\!-\!1} .
 \end{equation}

  Thus for even $N$
\begin{subequations}
 \begin{equation}
 \ket{\Phi(\xi)}=\Big({1+\tan\frac{\xi}{2}\,\sum_{j=1}^N c_j^\dagger}\Big)\prod_{m=0}^{m<\frac{N-1}{2}}
\left(\cos^2\frac{\xi}{2}+i\sin^2\frac{\xi}{2}\cot\frac{\pi(m+\frac{1}{2})}{N}
\tilde{c}^\dagger_m \tilde{c}^\dagger_{N\!-\!m\!-\!1}\right)\ket{\Omega},
 \end{equation}
 whereas for odd $N$
 \begin{equation}
\ket{\Phi(\xi)}=\Big({1+\tan\frac{\xi}{2}\,\sum_{j=1}^N c_j^\dagger}\Big)
\cos\frac{\xi}{2}\prod_{m=0}^{m<\frac{N-1}{2}}
\left(\cos^2\frac{\xi}{2}+i\sin^2\frac{\xi}{2}\cot\frac{\pi(m+\frac{1}{2})}{N}
\tilde{c}^\dagger_m \tilde{c}^\dagger_{N\!-\!m\!-\!1}\right)\ket{\Omega}.
 \end{equation}
 \end{subequations}
 Therefore, the overlap of $\ket{\Psi_{1/2}(r,h)}$ with $\ket{\Phi(\xi)}$ is
 \begin{equation}
 \ipr{\Psi_{1/2}(r,h)}{\Phi(\xi)}=\prod_{m=0}^{m<\frac{N-1}{2}}
\left(\cos\theta_m^{(1/2)}(r,h)\,\cos^2\frac{\xi}{2}+\sin\theta_m^{(1/2)}(r,h)\,\sin^2\frac{\xi}{2}\cot\frac{\pi(m+\frac{1}{2})}{N}\right)
 \end{equation}
 for even $N$ and
 \begin{equation}
 \ipr{\Psi_{1/2}(r,h)}{\Phi(\xi)}=\cos\frac{\xi}{2}\prod_{m=0}^{m<\frac{N-1}{2}}
\left(\cos\theta_m^{(1/2)}(r,h)\,\cos^2\frac{\xi}{2}+\sin\theta_m^{(1/2)}(r,h)\,\sin^2\frac{\xi}{2}\cot\frac{\pi(m+\frac{1}{2})}{N}\right)
 \end{equation}
 for odd $N$.

 Next, we discuss $b=0$ (odd-fermion) case. The lowest allowed state is
the one with one $\gamma^{(0)}_0=\tilde{c}^{(0)}_0$ fermion:
\begin{equation}
\ket{\Psi_{0}(r,h)}=\gamma^{(0)\dagger}_0\ket{G(r,h)}=\tilde{c}^{(0)\dagger}_0\ket{G(r,h)},
\end{equation}
where $\ket{G(r,h)}$ is the state with no $\gamma$ fermions:
\begin{equation}
\ket{G(r,h)}=\prod_{m=1}^{m<\frac{N}{2}}\Big[\cos\theta_m^{(0)}(r,h)
+i\sin\theta_m^{(0)}(r,h)\,\tilde{c}_m^{(0)\dagger}
\tilde{c}_{N-m}^{(0)\dagger} \Big]\ket{\Omega}.
\end{equation}
Similar to the case $b=1/2$, using
\begin{equation}
 \sum_{1\le j< l\le N} c_j^\dagger c_l^\dagger=i\sum_{m=1}^{m<\frac{N}{2}}
 \cot\frac{\pi m}{N}\tilde{c}^{(0)\dagger}_m \tilde{c}^{(0)\dagger}_{N\!-\!m},
 \end{equation}
we obtain that for even $N$
\begin{subequations}
 \begin{equation} \ket{\Phi(\xi)}=\Big({1+\sqrt{N}\tan\frac{\xi}{2}\,\tilde{c}_0^\dagger}\Big)\cos^2\frac{\xi}{2}\prod_{m=1}^{m<\frac{N}{2}}
\left(\cos^2\frac{\xi}{2}+i\sin^2\frac{\xi}{2}\cot\frac{\pi m}{N}
\tilde{c}^\dagger_m \tilde{c}^\dagger_{N\!-\!m}\right)\ket{\Omega},
 \end{equation}
 whereas for odd $N$
 \begin{equation}
\ket{\Phi(\xi)}=\Big({1+\sqrt{N}\tan\frac{\xi}{2}\,\tilde{c}_0^\dagger}\Big)
\cos\frac{\xi}{2}\prod_{m=1}^{m<\frac{N}{2}}
\left(\cos^2\frac{\xi}{2}+i\sin^2\frac{\xi}{2}\cot\frac{\pi m}{N}
\tilde{c}^\dagger_m \tilde{c}^\dagger_{N\!-\!m}\right)\ket{\Omega}.
 \end{equation}
 \end{subequations}
 Therefore, the overlap of $\ket{\Psi_{0}(r,h)}$ with $\ket{\Phi(\xi)}$ is
 \begin{equation} \ipr{\Psi_{0}(r,h)}{\Phi(\xi)}=\sqrt{N}\sin\frac{\xi}{2}\prod_{m=1}^{m<\frac{N}{2}}
\left(\cos\theta_m^{(0)}(r,h)\,\cos^2\frac{\xi}{2}+\sin\theta_m^{(0)}(r,h)\,\sin^2\frac{\xi}{2}\cot\frac{\pi
m}{N}\right)
 \end{equation}
 for even $N$ and
 \begin{equation} \ipr{\Psi_{0}(r,h)}{\Phi(\xi)}=\sqrt{N}\sin\frac{\xi}{2}\cos\frac{\xi}{2}\prod_{m=1}^{m<\frac{N}{2}}
\left(\cos\theta_m^{(0)}(r,h)\,\cos^2\frac{\xi}{2}+\sin\theta_m^{(0)}(r,h)\,\sin^2\frac{\xi}{2}\cot\frac{\pi
m}{N}\right)
 \end{equation}
 for odd $N$.
}

The formulas [in Eqs.~(\ref{eqn:overlap}) and~(\ref{eqn:f})]  contain all the
results that we shall explore shortly. By analyzing the structure of
Eq.~(\ref{eqn:overlap}), we find that the global entanglement does provide
information on the phase structure and critical properties of the quantum spin
chains.  Two of our key results, as captured in Figs.~\ref{fig:XYEnt10000} and
\ref{fig:Ent1000}, are: (i)~although the entanglement itself is, generically,
not maximized at the quantum critical line in the $(r,h)$ plane, {\it the
field-derivative of the entanglement diverges as the critical line $h=1$ is
approached\/}; and (ii)~the entanglement {\it vanishes\/} at the disorder line
$r^2+h^2=1$, which separates the oscillatory and ferromagnetic phases.

\begin{figure}
\psfrag{h}{$h$} \psfrag{r}{$r$} {\psfrag{E}{${\cal E}_{N} \ $}
\centerline{\psfig{figure=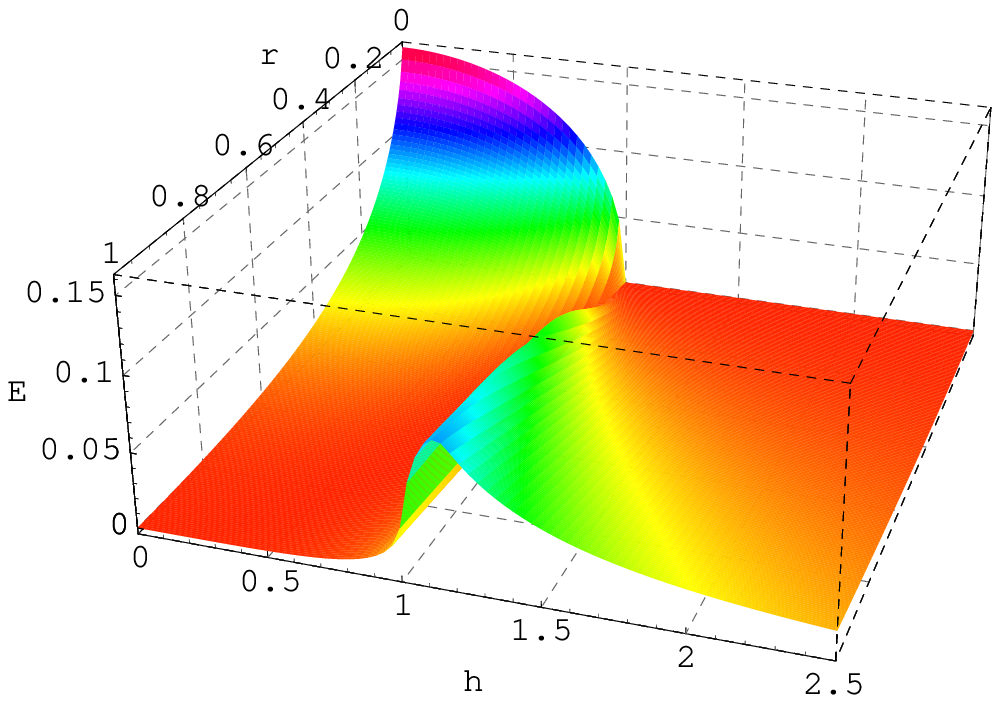,width=12cm,angle=0}}}
\vspace{0.5cm}
  \centerline{\psfig{figure=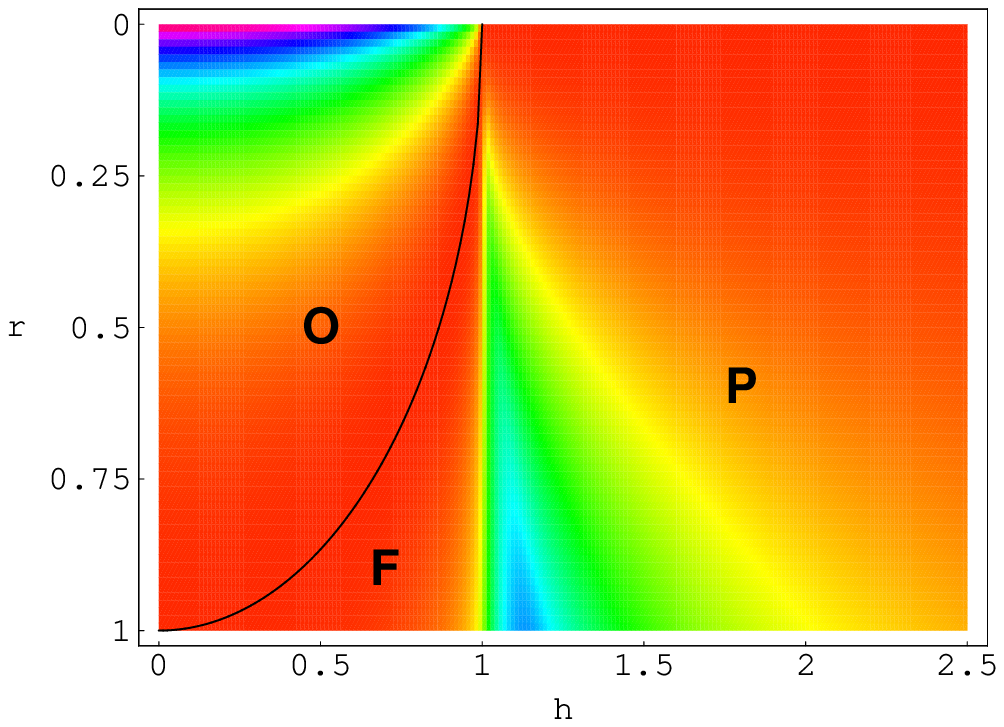,width=12cm,angle=0}}
\caption[Entanglement density and the phase diagram
 for the XY model]{
Entanglement density (upper) and phase diagram (lower) vs.~$(r,h)$ for the XY
model with $N=10^4$ spins, which is essentially in the thermodynamic limit.
There are three phases: {\bf O}: ordered oscillatory, for $r^2+h^2<1$ and
$r\ne 0$; {\bf F}: ordered ferromagnetic, between $r^2+h^2>1$ and $h<1$; {\bf
P}: paramagnetic, for $h>1$. As is apparent, there is a sharp rise in the
entanglement across the line $h=1$, which signifies a quantum phase
transition. The arc $h^2+r^2=1$, along which the entanglement density is zero
(see also Fig.~\ref{fig:Ent1000}), separates phases {\bf O} and {\bf F}. Along
$r=0$ lies the XX model, which belongs to a different universality class from
the anisotropic XY model. } \label{fig:XYEnt10000}
\end{figure}

As is to be expected, at finite $N$ the two lowest states $\ket{\Psi_0}$ and
$\ket{\Psi_{1/2}}$ featuring in Eq.~(\ref{eqn:overlap}) do not spontaneously
break the ${\rm Z}_2$ symmetry. However, in the thermodynamic limit they are
degenerate for $h\le 1$, and linear combinations are also ground states. The
question then arises as to whether linear combinations that explicitly break
${\rm Z}_2$ symmetry, i.e., the physically relevent states with finite
spontaneous magnetization, show the same entanglement properties. In fact, we
see from Eq.~(\ref{eqn:overlap}) that, in the thermodynamic limit, overlaps
for $\ket{\Psi_0}$ and $\ket{\Psi_{1/2}}$ are identical, up to the prefactors
$f^{(0)}_N$ and $f^{(1/2)}_N$. These prefactors do not contribute to the
entanglement density, and the entanglement density is therefore the same for
both $\ket{\Psi_0}$ and $\ket{\Psi_{1/2}}$. It  further follows that, in the
thermodynamic limit, the results for the entanglement density are insensitive
to the replacement of a symmetric ground state by a broken-symmetry one.

\begin{figure}
\psfrag{h}{$h$}\psfrag{E}{${\cal E}(h)$}\psfrag{D}{${\cal E}'(h)$}
\centerline{\psfig{figure=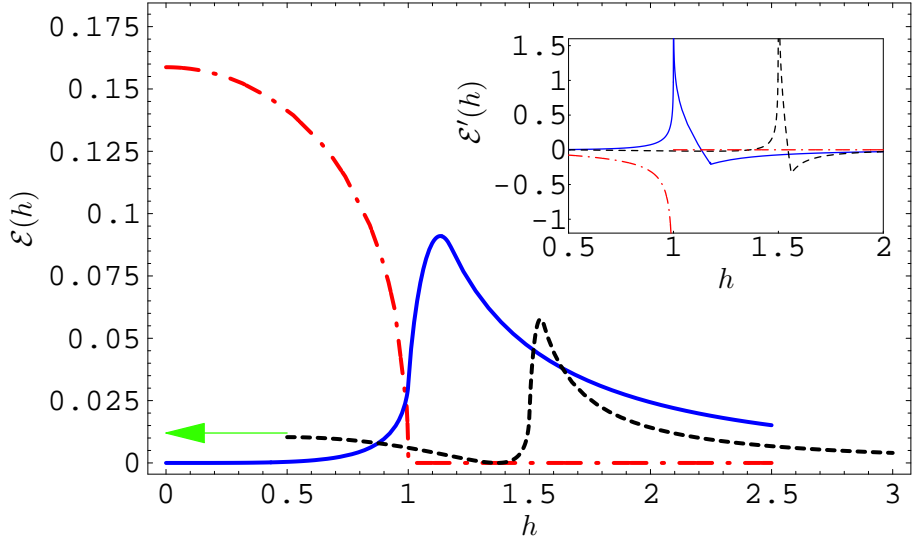,width=12cm,angle=0}}
\caption[Entanglement density and its $h$-derivative for the ground state of
three systems at $N=\infty$]{ Entanglement density and its $h$-derivative
(inset) for the ground state of three systems at $N=\infty$. Solid line: Ising
($r=1$) limit; dashed line: anisotropic ($r=1/2$) XY model; dash-dotted line:
($r=0$) XX model. For the sake of clarity, the XY-case curves are shifted to
the right by 0.5, indicated by the arrow. For the $r=1/2$ case, at  the
entanglement density vanishes at $h=\sqrt{1-r^2}$, which is a general property
for the anisotropic XY model. Note that whilst the entanglement itself has a
nonsingular maximum at $h\approx 1.1$ (Ising), $h\approx 1.04$ ($r=1/2$ XY),
and $h=0$ (XX), respectively, it has a singularity at the quantum critical
point at $h=1$, as revealed by the divergence of its derivative.}
\label{fig:Ent1000}
\end{figure}
\section{Entanglement and quantum criticality}
Before we discuss the thermodynamic limit of the entanglement density, we
compare the entanglement obtained via the results in Eq.~(\ref{eqn:overlap})
and that obtained via numerically diagonalizing the Hamiltonian and
calculating the maximal overlap. In Figure~\ref{fig:4plot} the results via
each method are shown for the Ising case ($r=1$) with small numbers of spins
($N=13$ through $22$), it is seen that our analytical results are exact even
for small $N$, both even and odd.

\begin{figure}
\centerline{\psfig{figure=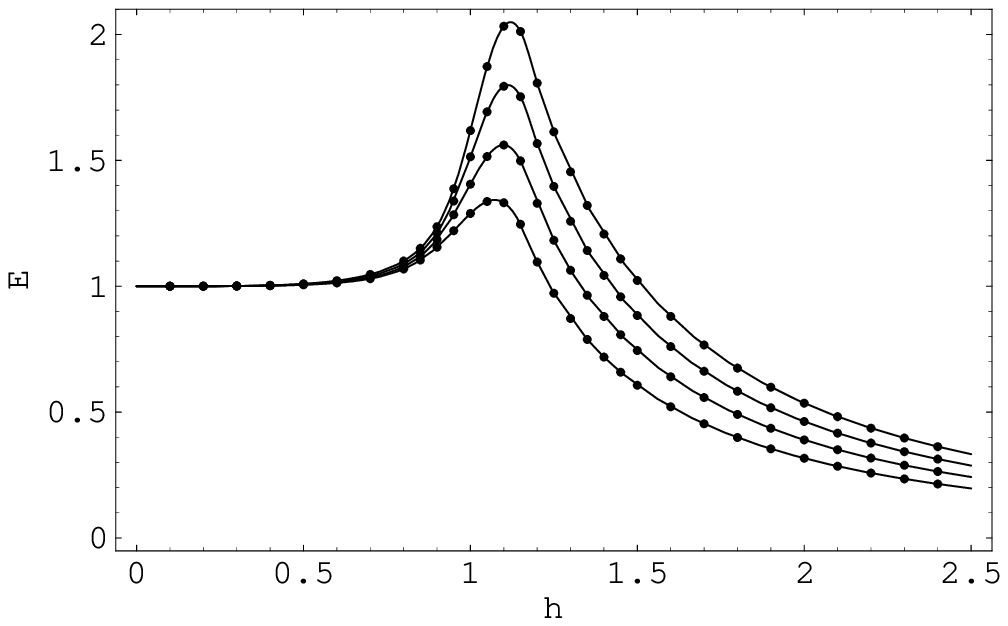,width=10cm,angle=0}}
\centerline{\psfig{figure=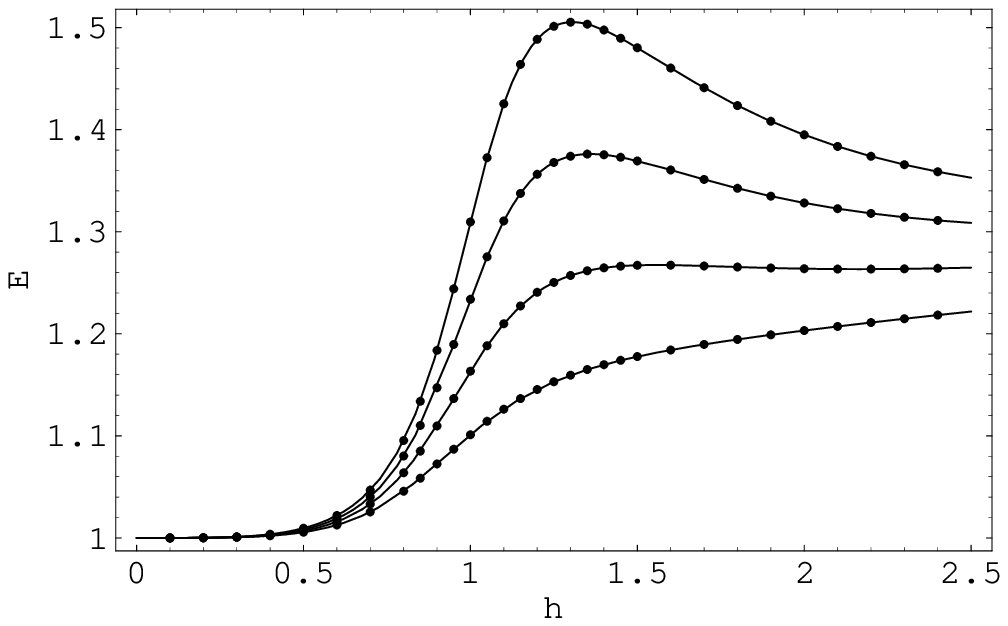,width=10cm,angle=0}}
\caption[Entanglement vs.~magnetic field $h$ for ground  and first excited
states with small of spins]{Entanglement vs.~magnetic field $h$ for ground
(upper panel) and first excited (lower panel) states with small
$N=13,16,19,22$ (from bottom to top) for the transverse Ising model ($r=1$).
The numerical results are shown as discrete points whereas the analytical
results are shown as lines. This demonstrates that the analytical results are
exact, even for small $N$, both even and odd. } \label{fig:4plot}
\end{figure}

From Eq.~(\ref{eqn:overlap}) it follows that the thermodynamic limit of the
entanglement density is given by
\begin{eqnarray}
\label{eqn:Erh} {\cal E}(r,h) = -\frac{2}{\ln2}\max_\xi
\int_0^{\frac{1}{2}}d\mu\, \ln\left[\cos\theta(\mu,r,h)\cos^2({\xi}/{2})
+\sin\theta(\mu,r,h)\sin^2({\xi}/{2})\cot\pi\mu\right], \label{eq:infNent}
\end{eqnarray}
where $\tan 2\theta(\mu,r,h)\equiv r\sin 2\pi\mu /(h-\cos 2\pi\mu)$.

Figure~\ref{fig:Ent1000} shows the thermodynamic limit of the entanglement
density ${\cal E}(r,h)$ and its $h$-derivative in the ground state, as a
function of $h$ for three values of $r$, i.e., three slices through the
surface shown in Fig.~\ref{fig:XYEnt10000}. As the $r=1$ slice shows, in the
Ising limit the entanglement density is small for both small and large $h$. It
increases with $h$ from zero, monotonically, albeit very slowly for small $h$,
then swiftly rising to a maximum at $h\approx 1.13$ before decreasing
monotonically upon further increase of $h$, asymptotically to zero.  The
entanglement maximum {\it does not\/} occur at the quantum critical point.
However, the derivative of the entanglement with respect to $h$ {\it does\/}
diverge at the critical point $h=1$, as shown in the inset. The slice at
$r=1/2$ (for clarity, shifted
 half a unit to the right) shows qualitatively similar
behavior, except that it is finite (although small) at $h=0$, and starts out
by decreasing to a shallow minimum of zero at $h=\sqrt{1-r^{2}}$. By
constrast, the slice at $r=0$ (XX) starts out at $h=0$ at a maximum value of
$1- 2 \gamma_C/(\pi \ln 2)\approx 0.159$. (where $\gamma_C\approx 0.9160$ is
the {\it Catalan\/} constant), the globally maximal value of the entanglement
over the entire $(r,h)$ plane.
For larger $h$ it falls monotonically until it vanishes at $h=1$, remaining
zero for larger $h$.

We find that along the line $r^{2}+h^{2}=1$ the entanglement density vanishes
in the thermodynamic limit. In fact, this line exactly corresponds to the
boundary separating the oscillatory and ferromagnetic phases; the boundary can
be characterized by a set of ground states with total entanglement of order
unity, and thus of zero entanglement density. The entanglement density is also
able to track the phase boundary ($h=1$) between the ordered and disordered
phases. Associated with the quantum fluctuations accompanying the transition,
the entanglement density shows a drastic variation across the boundary and the
field-derivative diverges all along $h=1$. The two boundaries separating the
three phases coalesce at $(r,h)=(0,1)$, i.e., the XX critical point.
Figures~\ref{fig:XYEnt10000} and \ref{fig:Ent1000} reveal all these features.

\def\XY{
\subsection{Divergence of entanglement-derivative in XY}
Define
\begin{equation}
F(\xi,r,h)\equiv\int_0^{\frac{1}{2}}d\mu\,
\ln\left[\cos\theta(\mu,r,h)\cos^2({\xi}/{2})
+\sin\theta(\mu,r,h)\sin^2({\xi}/{2})\cot\pi\mu\right].
\end{equation}
To find the stationarity condition:
\begin{equation}
\partial_\xi F(\xi,r,h)=-\frac{1}{2}\sin\xi \int_0^{\frac{1}{2}}d\mu\,
\frac{\cos\theta(\mu,r,h)-\sin\theta(\mu,r,h)\cot\pi\mu}{
\cos\theta(\mu,r,h)\cos^2({\xi}/{2})
+\sin\theta(\mu,r,h)\sin^2({\xi}/{2})\cot\pi\mu}=0.
\end{equation}
Denote $\xi^*(h)$ the solution for fixed $r$. Then the entanglement
field-derivative is
\begin{equation}
\partial_h {\cal E}(r,h)=- \frac{2}{\ln2}\partial_h F(\xi^*(h),h)
= - \frac{2}{\ln2}\left[ \frac{\partial \xi^*(h)}{\partial h}
\partial_{\xi}F(\xi,h)\Big|_{\xi^*}+\partial_h F(\xi^*,h)\right]=-
\frac{2}{\ln2}\partial_h F(\xi^*,h),
\end{equation}
where the first term in the square bracket vanishes identically. Thus
(dropping the * in $\xi$),
\begin{eqnarray}
&&\partial_h {\cal E}(r,h)=- \frac{2}{\ln2}\partial_h F(\xi^*,h)\\
&=&-\frac{2}{\ln2}\int_0^{\frac{1}{2}}d\mu\,
\frac{\partial_h\cos\theta(\mu,r,h)\cos^2({\xi}/{2})
+\partial_h\sin\theta(\mu,r,h)\sin^2({\xi}/{2})\cot\pi\mu}{
\cos\theta(\mu,r,h)\cos^2({\xi}/{2})
+\sin\theta(\mu,r,h)\sin^2({\xi}/{2})\cot\pi\mu}.
\end{eqnarray}
Recall that $\tan 2\theta(\mu,r,h)\equiv r\sin 2\pi\mu /(h-\cos 2\pi\mu)$ and
thus
\begin{equation}
\cos\theta=\sqrt{(1+\cos2\theta)/2}, \  \sin\theta=\sqrt{(1-\cos2\theta)}, \
\cos2\theta(\mu,r,h)= \frac{h-\cos 2\pi\mu}{\sqrt{(r\sin 2\pi\mu)^2 +(h-\cos
2\pi\mu)^2}}.
\end{equation}
Putting everything in, we get
\begin{eqnarray}
&&\partial_h {\cal E}(r,h)=-\frac{r}{\ln2}\int_0^{\frac{1}{2}}d\mu\,
\frac{\sin 2\pi\mu}{(r\sin 2\pi\mu)^2 +(h-\cos 2\pi\mu)^2} \nonumber\\
&&\quad\frac{\sqrt{\sqrt{\phantom{AAA}}-(h-\cos2\pi\mu)}\cos^2(\xi/2)-\sqrt{\sqrt{\phantom{AAA}}+(h-\cos2\pi\mu)}\sin^2(\xi/2)\cot\pi\mu}{\sqrt{\sqrt{\phantom{AAA}}+(h-\cos2\pi\mu)}\cos^2(\xi/2)+\sqrt{\sqrt{\phantom{AAA}}-(h-\cos2\pi\mu)}\sin^2(\xi/2)\cot\pi\mu},
\end{eqnarray}
where $\sqrt{\phantom{AAA}}\equiv\sqrt{(r\sin 2\pi\mu)^2 +(h-\cos
2\pi\mu)^2}$. Consider $h>1$ and $\epsilon\equiv h-1$. Make a change of
variable $t=h-\cos2\pi\nu$, giving lower and upper limits $\epsilon$ and
$2+\epsilon$, respectively. Shift the variable by $\epsilon$, giving
\begin{eqnarray}
&&\partial_h {\cal E}(r,h)=-\frac{r}{2\pi\ln2}\int_0^{{2}}d\,t
\frac{1}{(1-r^2)t^2+ 2(r^2+\epsilon)t+\epsilon^2}  \nonumber\\
&&\quad\frac{\sqrt{t}\sqrt{\sqrt{\phantom{AAA}}-(t+\epsilon)}\cos^2(\xi/2)-\sqrt{\sqrt{\phantom{AAA}}+(t+\epsilon)}\sin^2(\xi/2)\sqrt{2-t}}{\sqrt{t}\sqrt{\sqrt{\phantom{AAA}}+(t+\epsilon)}\cos^2(\xi/2)+\sqrt{\sqrt{\phantom{AAA}}-(t+\epsilon)}\sin^2(\xi/2)\sqrt{2-t}},
\end{eqnarray}
where $\sqrt{\phantom{AAA}}=\sqrt{(1-r^2)t^2+ 2(r^2+\epsilon)t+\epsilon^2}$.
As $h\rightarrow1$ or $\epsilon\rightarrow0$, the above expression diverges,
with the contribution coming from $t$ small, i.e., infrared divergence. Notice
further that only the second term in the numerator contributes to the
divergence. We proceed to evaluate the integral by separating into two parts:
\begin{equation}
\int_0^2=\int_0^\delta +\int_\delta^2,
\end{equation}
with $\delta\ll 1$. In the first region, we need only to keep $t$ to first
order at most. Also noting that for $t,\epsilon \ll 1$, the first term in the
denominator is much less than the second term, we get
\begin{eqnarray}
-\frac{r}{2\pi\ln2}\int_0^{{\delta}}d\,t
\frac{-1}{2r^2(t+\frac{\epsilon^2}{2r^2})}
\frac{\sqrt{\sqrt{2(r^2+\epsilon)t+\epsilon^2}+(t+\epsilon})}{\sqrt{\sqrt{2(r^2+\epsilon)t+\epsilon^2}-(t+\epsilon})}.
\end{eqnarray}
Simplify the second term ($\epsilon$ is ignored when there is no danger of
doing so)
\begin{equation}
\frac{\sqrt{\sqrt{2(r^2+\epsilon)t+\epsilon^2}+(t+\epsilon})}{\sqrt{\sqrt{2(r^2+\epsilon)t+\epsilon^2}-(t+\epsilon})}=\frac{\sqrt{t+\frac{\epsilon^2}{2r^2}}}{\sqrt{t}}+\frac{t+\epsilon}{\sqrt{2r^2
t}},
\end{equation}
and notice that only the first term on the right-hand side contributes to the
divergence. Thus the divergence part is
\begin{eqnarray}
-\frac{r}{2\pi\ln2}\int_0^{{\delta}}d\,t
\frac{-1}{2r^2(t+\frac{\epsilon^2}{2r^2})}
\frac{\sqrt{t+\frac{\epsilon^2}{2r^2}}}{\sqrt{t}}=
\frac{1}{4r\pi\ln2}\int_0^{{\delta}2 r^2/\epsilon^2}d\,t \frac{1}{\sqrt{t+1}}
\frac{1}{\sqrt{t}}.
\end{eqnarray}
The divergence part is then (for ${\delta}2 r^2/\epsilon^2\gg 1$)
\begin{equation}
\frac{1}{4r\pi\ln2}2\,{\sinh}^{-1}\sqrt{{\delta}2
r^2/\epsilon^2}\approx\frac{1}{2r\pi\ln2} \ln \left(2\sqrt{{\delta}2
r^2/\epsilon^2}\right).
\end{equation}
Therefore, the divergence in $\epsilon$ is
\begin{equation}
-\frac{1}{2r\pi\ln2} \ln \epsilon.
\end{equation}
Note that the part that involves $\delta$ has to be cancelled by the second
half of the integration $\int_\delta^2$, which we will not verify here.

In deriving the above divergence form, we have assumed that $r\ne 0$. Similar
consideration can be applied to the case when $h$ approaches 1 from below,
which we do not show here.

}

\begin{figure}
\psfrag{E}{$\partial{\cal E}_{N}/\partial h \vert_{h_{{\rm
max},N}}$}\psfrag{N}{$\ln N$}
\centerline{\psfig{figure=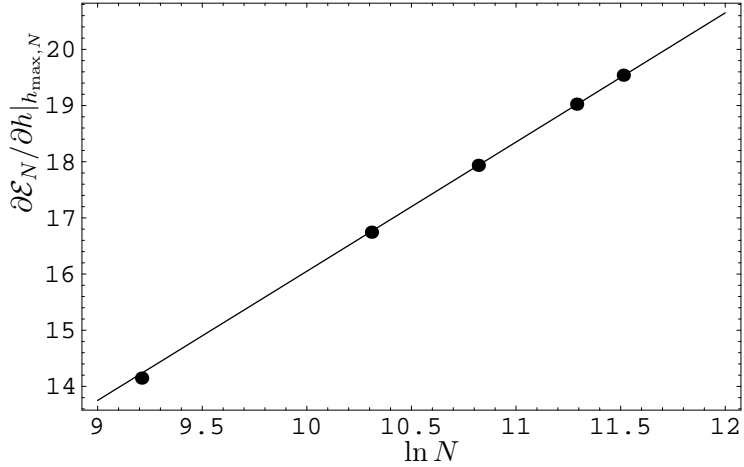,width=10cm,angle=0}}
\caption[Finte-size scaling]{Finte-size scaling. $\partial{\cal
E}_{N}/\partial h \vert_{h_{{\rm max},N}}$ vs. $\ln N$, for $N=10^4,
3\times10^4, 5\times10^4, 8\times10^4,$ and  $10^5$ (points) with $r=0.1$. The
solid line represents the fit $\partial{\cal E}_{N}/\partial h \vert_{h_{{\rm
max},N}}\approx 2.30 \ln N -6.95$.} \label{fig:dEvsNr01}
\end{figure}
In Appendix~\ref{app:DivXY} we analyze the singular behavior of the
field-derivative of the entanglement density~(\ref{eq:infNent}) in the
vicinity of the quantum critical line, and we find that the asymptotic
behavior (for $r\ne 0$)
\begin{equation}
\label{eqn:EntDiv} \frac{\partial{\cal E}}{\partial h} \approx -\frac{1}{2\pi
r \ln2}\ln|h-1|, \,\,\, \mbox{for $\vert{h-1}\vert\ll 1$}.
\end{equation}
From the arbitrary-$N$ results~(\ref{eqn:overlap}) of the entanglement we
analyze the approach to the thermodynamic limit, in order to develop further
connections with quantum criticality. We focus on the exponent $\nu$, which
governs the divergence at criticality of the correlation length: $L_c\sim
|h-1|^{-\nu}$. To do this, we compare the divergence of the slope
$\partial{\cal E}_{N}/\partial h$ (i)~near $h=1$ (at $N=\infty$), given above,
and (ii)~for large $N$ at the value of $h$ for which the slope is maximal
(viz.\ $h_{{\rm max},N}$), i.e., $\partial{\cal E}_{N}/\partial h
\vert_{h_{{\rm max},N}}\approx 0.230r^{-1}\ln N + {\rm const.}$,
 obtained by analyzing Eq.~(\ref{eqn:overlap})
for various values of $r$; see Fig.~\ref{fig:dEvsNr01} for the example of
$r=0.1$ case. Then, noting that $(2\pi\ln2)^{-1}\approx 0.2296$ and that the
logarithmic scaling hypothesis~\cite{Barber83} (or see
Appendix~\ref{app:finitesize}) identifies $\nu$ with the ratio of the
amplitudes of these divergences, $0.2296/0.230\approx 1$, we recover the known
result that $\nu=1$. Moreover, from Eq.~(\ref{eqn:znu}) we can extract the
value of the dynamical exponent: $z=1$.

\def\XX{
\subsection{Divergence of entanglement-derivative in XX}
The analysis for XX is much simpler. First note when $r=0$,
\begin{equation}
\cos2\theta(\mu,h)={\rm sgn}(h-\cos2\pi\mu).
\end{equation}
Let $\mu_0\equiv (\cos^{-1}h)/(2\pi)$. We have the expression for entanglement
density
\begin{eqnarray}
{\cal E}(h) &=&-\frac{2}{\ln2}\max_\xi \int_0^{\frac{1}{2}}d\mu\,
\ln\left[\cos\theta(\mu,r,h)\cos^2({\xi}/{2})
+\sin\theta(\mu,r,h)\sin^2({\xi}/{2})\cot\pi\mu\right]\\
&=&-\frac{2}{\ln2}\max_\xi\left[ \int_0^{\mu_0}d\mu\,
\ln\sin^2({\xi}/{2})\cot\pi\mu + \int_{\mu_0}^{\frac{1}{2}}d\mu\,
\ln\cos^2({\xi}/{2})\right].
\end{eqnarray}
Stationarity gives
\begin{equation}
\sin\xi[\mu_0- \frac{1}{4}(1-\cos\xi)]=0.
\end{equation}
The first case $\xi=0$ is good for large $h$ ($h\ge 1$). It is straightforward
to see that the entanglement density is identically zero. The second case
gives
\begin{equation}
\cos\xi=1-\frac{2}{\pi}\cos^{-1}h.
\end{equation}
Thus for $h\in[0,1]$
\begin{equation}
{\cal E}(h)
=-\frac{2}{\ln2}\left[\mu_0(h)\ln\frac{2\mu_0(h)}{1-2\mu_0(h)}+\frac{1}{2}\ln(1-2\mu_0(h))+\int_0^{\mu_0(h)}
d\mu\ln \cot\pi\mu\,\right].
\end{equation}
For $h>1$, ${\cal E}(h)=0$. The entanglement density at $h=0$ is the highest
among all XY models, which being ${\cal E}(0)=1- 2 \gamma_C/(\pi \ln 2)\approx
0.159$. (where $\gamma_C\approx 0.9160$ is the {\it Catalan\/} constant). The
entanglement density decreases monotonically as $h$ increases until $h=1$
beyond which it becomes zero identically. As the total z-component spin is
conserved, increasing $h$ simply increases the z-component spin of the ground
state until $h=1$ where all the spins are aligned with the field. The behavior
of entanglement is quite different from that of the transverse Ising model.

By directly taking the derivative w.r.t $h$, we get
\begin{equation}
\partial_h {\cal E}(h)=\frac{1}{\pi\ln2 \sqrt{1-h^2}} \ln\left[\frac{\cos^{-1}h}{\pi-\cos^{-1}h}\sqrt{\frac{1+h}{1-h}}\,\right].
\end{equation}
Near $h\approx 1$, we have (putting $1+h=2$ and evaluating the limit in the
argument of log function)
\begin{equation}
\frac{\partial}{\partial h}{\cal E}(0,h)\approx
-\frac{\log_2(\pi/2)}{\sqrt{2}\,\pi} \frac{1}{\sqrt{1-h}},
\qquad (h\to 1^{-}).
\end{equation}
}

We derive in Appendix~\ref{app:DivXX} the divergence behavior of the
field-derivative of the entanglement density for the isotropic ($r=0$) case.
Compared with $r\ne 0$ case, the nature of the divergence of $\partial{\cal
E}/\partial h$ at $r=0$  belongs to a different universality class:
\begin{equation}
\label{eqn:EntDivXX} \frac{\partial}{\partial h}{\cal E}(0,h)\approx
-\frac{\log_2(\pi/2)}{\sqrt{2}\,\pi} \frac{1}{\sqrt{1-h}},
\qquad (h\to 1^{-}).
\end{equation}
From this divergence, the scaling hypothesis, and the assumption that the
entanglement density is intensive, we can infer the known
result~\cite{Sachdev} that the critical exponent $\nu=1/2$ for the XX model.
Moreover, from Eq.~(\ref{eqn:znu}) we can extract the value of the dynamical
exponent $z=2$ for the XX model.

In keeping with the critical features of the XY-model phase diagram, for any
small but nonzero value of the anisotropy, the critical divergence of the
entanglement derivative is governed by Ising-type behavior. It is only at the
$r=0$ point that the critical behavior of the entanglement is governed by the
XX universality class. For small $r$, XX behavior ultimately crosses over to
Ising behavior.

We have mentioned that along the disorder line, $r^2+h^2=1$, the entanglement
density vanishes. One extreme limit is the Ising case, i.e., $r=1$ and $h=0$,
where the ground state is either
$\ket{\!\!\rightarrow\rightarrow\dots\rightarrow}$ or
$\ket{\!\!\leftarrow\leftarrow\dots\leftarrow}$, both of these being
unentangled. Any superposition of them is also a valid ground state, but it
has entanglement of order unity. In the thermodynamic limit the entanglement
per spin is identically zero. Is this a general feature along the disorder
line? Before we establish this, recall that the energies of the lowest two
levels are given in Eqs.~(\ref{eqn:E0}) and (\ref{eqn:E1/2}). Evaluating them
at $r^2=1-h^2$, we immediately find that  both are $-N$.

Now let us evaluate the expectation value of the Hamiltonian~(\ref{eqn:HXY})
with respect to a separable state with all spins pointing in the same
direction:
\begin{equation}
\langle{\cal H}\rangle =- N \left(\frac{1+r}{2}\langle\sigma^x\rangle^2
+\frac{1-r}{2}\langle\sigma^y\rangle^2+h\langle\sigma^z\rangle\right).
\end{equation}
Denoting $x\equiv\langle\sigma^x\rangle$, $y\equiv\langle\sigma^y\rangle$,
$z\equiv\langle\sigma^z\rangle$, we find that the above expression achieves
its minimum value $-N$ at $r^2+h^2=1$ when
\begin{equation}
(x,y,z)=\left(\pm\sqrt{\frac{2r}{1+r}},0,\sqrt{\frac{1-r}{1+r}}\right).
\end{equation}
Therefore, the separable state satisfying the above conditions is the ground
state, and there is a $Z_2$ degeneracy. Hence, along the disorder line the
entanglement density vanishes.

\section{Concluding remarks}
In summary, we have quantified the global entanglement of the quantum XY spin
chain.  This model exhibits a rich phase structure, the qualitative features
of which are reflected by this entanglement measure. Perhaps the most
interesting aspect is the divergence in the field-derivative of the
entanglement as the critical line ($h=1$) is crossed.  The behavior of the
divergence is dictated by the universality class of the model. Furthermore, in
the thermodynamic limit, the entanglement density vanishes on the disorder
line ($r^2+h^2=1$). The structure of the entanglement surface, as a function
of the parameters of the model (the magnetic field $h$ and the coupling
anistotropy $r$), is surprisingly rich.

We close by pointing towards a deeper connection between the global measure of
entanglement and the correlations among quantum fluctuations. The maximal
overlap~(\ref{eq:lambdamax}) can be decomposed in terms of correlation
functions (see Sec.~\ref{sec:correlations}):
\begin{equation}
\Lambda_{\max}^2 =\frac{1}{2^N}+ \frac{N}{2^N}\max_{|\vec{r}|=1}
\Big\{\langle\vec{r}\cdot\vec{\sigma}_{1}\rangle+\frac{1}{2}
\sum_{j=2}^N\langle\vec{r}\cdot \vec{\sigma}_{1} \otimes\vec{r}\cdot
\vec{\sigma}_{j}\rangle+ \cdots\Big\},\nonumber
\end{equation}
where translational invariance is assumed and the Cartesian coordinates of
$\vec{r}$ can be taken to be $(\sin\xi,0,\cos\xi)$. The two-point correlations
appearing in the decomposition are related to a bi-partite measure of
entanglement, namely, the concurrence, which shows similar singular
behavior~\cite{OsterlohAmicoFalciFazio02} to Eq.~(\ref{eqn:EntDiv}). It would
be interesting to establish the connection between the global entanglement and
correlations more precisely, e.g., by identifying which correlators are
responsible for the singular behavior in the entanglement and how they relate
to the better known critical properties.



\appendix

\chapter{Local hidden variable theories and Bell-CHSH inequality}
\label{app:Bell1} Local hidden variable (LHV) theories were conceived as
alternatives for quantum mechanics and were constructed in order to explain
the predictions of quantum mechanics in a statistical way without invoking the
nonlocal features that quantum mechanics allows. The most natural quantity for
two observers at a distance to measure is the correlation. Local hidden
variable theories dictate that the outcome of some measurement is
predetermined by the combination of measurement setting and some unknown local
hidden variable,  and that the result at one site should not be influenced by
that at the other. Suppose $A$ and $B$ are operators to be measured at the two
different sites, respectively. The correlation predicted by the LHV theories
is
\begin{equation}
E_L(a,b)\equiv\int d\lambda\, \rho(\lambda) A(a,\lambda) B(b,\lambda),
\end{equation}
where $A(a,\lambda)=\pm1$ and $B(b,\lambda)=\pm1$ are predetermined results
for the measurement settings $a$ for $A$ and $b$ for $B$ depending on the
local hidden variable $\lambda$; $\rho(\lambda)$ is the distribution for the
local hidden variable. Locality requires that the outcome $A(a,\lambda)$ does
not depend on $b$ and that of $B(b,\lambda)$ does not depend on $a$. For two
different settings at each site one arrives at the inequality
\begin{equation}
|E_L(a,b)+E_L(a,b')+E_L(a',b)-E_L(a',b')|\le 2.
\end{equation}
We shall see shortly that quantum mechanics can violate this inequality. To be
specific, the operators to be measured are the Pauli operators $\vec{\sigma}$.
The correlation
\begin{equation}
E_Q(a,b)\equiv \langle
\vec{\sigma}\cdot\vec{a}\otimes\vec{\sigma}\cdot\vec{b}\rangle,
\end{equation}
where $\vec{a}$ and $\vec{b}$ are unit vectors indicating the directions of
the Stern-Gerlach apparata at $A$ and $B$, respectively. Define
\begin{equation}
2B\equiv\vec{\sigma}\cdot\vec{a}\otimes\vec{\sigma}\cdot\vec{b}
+\vec{\sigma}\cdot\vec{a}\otimes\vec{\sigma}\cdot\vec{b'}+
\vec{\sigma}\cdot\vec{a'}\otimes\vec{\sigma}\cdot\vec{b}-
\vec{\sigma}\cdot\vec{a'}\otimes\vec{\sigma}\cdot\vec{b'}.
\end{equation}
It is straightforward to see that for a separable state
$\rho_s=\sum_i p_i \rho_A^i\otimes \rho_B^i$
\begin{equation}
\max_{a,a',b,b'}|2{\rm Tr}(B\rho_s)|\le 2,
\end{equation}
namely, it never violates the inequality. But for a singlet state
$(\ket{\uparrow\downarrow}-\ket{\downarrow\uparrow})/\sqrt{2}$, we have
\begin{equation}
\max_{a,a',b,b'}|\langle2B\rangle|=2\sqrt{2}.
\end{equation}
This can be achieved for the settings $\theta_a=0$, $\theta_a'=\pi/2$,
$\theta_b=\pi/4$, and $\theta_b'=-\pi/4$, where the angles are measured from
the $z$-axis in the $z-x$ plane.

Gisin~\cite{Gisin91} was the first to show that for any entangled pure state,
e.g., $\alpha\ket{\uparrow\uparrow}+\beta\ket{\downarrow\downarrow}$ with
$\alpha\beta\ne 0$, the inequality is also violated. Horodecki and
co-workers~\cite{Horodecki395} derived the maximal violation for any two-qubit
mixed states. It was then clear that there exist many mixed states that are
entangled but do not violate the CHSH inequality. The question remains open
whether there exists a Bell-like inequality that is necessary and sufficient
for a state being entangled.
\chapter{Schumacher's quantum data compression}
\label{app:compression}
\section{Quantum Data Compression}
To communicate quantum information by directly transmitting qubits may be
costly. The idea of quantum data compression (QDC) is to ask the question
whether we can compress the message into fewer qubits so as to minimize the
cost of transmission. Schmacher has provided an answer to achieve
this~\cite{Schumacher95}. One excellent review of quantum data compression is
by Preskill~\cite{Preskill}, which we follow here.

Let us go straight to the procedure of QDC. Suppose Alice needs to communicate
with Bob through some noiseless quantum channel as efficiently as possible,
that is, she hopes to compress her message using as few qubits  as possible.
The message consists of letters represented by some states $|\phi_x\rangle $.
Since on average, the frequency of each letter's appearance may not be equal
but is some probability $p_x$, the message can be said to be drawn from an
ensemble of states:
\begin{eqnarray}
\{|\phi_x\rangle ,p_x\}, \label{eq:ensemble}
\end{eqnarray}
so each letter has a density matrix
$\mbox{\boldmath $\rho$}=\sum_x p_x|\phi_x\rangle \langle \phi_x|$.
As we will see in the following, the lowest number of qubits per letter needed
to encode is set by the von Neumann entropy
  $S(\rho)=-\mbox{tr}(\rho\log_2\rho)$.
If we try to compress into fewer qubits, the fidelity of compression will be
ruined.

If the total length of the message is $n$, then the message has a density
matrix which is a direct product of $n$ letter density matrices
\begin{eqnarray}
\mbox{\boldmath$\rho$}^{\otimes n}\equiv\underbrace{ \mbox{\boldmath$\rho$}
\otimes\cdots\otimes\mbox{\boldmath$\rho$} }_{\mbox{$n$ $\rho$'s}}.
\end{eqnarray}
The procedure for quantum data compression goes as follows, \\
1)~Diagonalize {\boldmath$\rho$}. Work in the orthonormal basis in which
{\boldmath$\rho$} is diagonal. If {\boldmath $\rho$} has eigenvalues (arranged
decreasingly) ${\lambda_1\ge\lambda_2\ge\cdots\ge\lambda_d}$(the number $d$
depends on whether the state $|\phi\rangle $ is a qubit, tri-bit, or d-bit),
then $\mbox{\boldmath $\rho$}^{\otimes n}$  has eigenvalues of the form (i.e.,
the eigenvalues are obtained by choosing $n$ values from
$\lambda_1,\lambda_2,\ldots,\lambda_d$)
\begin{eqnarray}
\lambda(\{k_i\})\equiv\prod_{i=1}^{d}\lambda_i^{k_i}=\lambda_1^{k_1}\lambda_2^{k_2}\cdots\lambda_d^{k_d},
\end{eqnarray}
where $\sum_i^d k_i=n$, and each eigenvalue $\lambda(\{k_i\})$ occurs
$N(\{k_i\})={n!}/({k_1!k_2!\cdots k_d!})$
times. We will restrict ourselves to qubits, i.e., $d=2$: $\lambda(k_1,k_2=n-k_1)=\lambda_1^{k_1}\lambda_2^{k_2}$, and $N(k_1,k_2)=C^n_{k_1}$. \\
2)~Given a set of tolerances $\delta$ (tolerance for using slightly more
qubits than the asymptotically optimal case) and $\epsilon$ (tolerance for not
projecting onto the ``typical'' subspace), find the typical subspace $\Lambda$
and its dual subspace $\Lambda^\perp$:

2a)~First find out the smallest number $D(n)$ of necessary largest eigenvalues
(suppose they are $\lambda_{n,1}\ge\lambda_{n,2}\ge\cdots\ge\lambda_{n,D}$)
and corresponding eigenvectors($|\lambda_{n,1}\rangle ,|\lambda_{n,2}\rangle
,\cdots,|\lambda_{n,D}\rangle $) of ${\mbox{\boldmath $\rho$}}^{\otimes n}$
such that the sum of these eigenvalues is larger than $1-\epsilon$, but the
value $D(n)$ may be larger than $2^{n (S(\rho)+\delta)}$. Increase $n$ and
repeat the step until when $n>n_0$
\begin{eqnarray}
D(n)\le 2^{n (S(\rho)+\delta)},
\end{eqnarray}
where $n_0$ is the smallest number such that the above inequality is
satisfied. Note that it is sufficient to use at most some $n (S(\rho)+\delta)$
qubits to represent any state in $\Lambda$. This means there exists a unitary
transformation {\boldmath $U$} which takes any state $|\phi_\Lambda\rangle $
in $\Lambda$ to
\begin{eqnarray}
\mbox{\boldmath $U$}|\phi_\Lambda\rangle =|\phi_{\mbox{compressed}}\rangle
|0_{\mbox{rest}}\rangle ,
\end{eqnarray}
where $|\phi_{\mbox{compressed}}\rangle $ is a state of $n(S(\rho)+\delta)$
qubits, and $|0_{\mbox{rest}}\rangle $ is a state $|0\rangle
\otimes\cdots|0\rangle $ of ($n-n(S+\delta)$) qubits.

2b)~Those eigenvectors corresponding to the first $D(n)$ largest eigenvalues
span a typical subspace $\Lambda$, the remaining spanning a dual subspace
$\Lambda^\perp$. Note this division into two subspaces can be represented by a
projection operator {\boldmath $E$}
 which projects onto $\Lambda$ and the complement of which {\boldmath $1-E$}
to $\Lambda^\perp$. The condition that the sum of eigenvalues of eigenvectors
in $\Lambda$ is larger than $1-\epsilon$ can be rewritten as
\begin{eqnarray}
\mbox{tr}({\mbox{\boldmath $\rho$}}^n{\mbox{\boldmath $E$}})>1-\epsilon.
\end{eqnarray}
This means states in $\Lambda$ have much higher overlap with any state
drawn from the ensemble than those in $\Lambda^\perp$. \\
3)~Prepare the input state $|\psi\rangle =|\phi_1\rangle \cdots|\phi_n\rangle
$, where $|\phi_i\rangle $ belongs to the ensemble in the
Eq.(\ref{eq:ensemble}). Make the unitary transformation {\boldmath $U$} on
$|\psi\rangle $, and measure the state of the last ($n-n(S+\delta)$) qubits
mentioned above. If the result is $|0_{\mbox{rest}}\rangle $, Alice
successfully compresses $|\psi\rangle $ onto $|\psi_{\mbox{compressed}}\rangle
|0_{\mbox{rest}}\rangle $, and she simply sends
$|\psi_{\mbox{compressed}}\rangle $ to Bob. On the other hand, if Alice gets
the results other than $|0_{\mbox{rest}}\rangle $, she fails to compress her
message and the best she can do is send a state
$|0_{\mbox{compressed}}'\rangle $ which is the compressed state corresponding
to the largest eigenvector $|\lambda_{n,1}\rangle $ in $\Lambda$,
\begin{eqnarray}
\mbox{\boldmath $U$}|\lambda_{n,1}\rangle =|0_{\mbox{compressed}}'\rangle
|0_{\mbox{rest}}\rangle .
\end{eqnarray}
We note that the input state $|\psi\rangle $ has much higher overlap with
states in $\Lambda$ than any other states in $\Lambda^\perp$, the result for
Alice to get $|0_{\mbox{rest}}\rangle $ is of high probability
(larger than $1-\epsilon$).  \\
4)Bob, after receiving $|\psi_{\mbox{compressed}}\rangle $, appends
$|0_{\mbox{rest}}\rangle $ to it, and applies the inverse unitary
transformation {\boldmath $U^{-1}$}.  On average, Bob receives a density
matrix
\begin{eqnarray}
\mbox{\boldmath $\rho$'}^n=\mbox{\boldmath $E$}|\psi\rangle \langle
\psi|\mbox{\boldmath $E$}+
  |\lambda_{n,1}\rangle \langle \lambda_{n,1}|\langle \psi|(\mbox{\boldmath $1-E$})|\psi\rangle .
\end{eqnarray}

The averaged fidelity $\overline F$ of this procedure over the ensemble of
possible messages $\{|\psi_i\rangle ,p'_i\}$ can be shown to be larger than
$1-2\epsilon$. It can also be shown that if we try to compress the message
into $n(S(\rho)-\delta)$ qubits, the fidelity will be arbitrarily small for
sufficient large $n$.

Finally, we note quantum data compression cannot compress messages drawn from
a completely(maximally) mixed state, since $S(\rho_{\mbox{ completely
mixed}})=1$.

\section{An Example}
The example we will discuss shortly is for a small $n$. From previous
discussion, we know for any given $\delta$ and $\epsilon$, we can always find
a number $n_0$, such that for any $n> n_0$, the procedure succeeds with the
prescibed tolerances. In fact, for a given $n$, as $\epsilon$ becomes smaller,
the necessary $\delta$ increases, which means we need more qubits to compress,
as can be seen from the average number of qubits necessary to encode is $n
(S(\rho)+\delta)$. On the other hand, for a given $n$, as $\delta$ decreases
(we require fewer qubits to encode), $\epsilon$ increases, which means the
average fidelity decreases, as can be seen from $\overline F>1-2\epsilon$.
Hence, there is some tradeoff between $\delta$, and $\epsilon$ for a fixed
finite $n$.

Suppose the ensemble consists of $\{(|H\rangle ,p_H=\frac{1}{2}), (|D\rangle
,p_D=\frac{1}{2})\}$, where $|H\rangle $ is the state of horizontal
polarization while $|D\rangle $ is $45^\circ$,
\begin{eqnarray}
 |H\rangle = {1 \choose 0}, \
 |D\rangle = \frac{1}{\sqrt{2}}{1 \choose 1}.
\end{eqnarray}
The density matrix is
\begin{eqnarray}
  \rho=\frac{1}{2}|H\rangle \langle H|+\frac{1}{2}|D\rangle \langle D|=
     \left (\begin{array}{rr}
              \frac{3}{4} & \frac{1}{4} \\
              \frac{1}{4} & \frac{1}{4}
            \end{array}
     \right),
\end{eqnarray}
where the matrix is in $|H\rangle $ and $|V\rangle $ basis, and has two
eigenvectors and eigenvalues
\begin{eqnarray}
\begin{array}{l}
|Q\rangle =|22.5^\circ\rangle =\begin{pmatrix}{\cos\frac{\pi}{8}} \cr
{\sin\frac{\pi}{8} }\end{pmatrix},\
  \lambda_Q=\cos^2\frac{\pi}{8} \\
|\overline Q\rangle  =|112.5^\circ\rangle =\begin{pmatrix}{-\sin\frac{\pi}{8}}
\cr {\cos\frac{\pi}{8} }\end{pmatrix},\
  \lambda_{\overline Q} =\sin^2\frac{\pi}{8}.
\end{array}
\end{eqnarray}
The von Neumann entropy $S(\rho)$ is
\begin{eqnarray}
 S(\rho)=-\lambda_Q \log_2\lambda_Q - \lambda_{\overline Q}
       \log_2\lambda_{\overline Q}\approx 0.60088,
\end{eqnarray}
so the minimal number of qubits per letter needed to encode is about 0.6009.

Suppose Alice needs to send 3 letters to Bob, but she can afford only two
qubits. Since $3*S(\rho)\approx 1.8$, it's possible to compress 3 letters
using only 2 qubits with high fidelity. Note this means $\delta \approx 0.2$
and $D(n=3)=2$. The eigenvalues and eigenvectors of $\rho^3$ are
\begin{eqnarray}
\begin{array}{l}
\lambda_1  = \cos^3\frac{\pi}{8},\ \lambda_2=\lambda_3=
           \lambda_4=\cos^2\frac{\pi}{8}\sin\frac{\pi}{8},
 \nonumber \\
\lambda_5  =  \lambda_6=\lambda_7=\cos\frac{\pi}{8}
               \sin^2\frac{\pi}{8}, \
\lambda_8 =\sin^3\frac{\pi}{8}.
\end{array}
\end{eqnarray}
\begin{eqnarray}
\begin{array}{l}
|1\rangle  =  |QQQ\rangle ,  \ |2\rangle  =  |QQ\overline Q\rangle ,\
|3\rangle = |Q\overline QQ\rangle ,\
|4\rangle = |\overline QQQ\rangle ,  \\
|5\rangle  =  |Q\overline Q\overline Q\rangle , \ |6\rangle = |\overline
QQ\overline Q\rangle ,\ |7\rangle = |\overline Q\overline QQ\rangle ,\
|8\rangle  = |\overline Q\overline Q\overline Q\rangle .
\end{array}
\end{eqnarray}
The subspace $\Lambda$ is spanned by $\{|1\rangle ,|2\rangle ,|3\rangle
,|4\rangle \}$, while its dual subspace $\Lambda^\perp$ is by $\{|5\rangle
,|6\rangle ,|7\rangle ,|8\rangle \}$, and
\begin{eqnarray}
P_\Lambda &\equiv&\mbox{tr}(\rho^{\otimes3}\mbox{\boldmath $E$})=
 \sum_{i=1}^{4}\lambda_i\approx 0.9419 \nonumber \\
P_\Lambda^\perp &\equiv&\mbox{tr}(\rho^{\otimes 3}(\mbox{\boldmath $1-E$}))=
 \sum_{i=5}^{8}\lambda_i\approx 0.0581.
\label{eq:P_L}
\end{eqnarray}
Alice and Bob both agree on the form of the unitary transformation that they
will use,
\begin{eqnarray}
\mbox{\boldmath U}\left(\begin{array}{l}
                   |1\rangle  \\
                   |2\rangle  \\
                   |3\rangle  \\
                   |4\rangle
                  \end{array}\right) \rightarrow
                  \left(\begin{array}{l}
                   |HHH\rangle  \\
                   |HVH\rangle  \\
                   |VHH\rangle  \\
                   |VVH\rangle
                  \end{array}\right) \quad
\mbox{\boldmath U}\left(\begin{array}{l}
                   |5\rangle  \\
                   |6\rangle  \\
                   |7\rangle  \\
                   |8\rangle
                  \end{array}\right) \rightarrow
                  \left(\begin{array}{l}
                   |HHV\rangle  \\
                   |HVV\rangle  \\
                   |VHV\rangle  \\
                   |VVV\rangle
                  \end{array}\right);
\end{eqnarray}
this transformation is unitary since it is simply the transformation between
two sets of orthonormal bases of 3 qubits.

Alice prepares her message in a state $|\psi\rangle $, which can be expanded
in the basis of $|1\rangle \cdots|8\rangle $,
\begin{eqnarray}
|\psi\rangle =\sum_{i=1}^8 a_i|i\rangle ,
\end{eqnarray}
where from Eq.(\ref{eq:P_L}) we have
\begin{eqnarray}
\sum_{i=1}^4 |a_i|^2=P_\Lambda \gg \sum_{i=5}^{8}|a_i|^2=P_\Lambda^\perp.
\end{eqnarray}
Then Alice applies the unitary transformation {\boldmath $U$} on $|\psi\rangle
$ followed by a measurement on the third qubit. If the result is $|H\rangle $,
she sucessfully projects $|\psi\rangle $ into the likely subspace $\Lambda$.
At this stage, the total state is
\begin{eqnarray}
 a_1|HHH\rangle +a_2|HVH\rangle +a_3|VHH\rangle +a_4|VVH\rangle =|\psi_{\mbox{compressed}}\rangle |H\rangle ,
\end{eqnarray}
where $|\psi_{\mbox{compressed}}\rangle\equiv a_1|HH\rangle +a_2|HV\rangle
+a_3|VH\rangle +a_4|VV\rangle $. She simply sends this two-qubit state
$|\psi_{\mbox{compressed}}\rangle $ to Bob. Upon receiving
$|\psi_{\mbox{compressed}}\rangle $, Bob appends a third qubit $|H\rangle $ to
it, and does the inverse transformation {\boldmath $U^{-1}$} to get
\begin{eqnarray}
 |\psi'\rangle  = \mbox{\boldmath $U^{-1}$}(|\psi_{\mbox{compressed}}\rangle |H\rangle )
          = \sum_{i=1}^{4}a_i|i\rangle ,
\end{eqnarray}
which has high resemblance to the initial $|\psi\rangle $, that is
\begin{eqnarray}
F_1\equiv|\langle \psi|\psi'\rangle |^2=P_\Lambda\approx 0.9419.
\end{eqnarray}
On the other hand, if, when Alice measures the third qubit and gets $|V\rangle
$, she fails to project $|\psi\rangle $ into $\Lambda$ but $\Lambda^\perp$
instead, the best she can do is send a qubit state $|HH\rangle $. After Bob
receives it and decompresses it, he gets
\begin{eqnarray}
 |\psi''\rangle =\mbox{\boldmath $U^{-1}$}(|HH\rangle |H\rangle )=|1\rangle ,
\end{eqnarray}
which has overlap with the initial $|\psi\rangle $:
\begin{eqnarray}
F_2\equiv|\langle \psi|\psi''\rangle |^2=|a_1|^2=\lambda_1\approx 0.6219.
\end{eqnarray}
The fidelity of this procedure is $F=P_\Lambda F_1 + P_\Lambda^\perp
F_2=0.9234$.

How good is this? Let us compare it to the case when Alice sends the first two
letters without compressing and asks Bob to guess the third letter. Since both
$|H\rangle $ and $|D\rangle $ from the ensemble have higher overlap with
$|Q\rangle $ than with $|\overline Q\rangle $, the best guess he can make is
$|Q\rangle $. The fidelity of this procedure is
\begin{eqnarray}
  F=\frac{1}{2}|\langle H|Q\rangle |^2+\frac{1}{2}|\langle D|Q\rangle |^2=0.8535,
\end{eqnarray}
which is smaller than the case when we do compression.

\chapter{Wootters' formula}
\label{app:Wootters} \noindent The entanglement of formation defined in
Eq.~(\ref{eqn:Ef}) is, in general,
difficult to calculate. However, for two-qubit systems, 
Wootters~\cite{Wootters98} has provided and proved the following formula:
\begin{equation}
E_{\rm F}(\rho)=h\left( \frac{1}{2}[1 + \sqrt{1-C(\rho)^2}] \right),
\end{equation}
where $h(x) \equiv  -x \log_2 x - (1-x) \log_2(1-x)$,
and $C(\rho)$, the {\em concurrence} of the state $\rho$, is defined as
\begin{equation}C(\rho) \equiv \max \{0,\sqrt{\lambda_1}-\sqrt{\lambda_2}-
 \sqrt{\lambda_3}-\sqrt{\lambda_4}\},
\end{equation}
in which $\lambda_1,\ldots,\lambda_4$ are the eigenvalues of the matrix $\rho
(\sigma_y\otimes\sigma_y)\rho^{*}(\sigma_y\otimes\sigma_y)$ in nonincreasing
order and $\sigma_y$ is a Pauli spin matrix. $E_{\rm F}(\rho)$, $C(\rho)$ and
the  {\it tangle\/} $\tau(\rho)\equiv C(\rho)^2$ are equivalent measures of
entanglement, inasmuch as they are monotonic functions of one another. For
pure state $a\ket{00}+b\ket{01}+c\ket{10}+d\ket{11}$ the concurrence $C$ is
$2|ad-bc|$.

\chapter{Proof of entanglement monotone}
\label{app:Monotone} In this appendix we prove in detail that the geometric
measure of entanglement satisfies the criteria listed in Sec.~\ref{sec:Mixed},
and hence it is an entanglement monotone. For convenience we list those
criteria as follows
\begin{itemize}
\item[C1.](a)~$E(\rho)\!\ge\! 0$;
(b)~$E(\rho)\!=\!0$ if $\rho$ is not entangled.
\item[C2.]Local unitary transformations do not change $E$.
\item[C3.]Local operations and classical communication (LOCC)
(as well as post-selection) do not increase the expectation value of $E$.
\item[C4.]Entanglement is convex under the discarding of
information, i.e., $\sum_i p_i\,E(\rho_i)\ge E(\sum_i p_i\,\rho_i)$.
\end{itemize}

From the definition~(\ref{eqn:Emixed})
\begin{eqnarray}
E(\rho) \equiv \coe{\rm pure}(\rho) \equiv {\min_{\{p_i,\psi_i\}}}
\sum\nolimits_i p_i \, E_{\rm pure}(|\psi_i\rangle),\nonumber
\end{eqnarray}
it is evident that C1 and C2 are satisfied, provided that $E_{\rm pure}$
satisfies them, as it does for $E_{\rm pure}$ being any function of
$\Lambda_{\max}$ consistent with C1.  It is straightforward to check that C4
holds, by the convex hull construction. First, consider the case is which
$\rho=\sum_i p_i|\psi_i\rangle\langle\psi_i|$. From the
definition~(\ref{eqn:Emixed}) of $E(\rho)$, which is the {\it minimum\/} over
all decompositions, we have that $E(\rho)\le\sum_i p_iE_{\rm
pure}(|\psi_i\rangle)$. Hence we have that $E(\sum_i
p_i|\psi_i\rangle\langle\psi_i|)\le \sum_i  p_i
E(|\psi_i\rangle\langle\psi_i|)$, i.e., C4 is obeyed whenever the deomposition
is into {\it pure\/} states. Second, allow $\rho_i$ to be mixed.  To deal with
this case, express $\rho_i$ as its optimal decomposition: $\rho_i=\sum_k
q_{ik}|\psi_{ik}\rangle\langle\psi_{ik}|$, for which $E(\rho_i)=\sum_k
q_{ik}E_{\rm pure}(|\psi_{ik}\rangle)$. Inserting the above expression for
$\rho_i$ and $E(\rho_i)$ into the left hand side of the sought criterion, and
using the pure-state result just proved, we find $\sum_i p_i E(\rho_i)=
  \sum_{ik} p_i\,q_{ik} E(|\psi_{ik}\rangle\langle\psi_{ik}|)\ge
E(\sum_{ik} p_i\,q_{ik}|   \psi_{ik}\rangle\langle\psi_{ik}|)= E(\sum_i
p_i\rho_i)$. Thus we see that C4 is indeed obeyed.

 The consideration of C3 seems to be more delicate.
The reason is that our analysis of whether or not it holds depends on the
explicit form of $E_{\rm pure}$. For C3 to hold, it is sufficient to show that
the average entanglement is non-increasing under any trace-preserving,
unilocal operation~\footnote{What we mean by unilocal is that the operation is
performed by only one of the parties. All general multi-party operations can
be regarded as a sequence of unilocal operations}: $\rho\rightarrow \sum_k
V_k\rho V_k^\dagger$, where the Kraus operator $V_k$ has the form
$\openone\otimes\cdots \openone\otimes
V_k^{(i)}\otimes\openone\cdots\otimes\openone$ and obeys $\sum_k V_k^{\dagger}
V_k=\openone$. Furthermore, it suffices to show that C3 holds for the case of
a pure  initial state, i.e., $\rho=\ketbra{\psi}$.

We now prove that C3 holds for the particular (and by no means un-natural)
choice $E_{\rm pure}=E_{\sin^2}$. To be precise, for any quantum operation on
a pure initial state, i.e.,
\begin{equation}
|\psi\rangle\langle\psi| \rightarrow\sum\nolimits_k
V_k|\psi\rangle\langle\psi|V_k^\dagger,
\end{equation}
we aim to show that
\begin{equation}
\sum_k p_k \,E_{\sin^{2}} \left({V_k|\psi\rangle}/\!{\sqrt{p_k}}\right) \le
E_{\sin^{2}}(|\psi\rangle),
\end{equation}
where $p_k\!\equiv\! {\rm Tr}\,V_k|\psi\rangle\langle\psi|V_k^\dagger
=\langle\psi|V_k^\dagger V_k|\psi\rangle$, regardless of whether the operation
$\{V_k\}$ is state-to-state or state-to-ensemble. Let us  denote by $\Lambda$
and $\Lambda_k$ the respective entanglement eigenvalues corresponding to
$\ket{\psi}$ and the (normalized) pure state
${V_k|\psi\rangle}/{\sqrt{p_k}}\,$. Then our task is to show that $\sum_k
p_k\,\Lambda_k^2 \ge \Lambda^2$, of which the left hand side is, by the
definition of $\Lambda_k$, equivalent to
\begin{equation}
\sum_k p_k \max_{\scriptscriptstyle\xi_k\in D_s} \Vert {\langle\xi_k|
V_k|\psi\rangle}/\!{\sqrt{p_k}} \Vert^2=\sum_k
\max_{\scriptscriptstyle\xi_k\in D_s}\Vert \langle\xi_k| V_k|
\psi\rangle\Vert^2.
\end{equation}
Without loss of generality, we may assume that it is the first party who
performs the operation. Recall that the condition~(\ref{eqn:EigenForm}) for
the closest separable state
\begin{equation}
\ket{\phi}\equiv|\tilde{\alpha}\rangle_1\otimes|\tilde{\gamma}
\rangle_{2\cdots n}
\end{equation}
can be recast as
\begin{equation}
{}_{2\cdots n}\langle\tilde{\gamma}|\psi\rangle_{1\cdots n}=
\Lambda|\tilde{\alpha}\rangle_1.
\end{equation}
Then, by making the specific choice
\begin{equation}
\langle\xi_k|= ({\langle\tilde{\alpha}|V_k^{(1)\dagger}/\!{\sqrt{q_k}}})
\otimes\langle\tilde{\gamma}|,
\end{equation}
where $q_k \equiv \langle\tilde{\alpha}| V_k^{(1)\dagger}V_k^{(1)}
|\tilde{\alpha}\rangle$, we have the sought result
\begin{eqnarray}
\sum_k p_k\Lambda_k^2= \sum_k \max_{\xi_k\in D_s} \Vert\langle\xi_k|
V_k|\psi\rangle\Vert^2 \ge\Lambda^2\sum_k ({\langle\tilde{\alpha}|
V_k^{(1)\dagger}V_k^{(1)}| \tilde{\alpha}\rangle/\!\sqrt{q_k}} )^2=\Lambda^2.
\end{eqnarray}
  Hence, the form $1-\Lambda^2$, when generalized to
mixed states, is an entanglement monotone. We note that a different approach
to establishing this result has been used by Barnum and
Linden~\cite{BarnumLinden01}. Moreover, using the result that $\sum_k
p_k\Lambda_k^2\ge \Lambda^2$, one can further show that for any convex
increasing function $f_c(x)$ with $x\in[0,1]$,
\begin{equation}
\sum_k p_k\,f_c(\Lambda_k^{2})\ge f_c(\Lambda^{2}).
\end{equation}
Therefore, the quantity ${const.}-f_c(\Lambda^{2})$ (where the ${const.}$ is
to ensure the whole expression is non-negative), when extended to mixed
states, is also an entanglement monotone, hence a good entanglement measure.

\def\VW{
\chapter{The Vollbrecht-Werner technique}
\label{app:VW} In this Appendix, we briefly review a technique developed by
Vollbrecht and Werner~\cite{VollbrechtWerner01} for computing the entanglement
of formation for the generalized Werner states; this turns out to be
applicable to the computation of the sought quantity $E_{\sin^2}$.  We start
by fixing some notation. Let
\begin{itemize}
\item[(a)] $K$ be a compact convex set
(e.g., a set of states that includes both pure and mixed ones);
\item[(b)] $M$ be a convex subset of $K$ (e.g., set of pure states);
\item[(c)] $E:M\rightarrow R\cup\{+\infty\}$ be a function that maps
elements of $M$ to the real numbers (e.g., $E=E_{\sin^2}$); and
\item[(d)] $G$ be a compact group of symmetries, acting on $K$
(e.g., the group $U\otimes U^\dagger$) as $\alpha_g:K\rightarrow K$ (where
$\alpha_g$ is the representation of the element $g\in G$) that preserve convex
combinations.
\end{itemize}

We assume that $\alpha_g M\subset M$ (e.g., pures states are mapped into pure
states), and that $E(\alpha_g m)=E(m)$ for all $m\in M$ and $g\in G$ (e.g.,
that the entanglement of a pure state is preserved under $\alpha_g$). We
denote by ${\rm\bf P}$ the invariant projection operator defined via
\begin{equation}
{\rm\bf P}k= \int dg\,\alpha_g(k),
\end{equation}
where $k\in K$. Examples of ${\rm\bf P}$ are the operations ${\rm\bf P}_1$ and
${\rm\bf P}_2$ in the main text. Vollbrecht and Werner also defined the
following real-valued function $\epsilon$ on the invariant subset ${\rm\bf
P}K$:
\begin{equation}
\epsilon(x)= {\rm inf} \left\{E(m)\vert m\in M, {\rm \bf P}m=x\right\}.
\end{equation}

They then showed that, for $x\in{\rm\bf P}K$,
\begin{equation}
{\rm co}\,E(x)= {\rm co}\,\epsilon(x),
\end{equation}
and provided the following recipe for computing the function ${\rm co}\,E$ for
$G$-invariant states:
\begin{itemize}
\item[1.] For every invariant state $\rho$
(i.e., obeying $\rho={\rm\bf P}\rho$), find the set $M_\rho$ of pure states
$\sigma$ such that ${\rm \bf P}\sigma=\rho$.
\item[2.] Compute
$\epsilon(\rho)\equiv{\rm inf} \left\{E(\sigma)\vert \sigma\in
M_\rho\right\}$.
\item[3.] Then ${\rm co}\,E$ is the convex hull of this function $\epsilon$.
\end{itemize}
}

\chapter{Three $N$-qubit Bell inequalities}
\label{app:Bell2}
\section{Mermin-Klyshko-Bell inequality}
In Appendix~\ref{app:Bell1} we have discussed the Bell-CHSH inequality for two
qubits. Here we will discuss its generalization to $N$ qubits. The discussion
here follows D\"ur~\cite{Dur01}. First let us define a Pauli operator at
arbitrary direction
\begin{equation}
\sigma_{a_k}\equiv \vec{\sigma}\cdot \vec{a}_k,
\end{equation}
where $\vec{a}_k$ is a unit vector. The Mermin-Klyshko-Bell inequality is
conveniently defined via Bell operators in a recursive way:
\begin{equation}
B_k=\frac{1}{2}B_{k-1}\otimes
(\sigma_{a_k}+\sigma_{a_k'})+\frac{1}{2}B'_{k-1}\otimes
(\sigma_{a_k}-\sigma_{a_k'}),
\end{equation}
with $B_1\equiv\sigma_{a_1}$, $B_1'\equiv\sigma_{a_1'}$, and $B_k'$ is
obtained from $B_k$ by exchanging all $a_k$ with $a_k'$ and vice versa. For
example,
\begin{subequations}
\begin{eqnarray} B_2&=&(\sigma_{a_1}\otimes\sigma_{a_2}+\sigma_{a_1'}\otimes\sigma_{a_2}
+\sigma_{a_1}\otimes\sigma_{a_2'}-\sigma_{a_1'}\otimes\sigma_{a_2'})/2\\
B_3&=&(\sigma_{a_1}\otimes\sigma_{a_2}\otimes\sigma_{a_3'}+\sigma_{a_1}\otimes\sigma_{a_2'}\otimes\sigma_{a_3}
+\sigma_{a_1'}\otimes\sigma_{a_2}\otimes\sigma_{a_3}-\sigma_{a_1'}\otimes\sigma_{a_2'}\otimes\sigma_{a_3'})/2.
\end{eqnarray}
\end{subequations}

It is easy to see that for real $x,x',y,y'$ in the range $[-1,1]$, we have
that $|xy+xy'+x'y-x'y'|\le2$. As $\langle B_1\rangle$ and $\langle
B_1'\rangle$ are evidently in the range $[-1,1]$, local hidden variable
theories will then predict that $|\langle B_2\rangle|\le 1$, and by induction
$|\langle B_k\rangle|\le 1$. However, quantum mechanics can violate this
inequality. In particular, for the $N$-partite GHZ state
\begin{equation}
\ket{\rm GHZ}=(\ket{0\dots0}+\ket{1\dots1})/\sqrt{2},
\end{equation}
$|\langle B_N\rangle|$ can achieve the value $2^{(N-1)/2}$.
\section{Three-setting Bell inequality}
There are actually at least two approaches to define a three-setting Bell
inequality. These all involve selecting three settings for measurement at each
site. The first approach is to consider the linear combination of the joint
probabilities or correlations $E_(\xi_1,\dots,\xi_N)$, with $\xi_i$ being
chosen from three possible settings. The goal is to construct an inequality
that is satisfied by local hidden variable theores whereas it can be violated
by quantum mechanics. However, the number of such inequalities grows with the
number of parties. It is only possible to exhaust all inequivalent
inequalities for small number of parties, as investigated by Collins and
Gisin~\cite{CollinsGisin04} very recently.  Here we shall focus on a second
approach~\footnote{It would be interesting to compare the difference between
the two approaches.}.

The basic idea of the second approach is that, for two vectors $h$ and $q$, if
$|\langle h|q\rangle|< ||q||^2$, this means that $h\ne q$. Specifically, $h$
represents prediction from local hidden variables whereas $q$ represents that
from quantum mechanics. To be more precise, quantum mechanics predicts that
for a state $\rho_N$ the average outcome of the observables
$\{O_i,i=1,\dots,N\}$ (with possible outcomes being $\pm1$) at $N$ locations
is
\begin{equation}
E_Q(\xi_1,\dots,\xi_N)\equiv {\rm Tr}(O_1\cdots O_N\rho_N),
\end{equation}
where $\xi_k$ is the measurement setting at the $k$-th location. A local
hidden variable theory predicts
\begin{equation}
E_L(\xi_1,\dots,\xi_N)=\int d\lambda \,\rho(\lambda)\prod_{k=1}^N
I_k(\xi_k,\lambda),
\end{equation}
where $\rho(\lambda)$ is the probability distribution for the local hidden
variable $\lambda$, and $I_k(\xi_k,\lambda)$ is the outcome ascribed by the
local hidden variable $\lambda$ for the observable $O_k(\xi_k)$ measured with
the apparatus setting $\xi_k$.

 The inner product between $E(\xi_1,\dots,\xi_N)$ and $E'(\xi_1,\dots,\xi_N)$
 is defined as
  \begin{equation}
 \langle E |E'\rangle\equiv\sum_{\xi_1,\dots,\xi_N}  E(\xi_1,\dots,\xi_N)\,E'(\xi_1,\dots,\xi_N),
 \end{equation}
 where each $\xi$ has three different values.
If it can be shown that $|\langle E_Q |E_L\rangle|^2< ||E_Q||^2$ then the two
theories have different predictions. Zukowski and
Kaszlikowski~\cite{ZukowskiKaszlikowski97} considered the following basis to
be measured:
\begin{equation}
\label{eqn:basis} \ket{\pm,\xi_i}_i\equiv (\ket{0}_i\pm \ket{1}_i)/\sqrt{2}
\end{equation}
for party $i$. For the N-partite GHZ state they chose three values for
$\xi_1$: $(\pi/6,\pi/2,5\pi/6)$, and for $\xi_i$: $(0,\pi/3,2\pi/3)$ for
$i=2,\dots,N$. They showed that
\begin{equation}
|\langle E_L|E_Q\rangle|\le 2^{N-1}\sqrt{3} < \Vert E_Q\Vert^2=3^N/2.
\end{equation}
Thus, the two theories, LHV and quantum mechanics, have different predictions.

\section{Functional Bell inequality}
The idea of functional Bell inequality~\cite{SenSenZukowski02} is to consider
a continuous range of measurement settings, instead of a finite number.
 The inner product between $E(\xi_1,\dots,\xi_N)$ and $E'(\xi_1,\dots,\xi_N)$
 for the continuous setting is defined as
 \begin{equation}
 \langle E |E'\rangle\equiv\int[\prod_{i=1}^N d\xi_i] E(\xi_1,\dots,\xi_N)\,E'(\xi_1,\dots,\xi_N).
 \end{equation}
Similarly, if it can be shown that $|\langle E_Q |E_L\rangle|^2< ||E_Q||^2$
then the two theories have different predictions. Sen and co-workers
considered the same measurement basis as in Eq.~(\ref{eqn:basis}), with
$\xi_i\in[0,2\pi]$. They showed that
\begin{equation}
|\langle E_L|E_Q\rangle|\le 4^{N} < \Vert E_Q\Vert^2=(2\pi)^N/2.
\end{equation}
Thus again, the two theories, LHV and quantum mechanics, have different
predictions.

To compare the three-setting and the functional inequalities, we mention that
for the D\"ur's bound entangled state~(\ref{eqn:Dur}) discussed in
Sec.~\ref{sec:DurBound}, violation is achieved using the three-setting one
only when $N\ge 7$ whereas for the functional one $N\ge 6$, the latter being
stronger.

\chapter{Unextendible product bases and bound entangled states}
\label{app:UPB} To illustrate the idea of the unextendible product basis and
its connection to bound entanglement, we shall give an explicit example.
The discussion here follows Ref.~\cite{BennettDiVincenzoMorShorSmolinTerhal99}. 
 product
basis is a set $S$ of pure orthogonal product states, which span a subspace
${\cal H}_S$ of ${\cal H}$, a bi- or multi-partite quantum system. For
example, $S=\{\ket{0,1,+},\ket{1,+,0},\ket{+,0,1},\ket{-,-,-}\}$ in a
three-qubit system, where we have defined $\ket{\pm}\equiv
(\ket{0}\pm\ket{1})/\sqrt{2}$. It is straightforward to check that these basis
states are orthogonal to one another.

This set has the peculiar property that it is not possible to add an
additional basis vector that is a product state. Suppose we add $\ket{a,b,c}$.
This then requires that
\begin{equation}
\ipr{a,b,c}{0,1,+}=\ipr{a,b,c}{1,+,0}=\ipr{a,b,c}{+,0,1}=\ipr{a,b,c}{-,-,-}=0.
\end{equation}
But each of $\ket{a}$, $\ket{b}$, and $\ket{c}$ can be, at most, orthogonal to
any of the four states $\ket{0}$, $\ket{1}$, $\ket{+}$, and $\ket{-}$. Hence,
together, $\ket{a,b,c}$ can be orthogonal to three of the four states in $S$.
This means that the set $S$ cannot be extended, hence, the name unextendible
product basis (UPB). Therefore, the subspace that is orthogonal to the space
by the UPB contains no product states, nor mixture of them.

What is the use of UPB? It turns out that it can be used to construct a bound
entangled state. Suppose $S=\{\ket{\psi_1},\dots,\ket{\psi_n}\}$ contains a
UPB. Then the mixed state
\begin{equation}
\rho=\frac{1}{D-n}\left(\openone-\sum_{j=1}^n\ketbra{\psi_j}\right)
\end{equation}
is a bound entangled state, where $D$ is the total dimension (e.g., D=8 in the
above three-qubit example).

The state $\rho$ is entangled because it lies in the subspace ${\cal H}-{\cal
H}_S$, in which, by construction, there is no product state, and hence, it
cannot be written as a decomposition of pure product states. To see that its
entanglement is bound, we can look at its partial transpose with respect to
all bi-partite partitionings. As a product state is mapped to a product state
under partial transpose and the identity is unchanged, $\rho$, under partial
transpose, is mapped to a valid density matrix, which has non-negative
eigenvalues. Therefore, $\rho$ has PPT, and hence no entanglement can be
established across any bi-partite cut via local operations and classical
communication. We have then shown that the uniform mixture on the subspace
complementary to that spanned by a UPB is a bound entangled state. In
particular, the state
\begin{equation}
\rho=\frac{1}{4}\left(\openone-\ketbra{0,1,+}-\ketbra{1,+,0}-\ketbra{+,0,1}-\ketbra{-,-,-}\right)
\end{equation}
is bound entangled. We have shown previously in Sec.~\ref{sec:concludeBound}
that such a state cannot violate, e.g., the Mermin-Klyshko-Bell inequality.

\chapter{Derivation of overlap of the ground state with the
separable Ansatz state} \label{app:Derivation} In this appendix we derive the
results shown in Eqs.~(\ref{eqn:overlap}) and (\ref{eqn:f}). We first analyze
$b=1/2$ case, viz., the even-fermion case. The lowest state
$\ket{\Psi_{1/2}(r,h)}$ has zero number of Bogoliubov fermions. It is related
to the state that has no $c$-fermions, i.e.,
$\ket{\Omega}\equiv\ket{\!\uparrow\cdots\uparrow}$ via
\begin{subequations}
\begin{eqnarray}
\ket{\Psi_{1/2}(r,h)}&=&\prod_{m=0}^{m<\frac{N\!-\!1}{2}}\cos\theta_m^{(1\!/2)}(r,h)
e^{i\tan\theta_m^{(1\!/2)}(r,h)\,\tilde{c}_m^{(1\!/2)\dagger}
\tilde{c}_{N-m-1}^{(1\!/2)\dagger} }\ket{\Omega}\\
&=&\prod_{m=0}^{m<\frac{N\!-\!1}{2}}\Big[\cos\theta_m(r,h)
+i\sin\theta_m(r,h)\,\tilde{c}_m^{\dagger} \tilde{c}_{N-m-1}^{\dagger}
\Big]\ket{\Omega}.
\end{eqnarray}
\end{subequations}
The Ansatz state is then
\begin{subequations}
\begin{eqnarray}
\ket{\Phi(\xi)} &=&\prod_{j=1}^N
\Big[ \cos\frac{\xi}{2}+\sin\frac{\xi}{2}\prod_{1\le l<j}(1-2c_l^\dagger c_l) (c_j^\dagger-c_j) \Big] \ket{\Omega} \\
&=&\prod_{j=1}^N\big(\cos\frac{\xi}{2}+\sin\frac{\xi}{2}\,c_j^\dagger\big)
\ket{\Omega}\\
&=&\cos^{N}\frac{\xi}{2}\, e^{\tan\frac{\xi}{2}c_1^\dagger}\cdots e^{\tan\frac{\xi}{2}c_N^\dagger}\ket{\Omega}\\
&=&\cos^{N}\frac{\xi}{2}\, e^{\tan\frac{\xi}{2}\sum_{j=1}^N c_j^\dagger}
e^{\tan^2\frac{\xi}{2}\sum_{j< l}c_j^\dagger c_l^\dagger}\ket{\Omega},
\end{eqnarray}
\end{subequations}
where we have suppressed the index $(1/2)$. The term $\sum_{j< l}c_j^\dagger
c_l^\dagger$
 can be rewritten in momentum space as
 \begin{equation}
 \sum_{1\le j< l\le N} c_j^\dagger c_l^\dagger=i\sum_{m=0}^{m<\frac{N-1}{2}}
 \cot\frac{\pi(m+\frac{1}{2})}{N}\tilde{c}^\dagger_m \tilde{c}^\dagger_{N\!-\!m\!-\!1} .
 \end{equation}
  Thus, for even $N$
\begin{subequations}
 \begin{equation}
 \ket{\Phi(\xi)}=\Big({1+\tan\frac{\xi}{2}\,\sum_{j=1}^N c_j^\dagger}\Big)\prod_{m=0}^{m<\frac{N-1}{2}}
\left(\cos^2\frac{\xi}{2}+i\sin^2\frac{\xi}{2}\cot\frac{\pi(m+\frac{1}{2})}{N}
\tilde{c}^\dagger_m \tilde{c}^\dagger_{N\!-\!m\!-\!1}\right)\ket{\Omega},
 \end{equation}
 whereas for odd $N$
 \begin{equation}
\ket{\Phi(\xi)}=\Big({1+\tan\frac{\xi}{2}\,\sum_{j=1}^N c_j^\dagger}\Big)
\cos\frac{\xi}{2}\prod_{m=0}^{m<\frac{N-1}{2}}
\left(\cos^2\frac{\xi}{2}+i\sin^2\frac{\xi}{2}\cot\frac{\pi(m+\frac{1}{2})}{N}
\tilde{c}^\dagger_m \tilde{c}^\dagger_{N\!-\!m\!-\!1}\right)\ket{\Omega}.
 \end{equation}
 \end{subequations}
 Therefore, the overlap of the state $\ket{\Psi_{1/2}(r,h)}$ with $\ket{\Phi(\xi)}$ for even $N$ is
 \begin{subequations}
 \begin{equation}
 \ipr{\Psi_{1/2}(r,h)}{\Phi(\xi)}=\prod_{m=0}^{m<\frac{N-1}{2}}
\left(\cos\theta_m^{(1/2)}(r,h)\,\cos^2\frac{\xi}{2}+\sin\theta_m^{(1/2)}(r,h)\,\sin^2\frac{\xi}{2}\cot\frac{\pi(m+\frac{1}{2})}{N}\right),
 \end{equation}
whereas  for odd $N$
 \begin{equation}
 \ipr{\Psi_{1/2}(r,h)}{\Phi(\xi)}=\cos\frac{\xi}{2}\prod_{m=0}^{m<\frac{N-1}{2}}
\left(\cos\theta_m^{(1/2)}(r,h)\,\cos^2\frac{\xi}{2}+\sin\theta_m^{(1/2)}(r,h)\,\sin^2\frac{\xi}{2}\cot\frac{\pi(m+\frac{1}{2})}{N}\right).
 \end{equation}
 \end{subequations}

 Next, we discuss the $b=0$ (odd-fermion) case. The lowest allowed state is
the one with one $\gamma^{(0)}_0=\tilde{c}^{(0)}_0$ fermion:
\begin{equation}
\ket{\Psi_{0}(r,h)}\equiv\gamma^{(0)\dagger}_0\ket{G(r,h)}=\tilde{c}^{(0)\dagger}_0\ket{G(r,h)},
\end{equation}
where $\ket{G(r,h)}$ is the state with no $\gamma$ fermions:
\begin{equation}
\ket{G(r,h)}=\prod_{m=1}^{m<\frac{N}{2}}\Big[\cos\theta_m^{(0)}(r,h)
+i\sin\theta_m^{(0)}(r,h)\,\tilde{c}_m^{(0)\dagger}
\tilde{c}_{N-m}^{(0)\dagger} \Big]\ket{\Omega}.
\end{equation}
Similar to the $b=1/2$ case, by using
\begin{equation}
 \sum_{1\le j< l\le N} c_j^\dagger c_l^\dagger=i\sum_{m=1}^{m<\frac{N}{2}}
 \cot\frac{\pi m}{N}\tilde{c}^{(0)\dagger}_m \tilde{c}^{(0)\dagger}_{N\!-\!m},
 \end{equation}
we obtain that for even $N$
\begin{subequations}
 \begin{equation} \ket{\Phi(\xi)}=\Big({1+\sqrt{N}\tan\frac{\xi}{2}\,\tilde{c}_0^\dagger}\Big)\cos^2\frac{\xi}{2}\prod_{m=1}^{m<\frac{N}{2}}
\left(\cos^2\frac{\xi}{2}+i\sin^2\frac{\xi}{2}\cot\frac{\pi m}{N}
\tilde{c}^\dagger_m \tilde{c}^\dagger_{N\!-\!m}\right)\ket{\Omega},
 \end{equation}
 whereas
 for odd $N$
 \begin{equation}
\ket{\Phi(\xi)}=\Big({1+\sqrt{N}\tan\frac{\xi}{2}\,\tilde{c}_0^\dagger}\Big)
\cos\frac{\xi}{2}\prod_{m=1}^{m<\frac{N}{2}}
\left(\cos^2\frac{\xi}{2}+i\sin^2\frac{\xi}{2}\cot\frac{\pi m}{N}
\tilde{c}^\dagger_m \tilde{c}^\dagger_{N\!-\!m}\right)\ket{\Omega}.
 \end{equation}
 \end{subequations}
 Therefore, the overlap of $\ket{\Psi_{0}(r,h)}$ with $\ket{\Phi(\xi)}$ for
 even  $N$ is
 \begin{equation} \ipr{\Psi_{0}(r,h)}{\Phi(\xi)}=\sqrt{N}\sin\frac{\xi}{2}\prod_{m=1}^{m<\frac{N}{2}}
\left(\cos\theta_m^{(0)}(r,h)\,\cos^2\frac{\xi}{2}+\sin\theta_m^{(0)}(r,h)\,\sin^2\frac{\xi}{2}\cot\frac{\pi
m}{N}\right),
 \end{equation}
whereas for odd $N$
 \begin{equation} \ipr{\Psi_{0}(r,h)}{\Phi(\xi)}=\sqrt{N}\sin\frac{\xi}{2}\cos\frac{\xi}{2}\prod_{m=1}^{m<\frac{N}{2}}
\left(\cos\theta_m^{(0)}(r,h)\,\cos^2\frac{\xi}{2}+\sin\theta_m^{(0)}(r,h)\,\sin^2\frac{\xi}{2}\cot\frac{\pi
m}{N}\right).
 \end{equation}

\chapter{Analysis of singular behavior of entanglement density}
In this Appendix we investigate the singular behavior of the entanglement
density near the critical line $h=1$ for both anisotropic ($r\ne0$) and
isotropic ($r=0$) cases. We begin with $r\ne0$ case first.
\section{Divergence of entanglement-derivative for the anisotropic XY models}
\label{app:DivXY} The starting point is Eq.~(\ref{eqn:Erh}), in which there is
a maximization over the variable $\xi$. The function to be maximized is
\begin{equation}
F(\xi,r,h)\equiv\int_0^{\frac{1}{2}}d\mu\,
\ln\left[\cos\theta(\mu,r,h)\cos^2({\xi}/{2})
+\sin\theta(\mu,r,h)\sin^2({\xi}/{2})\cot\pi\mu\right].
\end{equation}

To find the stationarity condition, we demand the derivative with respect to
$\xi$ vanishes:
\begin{equation}
\label{eqn:stationarity}
\partial_\xi F(\xi,r,h)\Big|_{\xi=\xi*}=-\frac{1}{2}\sin\xi \int_0^{\frac{1}{2}}d\mu\,
\frac{\cos\theta(\mu,r,h)-\sin\theta(\mu,r,h)\cot\pi\mu}{
\cos\theta(\mu,r,h)\cos^2({\xi}/{2})
+\sin\theta(\mu,r,h)\sin^2({\xi}/{2})\cot\pi\mu}\Big|_{\xi=\xi*}=0.
\end{equation}
Denote by $\xi^*(h)$ the solution for fixed $r$. Then the  field-derivative of
the entanglement is
\begin{equation}
\partial_h {\cal E}(r,h)=- \frac{2}{\ln2}\partial_h F(\xi^*(h),h)
= - \frac{2}{\ln2}\left[ \frac{\partial \xi^*(h)}{\partial h}
\partial_{\xi}F(\xi,h)\Big|_{\xi^*}+\partial_h F(\xi^*,h)\right]=-
\frac{2}{\ln2}\partial_h F(\xi^*,h),
\end{equation}
where the first term in the square bracket vanishes identically due to the
condition~(\ref{eqn:stationarity}). Thus (dropping the * on $\xi$ for
convenience),
\begin{eqnarray}
\partial_h {\cal E}(r,h)&=&- \frac{2}{\ln2}\partial_h F(\xi^*,h)\\
\label{eqn:EB} &=&-\frac{2}{\ln2}\int_0^{\frac{1}{2}}d\mu\,
\frac{\partial_h\cos\theta(\mu,r,h)\cos^2({\xi}/{2})
+\partial_h\sin\theta(\mu,r,h)\sin^2({\xi}/{2})\cot\pi\mu}{
\cos\theta(\mu,r,h)\cos^2({\xi}/{2})
+\sin\theta(\mu,r,h)\sin^2({\xi}/{2})\cot\pi\mu}.
\end{eqnarray}
Recall that $\tan 2\theta(\mu,r,h)\equiv r\sin 2\pi\mu /(h-\cos 2\pi\mu)$ and
thus
\begin{subequations}
\begin{eqnarray}
&&\cos\theta=\sqrt{(1+\cos2\theta)/2}, \  \sin\theta=\sqrt{(1-\cos2\theta)/2}, \\
&&\cos2\theta(\mu,r,h)= \frac{h-\cos 2\pi\mu}{\sqrt{(r\sin 2\pi\mu)^2 +(h-\cos
2\pi\mu)^2}}.
\end{eqnarray}
\end{subequations}
Putting everything in Eq.~(\ref{eqn:EB}), we get
\begin{eqnarray}
\partial_h {\cal E}(r,h)&=&-\frac{r}{\ln2}\int_0^{\frac{1}{2}}d\mu\,
\frac{\sin 2\pi\mu}{(r\sin 2\pi\mu)^2 +(h-\cos 2\pi\mu)^2} \nonumber\\
&&\!\!\!\!\!\!\!\!\!\!\!\!\!\!\!\!\!\!\!\!\!\!\!\!\frac{\sqrt{\sqrt{\phantom{AAA}}-(h-\cos2\pi\mu)}\cos^2(\xi/2)-\sqrt{\sqrt{\phantom{AAA}}+(h-\cos2\pi\mu)}\sin^2(\xi/2)\cot\pi\mu}{\sqrt{\sqrt{\phantom{AAA}}+(h-\cos2\pi\mu)}\cos^2(\xi/2)+\sqrt{\sqrt{\phantom{AAA}}-(h-\cos2\pi\mu)}\sin^2(\xi/2)\cot\pi\mu},
\end{eqnarray}
where $\sqrt{\phantom{AAA}}\equiv\sqrt{(r\sin 2\pi\mu)^2 +(h-\cos
2\pi\mu)^2}$.

We aim to explore the behavior near $h=1$. First consider $h>1$ and define
$\epsilon\equiv h-1$, which is the deviation from the critical point. Make the
change of variables $t=h-\cos2\pi\nu$, giving lower and upper limits
$\epsilon$ and $2+\epsilon$, respectively. We further shift the integration
variable by $\epsilon$, arriving at
\begin{eqnarray}
\label{eqn:EB2}
\partial_h {\cal E}(r,h)&=&-\frac{r}{2\pi\ln2}\int_0^{{2}}d\,t
\frac{1}{(1-r^2)t^2+ 2(r^2+\epsilon)t+\epsilon^2}  \nonumber\\
&&\!\!\!\!\!\!\!\!\!\!\!\!\frac{\sqrt{t}\sqrt{\sqrt{\phantom{AAA}}-(t+\epsilon)}\cos^2(\xi/2)-\sqrt{\sqrt{\phantom{AAA}}+(t+\epsilon)}\sin^2(\xi/2)\sqrt{2-t}}{\sqrt{t}\sqrt{\sqrt{\phantom{AAA}}+(t+\epsilon)}\cos^2(\xi/2)+\sqrt{\sqrt{\phantom{AAA}}-(t+\epsilon)}\sin^2(\xi/2)\sqrt{2-t}},
\end{eqnarray}
where $\sqrt{\phantom{AAA}}=\sqrt{(1-r^2)t^2+ 2(r^2+\epsilon)t+\epsilon^2}$.

As we inspect the limit $h\rightarrow1$ or $\epsilon\rightarrow0$, we see that
the above expression diverges, with the contribution coming from $t$ small,
i.e., infrared divergence. Large $t$($\le2$) does not contribute to the
divergence. Note further that only the second term in the numerator
contributes to the divergence. We then proceed to evaluate the integral by
separating it into two parts:
\begin{equation}
\int_0^2=\int_0^\delta +\int_\delta^2,
\end{equation}
with $\delta\ll 1$. In the first region we only need to keep $t$ to first
order at most. Also noting that for $t,\epsilon \ll 1$, the first term in the
denominator is much smaller than the second term, we get
\begin{eqnarray}
-\frac{r}{2\pi\ln2}\int_0^{{\delta}}d\,t
\frac{-1}{2r^2(t+\frac{\epsilon^2}{2r^2})}
\frac{\sqrt{\sqrt{2(r^2+\epsilon)t+\epsilon^2}+(t+\epsilon})}{\sqrt{\sqrt{2(r^2+\epsilon)t+\epsilon^2}-(t+\epsilon})}.
\end{eqnarray}
Next, we simplify the second term (ignoring $\epsilon$ when there is no danger
in doing so)
\begin{equation}
\frac{\sqrt{\sqrt{2(r^2+\epsilon)t+\epsilon^2}+(t+\epsilon})}{\sqrt{\sqrt{2(r^2+\epsilon)t+\epsilon^2}-(t+\epsilon})}=\frac{\sqrt{t+\frac{\epsilon^2}{2r^2}}}{\sqrt{t}}+\frac{t+\epsilon}{\sqrt{2r^2
t}}.
\end{equation}
Observing that only the first term on the right-hand side contributes to the
divergence, we have that the divergent part is
\begin{eqnarray}
-\frac{r}{2\pi\ln2}\int_0^{{\delta}}d\,t
\frac{-1}{2r^2(t+\frac{\epsilon^2}{2r^2})}
\frac{\sqrt{t+\frac{\epsilon^2}{2r^2}}}{\sqrt{t}}=
\frac{1}{4r\pi\ln2}\int_0^{{\delta}2 r^2/\epsilon^2}d\,t \frac{1}{\sqrt{t+1}}
\frac{1}{\sqrt{t}}.
\end{eqnarray}
The divergence part is then (for ${\delta}2 r^2/\epsilon^2\gg 1$)
\begin{equation}
\frac{1}{4r\pi\ln2}2\,{\sinh}^{-1}\sqrt{{\delta}2
r^2/\epsilon^2}\approx\frac{1}{2r\pi\ln2} \ln \left(2\sqrt{{\delta}2
r^2/\epsilon^2}\right)=-\frac{1}{2r\pi\ln2} \ln \epsilon+\frac{\ln
\big(2\sqrt{\delta 2 r^2}\big)}{2r\pi \ln2}.
\end{equation}
As the integral~(\ref{eqn:EB2}) does not depend on the choice of $\delta$, the
part that involves $\delta$ must be cancelled by the second half of the
integration $\int_\delta^2$, which can be verified by direct evaluation.
Therefore, for $h$ very close to $h_c=1$, we have
\begin{equation}
\frac{\partial{\cal E}}{\partial h} \approx -\frac{1}{2\pi r \ln2}\ln|h-1|.
\end{equation}

In deriving the above divergence form, we have assumed that $r\ne 0$. Similar
consideration can be applied to the case when $h$ approaches 1 from below, and
the behavior is the same.

\section{Divergence of entanglement-derivative for the  XX limit of the model}
\label{app:DivXX} We now analyze the $r=0$ isotropic case. It turns out that
the analysis for this case is much simpler. To see this, we first note when
$r=0$ we have the simplification
\begin{equation}
\cos2\theta(\mu,h)={\rm sgn}(h-\cos2\pi\mu).
\end{equation}
The above expression changes sign when $h=\cos2\pi\nu$. So let us introduce
the variable $\mu_0\equiv (\cos^{-1}h)/(2\pi)$. The expression for the
entanglement density Eq.~(\ref{eqn:Erh}) becomes
\begin{subequations}
\begin{eqnarray}
{\cal E}(h) \label{eqn:EB3} &=&-\frac{2}{\ln2}\max_\xi
\int_0^{\frac{1}{2}}d\mu\, \ln\left[\cos\theta(\mu,r,h)\cos^2({\xi}/{2})
+\sin\theta(\mu,r,h)\sin^2({\xi}/{2})\cot\pi\mu\right]\\
&=&-\frac{2}{\ln2}\max_\xi\left[ \int_0^{\mu_0}d\mu\,
\ln\sin^2({\xi}/{2})\cot\pi\mu + \int_{\mu_0}^{\frac{1}{2}}d\mu\,
\ln\cos^2({\xi}/{2})\right].
\end{eqnarray}
\end{subequations}
Demanding stationarity with respect to $\xi$ gives the condition
\begin{equation}
[\mu_0- \frac{1}{4}(1-\cos\xi)]\sin\xi=0.
\end{equation}
The solution $\xi=0$ gives the entanglement density for $h\ge 1$. It is
straightforward to see from Eq.~(\ref{eqn:EB3}) that the entanglement density
is identically zero. The solution $\mu_0- \frac{1}{4}(1-\cos\xi)=0$ gives
\begin{equation}
\cos\xi=1-\frac{2}{\pi}\cos^{-1}h.
\end{equation}
This in turn gives the entanglement density for $0\le h\le 1$:
\begin{equation}
\label{eqn:Eh2} {\cal E}(h)
=-\frac{2}{\ln2}\left[\mu_0(h)\ln\frac{2\mu_0(h)}{1-2\mu_0(h)}+\frac{1}{2}\ln(1-2\mu_0(h))+\int_0^{\mu_0(h)}
d\mu\ln \cot\pi\mu\,\right].
\end{equation}
Thus we have derived the entanglement density as a function of the magnetic
field $h$ in the XX limit. The result is shown in Fig.~(\ref{fig:Ent1000}).

 We see from the figure that the entanglement density at $h=0$ is
the highest, which being ${\cal E}(0)=1- 2 \gamma_C/(\pi \ln 2)\approx 0.159$
by evaluating Eq.~(\ref{eqn:Eh2}) at $h=0$. The constant $\gamma_C\approx
0.9160$ is the {\it Catalan\/} constant. The entanglement density decreases
monotonically as $h$ increases until $h=1$ beyond which it becomes zero
identically. This qualitative behavior can be understood as follows. As the
total $z$-component spin is conserved, increasing $h$ simply increases the
z-component spin of the ground state until $h=1$ where all the spins are
aligned with the field, hence there is no entanglement beyond this value of
$h$.

By directly taking the derivative with respect to $h$, we get
\begin{equation}
\partial_h {\cal E}(h)=\frac{1}{\pi\ln2 \sqrt{1-h^2}} \ln\left[\frac{\cos^{-1}h}{\pi-\cos^{-1}h}\sqrt{\frac{1+h}{1-h}}\,\right].
\end{equation}
Near $h\approx 1$, we have (putting $1+h=2$ and evaluating the limit in the
argument of log function)
\begin{equation}
\frac{\partial}{\partial h}{\cal E}(0,h)\approx
-\frac{\log_2(\pi/2)}{\sqrt{2}\,\pi} \frac{1}{\sqrt{1-h}},
\qquad (h\to 1^{-}).
\end{equation}
This completes our derivation of the singular behavior of the entanglement
density near the critical points.
\chapter{Finite-size scaling}
\label{app:finitesize} The discussion here follows Barber~\cite{Barber83}.
In the vicinity of the bulk critical temperature $T_C$ the behavior of a
system should depend on $y\equiv L/\xi(T)$, where $\xi(T)$ is the bulk
correlation length and $L$ is the characteristic length of the system. How
does the divergence of certain thermodynamic quantities emerge as the system
size $L$ grows?

\section{Algebraic divergence}
Assume that some thermodynamic quantity at $L\rightarrow\infty$ diverges as
$t\equiv(T-T_C)/T_C\rightarrow\infty$:
\begin{equation}
P_\infty(T)\sim C_\infty t^{-\rho}.
\end{equation}
Finite-size scaling hypothesis asserts that for finite $L$ and $T$ near $T_C$,
\begin{equation}
P_L(T)\sim l^\omega Q_P(l^{{1}/{\nu}}\tilde{t}), \ \ l\rightarrow\infty,\
\tilde{t}\rightarrow 0,
\end{equation}
where $l\equiv L/a$ ($a$ is some microscopic length), $\tilde{t}\equiv
[T-T_C(L)]$. The exponent $\omega$ can be determined by the requirement that
the $P_L(T)$ reproduces $P_\infty(T)$ as $l\rightarrow\infty$. Thus,
\begin{equation}
Q_P(x)\sim C_\infty x^{-\rho}, \ \ x\rightarrow\infty,
\end{equation}
and $\omega=\rho/\nu$. We consider the case that the finite system does not
exhibit a true transition, then
\begin{equation}
Q_P(x)\rightarrow Q_0, \ \ x\rightarrow 0.
\end{equation}
From this we have that at the peak or rounding temperature $T_m^*(l)$ (where
$P_L$ reaches the maximum or deviates significantly from the bulk value)
\begin{equation}
P_L(T_m^*(l))\sim Q_0 l^{\rho/\nu}, \ \ l\rightarrow\infty.
\end{equation}
This means that the behavior of a thermodynamic quantity varies with the
system size is determined by the bulk critical exponent.
\section{Logarithmic divergence}
Now assume the thermodynamic quantity $P(T)$ diverges logarithmically as
\begin{equation}
P_\infty(T)\sim C_\infty \ln t, \ \ t\rightarrow0,
\end{equation}
as in the field-derivative of the entanglement density for anisotropic XY spin
chains. The finite-size scaling hypothesis in this case is to assume
\begin{equation}
P_L(T)-P_L(T_0)\sim Q_P(l^{1/\nu}\tilde{t})-Q_P(l^{1/\nu}\tilde{t}_0),
\end{equation}
where $T_0$ is some non-critical temperature and $\tilde{t}_0\equiv
(T_0-T_C(L))/T_C$. The hypothesis has to recover the $l\rightarrow\infty$
limit at fixed $T$, which requires
\begin{equation}
Q_P(x)\sim C_\infty\ln x, \ \ x\rightarrow\infty.
\end{equation}
Thus in the limit $\tilde{t}\rightarrow0$ at fixed large $l$, we have
\begin{equation}
P_L(T_C(L))\sim -\frac{C_\infty}{\nu}\ln l + O(1), 
\end{equation}
if $Q_P(x)=O(1)$ as $x\rightarrow 0$. This allows us to obtain the exponent
$\nu$ by analyzing how the divergence develops as the system size $l$ (which
is $N$ in our spin-chain entanglement problem) increases.

\backmatter


\begin{thebibliography}{999}
\bibitem{Note}
As we shall show later, this state has $E_{\rm R}=\log_2(9/4)$, exactly the
value cited in Ref.~\cite{PlenioVedral01} from a numerical result.
\bibitem{footnote}
In arriving at this result we have made a hypothesis about the form of the
closest separable pure state. The result has been verified numerically for the
$N=4$ case.
\bibitem{footnote:ansatz}
We have confirmed this Ansatz numerically for small numbers of spins.
\end{thebibliography}


\vita
Tzu-Chieh Wei was born in Taichung, Taiwan on February 25, 1973. He received
his B.S. in Physics with a minor in Mathematics in 1994 and M.S.
in Physics in 1996, both from National
Taiwan University. He fulfilled his obligatory military service from
1996 to 1998. He entered the Ph.D. program in Physics at the University of
Illinois in 1999.
\chapter{List of Publications}
\begin{enumerate}
\item ``Two-qubit mixed states and the entanglement-entropy frontier",
T.-C. Wei, K. Nemoto, P.M. Goldbart, P.G. Kwiat, W.J. Munro, and F. Verstraete,
Proceedings of the 6th international conference on quantum communication,
measurement and computing, July 22-26, 2002, p.37-40, ed. J.H. Shapiro and
O. Hirota, Rinton Press, 2003.

\item ``Taming entanglement", P.G. Kwiat, J.B. Altepeter, D. Branning, E. Jeffrey,
N. Peters, and T.-C. Wei, Proceedings of the 6th international conference on
quantum communication, measurement and computing, July 22-26, 2002, p.117-122,
ed. J.H. Shapiro and O. Hirota, Rinton Press, 2003.

\item ``Quantum information with optics", T.-C. Wei, Physics Bimonthly, {\bf 25},
p.555-564 (2003), published by the Physical Society of the Republic of
China, Taipei, Taiwan.

\item ``Maximal entanglement versus entropy for mixed quantum states",
T.-C. Wei, K. Nemoto, P.~M.~Goldbart, P.~G.~Kwiat, W.~J.~Munro, and F.~Verstraete,
Phys. Rev A {\bf 67}, 022110 (2003).

\item ``Ancilla-assisted quantum process tomography", J. B. Altepeter, D. Branning,
E. Jeffrey, T.-C. Wei, P. G. Kwiat, R. T. Thew, J. L. O'Brien, M. A. Nielsen, and
A. G. White, Phys. Rev. Lett. {\bf 90}, 193601 (2003).

\item ``Geometric measure of entanglement for bipartite and multipartite quantum
states", T.-C. Wei and P. M. Goldbart, Phys. Rev. A {\bf 68}, 042307 (2003).

\item ``Maximally entangled mixed states: creation and concentration", N. A. Peters,
J. B. Altepeter, D. A. Branning, E. R. Jeffrey, T.-C. Wei, and P. G. Kwiat,
Phys. Rev.  Lett. {\bf 92}, 133601 (2004).

\item ``Benchmarking and procrustean noise reduction of entangled mixed states",
N. A. Peters, T.-C. Wei, P. G. Kwiat,
Proc. SPIE Vol. {\bf 5468}, p.269-281, Fluctuations and Noise in
Photonics and Quantum Optics II; Ed. Peter Heszler, 2004.

\item ``h/e magnetic flux modulation of the energy gap in nanotube quantum dots",
U. C. Coskun, T.-C. Wei, S. Vishveshwara, P. M. Goldbart, and
A. Bezryadin, Science {\bf 304}, 1132 (2004).

\item ``Connections between relative entropy of entanglement and geometric measure of entanglement", T.-C. Wei, M. Ericsson, P. M. Goldbart, and
W. J. Munro, Quantum Info. Comput. {\bf 4}, 252 (2004).

\item ``Measures of entanglement in bound entangled states", T.-C. Wei,
J. B. Altepeter, P. M. Goldbart, and W. J. Munro, Phys. Rev. A {\bf 70}, 022322 (2004).

\item ``Mixed state sensitivity of several quantum information benchmarks",
N. A. Peters, T.-C. Wei, and P. G. Kwiat, to appear in
Phy. Rev. A {\bf 70}; e-print quant-ph/0407172.

\item ``Synthesizing arbitrary two-photon polarization
           mixed states",
T.-C. Wei, J. B. Altepeter, D. A. Branning, P. M. Goldbart,
D. F. V. James, E. Jeffrey, P. G. Kwiat, S. Mukhopadhyay,
and N. A. Peters, submitted to Phys. Rev. A.

\item ``Global entanglement and quantum criticality in spin chains", T.-C. Wei,
D. Das, \\ S. Mukhopadyay, S. Vishveshwara, and P. M. Goldbart, e-print quant-ph/0405162.

\item ``Quantifying multipartite entanglement'', T.-C. Wei, J. B. Altepeter,
D. Das, M. Ericsson,\\ P. M. Goldbart, S. Mukhopadyay, W. J. Munro, and
S. Vishveshwara, to appear in the Proceedings of the 7th international conference on
quantum communication, measurement and computing.

\end{enumerate}
\end{document}